\documentclass[11pt,a4paper]{article}
\pdfoutput = 1
\usepackage{jheppub}

\usepackage{texnames}
\usepackage[T1]{fontenc}
\usepackage{hepnames}
\usepackage{siunitx}
\usepackage{nicefrac}
\usepackage{setspace}

\newcommand{\sfrac}[2]{\nicefrac{\displaystyle #1}{#2}}
\newcommand{\ssfrac}[2]{\nicefrac{#1}{#2}}

\title{Polarization and Centre-of-mass Energy Calibration at FCC-ee}

\author[]{The FCC-ee Energy and Polarization Working Group:\break} 
\author[1,2,3]{Alain~Blondel,} 
\author[2]{Patrick~Janot,}
\author[2]{J\"org~Wenninger}
\author[]{(Editors)\break}
\author[4]{Ralf~A{\ss}mann,}
\author[2]{Sandra~Aumon,}
\author[5]{Paolo~Azzurri,}
\author[4]{Desmond~P.~Barber,}
\author[2]{Michael~Benedikt,}
\author[6]{Anton~V.~Bogomyagkov,}
\author[7]{Eliana~Gianfelice-Wendt,}
\author[2]{Dima~El~Kerchen,}
\author[6]{Ivan~A.~Koop,}
\author[8]{Mike~Koratzinos,}
\author[6]{Evgeni~Levitchev,}
\author[2]{Thibaut~Lefevre,}
\author[2]{Attilio~Milanese,}
\author[6]{Nickolai~Muchnoi,}
\author[6]{Sergey~A.~Nikitin,}
\author[2]{Katsunobu~Oide,}
\author[2]{Emmanuel~Perez,} 
\author[4]{Robert~Rossmanith,}
\author[9]{David~C.~Sagan,}
\author[5]{Roberto~Tenchini,}
\author[2]{Tobias~Tydecks,}
\author[6]{Dmitry~Shatilov,}
\author[2]{Georgios~Voutsinas,} 
\author[10]{Guy~Wilkinson,}
\author[2]{Frank~Zimmermann.}

\affiliation[1]{DPNC, University of Geneva, CH-1205 Geneva, Switzerland}
\affiliation[2]{CERN, EP Department, 1 Esplanade des Particules, CH-1217 Meyrin, Switzerland} 
\affiliation[3]{CNRS/IN2P3 Laboratoire de Physique Nucl\'eaire et de Hautes Energies, 4 place Jussieu, 75252 Paris Cedex 05, FRANCE}
\affiliation[4]{DESY, Notkestra{\ss}e 85, D-22607 Hamburg, Germany}
\affiliation[5]{INFN, Sezione di Pisa, Largo Bruno Pontecorvo, 3, 56127 Pisa, Italy}
\affiliation[6]{Budker Institute of Nuclear Physics  (BINP SB RAS), 11, Acad. Lavrentieva Pr., Novosibirsk, 630090, Russian Federation}
\affiliation[7]{Fermi National Accelerator Laboratory, Fermilab, PO Box 500, Batavia, IL~60510-5011, USA} 
\affiliation[8]{Massachusetts Institute of Technology, 77 Massachusetts Ave, Cambridge, MA 02139, USA}
\affiliation[9]{Wilson Laboratory, Cornell University, Ithaca, NY 14853, USA}
\affiliation[10]{University of Oxford, Particle Physics Department, Oxford OX1 3RH, United Kingdom} 

\emailAdd{Alain.Blondel@cern.ch}
\emailAdd{Patrick.Janot@cern.ch}
\emailAdd{Jorg.Wenninger@cern.ch}

\begin{abstract}
{The proposed Frontier Circular Collider (FCC) project integrates in sequence ${\rm e^+ e^-}$ and hadron colliders in the same 100\,km infrastructure. The FCC provides a most effective and comprehensive exploration of open questions in modern particle physics, by a combination of much increased precision, sensitivity, and centre-of-mass energy. The first stage is a high-luminosity electron-positron storage ring collider  (FCC-ee) with centre-of-mass energy ranging from 88 to 365\,GeV, to study with high precision the Z, W, Higgs and top particles, with samples of $5 \times 10^{12}$ Z bosons, $10^8$ W pairs, $10^6$ Higgs bosons and $10^6$ top quark pairs.  A cornerstone of the FCC-ee physics program lays in the precise (ppm) measurements of the W and Z masses and widths, as well as forward-backward asymmetries. To this effect, the centre-of-mass energy and its distribution should be determined with the highest feasible precision. \\ \\
This document describes the capacity offered by FCC-ee, starting with the possibility to obtain transverse polarization of the beams around both the Z pole and the W pair threshold. A running scheme based on a regular (several times per hour) measurement of the beam energy by means of resonant depolarization of pilot bunches, during physics data taking, is proposed.  Feasible designs for polarization wigglers, polarimeters and RF depolarizer are outlined, resulting in the ability to monitor the beam energies $E_{\rm b}^\pm$ of the ${\rm e}^\pm$ beams with a relative precision of around $10^{-6}$. The centre-of-mass energy, $\sqrt{s} = 2 \sqrt{E_{\rm b}^+E_{\rm b}^-}\cos\sfrac{\alpha}{2}$, is derived subject to further corrections, related to the beam acceleration and to the energy losses due to synchrotron radiation and beamstrahlung; these effects are identified and evaluated. Dimuon events ${\rm e^+ e^-} \to \mu^+\mu^-$ recorded in the detectors, provide  with great precision the average beam crossing angle $\alpha$, the centre-of-mass energy spread, and the difference between ${\rm e^+}$ and ${\rm e^-}$ beam energies. Monitoring methods to minimize both the absolute error and the relative uncertainties of the different energy settings with each other are discussed. The final impact on the physics measurements is given.  Elements of a programme of further simulations, design, monitoring and {R\&D} are outlined.}  
\end{abstract}

\begin{document}
\maketitle 




\section{Overview}
\label{sec:overview}

This document presents a status report on the present understanding of energy calibration and beam polarization issues for the 100\,km ${\rm e^+e^-}$ Future Circular Collider (FCC-ee)~\cite{Abada2019}. The knowledge of the centre-of-mass energy is a key characteristic of ${\rm e^+ e^-}$ colliders, and more particularly of storage ring colliders, as was demonstrated at LEP by the determination of the Z mass with a precision of $2\times 10^{-5}$~\cite{ALEPH:2005ab} and more recently at VEPP4 with the determination of the $J / \psi$ mass with a relative precision of $2\times 10^{-6}$~\cite{Anashin2015}. The cornerstone of this exceptional precision is the possibility to measure -- by resonant depolarization  -- the spin precession frequency of the beams, which in a planar magnetic ring is closely related to their energy, averaged over the ring.

  \begin{figure}[htbp]
	\begin{center}
	\centering

		\includegraphics*[width=0.6\textwidth]{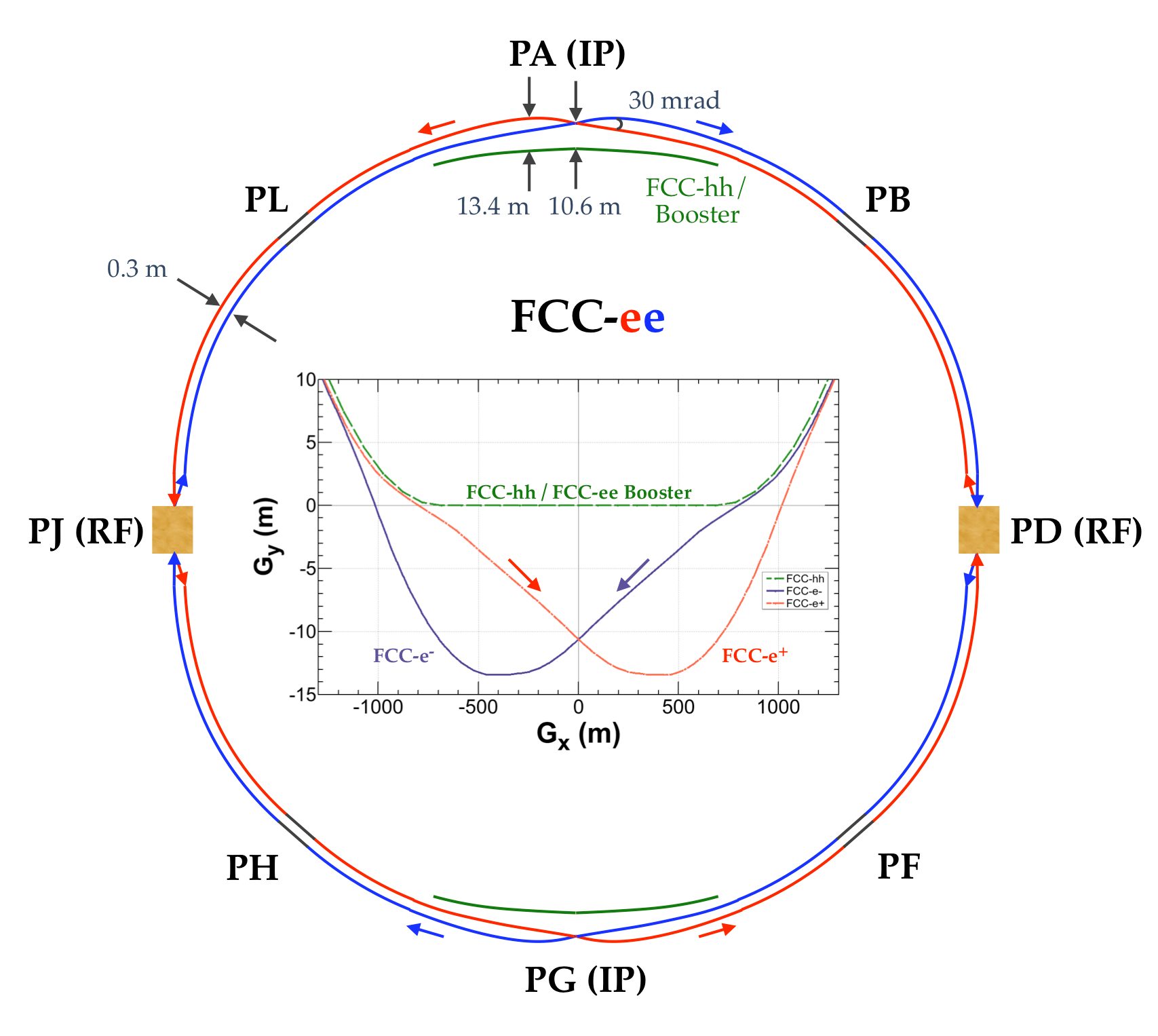}

\caption{Overall layout of the FCC-ee collider 
 layout with a zoomed view of the trajectories across interaction point PG \protect\cite{Abada2019}.  
The FCC-ee rings are placed 1\,m outside the FCC-hh footprint in the arc. 
The main booster follows the footprint of FCC-hh. 
In the arc the $\rm e^+$ and $\rm e^-$ rings are horizontally separated by 30\,cm. The interaction points are shifted by 10.6\,m towards the outside of FCC-hh. The beam trajectories toward the IP are straighter than the outgoing ones in order to reduce the synchrotron radiation at the IP
\protect\cite{Boscolo:2281384}. 
}
\label{fig0:layout}
\end{center}
\end{figure} 

\begin{table}[htbp]
\caption{Machine parameters of FCC-ee for different beam energies~\protect\cite{Abada2019} \vspace{3mm}}.
\label{tab0:params}
\centering
\footnotesize
\begin{tabular}{l|c|c|c|c|c}
\hline
 & \bf Z & \bf WW & \bf ZH & 
\multicolumn{2}{c}{\textbf{t}$\overline{\mbox{\bf{t}}}$} \\ 
\hline
Circumference [km]& \multicolumn{5}{c}{97.756}\\ 
Bending radius [km]& \multicolumn{5}{c}{10.760}\\ 
Free length to IP ${\it l}^*$ [m]&  \multicolumn{5}{c}{2.2}\\
Solenoid field at IP [T]& \multicolumn{5}{c}{2.0}\\
Full crossing angle at IP $\alpha$  [mrad] &\multicolumn{5}{c}{30}\\
SR power / beam [MW]&\multicolumn{5}{c}{50}\\
\hline
Beam energy [GeV]& 45.6 & 80 & 120 &175 & 182.5\\ 
\hline
Beam current [mA]& 1390 & 147 & 29 & 6.4 & 5.4\\
\hline
Bunches / beam  & 16640 & 2000 & 328 & 59 & 48 \\
\hline
Average bunch spacing [ns]& 19.6 & 163 & 994 &  2763 & 3396  \\
\hline
Bunch population [$10^{11}$]& 1.7&1.5 &1.8  &2.2 &2.3\\
\hline
Horizontal emittance $\varepsilon_x$ [nm]& 0.27 & 0.84 & 0.63 & 1.34 & 1.46\\
Vertical emittance $\varepsilon_y$ [pm]& 1.0 & 1.7 & 1.3 & 2.7 & 2.9\\
\hline
Arc cell phase advances [deg]&\multicolumn{2}{c|}{60/60}& \multicolumn{3}{c}{90/90} \\
\hline
 Momentum compaction $\alpha_p$ [$10^{-6}$]& \multicolumn{2}{c|}{14.8}&  \multicolumn{3}{c}{7.3} \\
\hline
Horizontal $\beta^*_x$ [m] & 0.15 & 0.2 & 0.3 & \multicolumn{2}{c}{1.0} \\
Vertical $\beta^*_y$ [mm] & 0.8 & 1.0 & 1.0 &  \multicolumn{2}{c}{1.6}  \\
\hline
Horizontal size at IP $\sigma^*_x$ [$\mu$m]& 6.4 & 13.0 & 13.7 &  36.7 &  38.2\\
Vertical size at IP $\sigma^*_y$ [nm]& 28 &41 & 36 &66 & 68\\
\hline
Natural Energy spread $\sigma_\delta$ [\%]/MeV& 0.038/17 & 0.066/53 & 0.099/119 & 0.144/252 & 0.150/274\\
\hline
Energy spread in collision $\sigma_{\delta_c}$ [\%]& 0.132 & 0.131 & 0.165 & 0.186 & 0.192\\
\hline
Bunch length in collision $\sigma_z$ [mm]& 12.1 & 6.0 & 5.3 & 2.62 & 2.54 \\
\hline
Piwinski angle (SR/BS) $\phi$  & 8.2/28.5 &3.5/7.0 &3.4/5.8 &0.8/1.1 & 0.8/1.0\\
\hline
Energy loss / turn [GeV]& 0.036 & 0.34 & 1.72 & 7.8 & 9.2\\
\hline
RF frequency [MHz] &  \multicolumn{3}{c|}{400}&\multicolumn{2}{c}{400 / 800} \\
\hline
RF voltage  [GV] & 0.1&0.75&2.0&4.0 / 5.4 &4.0 / 6.9\\
\hline
Longitudinal damping time [turns] & 1273 & 236 & 70.3 & 23.1&20.4\\
\hline
Energy acceptance (DA) [\%]& $\pm$1.3 & $\pm$1.3 & $\pm$1.7 & \multicolumn{2}{c}{$-2.8$, $+2.4$} \\
\hline
Polarization time $t_p$ [min]& 15000 & 900 & 120 & 18.0 & 14.6 \\
\hline
Luminosity / IP [$10^{34}$/cm$^2$s] & 230 & 28 & 8.5 & 1.8 & 1.55\\
\hline
Vertical beam-beam parameter $\xi_y$  & 0.133 & 0.113 & 0.118 & 0.128 & 0.126\\
\hline
Beam lifetime [min]& $>200$ & $>200$ & 18 & 24 & 18\\
\hline
\end{tabular}
\end{table}

The Future Circular Collider (FCC)~\cite{Benedikt:2653673} is based on a large  common infrastructure, a tunnel in the Geneva area of 100\,km circumference, shared in succession by several colliders. First, a luminosity-frontier electron-positron collider (FCC-ee) spanning the energies corresponding to the Z resonance, the WW threshold, the HZ production maximum, and the ${\rm t\bar t}$ threshold and 20\,GeV above. The FCC-ee is a precision instrument to study the Z, W, Higgs and top particles, and offers unprecedented sensitivity to signs of new physics. Most of the FCC-ee infrastructure can be reused later for the subsequent hadron collider, FCC-hh~\cite{Abada2019hh}. The layout of FCC-ee is shown in~Fig.~\ref{fig0:layout}, and the parameter list in Table~\ref{tab0:params}. 

The document first recalls the experience from LEP (Section~\ref{sec:LEP}), where transverse polarization was available at 45\,GeV beam energy (the Z pole) but was achievable only up to a beam energy of 60\,GeV, due to the increase of energy spread. Instantaneous beam energy measurement with a precision of 100\,keV could be achieved, but was limited to about 2\,MeV uncertainty on the average luminosity-weighted centre-of-mass energy. This uncertainty was due to the fact that energy calibrations were performed at the end of physics coasts, making it sensitive to a number of time-dependent systematic effects that have been identified and understood only after completion of data taking. 

Already in the early studies~\cite{Gomez-Ceballos:2013zzn}, the importance of beam polarization and the possibility of precise energy calibration had been stressed. It is expected that beam polarization will build up naturally for  energies where the natural beam energy spread $\sigma_E / E \simeq E^4/\rho $ is small (<60\,MeV) compared to the spacing of 440\,MeV in beam energy between the integer spin resonances ($\rho$ is the radius of curvature of the storage ring). As can be seen in Table~\ref{tab0:params}, this is the case for non-colliding beams at the Z and WW runs, where precision will be most essential for the measurements of the Z and W mass and width, and for the precise measurement of forward-backward asymmetries, which all involve measuring quantities that depend strongly on the centre-of-mass energy. The requirements on precision at the Z and WW energies are elaborated in Sections~\ref{sec:Zrequirements} and~\ref{sec:Wrequirements}, where it is shown that in addition to the average beam energy, the energy spread must be known with great precision. At higher energies, the ${\rm e^+e^- \rightarrow Z\gamma, WW, ZZ}$ processes provide centre-of-mass energy calibration from the knowledge of the Z and W masses.    

The physics of beam polarization and the prospects for obtaining it in FCC-ee are described in Section~\ref{sec:polar}, including simulations of spin motion in the FCC-ee optics, the use of wigglers, and  the necessary corrections for optics imperfections and for the  detector solenoids correction system. 

With hindsight, it is possible to design the data taking to avoid a great number of the LEP sources of systematics. It is proposed to organize the luminosity runs with a sufficient number (250 out of 16600 at the Z pole) non-colliding pilot bunches on which the resonant depolarization can be performed a short intervals of 10 to 15 minutes. The long polarization time at the Z pole energies requires the use of polarization wigglers, to be powered only for about one to two hours at the beginning of each data taking fill before the full current is injected. The scheme and the (limited) necessary equipment are described in Section~\ref{sec:running-scheme}; a novel feature is the fact that the polarimeter, based on inverse Compton scattering, detects both the recoil photon and the recoil electron, thus allowing simultaneously the beam polarization to be monitored and a continuous (relative) beam energy measurement to be performed (Section~\ref{sec:polarimeter}). 

In 1992, the four LEP experiments reported~\cite{Quast:1992et} a large discrepancy among their observed Z boson masses from the 1991 data taking. The cause was soon understood as the effect of the RF acceleration of the beams, which was unevenly distributed around the ring. This example illustrates that the extraction of the centre-of-mass energies from the determination of the spin precession frequency requires several corrections, and associated sources of uncertainties. These corrections and uncertainties are addressed in Sections~\ref{sec:energy-systematics} and~\ref{sec:cm-energy}.

Considerable information will be available from the experiments, on one hand to measure the beam crossing angle $\alpha$, needed to determine the centre-of-mass energy $\sqrt{s} = 2 \sqrt{E_{\rm b}^+E_{\rm b}^-} \cos\alpha/2$; and on the other, to reduce two important elements of systematic uncertainties: the beam-energy spread and the relative point-to-point energy scale uncertainty. These elements are described in Sections~\ref{sec:EnergySpreadAndCrossingAngle} and~\ref{sec:point-to-point}. 
The updated estimates of the related systematic uncertainties on the relevant precision electroweak observables are given in Section~\ref{sec:status} and Table~\ref{tab:Z-Ecal-errors-final}. Monitoring of accelerator parameters (orbits, magnetic fields, etc..) are of crucial importance for ensuring the result reproducibility. The requirements in terms of simulation tools and monitoring devices and procedures, as well as an overview of future work, are given in Sections~\ref{sec:sim-tools} and ~\ref{sec:future-studies}.


\section{Experience from LEP}
\label{sec:LEP}

\subsection{Measuring \texorpdfstring{$m_{\rm Z}$}{mZ} and \texorpdfstring{$\Gamma_{\rm Z}$}{GZ} at \texorpdfstring{LEP\,1}{LEP1}}
\label{sec:LEPscans}
The precise determination of the Z lineshape parameters $m_{\rm Z}$ and $\Gamma_{\rm Z}$, with uncertainties of 2.1\,MeV and 2.3\,MeV,
respectively~\cite{ALEPH:2005ab}, was one of the most important achievements of the LEP\,1 physics programme. The beam-energy calibration, fully described in Ref.~\cite{Assmann:1998qb}, was central to this accomplishment. The lineshape measurements were based on two Z resonance scans conducted in 1993 and 1995, each corresponding to approximately 40\,${\rm pb^{-1}}$ of integrated luminosity, and one run in 1994 of around 60\,${\rm pb^{-1}}$, taken solely at the peak of the resonance.  During the scans,
collisions were delivered at three energies: the peak itself (`P') and approximately $1.8$\,GeV below (`P-2') and above (`P+2') the peak.   These off-peak energies were close to optimal for the measurement of $\Gamma_{\rm Z}$ and were also at half-integer spin tunes\footnote{The ``spin tune'' is the number of spin precessions that an electron spin undergoes when the electron does one revolution around the ring; in other words it is the ratio between the electron spin precession frequency in the magnetic field and the electron revolution frequency around the ring.}, which facilitated beam polarization, necessary for the energy calibration.   The
data collection occurred in coasts of varying lengths, but typically lasting 10 hours each.  During the scans, an effort was made to alternate between the two off-peak energies in adjacent coasts, interspersed with measurements at the peak, in order to reduce systematic biases.

\subsubsection{Resonant depolarization at LEP}
\label{sec:LEPrdp}
\begin{figure}[tbh]
\centerline{\includegraphics[width=0.6\textwidth]{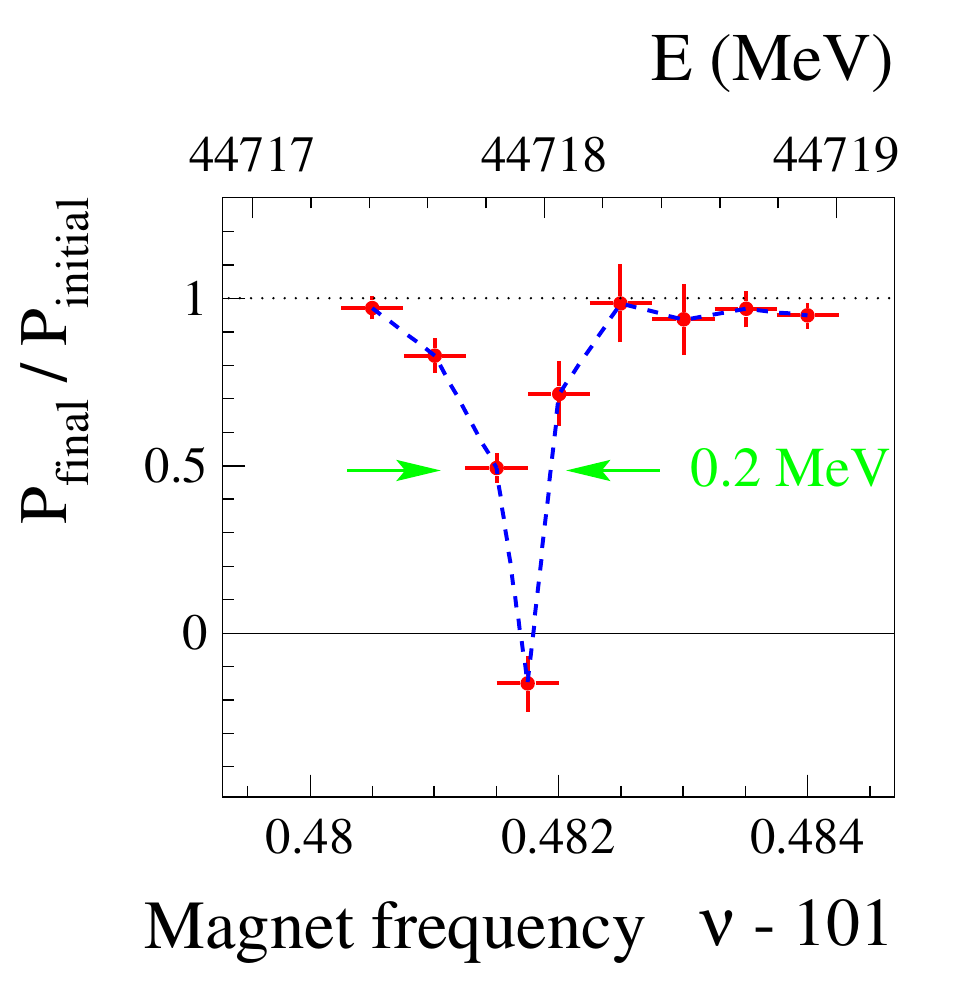}}
\caption{Example of RDP measurement at LEP, showing the variation of the relative polarization plotted against the frequency of the perturbing field.}
\label{fig:rdp}
\end{figure}

The average beam energy, $E_{\rm b}$, was measured by resonant depolarization (RDP). The level of polarization was monitored from Compton-scattered polarized laser light, and depolarization was achieved by exciting the beam with a transverse oscillating magnetic field, the frequency of which allowed the spin tune to be determined, and hence $E_{\rm b}$ to
be measured with an instantaneous precision of 200$\,$keV~\cite{Arnaudon:1994zq}, as is shown in Fig.~\ref{fig:rdp}.  Transverse polarization was first observed at LEP in 1990~\cite{Knudsen:1991cu} and was exploited for measurements of the lineshape parameters the following year~\cite{Arnaudon:1992rn}.  However, RDP calibration only became a regular operational tool in 1993, when the measurement became routinely possible with separated beams.  Transverse polarization with colliding beams
was obtained only in special conditions, far from the physics-operation mode.  This limitation had important consequences for the lineshape measurement, imposing that a precise determination of $E_{\rm b}$ could only be performed outside physics conditions.

At the end of each off-peak coast in the 1995 scan, the beams  were separated and a RDP measurement was attempted, usually on the electron beam.  A few of these calibrations also happenned at the start, prior to collisions. In total, around half the off-peak coasts, corresponding to approximately 65\% of  the integrated luminosity at these energies, were successfully calibrated, with others being lost before this procedure could be
performed.  In 1993, around 40\% of the off-peak integrated luminosity was calibrated. Fewer calibrations were performed at the peak itself, as knowledge of this energy was less critical for the determination of the lineshape parameters.

\subsubsection{Time variations in the beam energy:  the LEP \texorpdfstring{$E_{\rm b}$}{Eb} model}
\label{LEPvar}

The distribution of RDP values obtained at each energy point  exhibited an RMS of around 5\,MeV, which indicated that there were underlying mechanisms of significant energy variation between coasts.  This spread led to an unacceptably large uncertainty in the estimated mean energy of those coasts that had not been calibrated, and hence understanding of this inter-fill variation was necessary in order not to compromise the
goals of the lineshape campaign.   This understanding was gradually acquired, and an {\it $E_{\rm b}$ model}~\cite{Assmann:1998qb} developed to track the variation in beam energy between measurements. Significant machine time was devoted to calibration experiments which validated the various components of this model. The RMS of the difference between the RDP
measurements and the predictions of the final $E_{\rm b}$ model are less than half those of the spread in raw RDP measurements.

\begin{figure}[tbh]
\centering
\begin{tabular}{cc}
\includegraphics[width=0.54\textwidth]{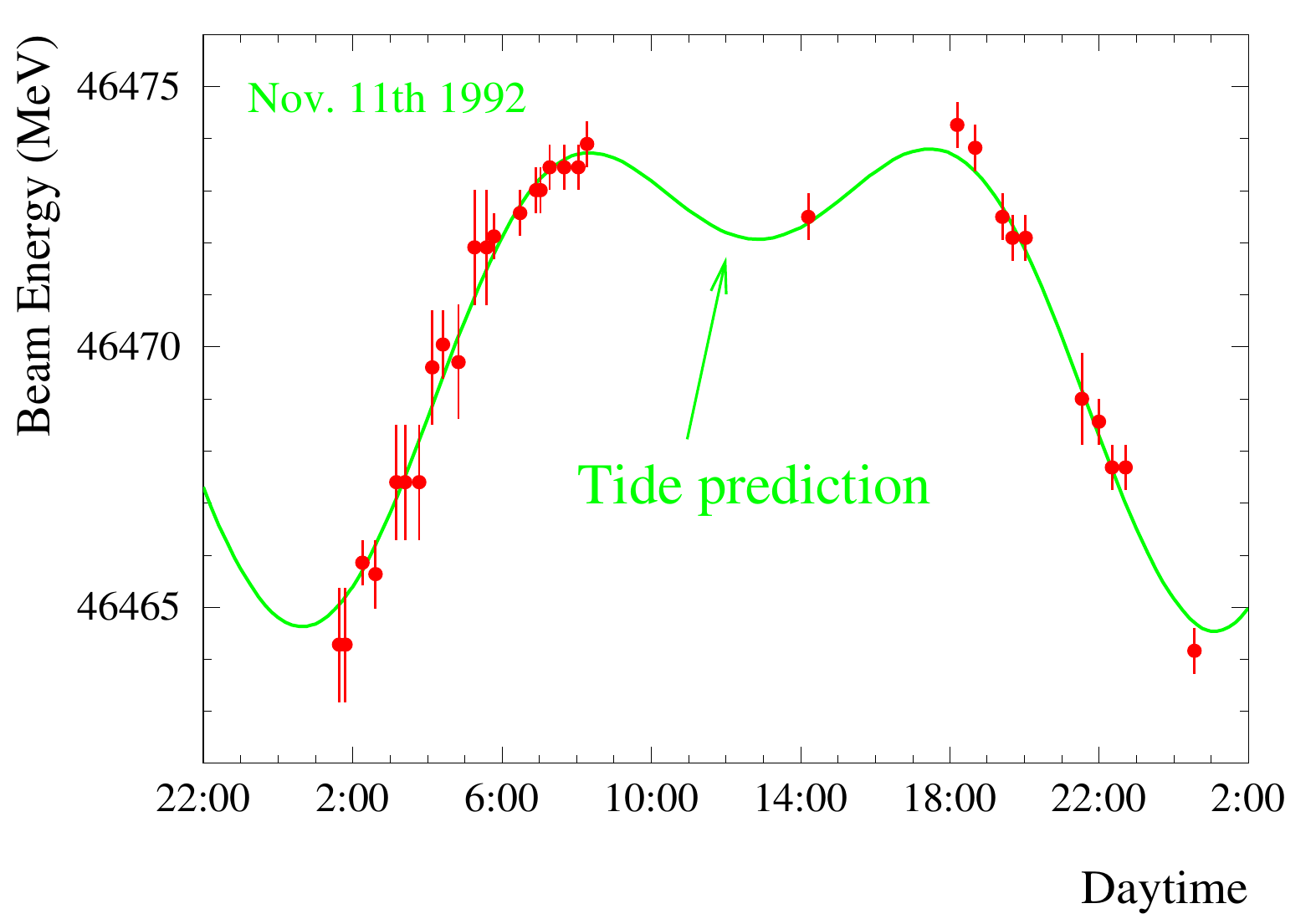}
&
\includegraphics[width=0.43\textwidth]{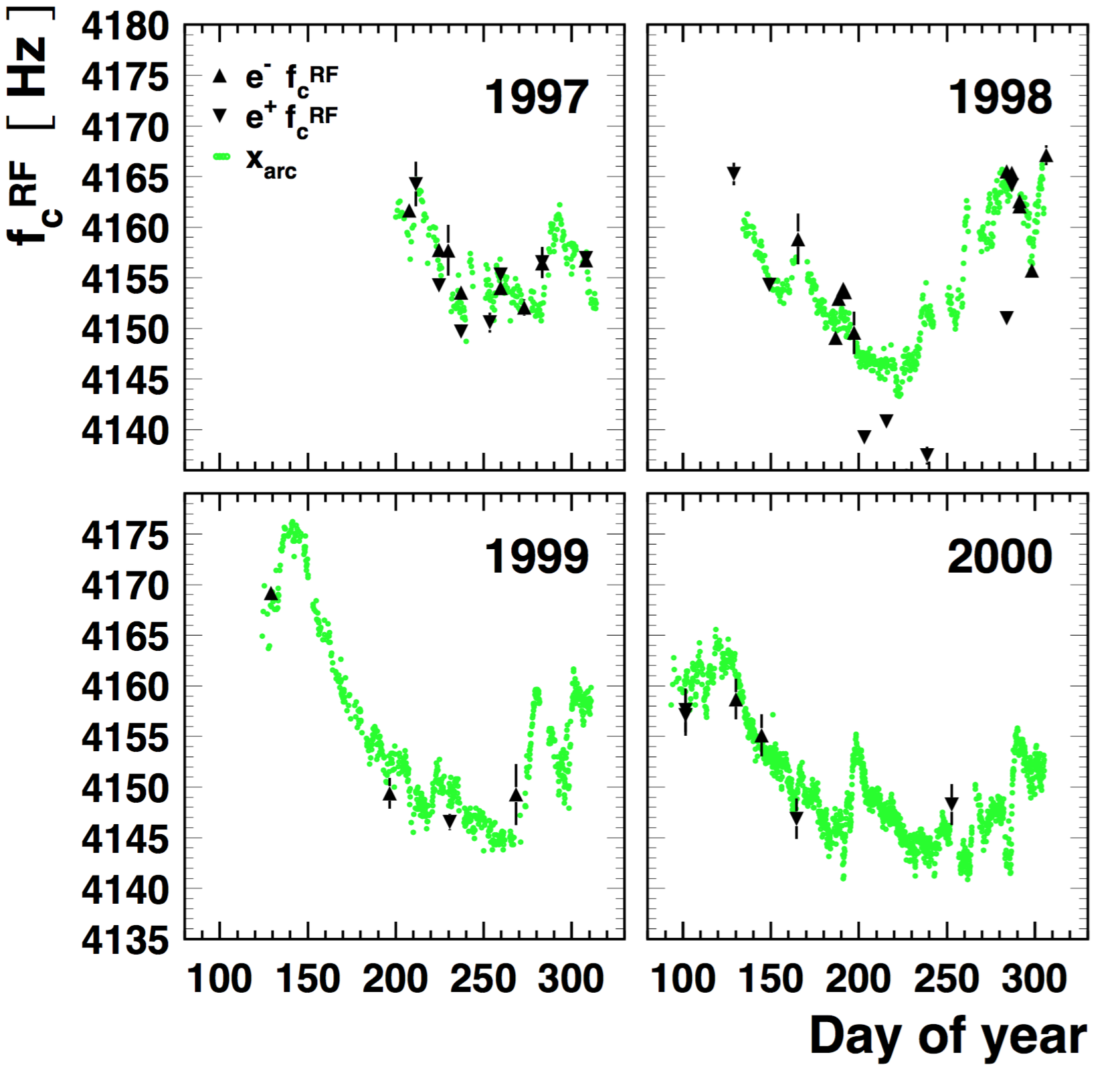}
\end{tabular}
\caption{Left: comparison between RDP measurements (red points) and prediction of tide model (green line) over a four hour period.   Right: central frequency measurements (black triangles) and BOM measurements (green points) showing slowly varying distortions in ring circumference.}
\label{fig:tidesandfrfc}
\end{figure}

 The most significant component of the model is the effect of relative changes in the ring circumference $\Delta C/C$, which leads to a relative change in the beam energy of $-\frac{1}{\chi}\frac{\Delta C}{C}$, where $\chi$ is the momentum compaction factor. At LEP $\frac{1}{\chi}\sim 5000$, which meant that 1\,mm distortions led to ${\cal O}({10}\,{\rm MeV})$ effects during the Z scans. Distortions of this magnitude did indeed occur, driven both by Earth tides, which made the beam energy vary over the course of each coast, and by longer-term variations most likely driven by seasonal changes in both the local water
table and also the water level in Lac L\'{e}man.  The tidal contribution was modelled analytically, whereas the slowly-varying component was tracked by central-frequency measurements and the beam-orbit monitor (BOM) system of the accelerator. Figure~\ref{fig:tidesandfrfc} (left) shows good agreement between RDP measurements and a tidal model throughout an experiment lasting several hours~\footnote{These data were collected after the dipole field rise, discussed below, had saturated, allowing all energy variation observed during the experiment to be described by the tidal model alone.},  and also (right) central-frequency and BOM measurements that exhibit slowly-varying ring distortions.

\begin{figure}[tbh]
\centering
\begin{tabular}{cc}
\includegraphics[width=0.47\textwidth]{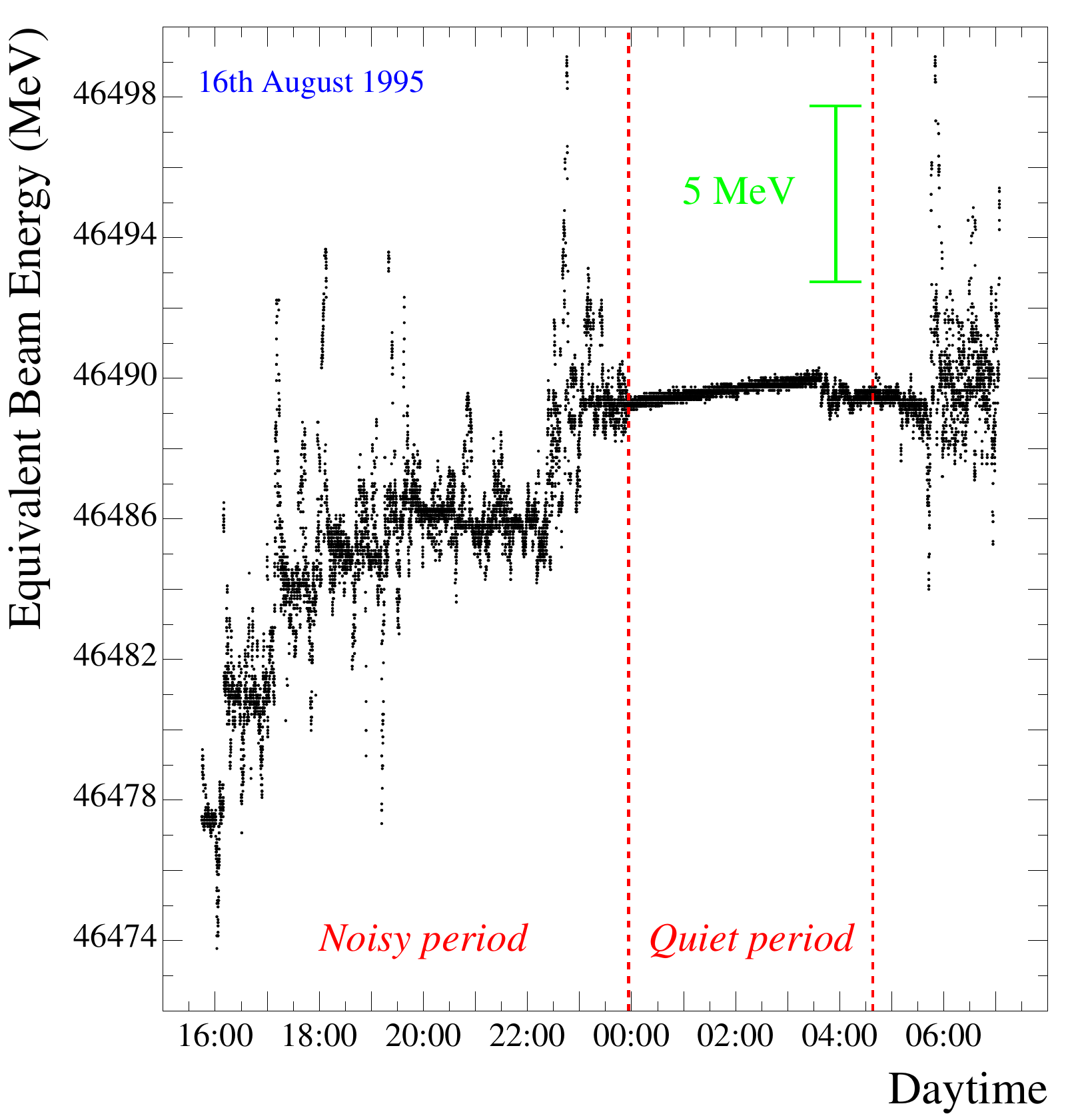}
&
\includegraphics[width=0.485\textwidth]{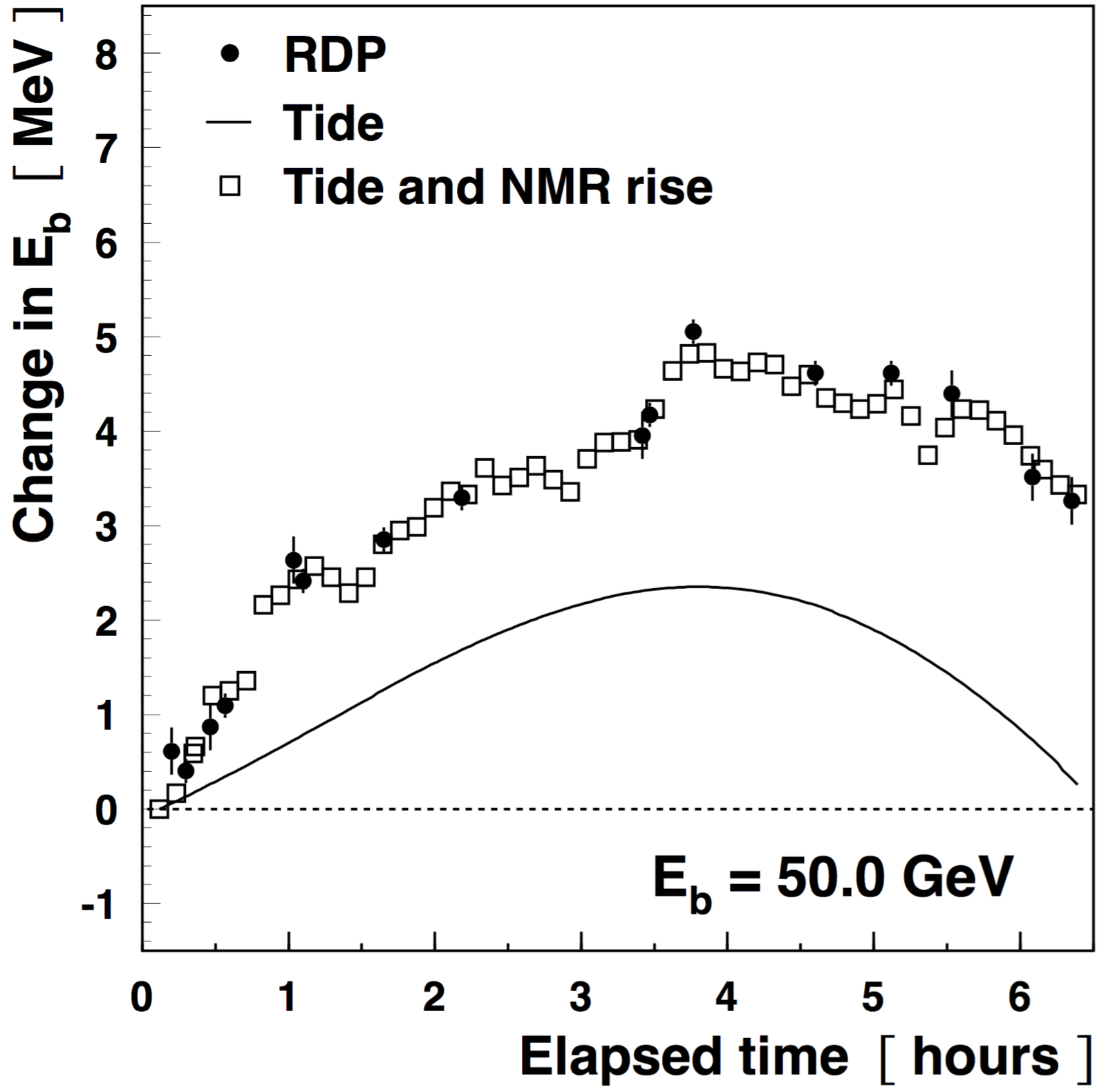}
\end{tabular}
\caption{Left: dipole field as sampled in an NMR probe, expressed as equivalent beam energy, plotted against time of day.  Right: energy rise during a fill measured by RDP, compared withpredictions of the energy model including both the tidal component and NMR ({\it i.e.} dipole field)  rise.}
\label{fig:nmrrise}
\end{figure}

An important source of energy variation during the coast itself was that of a rising dipole field. This effect was first discovered in 1995 after two dipoles in opposite octants were equipped with NMR probes to provide local field readings.  More dipoles in the tunnel were instrumented the following year and showed similar behaviour. Complementary information came from a series of measurements performed on a magnet in the laboratory.  The observations from the tunnel indicated an average field rise of around
$1 \times 10^{-4}$ during a typical coast.  This rise was accompanied by a characteristic noise or fluctuation in the field values.  Both the gradient of the field rise and amplitude of the noise was found to vary with time of day, time into coast, and location around the ring.  Figure~\ref{fig:nmrrise}~(left) shows the response of one NMR probe as a function of time of day. Eventually the noise was correlated with the measurement of a current flowing on the beam-pipe, which arose from electrical trains passing along the Geneva-Bellegarde railway line.  This current stimulated the magnetization status of the
iron in the dipoles, provoking a field rise that gradually saturated.  An additional contribution to the field rise came from temperature effects, which had a complicated dependence on the thermal history of the dipole during the fill.   An empirical model of the field rise was developed, which was applied both in 1995 and also retrospectively to
the data sets of 1993 and 1994.  Figure~\ref{fig:nmrrise}~(right) shows excellent agreement between RDP measurements and a model including both the dipole field rise and the tidal variations for data collecting at LEP\,2.
This correction shifted the overall energy scale, and hence $m_{\rm Z}$, {\it downwards} by a few MeV, since the bulk of  RDP measurements had been made at the end of coast, after the saturation of the field rise.  

\subsubsection{Centre-of-mass energy (\texorpdfstring{$\sqrt{s}$}{ECM}) corrections }
\label{LEPEcmcorr}

The energy loss from synchrotron radiation, and its replenishment from the RF system, led to a characteristic {\it sawtooth} in the local beam energy.  A good understanding of this sawtooth was necessary to correct the centre-of-mass energy $\sqrt{s}$ at each of the four LEP interaction points (IP) from the naive expectation of $\sqrt{s} = 2E_{\rm
b}$.  Specifically, $\sqrt{s}$ could be shifted differently at each IP depending on the relative misalignment of the RF cavities in use, the phase errors on these cavities, and the precise configuration of cavities being operated.  These considerations were particularly important during 1995, when the commissioning of a large number of
superconducting cavities required for LEP\,2 operation led at times to a very asymmetric distribution in the accelerating voltage.  Care was taken to log the status of each cavity throughout the coast, so that voltage trips and trip recoveries were recorded.  An
{\it RF model} was developed to calculate the local energy shift at each IP as a function of time, and this model was validated against measured values of the separation between the electron and positron orbits from the BOM system, and also the synchrotron tune. Shifts in $\sqrt{s}$ of up to 20\,MeV were found for certain IPs and energy points, but
the overall uncertainty on the lineshape parameters from this source was small ({\it e.g.} 0.4\,MeV in $m_{\rm Z}$) due to the reliability of the RF model, and
anti-correlations between IPs.

\begin{figure}[tbh]
\centering
\begin{tabular}{cc}
\includegraphics[width=0.35\textwidth]{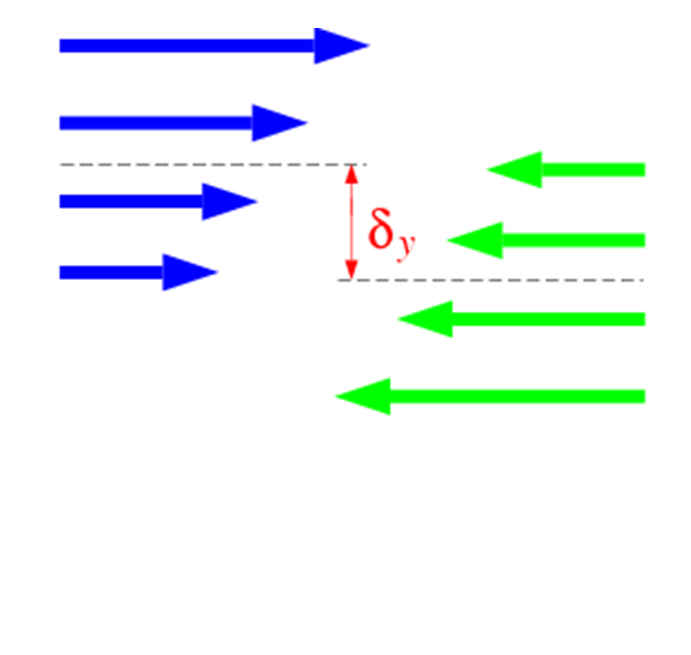}\hspace*{0.3cm}
&
\includegraphics[width=0.5\textwidth]{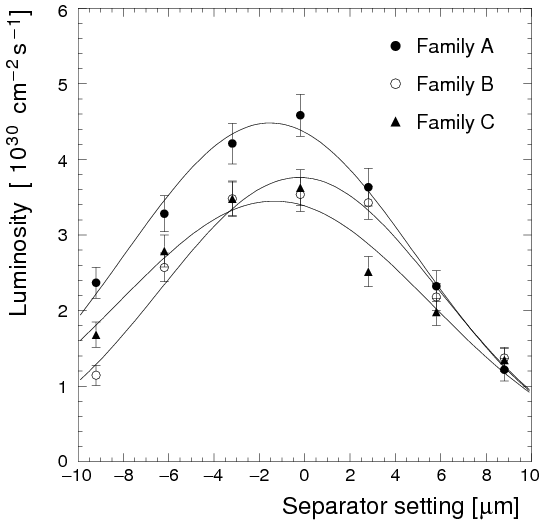}
\end{tabular}
\caption{Opposite-sign vertical dispersion in 1995.  Left: schematic indicating, by the length of the arrows, the variation in beam energy with vertical position within the bunch, and the offset $\delta_y$ between the colliding bunches.  Right: a separator scan showing the variation of luminosity with separator setting for each family of bunches within the train.}
\label{fig:vernier}
\end{figure}

An additional source of IP-specific shift in $\sqrt{s}$ arose in 1995. That year, a new mode of bunch-train operation was deployed, which necessitated the application of vertical bumps in the straight sections of LEP to avoid unwanted collisions.  These bumps induced a non-negligible vertical dispersion ({\it i.e.} a momentum ordering of the particles in the vertical plane).   Such a dispersion, combined with any vertical offset
between the colliding bunches, leads to a shift in $\sqrt{s}$.   These shifts were minimised through the regular use of separator scans at each IP.  Vertical scans in steps of known offsets were applied to each beam and the luminosity was measured per bunch at each setting.  A schematic to illustrate the dispersion, and an example of a separator scan are shown in Fig.~\ref{fig:vernier}. From these scans the settings were determined that minimised the net collision offset and hence maximised the luminosity.
  These settings were then applied,
thereby minimising any  energy shift through dispersion effects.  The scans also enabled the size of any residual bias to be estimated.  Through this approach the luminosity-weighted vertical offsets were restricted to $<0.1\,{\rm \mu m}$ and the uncertainty on $m_{\rm Z}$ and $\Gamma_{\rm Z}$ to 0.1\,MeV and 0.2\,MeV, respectively.

\subsubsection{\texorpdfstring{LEP\,1}{LEP1} results}
\label{sec:LEPresults}

The final uncertainty on $m_{\rm Z}$ was 2.1\,MeV, of which the LEP  energy contribution was 1.7\,MeV.  The corresponding values for $\Gamma_{\rm Z}$ were 2.3\,MeV and 1.2\,MeV, respectively~\cite{ALEPH:2005ab}.  The dominant components to the energy uncertainties for both parameters came from the residual scatter between the $E_{\rm b}$ model and the RDP measurements, which limits the knowledge of the energy scale in the un-calibrated coasts, and the understanding of the energy rise within a coast.

\subsection{Measuring \texorpdfstring{$m_{\rm W}$}{mW} at \texorpdfstring{LEP\,2}{LEP2}}
\label{sec:LEPmw}

During the LEP\,2 period (1996-2000) data corresponding to almost 700\,${\rm pb{^{-1}}}$ of  integrated luminosity were collected by each experiment at and above the W-pair production threshold.  This sample was used to determine the mass of the W boson, $m_{\rm W}$, through the methods of a threshold scan, which was performed in 1996, and through direct
reconstruction of the W-pair system, using the full data set.  For both approaches the relative uncertainty on the collision energy induces a corresponding relative uncertainty on $m_{\rm W}$, essentially common between experiments. The statistical uncertainty of the full LEP\,2 sample was around 30\,MeV, hence setting a goal on the relative knowledge of $\sqrt{s}$ of $1$--$2 \times 10^{-4}$.  The LEP\,2 energy-calibration campaign is fully described in Ref.~\cite{Assmann:2004gc}.

\subsubsection{Setting the absolute energy scale above the W-pair threshold}
\label{sec:LEPWEcal}

The high-energy operation of LEP\,2, spanning the range $80.5 < E_{\rm b} < 104.6$\,GeV, presented a new challenge for the energy calibration.  The polarization level of the machine dropped rapidly with $E_{\rm b}$, because of the depolarization effects of synchrotron side-band resonances, which grew in importance due to the increasing energy spread of the beam. A detailed discussion of polarization at LEP\,2 can be found
in Ref.~\cite{Assmann:535999}.  Techniques such as harmonic-spin matching~\cite{Sonnemann:357577}, k-modulation studies~\cite{Dehning:1998bq} and the development of a dedicated polarization optics~\cite{Uythoven:341927} were employed to ameliorate the problem, but RDP was limited to
beam energies of 61\,GeV and below.  Therefore RDP calibrations were performed at a variety of energies over the interval $41 < E_{\rm b} < 61$\,GeV and these measurements were used as input to a linear fit of the local $B$-field readings provided by 16  NMR
probes in  dipoles around the ring.  This {\it NMR model} was then applied at high energy to predict $E_{\rm b}$ in the physics regime.   In order to test the validity of the NMR model at these higher energies three complementary sets of tests were performed.
\begin{itemize}
\item{{\bf Flux loop}\\
The flux-loop was a sequence of copper loops which were embedded  in the dipole cores, and which sensed the change of flux as the magnets were ramped.  Although the flux loop provided no direct energy measurement it was sensitive to 96.5\% of the total bending field of LEP.  Hence comparison of the flux loop with the NMR model, particularly
during the evolution from calibration energies to the physics regime, enabled the representability of the NMR sampling to be validated. }
\item{{\bf LEP spectrometer}\\
The spectrometer was a device installed and commissioned in 1999 and used in  dedicated calibration periods throughout the 2000 run.  It consisted of a steel dipole with a precisely known integrated field~\cite{Chritin:2004km}, and triplets of beam-position monitors (BPMs) on either side which enabled the beam deflection to be be measured, and thus the energy to be determined.  The spectrometer dipole was part of the LEP lattice
and ramped together with the rest of the machine.  Hence the bending angle, which was approximately 3.8\,mrad, changed very little with energy. The design goal was to achieve $\sim\,10^{-4}$ precision at $E_{\rm b} \sim 100$\,GeV.  A {\it relative} measurement was performed, in which the change in integrated dipole field and and any small evolution in bending angle were determined over a ramp from an energy well known
from RDP calibration.  Shielding of the BPM system and water cooling of the coils suppressed thermal effects driven by the large increase in synchrotron radiation over the energy step.  As the spectrometer determined the local beam energy at its location between IPs 2 and 4 it was necessary to use the RF model to relate this measurement to the average beam energy $E_{\rm b}$. }
\item{{\bf  Synchrotron tune analysis}\\
The synchrotron tune $Q_s$ depends on the beam energy, the energy loss per  turn and the total RF voltage, $V_{\rm RF}$.  Since the energy loss itself depends on the beam energy, an analysis of the variation of $Q_s$ with $V_{\rm RF}$ can be used to infer $E_{\rm b}$.  Experiments were conducted in 1998, 1999 and 2000 to exploit this method in which $V_{\rm RF}$ was varied over as wide a range as beam stability permitted, and
data were taken around 50\,GeV  as well as at higher energies.  The low energy data were required to fix the RF voltage scale in a regime where $E_{\rm b}$ was well known. The model used in the fit was refined to account for higher-order effects such as synchrotron loss in machine components other than the dipoles. }
\end{itemize}

\subsubsection{\texorpdfstring{LEP\,2}{LEP2} results}
\label{sec:LEPWresults}

The three methods employed to validate the NMR model yielded  consistent results and demonstrated that the model gave an unbiased determination of the beam energy in the W-pair regime, within an uncertainty of $\pm 10$\,MeV at  $E_{\rm b}=100$\,GeV. The same $E_{\rm b}$ and RF models developed for LEP\,1 were required to track the evolution of the beam energy and the IP-specific corrections to the collision energy, respectively.
The relative uncertainty on the calculated value of $\sqrt{s}$ varied from year to year, but was typically $1.1 - 1.2 \times 10^{-4}$ for the bulk of data taking, which satisfied the goals of the LEP\,2 energy calibration.  The corresponding uncertainty on the global LEP energy measurement of $m_{\rm W}$ is 9\,MeV~\cite{Schael:2013ita}.

\subsection{Conclusions on the LEP energy calibration program}
\label{sec:LEPsummary}
Three main reasons contributed to the success of the LEP energy-calibration campaigns. Firstly was the existence of the LEP Energy Calibration Working Group, a committed team of physicists and engineers drawn from both the machine and the detectors, who led the work.  Secondly was the willingness of the LEP community to devote significant machine time to energy-calibration related studies:  these occupied more than 50 full days of
operation from 1993 onwards.  Thirdly was the investment in high quality instrumentation, and the commitment to mundane tasks such as the continuous logging of accelerator parameters, which was essential for the understanding and modelling of almost all effects.

The calibration of the collision energy at LEP was performed to a relative  precision of $2 \times 10^{-5}$ at LEP\,1 and around $1 \times 10^{-4}$ at LEP\,2, which, in both cases, matched well the goals of the physics programme.  At LEP\,1, the sensitivity was limited by the fact that RDP could only be performed in selected coasts with separated beams.  The main limitation on the LEP\,2 calibration arose from the inability to obtain polarization for RDP measurements above beam energies of 61\,GeV.  These restrictions do not apply at FCC-ee.


\section{Requirements for precision measurements at FCC-ee}

\subsection{Requirements for precision measurements at and around the Z peak}
\label{sec:Zrequirements}

One of the main physics goals of FCC-ee is a set of precision measurements with statistical precision improved by more than two orders of magnitude with respect to the state-of-the-art. A sample of these measurements is presented in Table~\ref{tab:EWPO}. The great FCC-ee experimental challenge will be to reach systematic uncertainties at the same level or smaller than the statistical ones.  

\setlength{\tabcolsep}{1.5pt}
\begin{table}[htbp]
\centering
\caption{\small Measurement of selected electroweak quantities at FCC-ee, compared with the present precision. The systematic uncertainties are initial estimates and might change with further examination.
This set of measurements, together with those of the Higgs properties, achieves indirect sensitivity to new physics up to a scale $\Lambda$ 
of 70 TeV in a description with dim 6 operators, and possibly much higher in some specific new physics models.\vspace{1mm}\label{tab:EWPO}}
\begin{tabular}{|l|rcl|c|c|r|}
\hline
Observable  & present &  &          &  FCC-ee  &  FCC-ee  &  Comment and   \\ 
            & value  &$\pm$& error  &  Stat. &   Syst.     &  dominant exp. error \\ 
\hline\hline  
$ \mathrm{ m_Z  ~(keV) } $  &  91186700   & $\pm$ &  2200    & 4  & 100  & From Z line shape scan  \\ 
$  $  &  & &    &   &  & Beam energy calibration \\
\hline
$ \mathrm{  \Gamma_Z  ~(keV) } $  & 2495200   & $\pm$ &  2300    & 4  & 100  & From Z line shape scan  \\
$  $  &  & &    &   &   &  Beam energy calibration  \\
\hline
$ \mathrm{  R_{\ell}^{Z}} ~(\times 10^3) $  & 20767 & $\pm$ &  25   & 0.06   & 0.2-1   &  ratio of hadrons to leptons \\
$  $  &  & &    &   &   &  acceptance for leptons  \\
\hline
$ \mathrm{ \alpha_{s} (m_Z^2) } ~(\times 10^4) $  & 
 1196 & $\pm$ &  30  &  0.1  &  0.4-1.6  &   
from $\mathrm{  R_{\ell}^{Z}}$ above\\
\hline
$ \mathrm{  R_b} ~(\times 10^6) $  & 216290 & $\pm$ &  660   & 0.3   &  <60  &  ratio of $\rm{ b\bar{b}}$  to hadrons  \\
$  $  &  & &    &   &   &  stat. extrapol. from SLD
\\
\hline
$ \mathrm{\sigma_{had}^0} ~(\times 10^3)$ (nb) & 41541 & $\pm$ &  37   & 0.1  &  4  &  peak hadronic cross section  \\
$  $  &  & &    &   &   &  \small luminosity measurement  \\
$ \mathrm{  N_{\nu}}  (\times 10^3) $  & 2992  & $\pm$ &  8   & 0.005   &  1  &  Z peak cross sections \\
$  $  & \cite{LEP1EWWG}  & \cite{Voutsinas:2019hwu} &    &   &  &   Luminosity measurement \\
\hline
$ \mathrm{ sin^2{\theta_{W}^{\rm eff}}} (\times 10^6) $  & 231480   & $\pm$ &  160   & 3   &  2 - 5  &   
from $ \mathrm{ A_{\rm FB}^{{\mu} {\mu}}}$  at Z peak\\
$  $  &  & &    &   &   &  Beam energy calibration  \\
\hline
$ \mathrm{ 1/\alpha_{QED} (m_{\rm Z}^2) } (\times10^3) $  & 128952 
  & $\pm$ &  14   & 3   &  small  &   
from $ \mathrm{ A_{\rm FB}^{{\mu} {\mu}}}$ off peak\\
$  $  &  & &    &   &  &  QED\&EW errors dominate  \\\hline
$ \mathrm{  A_{\rm FB}^b,0} ~(\times 10^4) $  & 992 & $\pm$ &  16   & 0.02   &  1-3  &  b-quark asymmetry at Z pole  \\
$  $  &  & &    &   &   &  from jet charge \\
\hline
$ \mathrm{A_{\rm FB}^{pol,\tau} ~(\times 10^4)} $  & 1498 & $\pm$ &  49   & 0.15   &  <2  &  $\tau$ polarization asymmetry  \\
$  $  &  & &    &   &   &  $\tau$ decay physics \\
\hline  
$ \mathrm{ m_W  ~(MeV) } $  &  80350   & $\pm$ &  15    & 0.5  & 0.3  & From WW threshold scan \\ 
$  $  &  & &    &   &  &  Beam energy calibration  \\
\hline
$ \mathrm{  \Gamma_W  ~(MeV) } $  & 2085   & $\pm$ &  42    & 1.2  & 0.3  & From WW threshold scan \\
$  $  &  & &    &   &   &  Beam energy calibration  \\
\hline
$ \mathrm{ \alpha_{s} (m_W^2) }  (\times 10^4)$  & 
 1170   & $\pm$ & 420 &  3  & small  &   
  from $ \mathrm{R_{\ell}^{W} }$\\
\hline
$ \mathrm{  N_{\nu}}  (\times 10^3) $  & 2920 & $\pm$ &  50   & 0.8   & small   &   ratio of invis. to leptonic \\
$  $  &  & &    &   &  & in radiative Z returns  \\
\hline
$ \mathrm{ m_{top}  ~(MeV/c^2) } $  &  172740   & $\pm$ &  500    & 17  & small  & From $\mathrm {t\bar{t}}$ threshold scan \\ 
$  $  &  & &    &   &  &  QCD errors dominate  \\
\hline
$ \mathrm{ \Gamma_{top}  ~(MeV/c^2) } $  &  1410   & $\pm$ &  190    & 45  & small  & From $\mathrm {t\bar{t}}$ threshold scan \\ 
$  $  &  & &    &   &  &  QCD errors dominate \\
\hline
$ \mathrm{ \lambda_{top}/\lambda_{top}^{SM}   } $  &   1.2 
    & $\pm$ &  0.3    & 0.10  & small  & From $\mathrm {t\bar{t}}$ threshold scan \\ 
$  $  &  & &    &   &  &  QCD errors dominate \\
\hline
$ \mathrm{ ttZ ~couplings   } $  &   
   & $\pm$ &  30\%   & 0.5 -- 1.5\%  & small  & From $\sqrt{s}=365$\,GeV run  \\ 
\hline
\end{tabular} 
\end{table}

The proposed method is to use the natural polarization of the beams, as was done in LEP, and to perform precise energy calibrations with the resonant depolarization technique already described in the previous section. This choice limits the data taking to half-integer spin-tune energy points $E_{\rm b} = 0.4406486\,(n+\ssfrac{1}{2})$\,GeV for resonant depolarization. Given that the energy dependence of the Z line shape is much broader than the interval between these quantized energies, essentially no loss of precision is to be expected. 

A number of shortcomings with respect to the  LEP procedures should be fixed: 
\begin{itemize}
    \item Unlike at LEP where measurements were performed outside of the normal data taking, the RD measurements will take place parasitically during normal data taking, using special non-.colliding beams. A frequency of about four times per hour should be feasible. 
    \item The measurements should be performed both on the electron and positron beams, especially since the two beams circulate in different rings. 
    \item To mitigate the large effect of earth tides and other ground motion, the RF frequency should be adjusted at short intervals to maintain the beam in position within the optical elements (as is done nowadays at the Large Hadron Collider~\cite{PhysRevAccelBeams.20.081003}).
    \item To monitor and mitigate opposite sign dispersion effects at the IRs, luminosity scans  or  Van der Meer scans) must be performed regularly in the vertical plane. In the horizontal plane transverse scans cannot be applied due to the large crossing angle, the longitudinal overlap must be scanned using the relative RF phase between electron and positron beams.
    \item Monitoring of all diagnostics and of changes in the optics and machine settings should be foreseen. 
    \item All possible means of monitoring energy spread and centre-of-mass energy will be welcome. This applies in particular i) to the ability of the polarimeter to be used as a spectrometer; ii) of the experiments to monitor quantities related to beam size, energy difference between e+ and e- beams, energy spread and centre-of-mass energy.  
\end{itemize}

\begin{figure}[htbp]
	\begin{center}
	\centering
	\vspace{-2cm}
 \begin{minipage}[t]{0.49\textwidth}
		\includegraphics*[width=1.0\textwidth]{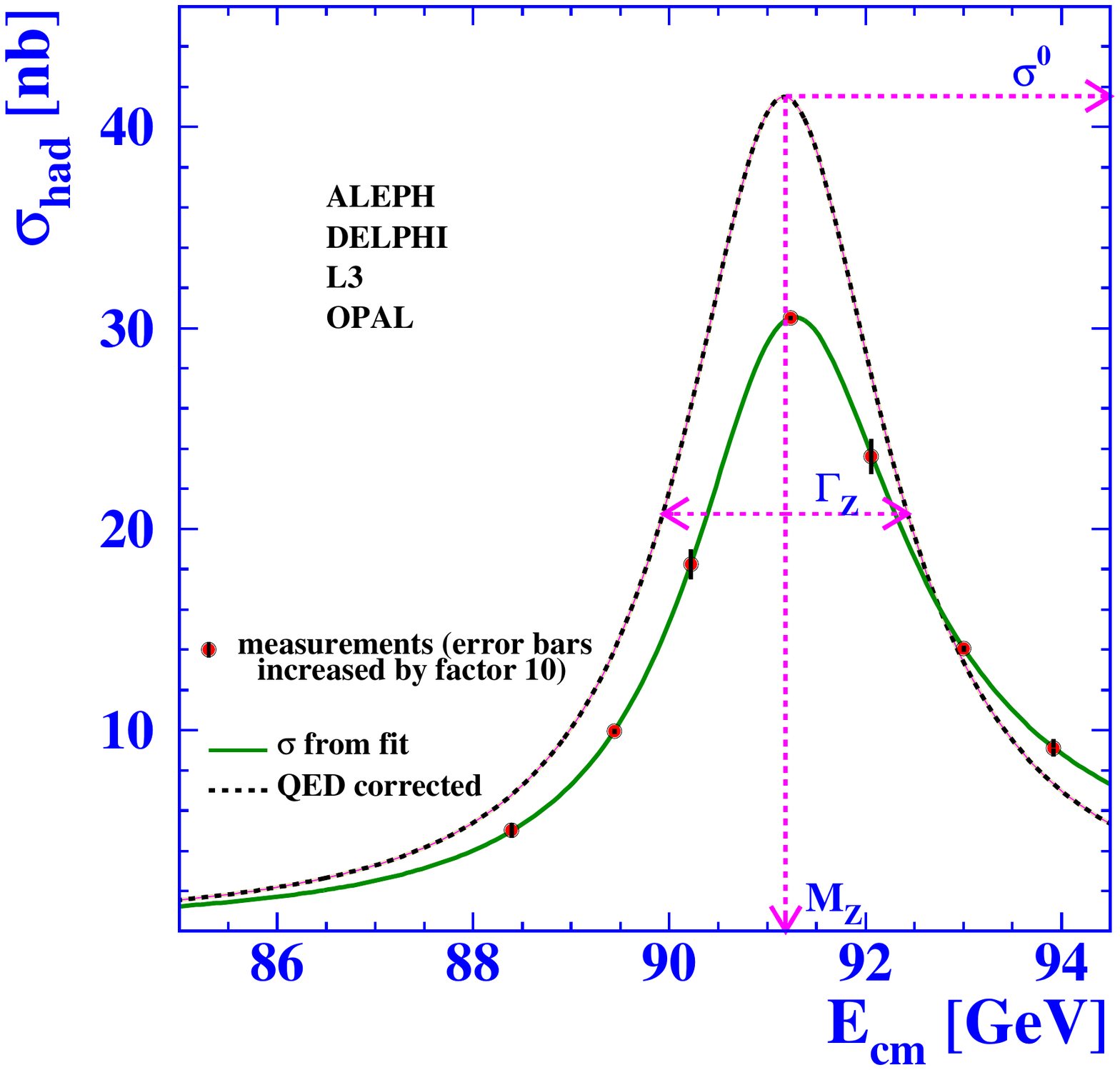}
\end{minipage}
\begin{minipage}[t]{0.49\textwidth}
		\includegraphics*[width=1.\textwidth]{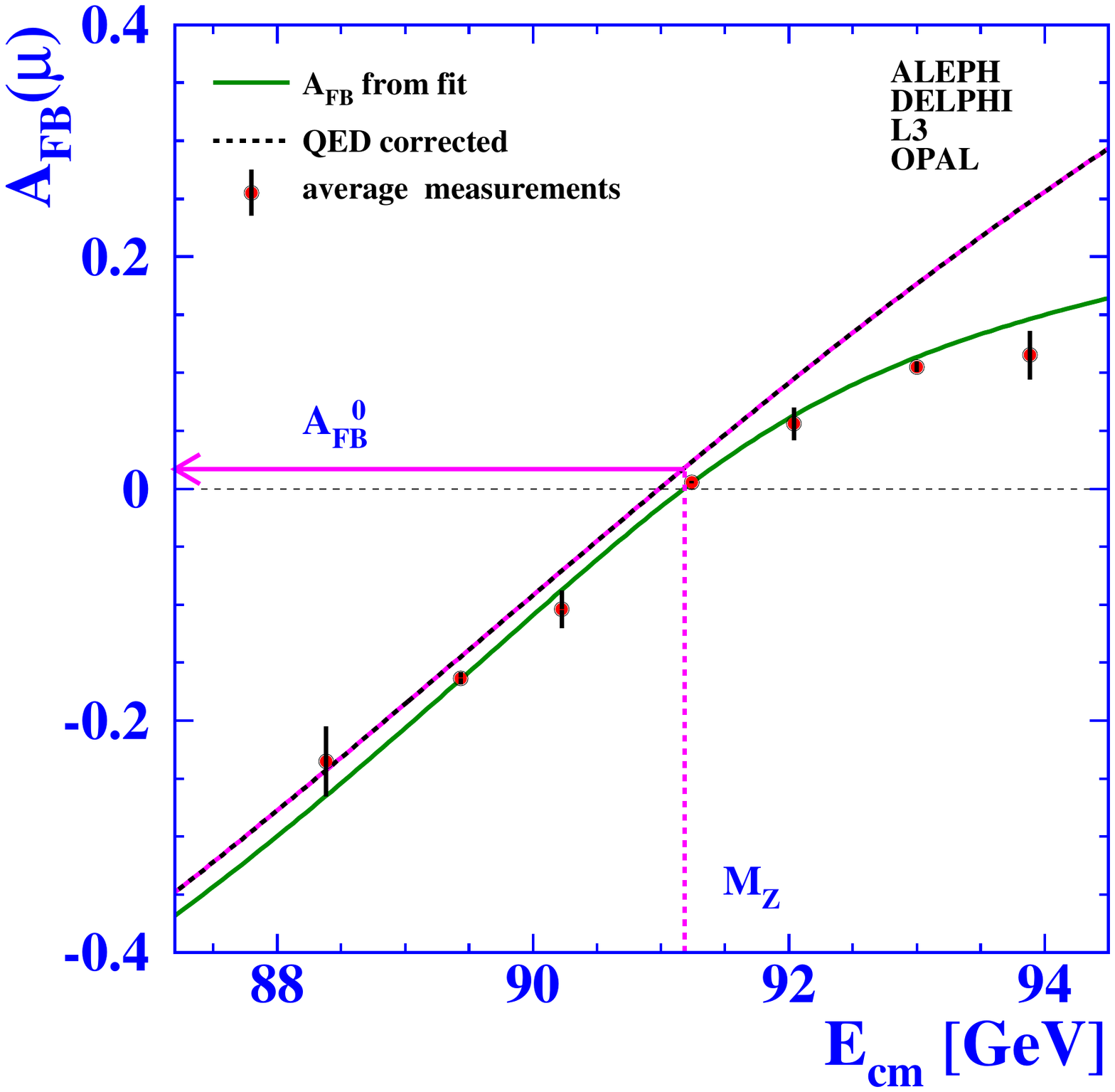}
\end{minipage} 
\caption{The total hadronic  cross section (Left) and the muon forward-backward charge asymmetry  (right) as  function of the centre-of-mass energy around the Z peak, from~\cite{ALEPH:2005ab}. 
}
\label{fig:ZlineAFB}
\end{center}
\end{figure}

The measurements most affected by the energy calibration uncertainties are those associated with a phenomenon having a strong variation with $\sqrt{s}$. This is the case particularly of the Z line shape and of the forward backward asymmetry for leptons, shown on Fig.~\ref{fig:ZlineAFB}, where the measurement points from LEP are also shown. From these measurements are extracted the Z mass $\rm{m_Z}$, the Z width $\rm{\Gamma_Z}$, the Z peak cross section $\mathrm{\sigma_{had}^0}$, the forward-backward asymmetry for leptons $A_{\rm FB}^{\mu \mu}$ both on resonance from which is extracted the weak mixing angle $\sin^2{\theta_{W}^{\rm eff}}$ and off resonance, where the energy dependence of the asymmetry is used to extract the value of  the electromagnetic constant $\alpha_{QED} (m_{\rm Z}^2)$ ~\cite{Janot:2015gjr}. 

A three-point scan of the resonance, with centre-of-mass energies $\sqrt{s_-}$, $\sqrt{s_0}$, and $\sqrt{s_+}$ as discussed below, at and above the Z peak (defined as the point of maximal cross section) is sufficient to provide the necessary information. It should be optimized around the quantity whose precision will be most limiting for the electroweak fits; this, it turns out, is the measurement of the off-peak forward-backward asymmetries from which $\mathrm{\alpha_{QED} (m_Z)}$ is extracted. The optimal energy points~\cite{Janot:2015gjr} being  $\sqrt{s_-} = 87.9$\,GeV, $\sqrt{s_0}=91.2$\,GeV, and $\sqrt{s_+}= 94.3$\,GeV, the set of energies can be chosen as follows. 

The requirement for the energy calibration on a three point scan around the line shape at the three centre-of-mass energies can be phrased in the following quantities, following the description given for LEP~\cite{Blondel:251734}, which gives the resulting uncertainties for a given set of centre-of-mass energy errors:
\begin{equation}
\begin{array}{r@{}l}
\frac{\Delta m_{\rm Z}}{m_{\rm Z}} &{} = 
\left \{ \frac{\Delta \sqrt{s}}{\sqrt{s}} \right \}_{\rm abs} 
\oplus   \left \{\frac{\Delta (\sqrt{s_+}  + \sqrt{s_-} )}{\sqrt{s_+} +\sqrt{s_-}}\right \}_{\rm ptp-syst} 
\oplus_i  \left \{\frac{\Delta \sqrt{s_\pm^i}}{\sqrt{s_\pm^i N_\pm^i} }\right \}_{\rm sampling} ,
\\
 \frac{\Delta{\rm \Gamma_Z}}{{\rm \Gamma_Z}} &{} = 
\left \{ \frac{\Delta \sqrt{s}}{\sqrt{s}} \right \}_{\rm abs} 
\oplus   \left \{\frac{\Delta (\sqrt{s_+}  - \sqrt{s_-} )}{\sqrt{s_+} -\sqrt{s_-}}\right \}_{\rm ptp-syst} 
\oplus_i  \left \{\frac{\Delta \sqrt{s_\pm^i}}{\sqrt{s_\pm^i N_\pm^i} }\right \}_{\rm sampling} ,
\\
{\scriptstyle \Delta A_{\rm FB}^{\mu \mu} {\rm (pole)}}  &{} = \frac{\partial A_{\rm FB}^{\mu \mu} }{\partial \sqrt{s} }  {\scriptstyle \left \{ \Delta (\sqrt{s_0} -  0.5 ( \sqrt{s_+} + \sqrt{s_-}) )\right \}}_{\rm ptp-syst}  \oplus_i  \frac{\partial A_{\rm FB}^{\mu \mu} }{\partial \sqrt{s}} \left \{ \frac{\Delta \sqrt{s_{0,\pm}^i}}{\sqrt{N_{0,\pm}^i} }\right \}_{\rm sampling} ,
\\
\frac{\Delta \alpha_{\rm QED} (m_{\rm Z}^2)}{\alpha_{\rm QED} (m_{\rm Z}^2)}  &{} = 
\left \{ \frac{\Delta \sqrt{s}}{\sqrt{s}} \right \}_{\rm abs} 
\oplus   \left \{\frac{\Delta (\sqrt{s_+}  - \sqrt{s_-} )}{\sqrt{s_+} -\sqrt{s_-}}\right \}_{\rm ptp-syst} 
\oplus_i  \left \{\frac{\Delta \sqrt{s_\pm^i}}{\sqrt{s_\pm^i N_\pm^i} }\right \}_{\rm sampling} ,
\end{array}
\label{eq:ECM-errors}
\end{equation}
with $\frac{\partial A_{\rm FB}^{\mu \mu} }{\partial \sqrt{s}} \simeq 0.09 / {\rm GeV} $.  Equation~\ref{eq:ECM-errors} can be interpreted as follows. 
\begin{itemize}
    \item The absolute energy scale uncertainty,  denoted ``abs'', represents a global energy scale error, and absorbs all uncertainties that are common to the scan points, such as e.g energy-independent systematic offsets of the collision point, the bulk of the beamstrahlung errors, the common errors in the DP technique, etc;  
    \item The point-to-point energy uncertainties, denoted ``ptp-syst'', affect each of the lumino\-sity-averaged centre-of-mass energies of the scan points in independent fashion. They concern the statistical errors for all quantities determined from the experiments data themselves, as well as the systematic variations with energy of, e.g., the collision offsets, the dispersion at IR, and  spin-tune- (thus energy-) dependent interference of the depolarizing resonance with synchrotron or betatron resonances, etc; 
    \item The energy calibration statistical uncertainty, denoted ``sampling'', represents the precision of each energy measurement, scaled with the number of measurements as $1/\sqrt{N^i}$, where $i$ indicates that  a quadratic sum over the centre-of-mass energies should be performed, with the appropriate coefficients. Given a sampling of roughly $10^4$ energy measurements on each of the scan points and a beam energy measurement error of $\pm 100$<,keV, the statistical uncertainty is of the order of 1\,keV and has a negligible effect. This precision however is of great importance as it allow a great number of systematic checks.   
\end{itemize}

In addition, the centre-of-mass energy spread in collision, denoted $\sigma_{\sqrt{s}}$, affects the quantities with a strong quadratic dependence on centre-of-mass energy~\cite{Assmann:1998qb}, such as the measured cross section at the top of the resonance, which affects mainly the Z width $\Gamma_{\rm Z} $ and the Z peak cross section $\sigma_{\rm had}^0$. The energy spread affects the measured cross section as described in Ref.~\cite{Assmann:1998qb}, Section 12: 
$$ 
\delta \sigma^0_{\rm had} = 0.5 \frac{d^2\sigma_{\rm had} }{d\sqrt{s}^2} \sigma^2_{\sqrt{s}}
$$
The main effect is to smear the cross sections, adding in quadrature the energy spread to the Z width.  

Table~\ref{tab:Zscan} summarizes possible scan energies. For the peak point, nature is kind to provide directly an half-integer spin tune precisely at the requested energy. For the other two points, it is possible to find  energies with half-integer spin tune that are close, but not exact. If requested, it is possible to take data both above and below the requested value, so that the luminosity-averaged energies are the requested ones. Five energy points might actually help controlling and reducing the effect of possible point-to-point energy calibration systematic uncertainties, which turn out to be dominant.

\begin{table}[htbp]
\centering
\caption{Centre-of-mass energies for the proposed Z scan. The points noted A and B are half integer spin tune points with energies closest to the requested energies.\vspace{3mm}     
   \label{tab:Zscan}}
\begin{tabular}{|l|r|r|r|}
\hline\hline
Scan point & $\sqrt{s}$ (GeV) & $E_{\rm b}$ (GeV) & Spin tune \\ \hline\hline
$\sqrt{s_-}$ A &  87.69 & 43.85 & 99.5 \\
$\sqrt{s_-}$ Request  &  87.9 & 43.95 & 99.7 \\
$\sqrt{s_-}$ B  &  88.57 & 44.28 & 100.5 \\
\hline
$\sqrt{s_0}$  &  91.21  & 45.61 & 103.5 \\
\hline
$\sqrt{s_+}$ A &  93.86 & 46.93 & 106.5 \\
$\sqrt{s_+}$ Request  &  94.3 & 47.15 & 107.0 \\
$\sqrt{s_+}$ B  &  94.74 & 47.37 & 107.5 \\
\hline\hline
\end{tabular} 
\end{table}

Table~\ref{tab:Z-Ecal-errors} compiles the uncertainties on the most sensitive quantities 
$m_{\rm Z}$, $\Gamma_{\rm Z}$, $\sin^2{\theta_{\rm W}^{\rm eff}}$ from  $A_{\rm FB}^{\mu \mu}$ at the Z peak, and  $1/\alpha_{\rm QED} ({\rm m}_Z^2) $ from   $A_{\rm FB}^{\mu \mu}$ at the optimal off peak. The absolute measurement centre-of-mass energy calibration error is assumed to amount to $\pm 100$\,keV, the point-to-point errors are assumed to be also equal to $\pm 100$\,keV, and the centre-of-mass energy spread is assumed to be known to 500\,keV. As for statistics, the measurements of beam energy are assumed to take place with a resonant depolarization (RD) every 1000 seconds (15 minutes) with a beam energy precision of $\pm 100$\,keV. During that time the spectrometer part is able to measure the beam energy with a statistical precision of around 120\,keV ($4/\sqrt{1000}$\,MeV); we assume an overall $\sqrt{s}$ determination at a precision of 200\,keV every 15 minutes.  The number of measurements per scan point is then taken to be spread over the $10^7$ seconds of data taking at each off-peak point and twice as many on the peak.  

\begin{table}[htbp]
\centering
\caption{Calculated uncertainties on the quantities most affected by the centre-of-mass energy uncertainties, under the initial systematic assumptions.\vspace{3mm}
   \label{tab:Z-Ecal-errors}}
\begin{tabular}{|l|c|c|c|c|c|}
\hline\hline
  & statistics & $\Delta \sqrt{s}_{\rm abs} $ & $\Delta \sqrt{s}_{\rm syst-ptp}$ & calib. stats. & $\sigma_{\sqrt{s}}$ \\
Observable &  & 100\,keV&100\,keV & 200\,keV/$\sqrt{N^i}$& $85 \pm 0.5$\,MeV \\ \hline
$m_{\rm Z}$ (keV) & 4  & 100 & 70 & 1 & --  \\
$\Gamma_{\rm Z}$ (keV) & 7  & 2.5 & 55 & 1 & 100   \\
$\sin^2{\theta_{\rm W}^{\rm eff}}\times 10^6 $  from  $A_{\rm FB}^{\mu \mu}$ & 2  & -- & 6 & 0.1  & --\\
$\frac{\Delta \alpha_{\rm QED} (m^2_{\rm Z})}{\alpha_{\rm QED} (m^2_{\rm _Z})} \times 10^5$ &  3 & 0.1 & 2.2 & -- & 1\\ 
\hline\hline
\end{tabular} 
\end{table}

The conclusions from this table are very clear: the frequent RD measurements of the proposed scan should ensure that the absolute energy scale and its reproducibility are at or below the 100\,keV mark. However, the energy spread and the point-to-point systematic uncertainties will dominate the Z width and the $\sin^2{\theta_{\rm W}^{\rm eff}}$ accuracy. It is therefore necessary to find independent measurements either from accelerator diagnostics or from the measurements of, e.g., muon pairs in the detectors, to extract the energy spread and a relative measure of the centre-of-mass energy to reduce these uncertainties to a level that is commensurate with the available statistical precision.  


\subsection{Requirements for precision measurements of the W mass and width}
\label{sec:Wrequirements}

\begin{figure}[tbh]
\centerline{\includegraphics[width=0.8\textwidth]{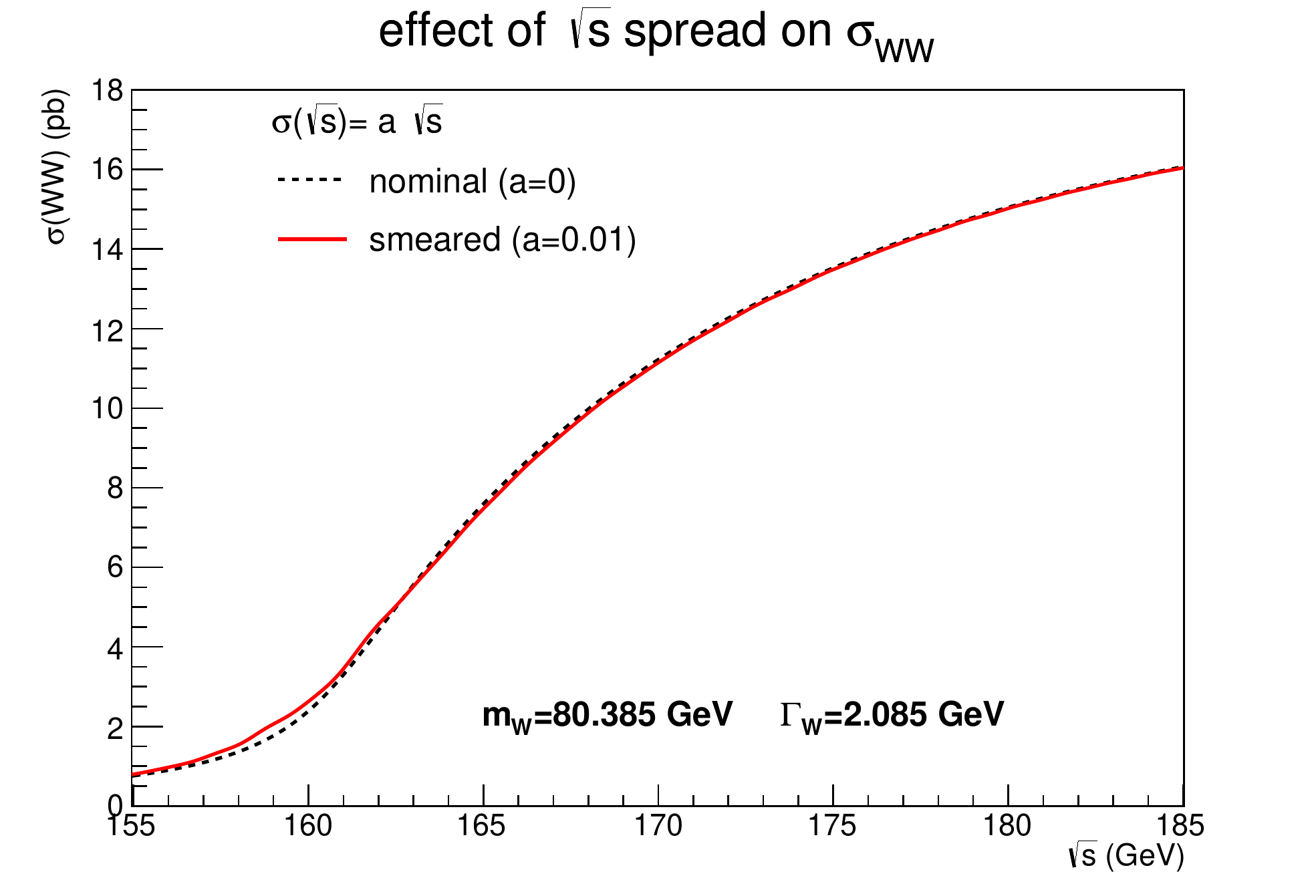}}
\caption{W-pair production cross section as a function of
  the $\Pep\Pem$ collision energy $\sqrt{s}$ as evaluated with YFSWW3 1.18~\cite{Jadach:2001uu}.
  The central curve corresponds to the predictions obtained with
  $m_{\rm W}=80.385$\,GeV and $\Gamma_{\rm W}=2.085$\,GeV.
  The dashed curve represents the effect of a 1\% centre-of-mass energy spread.}
\label{fig:sww_espread}
\end{figure}

The FCC-ee precision measurements of the W mass and width make use of the total W-pair cross section lineshape near the kinematic production threshold (Fig.~\ref{fig:sww_espread}) and have been presented and discussed in Ref.~\cite{Azzi:2017iih}. 
A minimal and optimal threshold-scan strategy sharing a total luminosity of 12\,ab$^{-1}$ on two energy points  $\sqrt{s_1}=157.1$\,GeV and $\sqrt{s_2}=162.3$\,GeV provide a statistical uncertainty on the W mass and width measurements of $\Delta m_{\rm W} ({\rm stat})= 0.45$\,MeV and $\Delta \Gamma_{\rm W} ({\rm stat})= 1.2$\,MeV. When limiting the data taking to half-integer spin-tune energy points, adequate for energy calibration with resonant depolarization, the optimal data taking shifts to $\sqrt{s_1}=157.3$\,GeV and $\sqrt{s_2}=162.6$\,GeV, with a small degradation ($<10\%$) of the statistical precision.

\begin{figure}[htbp]
	\begin{center}
	\centering
 \begin{minipage}[t]{0.49\textwidth}
		\includegraphics*[width=1.0\textwidth]{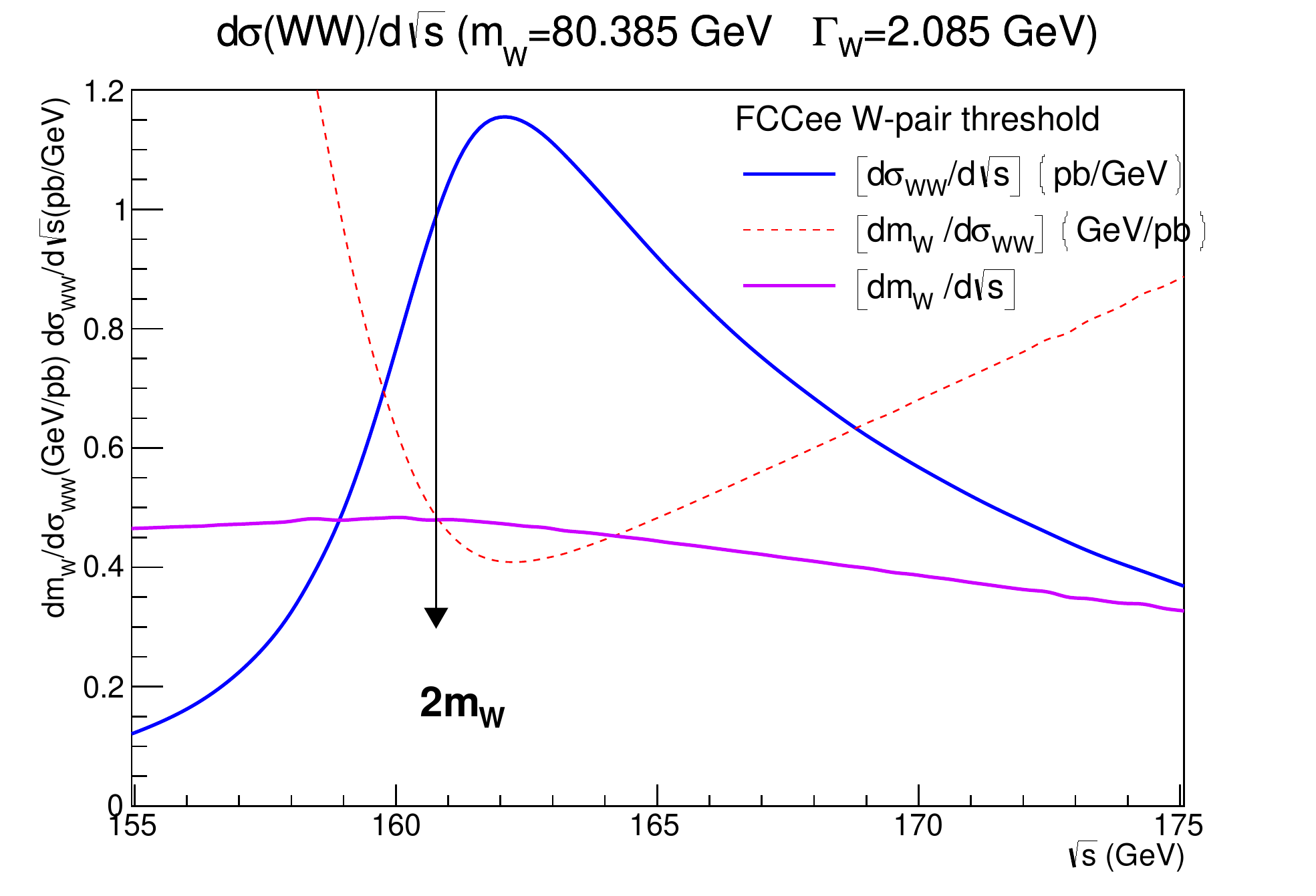}
\end{minipage}
\begin{minipage}[t]{0.49\textwidth}
		\includegraphics*[width=1.\textwidth]{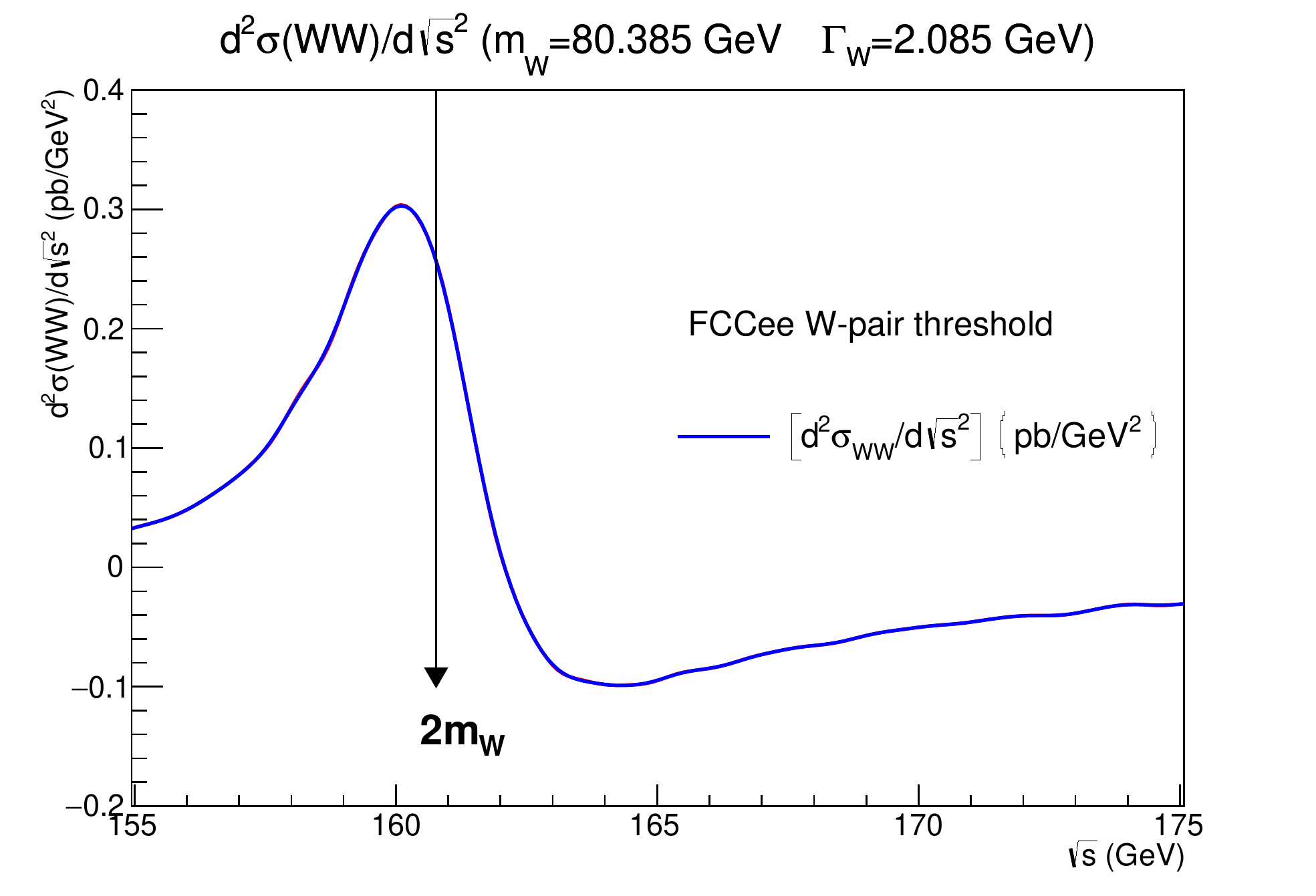}
\end{minipage} 
\caption{W-pair cross section first-order (left) and second-order (right) differential functions
  with respect to the centre-of-mass energy, evaluated with YFSWW3 1.18~\cite{Jadach:2001uu}, and as a function of $\sqrt{s}$ around and above the kinematic production threshold.
  Central mass and width values are set to
  $m_{\rm W}=80.385$\,GeV and $\Gamma_{\rm W}=2.085$\,GeV. 
}
\label{fig:dsWWdE}
\end{center}
\end{figure}

Any uncertainty on the centre-of-mass energy propagates directly to the W mass determination as
\begin{equation}
    \delta m_{\rm W} = F(\sqrt{s}) \delta \sqrt{s} = 
    \frac {dm_{\rm W}}{d\sigma_{\rm WW}} \frac{d\sigma_{\rm WW}}{d\sqrt{s}} \delta \sqrt{s}.
\end{equation}
The derivatives $d\sigma_{\rm WW}/d\sqrt{s}$
and $dm_{\rm W}/d\sigma_{\rm WW}$, together with their product $F(\sqrt{s})$ are shown in 
Fig.~\ref{fig:dsWWdE} as a function of $\sqrt{s}$. The function $F$ reaches a maximum at $\sqrt{s} \simeq 2m_{\rm W}-1$\,GeV, and in general $F(\sqrt{s}) \leq  1/2$.  
Therefore, the uncertainty on the average beam energy translates into an uncertainty on the W mass of  $\delta m_{\rm W} \leq  \delta \sqrt{s} /2$, such that a beam-energy uncertainty of $ \sim 0.2$\,MeV or better is necessary to have a small impact on the W mass measurement.

The effect of the centre-of-mass energy spread on the total W-pair cross-section lineshape is given in Eq.~\ref{eq:swwspread}
\begin{equation} \label{eq:swwspread} 
    \sigma'_{\rm WW}(\sqrt{s_0}) = \int  \sigma_{\rm WW}(\sqrt{s}) G(\sqrt{s}-\sqrt{s_0})d\sqrt{s} 
\end{equation}
where the energy spread distribution can be approximated by a Gaussian: 
\begin{equation}
    G(x) = \frac{1}{\sqrt{2\pi}\sigma_{\sqrt{s}}} \exp \left(-\frac{x^2}{2\sigma^2_{\sqrt{s}}}\right)
\end{equation}
As shown in Table~\ref{tab0:params}, the centre-of mass energy spread is estimated to be $\sim 150$\,MeV at the WW energies. The effect of a ten times larger energy spread ($\sim 1.6$\,GeV) on the cross-section lineshape is shown in Figure~\ref{fig:sww_espread}. To first order, the cross-section variation amounts to
\begin{equation} 
\label{sww_espread2}
    \delta\sigma_{\rm WW} \simeq \frac{1}{2}\frac{d^2\sigma_{\rm WW}}{d\sqrt{s}^2}  \sigma^2_{\sqrt{s}}.
\end{equation}  
Figure~\ref{fig:dsWWdE} shows dependence of the first and second order derivatives of the W-pair cross-section lineshape on $\sqrt{s}$. From the second derivative plot, it can be inferred that the maximum (positive) cross-section variations are expected at $\sqrt{s}=160.1$\,GeV. Effects above and below the threshold region ($\sqrt{s}>170$\,GeV and $\sqrt{s}<155$\,GeV) are much smaller, and it essentially vanishes at $\sqrt{s} = 162.3$\,GeV.
With a minimal two-point data taking configuration at the $\sqrt{s_1}=157.3$\,GeV and $\sqrt{s_2}=162.6$\,GeV half-integer spin-tune energy points, 
the full impact of the energy spread effects on the measured cross sections are 
$$
\delta\sigma_{\rm WW_1} = +0.82~{\rm fb} \qquad \delta\sigma_{\rm WW_2} =-0.56~{\rm fb}
$$
with corresponding induced shifts on the measured W mass and width of
$$
\delta m_{\rm W} = -0.28\,{\rm MeV} \qquad
\delta \Gamma_{\rm W} = +2.0\,{\rm MeV}.
$$
These values are comparable to the statistical uncertainties on the W mass and width of 0.45\,MeV and 1.3\,MeV, respectively, expected with an integrated luminosity of $12\,{\rm ab}^{-1}$. A measurement of $\sigma_{\sqrt{s}}$ with a 10\% accuracy would suffice to   
reduce the impact of the centre-of-mass energy spread to negligible levels, with a  degradation smaller than 5\% of the W width precision. 

\subsection{Requirements at Higher Energies}
\label{sec:TOPrequirements}

At energies well above the W pair production threshold, the levels of transverse polarization are expected to be way too small to be suitable for energy calibration. The statistical accuracy with which the Higgs boson and the top quark masses will be measured at FCC-ee, typically 10\,MeV and 17\,MeV, respectively, sets the requirement on the beam energy calibration. Following the reasoning in Section~\ref{sec:Wrequirements}, the beam energy spread (325\,MeV) needs to be evaluated around the ${\rm t \bar t}$ threshold with a moderate precision of 35\% for the measurement of the top-quark width. 

At LEP\,2~\cite{Barate:1999tg, Achard:2003tw,Abbiendi:2004zw, Abdallah:2006yy}, it was shown that radiative fermion pair events ${\rm e^+e^-\to Z}\gamma$ with ${\rm Z \to f\bar f}$ (with f being a charged lepton or a quark, and where the photon escapes undetected along one of the beams), over-constrained by the precise measurement of the fermion angles, the knowledge of the Z boson mass 
and the total energy-momentum conservation, offered an in-situ determination of the centre-of-mass energy with a statistical precision of a few tens of MeV. 

With the 12\,${\rm ab}^{-1}$ run of FCC-ee at the W pair production threshold ($\sqrt{s} \simeq 160$\,GeV), almost a billion fermion pairs are selected in the detector acceptance (with a polar angle in excess of 10 degrees for the charged leptons and 30 degrees for the hadronic jets) and with a mass within $\pm 10$\,GeV from the Z mass. The method yields a statistical precision of 300\,keV on the centre-of-mass energy, and can be calibrated with a similar accuracy from resonant depolarization in view of its use at higher energies. The knowledge of the absolute angular scale and, to a lesser extent, of the angular resolution and the beam energy spread, were shown to be the dominant systematic uncertainties at LEP\,2. 
The large beam crossing angle and luminosity at FCC-ee allow the absolute angular scale and the beam energy spread to be monitored continuously with a virtually infinite precision, as explained in Section~\ref{sec:EnergySpreadMuons}. The muon angular resolution of the FCC-ee detectors, of the order of 0.1\,mrad, is good enough to have an impact smaller than 1\,MeV on the centre-of-mass energy determination, and can anyway be measured with di-muon events with a more-than-adequate precision over the whole acceptance (Section~\ref{sec:EnergySpreadMuons}). The jet angular resolution is 150 times larger than the muon angular resolution ($\sim 15$\,mrad), but its impact can be predicted  in-situ from its macroscopic effects on the ${\rm q \bar q}$ mass resolution in the very large samples of ${\rm q \bar q} (\gamma)$ events, either with quasi back-to-back di-jet events, or with a photon in the detector acceptance. The same is true for fragmentation and hadronization effects.

At $\sqrt{s} = 240$\,GeV, the same method still enjoys 70 million Z$\gamma$ events (with a twice smaller luminosity, a twice smaller cross section, and a twice smaller acceptance due to the larger longitudinal boost than at the WW threshold), which suffice for a determination of the average centre-of-mass energy with a precision of 1.7\,MeV. This finite accuracy, which comprises the systematic uncertainty arising from  the accuracy of the calibration at the WW threshold, has a negligible impact on the Higgs boson mass experimental precision of 10\,MeV. 

The scan of the ${\rm t \bar t}$ threshold proceeds with a centre-of-mass energy around 350\,GeV and an integrated luminosity of $0.2\,{\rm ab}^{-1}$. The even larger longitudinal boost causes the ${\rm q \bar q}\gamma$ selection efficiency to vanish, and the $\ell^+\ell^-\gamma$ selection efficiency to be substantially reduced, which limits -- with only 200\,000 events -- the statistical power of the method to a precision of 30\,MeV on the centre-of-mass energy. The direct reconstruction of the two million ${\rm e^+e^- \to W^+W^-}$ events in the fully hadronic and semi-leptonic final states, constrained by the precise knowledge of the W mass 
and the total energy-momentum conservation, provides an instrumental alternative. At $\sqrt{s} = 350$\,GeV, these events allow the mean centre-of-mass energy to be determined with a statistical precision of 5\,MeV~\cite{PaoloMarina:2019}, with a negligible impact on the top-quark mass experimental precision of 17\,MeV. The systematic uncertainties (quark fragmentation and hadronization, colour reconnection, etc.) can be controlled and calibrated to better than 1\,MeV with the 100 million ${\rm e^+e^- \to W^+W^-}$ events recorded at $\sqrt{s} = 240$\,GeV and the measurement of the W mass at $\sqrt{s} = 160$\,GeV. The same method yields a precision of 2\,MeV on the mean centre-of-mass energy at $\sqrt{s} = 365$\,GeV. A marginal improvement is expected from the use of ${\rm e^+e^- \to ZZ}$ events at 240, 350, and 365\,GeV. 

In summary, the centre-of-mass energy can be measured with an adequate precision at energies above the WW threshold, for the measurement of the Higgs boson and top quark masses, once the Z and W masses are measured at FCC-ee at the Z pole and the WW threshold as displayed in Table~\ref{tab:EWPO}. The beam energy spread can be measured continuously at all energies as described in Section~\ref{sec:EnergySpreadMuons}. 

\section{Beam Polarization prospects at FCC-ee}
\label{sec:polar}
\subsection{Expectations based on LEP experience}
\begin{figure}[htbp]
	\begin{center}
	\centering
 \begin{minipage}[t]{0.535\textwidth}
 \vspace*{0.45cm}
		\includegraphics*[width=1.0\textwidth]{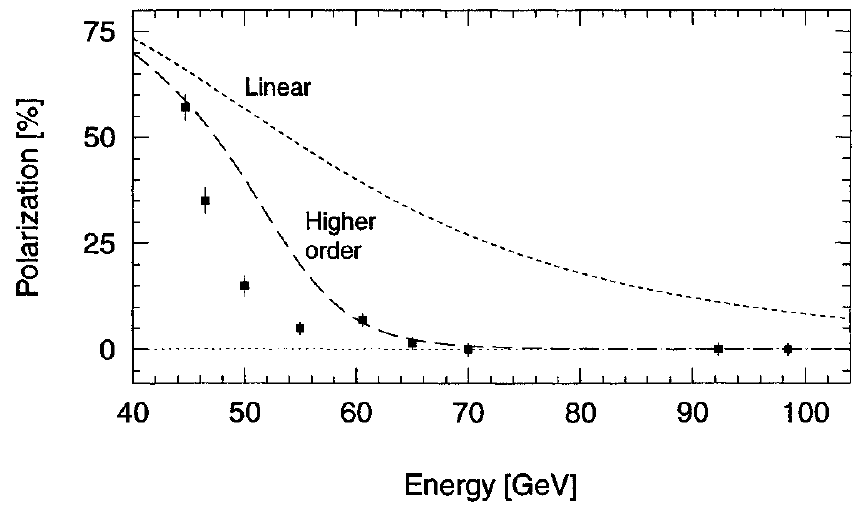}
\end{minipage}
\begin{minipage}[t]{0.45\textwidth}
\vspace*{-0.1cm}
		\includegraphics*[width=1.\textwidth]{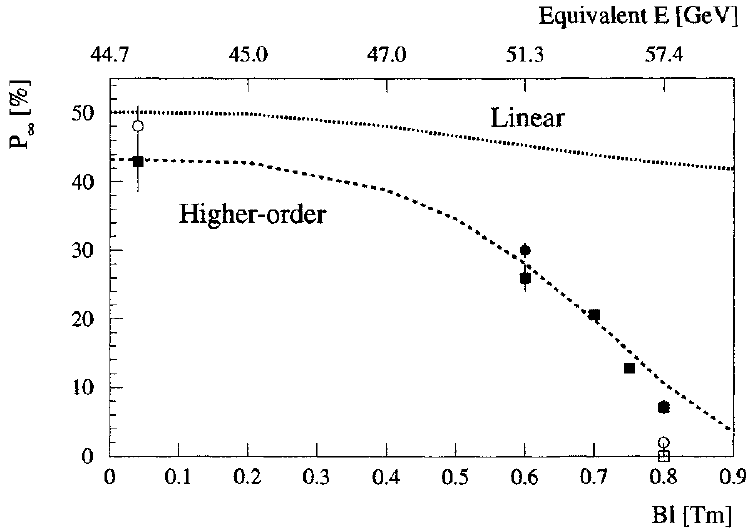}
\end{minipage} 
\caption{Left: Measured transverse polarization of the LEP electron beam as function of the beam energy. Only the points at 45\,GeV and 60.5\,GeV were fully corrected with harmonic spin matching. Right: Measured transverse polarization at 45\,GeV as a function of the damping wiggler field, or equivalently, of the beam-energy spread. }
\label{fig:polar-at-LEP}
\end{center}
\end{figure}

Transverse beam polarization builds up naturally in a storage ring by the Sokolov-Ternov effect. Spin dynamics experience at LEP is summarized in Ref.~\cite{Assmann:535999}.  A transverse polarization in excess of 5--10\% was  sufficient in LEP for beam energy calibration purposes. This was obtained at the Z pole energies with no special tuning, apart from turning off the experiments solenoids and the ${\rm e^+ e^-}$ collisions.  This level of polarization could be significantly increased by vertical dispersion correction; deterministic spin matching based on correction of the  harmonics of the measured vertical orbit; and empirical harmonic spin matching by optimization of the observed polarization by tuning four orthogonal harmonics by steps.  With these methods, beam polarization  was obtained for beam energies up to 61\,GeV at LEP. It is generally accepted that this upper limit is determined by the energy spread, which increases with the beam energy as $\sigma_{E_{\rm b}} \propto {E_{\rm b}}^2 / \rho$ where $\rho$ is the bending radius of the accelerator. When $\sigma_{E_{\rm b}}$ becomes commensurate with the spacing (440\,MeV), the depolarization of the beam is significant. This depolarizing effect was verified experimentally by exciting the ``damping wigglers'' situated in a nominally dispersion free region, to generate an increase of the beam-energy spread, as can be seen in Fig.~\ref{fig:polar-at-LEP}. 

This observation leads to the following rule of thumb: sufficient polarization can be obtained as long as the beam-energy spread is less than around 55\,MeV. Given the scaling of the beam-energy spread with the ring radius, it is expected that beam polarization sufficient for beam energy calibration should be readily available up to and above the WW threshold (i.e., 81.3\,GeV per beam) at FCC-ee. A new machine with a better control of the orbit should be able to increase this limit. These expectations were verified with a spin simulation code, as is shown in the next section. 

\subsection{Beam polarization at FCC-ee} 
Beam polarization at FCC-ee may be obtained by the Sokolov-Ternov effect~\cite{Sokolov:1963zn}. In an ideally planar ring
\footnote{with \emph{ideally planar}  it is meant  that the design orbit is contained in a plane
and the magnets are  perfectly aligned.  To be more precise we shall also exclude the presence
of dipole fields pointing in the direction opposite to the guiding field.}
the asymptotic polarization is
$P_{\mathrm ST}=92.4\% $ and the polarization build-up rate is given by
\begin{equation}
\tau_{p} ^{-1}= \frac{5 \sqrt{3}}{8}\frac{r_e\gamma^5\hbar}{m_0 C}\oint \frac{ds}{|\rho|^3} 
\end{equation}

The expectation value $\vec S$ of the
spin operator obeys the Thomas-Bargmann-Michel-Telegdi
(Thomas-BMT) equation \cite{Thomas:1927yu}\cite{Bargmann:1959gz}

\begin{equation}
\frac{d\vec S}{dt} = \vec \Omega  \times \vec S  \label{eq:thomas}
\end{equation}
$\vec \Omega$ depends on machine azimuth, $s$, and the phase-space position, $\vec u$.
 In the laboratory frame and MKS units it is given by
\begin{equation}
\vec \Omega(\vec{u};s) = -\frac{e}{m_0}\Bigl[\Bigl(a+\frac{1}{\gamma}\Bigr)\vec{B}
-\frac{a\gamma}{\gamma +1}\vec{\beta}\cdot \vec{B}\vec{\beta}
-\Bigl(a+\frac{1}{\gamma +1}\Bigr)\vec{\beta} \times \frac{\vec{E}}{c} \Bigr]
\label{eq:Omega_Thomas}
\end{equation}
with $\vec\beta\equiv \vec v/c$ and `
$a=(g-2)/2$=0.0011597 (for $e^\pm$).

In an ideally planar machine the \emph{periodic}
solution, $\hat n_0(s)$, to equation~(\ref{eq:thomas}) on the closed orbit is vertical and, neglecting the electric field,
the number of spin precessions around $\hat n_0$ per turn, the naive ``spin tune'', in the rotating frame is $a\gamma$.
Photon emission results in a randomization of the particle spin directions (\emph{spin diffusion}).
Polarization will be therefore the result of the  competing process, the  Sokolov-Ternov effect and the spin diffusion
caused by stochastic photon emission. See \cite{Barber:1999xm} for a succinct summary of the definitions, theory and phenomenology.

Using a semiclassical approach, Derbenev and Kondratenko\cite{Derbenev:1973ia} found 
that the polarization is oriented along 
$\hat n_0$ and
its asymptotic value is 

\begin{equation}
P_\mathrm{DK}= P_\mathrm{ST} \frac{\oint ds < \frac{1}{|\rho|^3}\hat b \cdot
(\hat n - \frac{\partial \hat n}{\partial \delta})>}{
\oint ds < \frac{1}{|\rho|^3}\Bigl[1-\frac{2}{9}(\hat n\cdot \hat v)^2+
\frac{11}{18}
(\frac{\partial\hat n}{\partial \delta})^2\Bigr]>} \label{eq:derbenev}
\end{equation}
with 
$\hat b\equiv \hat{v} \times \dot{\hat{v}}/
|\dot{\hat{v}}| $ and $\delta\equiv \delta E/E$.

The vector $\hat n$ is the \emph{invariant spin field} \cite{Hoffstaetter:2000cf,Hoffstaetter:2006zu},
i.e. a solution of equation~(\ref{eq:thomas}) satisfying the condition $\hat n(\vec{u};s) = \hat n(\vec{u};s+C)$,
$C$ being the machine length. 
The $<>$ brackets indicate averages over the phase space.
The term with $ (\partial \hat{n}/ \partial \delta)^2$
quantifies the depolarizing effects resulting from the trajectory
perturbations due to photon emission.

The corresponding polarization rate is
\begin{equation}
\tau_{p} ^{-1}=P_\mathrm{ST}\frac{r_e\gamma^5\hbar}{m_0 C}\oint <\frac{1}{|\rho|^3} \Big[1-\frac{2}{9}(\hat n \cdot \hat v)^2+
\frac{11}{18}\Bigl(\frac{\partial \hat n}{ \partial \delta}\Bigr)^2\Big]>  
\end{equation}
In a perfectly planar machine $\partial \hat{n}/ \partial \delta$=0 and $P_\mathrm{DK}$=$P_\mathrm{ST}$.  In the presence of quadrupole 
vertical misalignments (and/or spin rotators) $\partial \hat{n}/ \partial \delta\neq$0 and it is particularly large when
spin and orbital motions are in resonance
\begin{equation}
{\nu}_0 \approx  k_0 + k_1 Q_1 + k_2 Q_2 + k_3 Q_3
\end{equation}
for integers $k$ and tunes $Q$ and where $\nu_0$ is the actual spin tune on the closed orbit.
See Section 12 for a detailed discussion around the concept of spin tune.
For FCC-ee with $\rho\simeq$ 10424 m, fixed by the maximum attainable dipole field for the hadron collider,
the polarization times at 45 and 80\,GeV are 256 and 14 hours respectively.

Here it is assumed that beam polarization of about 10\% is sufficient for an accurate 
depolarization measurement. The time, $\tau_{10\%}$, needed for the beam to reach this polarization level  is 
given by
\begin{equation}
\tau_{10\%}=-\tau_p \times \ln(1-10/P_{\infty}) 
\end{equation}
with $P_{\infty}$ being the asymptotic polarization value in \%.
At 80\,GeV it is $\tau_{10\%}$=1.6 hours, but $\tau_{10\%}$= 29 hours at 45\,GeV. 

At low energy the polarization time may be reduced by introducing properly designed wiggler magnets
i.e. a sequence of vertical dipole fields, $\vec B_w$,  with alternating signs.

The  maximum synchrotron radiation power at FCC-ee is set to 50 MW per beam and
the beam current at the various energies has been scaled accordingly.  
This limits the integrated wiggler strength.
Moreover wigglers increase the beam energy spread whose effect on polarization 
in the presence of machine imperfections must be investigated.

At 80\,GeV wigglers are not needed. However the energy dependence of the spin motion
makes the attainable polarization level more sensitive to machine errors.

\subsection{Polarization wigglers}
\label{sec:polar-wigglers}
The use of wiggler magnets for reducing the polarization time was first proposed
for LEP~\cite{Blondel:369485}.

Since $\hat n_0 \equiv \hat y $ in a perfectly planar ring and the field is piecewise constant in the
wiggler,
the asymptotic polarization and the polarization build-up rate write
\begin{equation}
P_{\infty}=
\frac{8 F \gamma^5}{5\sqrt{3}} \hskip 1mm \tau_p \Bigl[
\int_{dip} ds \hskip 1 mm \frac{\hat B_d \cdot \hat n_0}{|\rho_d|^3}+
\frac{L^+}{|\rho^+|^3}\bigl(1-\frac{1}{N^2}\bigr)
\Bigr]
\end{equation}
\begin{equation}
\tau_{p} ^{-1}=F\gamma^5\Bigl[
{\int_{dip}  \frac{ds}{|\rho_d|^3} +  \int_{wig} \frac{ds}{|\rho_w|^3}} \Bigr]=
F\gamma^5\Bigl[
\int_{dip}  \frac{ds}{|\rho_d|^3} + \frac{L^+}{|\rho^+|^3}\bigl(1+\frac{1}{N^2}\bigr)
\Bigr]
\end{equation}
where $N\equiv L^{-}/L^{+}=B^{+}/B^{-}$.

The particle energy lost per turn and the energy spread are 
\begin{equation}
U_{loss} = \frac{C_{\gamma} E^4}{2\pi} \oint  \frac{ds} {\rho^2}  \hskip 20mm 
(\sigma_E/E)^2= \frac{C_q}{J_{\epsilon}} \gamma^2   \oint  \frac{ds}{|\rho|^3}/  \oint \frac{ds}{\rho^2} 
\end{equation}

The presence of wigglers increases $U_{loss}$ and $ \sigma_E/E $.
The following equation
\begin{equation}
(\sigma_E/E)^2=\frac{C_q C_{\gamma}   E^4}{2\pi J_{\epsilon} F \gamma^3} \frac{1}{\tau_p U_{loss} }
\end{equation}
indicates that a smaller $\tau_p$ comes at the price of a higher $U_{loss}$ and/or $\sigma_E$.

\begin{figure}[htb]
\centering\includegraphics*[width=102mm,height=72mm]{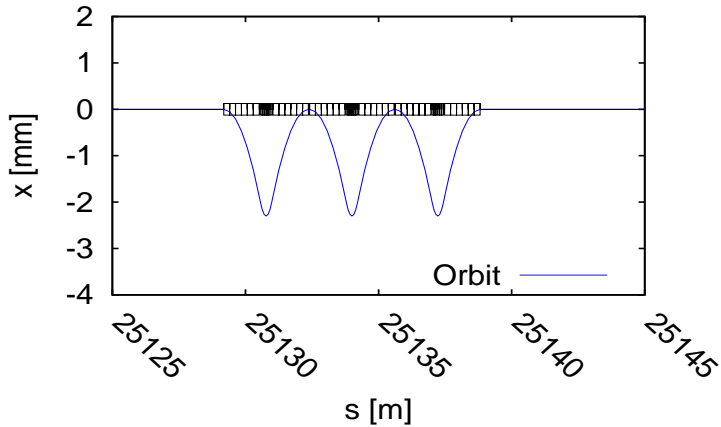}
\vspace*{-10mm}
\caption{\label{eliana-f1} Beam trajectory through the wiggler
for $B^+$=2T; the amplitude scales proportionally to the wiggler excitation.}
\end{figure}

As shown in \cite{Gianfelice-Wendt:2016jgk} there is no ``magic'' set of wiggler parameters, all cases with $N>$2 are
almost equivalent. To limit the impact on horizontal emittance
the wiggler period has been reduced wrt to the original LEP design.
The corresponding beam trajectory through the wiggler is shown in
Fig.~\ref{eliana-f1} for $B^+$=2 T and $N$=6. With 8 of such wigglers and $B^+$=0.67 T 
$\tau_{10\%}\simeq$1.8 h and $\sigma_E$= 60\,MeV.
The critical energy is 902\,keV.

\subsection{Effect of machine misalignments}
\label{sec:misalignments}


In the presence of misalignments the polarization may be evaluated 
analytically if orbital and spin motion are linearised. Evaluation of equation~(\ref{eq:derbenev}) for the general case is extremely challenging. We resorted to the tracking code SITROS\cite{Kewisch:1985nf} whose results for HERAe were in fair agreement with observations. SITROS tracks particles in the presence of quantum excitation and damping, 
photons being emitted at user selected dipoles.
Once the equilibrium distribution is reached, the particle spins initially aligned to the periodic solution, $\hat n_0$,  of~(\ref{eq:thomas}) are tracked
and from the fit of $P(t)$ the depolarization rate and 
asymptotic polarization are evaluated. The package also contains the code SITF for computing polarization in the approximation of
linearised  spin motion.

Due to the strong IRs quadrupoles at large  $\beta_{x,y}$ locations, and to the large number of magnets, the FCCee closed orbit
is very sensitive to magnet misalignments. 
The SY sextupoles used for correcting the IR chromaticity introduce large tune shifts and coupling. 
In addition the machine may be anti-damped due to the IR quadrupole offsets.
To achieve the required collider performance, orbit,
spurious vertical  dispersion and betatron coupling
must be extremely well corrected and maintained. In this section we assume that those issues have been understood and solved and
we look at the achievable polarization level for a well corrected machine.
For the simulation reasonable misalignments were introduced in small steps, at each step the orbit was corrected with a large number of correctors. 

We have introduced a beam position monitor (BPM) close to each quadrupole and to the SY sextupoles, as well as a horizontal 
and a vertical corrector close to each horizontally and vertically focusing quadrupole respectively. The IR quadrupoles are equipped with both kinds of correctors.
For correcting betatron coupling and spurious vertical dispersion one skew quadrupole has been
introduced into each 10th FODO cell. The correction formalism is described in\cite{Alexahin:2006nw}.  

When necessary, a correction of the periodic solution to~(\ref{eq:thomas}) has been applied as done
for instance at 
HERAe\cite{Barber:1993ui}
and LEP\cite{Arnaudon:1994ej}. 

In the few years of FCC design study, the optics has evolved and
it is impossible to present all polarization results in a consistent way. In summary we can say 
that polarization studies have shown that the goal of 10\% polarization is within reach at 45 as well as at
80\,GeV even in the presence of BPM position and calibration errors. However for the latest optics (60$^0$/60$^0$ FODO cells
with $\beta_y^*$=0.8 mm and 1 mm at 45\,GeV and 80\,GeV respectively)  we had to reduce the
quadrupole misalignments in order to get a stable machine. The errors considered are summarized in 
Table~\ref{eliana-tab-errors} 

The available spin simulation codes are not integrated in the code with which the orbit correction algorithms have been studied. As a result, the level of initial misalignments that are assumed here are smaller than the $100\,\mu$m normally assumed.  
They were chosen so that the corrected orbits have similar values for vertical dispersion  and vertical emittance than those used for the luminosity optimization~\cite{Charles}. The future action required is to integrate the codes, as explained in Section~\ref{sec:sim-tools}.

\begin{table}[htb]
\centering
  \caption{Assumed errors.}
\vspace*{2mm}
\begin{tabular}{ c c c c c }  
\hline
 & IR Quads & IR BPMs & other Quads & other BPMs \\
$\delta x$ ($\mu$m) & 10 & 10 & 30 & 30  \\
$\delta y$ ($\mu$m) & 10 & 10 & 30 & 30 \\
$\delta \theta$ ($\mu$rad) & 10 & 10 & 30 & 30  \\
calibration & - & 1\% & - & 1\% \\
 \hline
\end{tabular}
  \label{eliana-tab-errors}
\end{table}

\begin{figure}[htb]
\vspace*{-10mm}
\centering\includegraphics*[width=102mm,height=72mm]{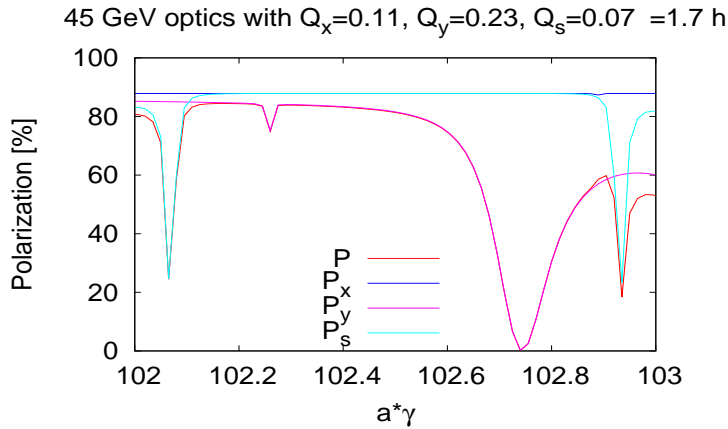}
\vspace*{-10mm}
\caption{\label{eliana-f3} Linear beam polarization vs.~a$\gamma$ for the $\beta^*_y$= 0.8 mm and 90$^0$/90$^0$ FODO optics. The misalignments are shown in Table~\ref{eliana-tab-errors}. }
\end{figure}

\begin{figure}[htb]
\vspace*{-10mm}
\centering\includegraphics*[width=102mm,height=72mm]
{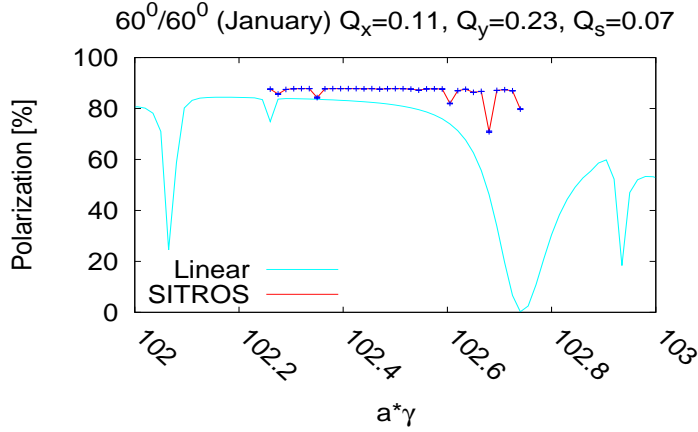}
\vspace*{-10mm}
\caption{\label{eliana-f4} Polarization for the perturbed $Z$ optics.}
\end{figure}

Fig.~\ref{eliana-f3} shows the expected polarization for the last 60$^0$/60$^0$ optics at 45\,GeV
computed in linear spin motion approximation, while Fig.~\ref{eliana-f4}
shows the result of the tracking.

The blue, magenta and cyan lines refer to the polarization for purely radial, vertical and longitudinal motion respectively.
The lattice includes 8 wigglers with $B^+$=0.66 T ($\tau_{10\%}\simeq$1.7 h).
The beam parameters after orbit correction are summarized in Table~\ref{eliana-tab-corrected45}.

\begin{table}[htb]
\centering
  \caption{Beam parameters after orbit correction at 45\,GeV.}
\vspace*{2mm}
\begin{tabular}{ c c c c c c c }  
\hline
 & $x_{rms}$ & $y_{rms}$ & $D^y_{rms}$ & $\epsilon_x$ & $\epsilon_y$ & $|C^-|$ \\
 & ($\mu$m)  & ($\mu$m)  & (mm) &  (nm)  &  (pm)  & \\
no skews & 26 & 11 & 2 & 0.222 & 0.5 & 0.0014 \\
 \hline
\end{tabular}
  \label{eliana-tab-corrected45}
\end{table}

The resulting rms value of $|\delta\hat{n_0}|$ is 0.05 mrad at $a\gamma$=102.5.
No attempt has been made to improve betatron coupling and spurious vertical dispersion by the skew quadrupoles as they are already
small. Despite that, polarization is limited by the vertical motion. We shall
recall the \emph{spin-orbit coupling integrals} relating in 
linear approximation spin diffusion to orbital motion~\cite{ChaoYokoya:1982, Barber:1999xm}
\begin{equation}
\frac{\partial\hat n}{\partial \delta}(\vec u;s)=\vec d (s)=\frac{1}{2} \Im{\Bigl\{(\hat m_0+i\hat l_0)^* \sum_{k=\pm x,\pm y,\pm s}  \Delta_k \Bigr\}}
\end{equation}
with
\begin{equation}
\Delta_{\pm x,\pm y}=(1+a\gamma)\frac{e^{\mp i\mu_{x,y}} } { e^{2i\pi(\nu\pm Q_{x,y)}}-1}
\frac{ [-D\pm i(\alpha D+\beta D')]_{x,y}}{\sqrt{\beta_{x,y}}} \hspace*{2mm} J_{x,y}
\end{equation}
\begin{equation}
\Delta_{\pm s}=(1+a\gamma)\frac{e^{\pm i\mu_{s}} } { e^{2i\pi(\nu\pm Q_{s)}}-1}\hspace*{2mm} J_{s}
\end{equation}

\begin{equation}
J_{\pm x,\pm y}=\int_s^{s+L} \hspace*{-1mm} ds' (\hat m_0+i\hat l_0)\cdot \left\{\begin{array}{c} \hat y \sqrt{\beta_x} \\ \hat x \sqrt{\beta_y} \end{array} \right\}
\hspace*{1mm} K e^{\pm i \mu_{x,y}}
\end{equation}

\begin{equation}
J_{s}=\int_s^{s+L} \hspace*{-1mm} ds' (\hat m_0+i\hat l_0)\cdot (\hat y D_x + \hat x D_y) 
\hspace*{1mm} K
\end{equation}
Plotting the factor $f_y\equiv[-D_y\pm i(\alpha_y D_y+\beta_y D'_y)]/\sqrt{\beta_y}$ 
vs. position for the perturbed optics, we notice that in some short regions
$f_y$ is much larger than in the rest of the ring. 
It is worth noting that 
the tracking shows larger polarization than the linear calculations. It could be
some ``offending dipoles'' have been missed in choosing the dipoles where
emission takes place, although the dipoles at largest $f_y$ have been included.

\begin{figure}[htb]
\centering\includegraphics*[width=102mm,height=72mm]{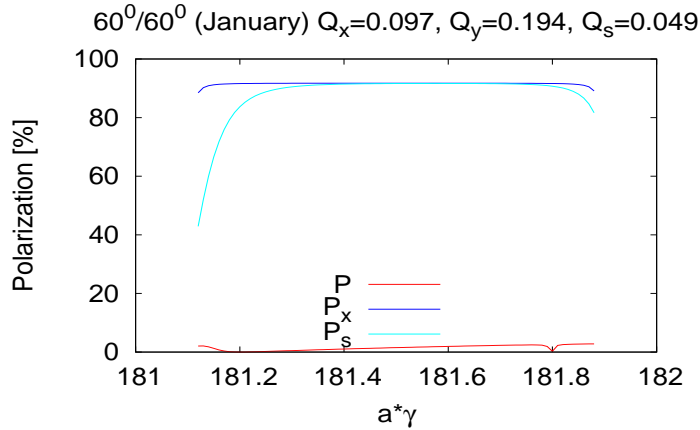}
\caption{\label{eliana-f5} Linear beam polarization vs.~a$\gamma$ for the $WW$ optics. The assumed misalignments are the same as for the $Z$ optics. 
Polarization is limited to few percent by vertical betatron motion.}
\end{figure}

\begin{figure}[htb]
\centering\includegraphics*[width=102mm,height=72mm]{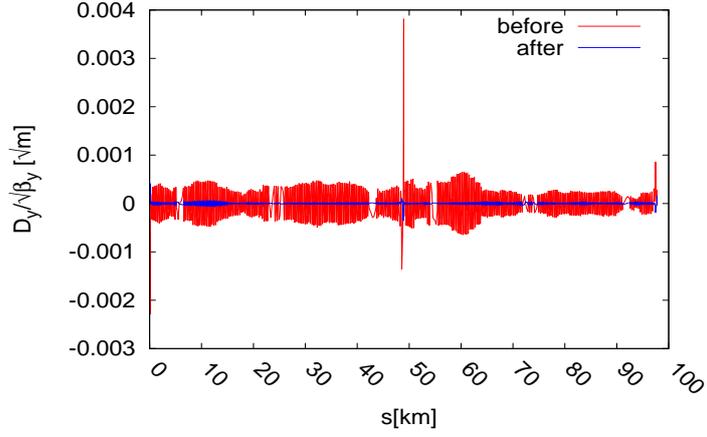}
\caption{\label{eliana-f6} The real part of $f_y$ before (red) and after (blue)
spurious vertical dispersion for the $WW$ optics.}
\end{figure}

At 80\,GeV the linear polarization is limited to few percent 
(see Fig.~\ref{eliana-f5}). Again polarization is limited by
the vertical betatron motion and large values of $f_y$ are visible 
at a few ring locations (see Fig.~\ref{eliana-f6}). 

The beam parameters are shown in Table~\ref{eliana-tab-corrected80}.

\begin{table}[htb]
\centering
  \caption{Beam parameters after orbit correction at 80\,GeV.}
\vspace*{2mm}
\begin{tabular}{ c c c c c c c }  
\hline
 & $x_{rms}$ & $y_{rms}$ & $D^y_{rms}$ & $\epsilon_x$ & $\epsilon_y$ & $|C^-|$ \\
 & ($\mu$m)  & ($\mu$m)  & (mm) &  (nm)  &  (pm)  & \\
no skews & 144 & 11 & 2 & 0.792 & 0.1 & $<$ 0.001 \\
 \hline
\end{tabular}
  \label{eliana-tab-corrected80}
\end{table}

\begin{figure}[htb]
\centering\includegraphics*[width=102mm,height=72mm]{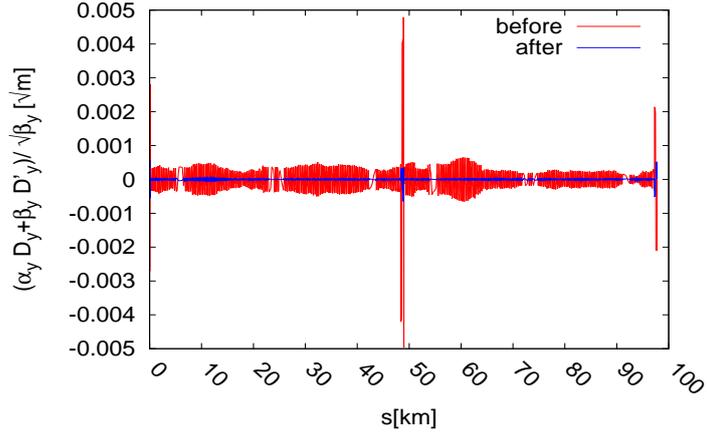}
\caption{\label{eliana-f7} The imaginary part of $f_y$ before (red) and after (blue)
spurious vertical dispersion for the $WW$ optics.}
\end{figure}

\begin{figure}[htb]
\centering\includegraphics*[width=102mm,height=72mm]{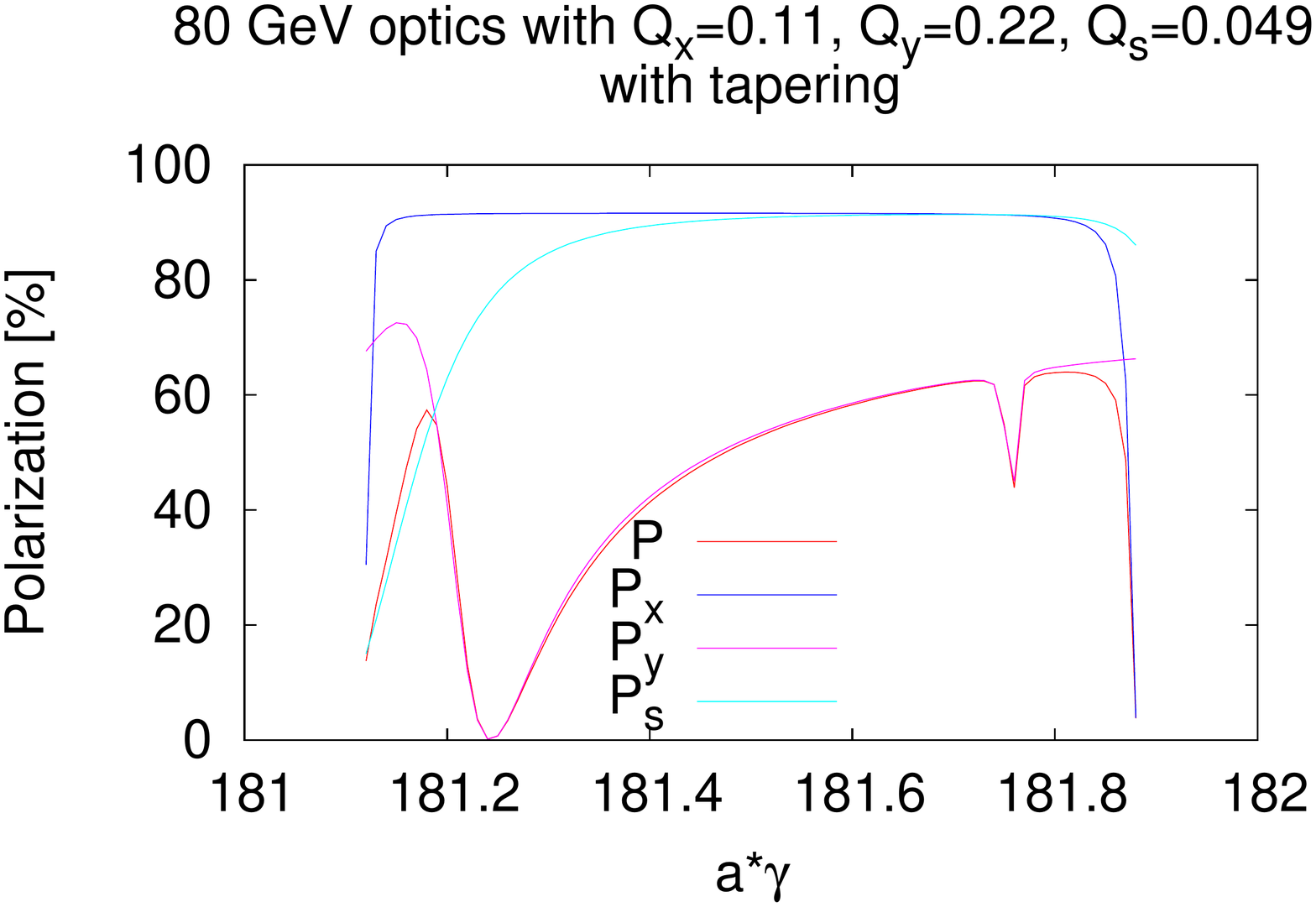}
\vspace*{-5mm}
\caption{\label{eliana-f8} Linear beam polarization vs.~a$\gamma$ for the $WW$ optics after correcting spurious vertical dispersion. }
\end{figure}

It is worth noting that at 80\,GeV the sawtooth orbit
and sextupole feed-down effect on tunes are already evident. Tapering of the dipole magnets and compensation of the tune changes using the FODO circuits did not improve the polarization level.
Attempts to correct the $f_y$ spikes with the skew quadrupoles were  unsuccessful. The vertical correctors were therefore used to correct the vertical dispersion. The real and imaginary parts of the factor $f_y$  before and after correction and the linear polarization are shown in Fig.~\ref{eliana-f6}, ~\ref{eliana-f7} and ~\ref{eliana-f8} respectively.

\begin{figure}[htb]
\centering\includegraphics*[width=102mm,height=72mm]{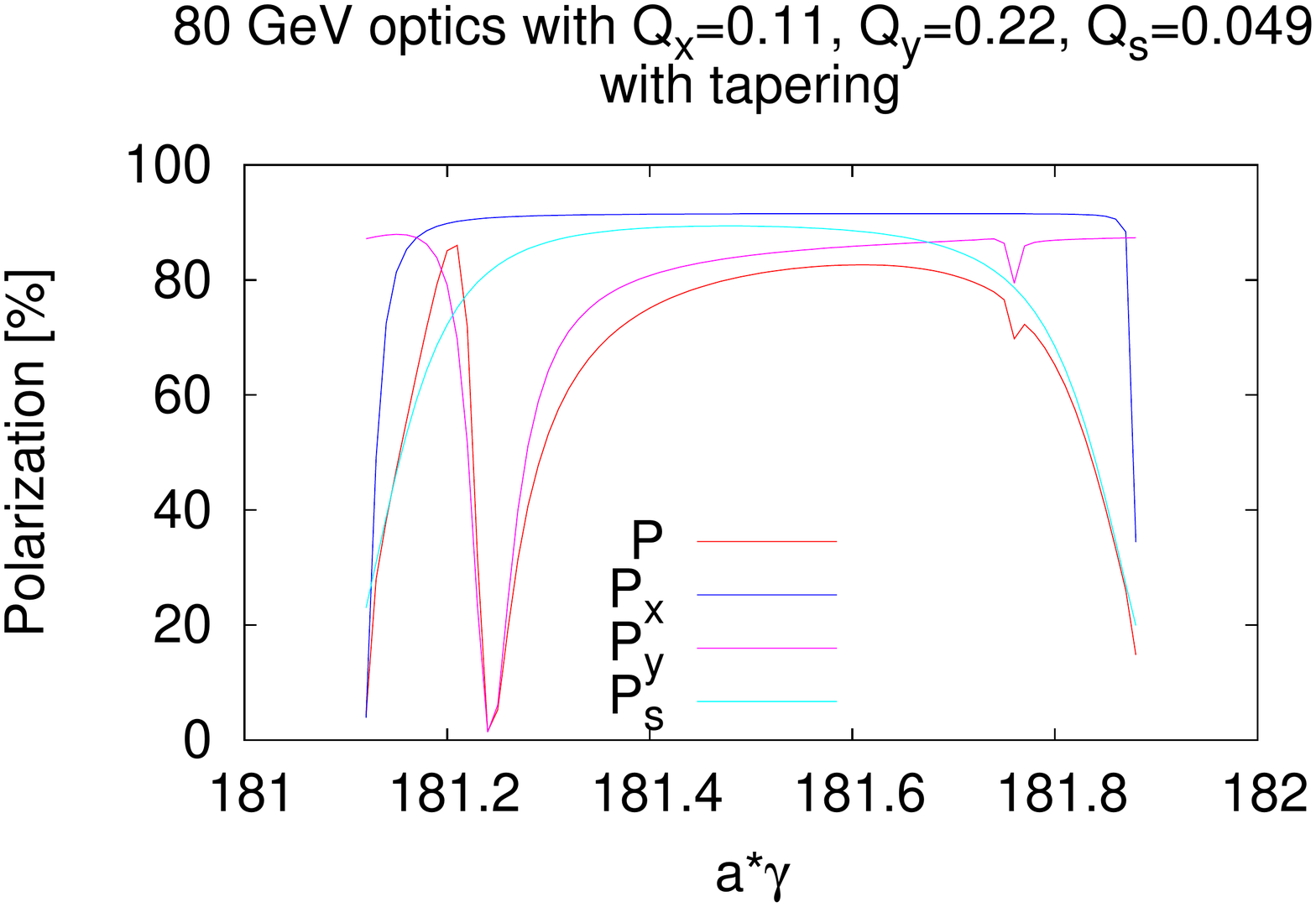}
\vspace*{-5mm}
\caption{\label{eliana-f9} Linear beam polarization vs.~a$\gamma$ for the $WW$ optics after correcting spurious vertical dispersion and $\delta\hat n_0$.}
\end{figure}

\begin{figure}[htb]
\centering\includegraphics*[width=102mm,height=72mm]{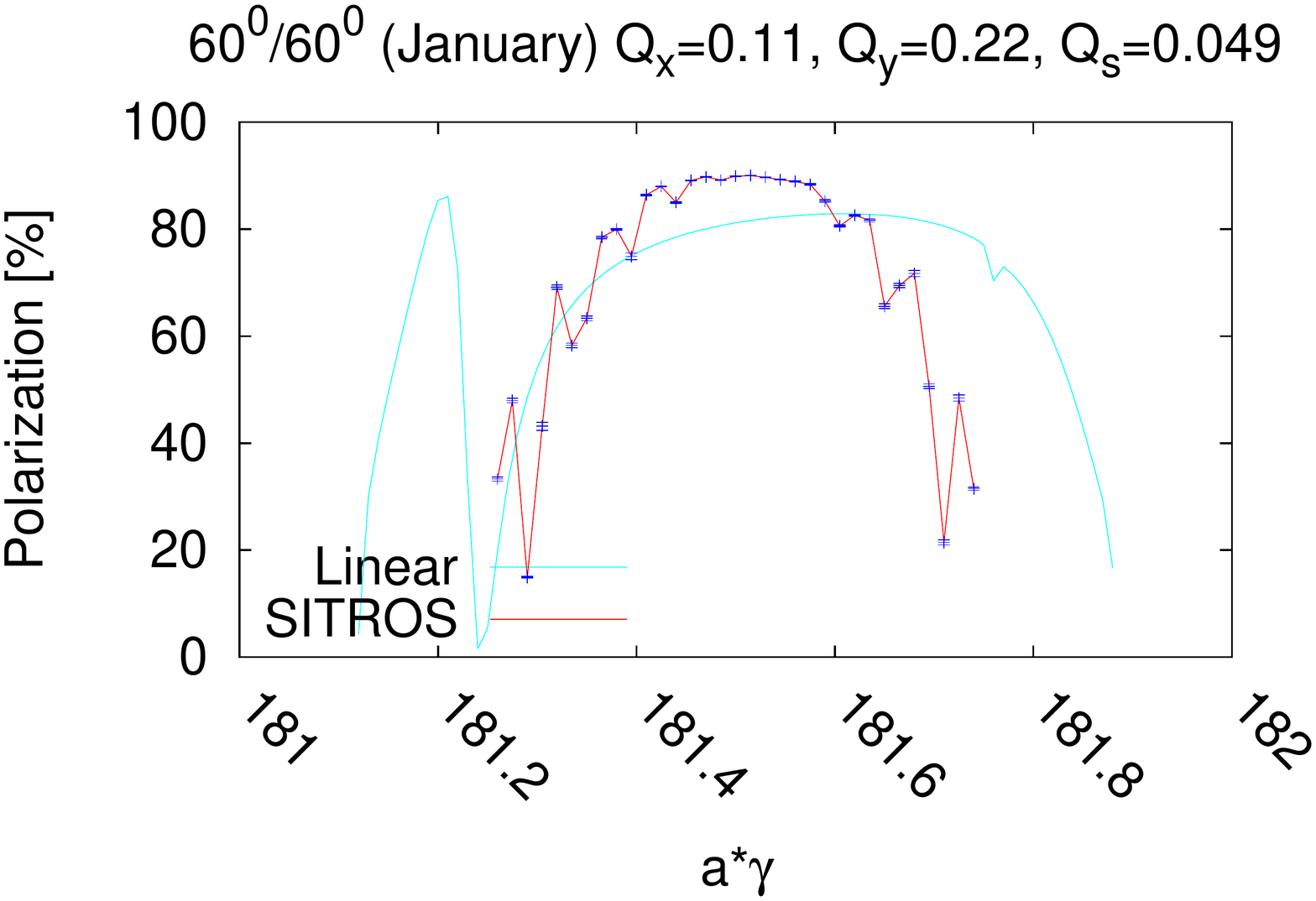}
\caption{\label{eliana-f10} Polarization after correction of spurious vertical dispersion and optimization of the harmonic bumps.}
\end{figure}

An additional correction of $\delta\hat n_0$ resulted in further improvement of the linear polarization visible in Fig.~\ref{eliana-f9}. The SITROS polarization for the final correction shown in Fig.~\ref{eliana-f10}.

\subsection{Effect of detector solenoids}
\label{sec:solenoids}

A detector's  solenoid field $B_s$ rotates the nominally vertical 
polarization axis $\hat n_0$ around the
longitudinal coordinate and shifts the spin tune, breaking the proportionality between spin tune $\nu_0$ away and $a \gamma$.

\begin{figure}[htb]
\centering\includegraphics[width=.9\linewidth]{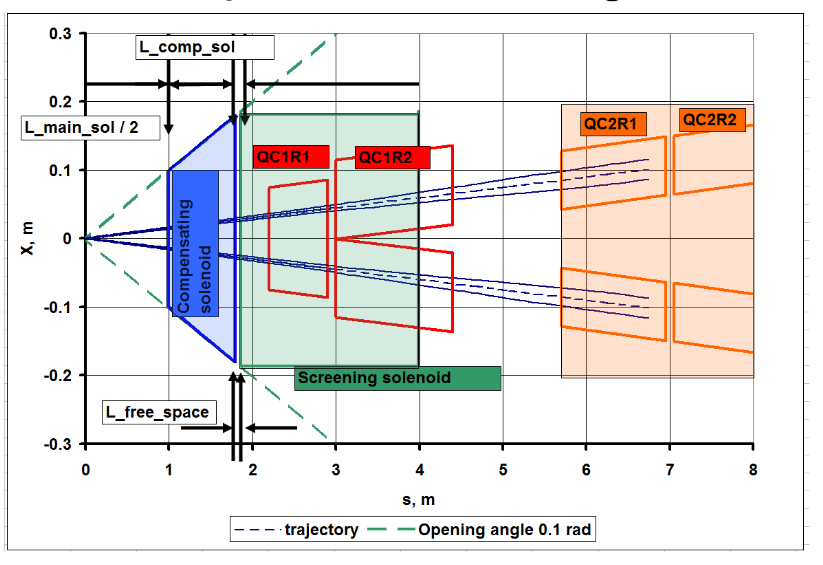}
\caption{\label{eliana-f2} Schematic layout of the right side of one IP, showing the solenoids layout.}
\end{figure}

The FCC-ee solenoid layout is shown in Fig.~\ref{eliana-f2} for the
right hand side of one IP. The detector solenoids are compensated by screening solenoids. The crossing angle is 30 mrad
and the detector field $B_s$ is 2~T. The layout is symmetric with respect to the IP.

The effect of the FCC-ee detector solenoids has been evaluated
with SLIM\cite{Chao:1978us} which can handle solenoids tilted around the vertical axis. Thanks to the screening solenoids,
$|\delta\hat n_0|_{rms}$ is only 10\,$\mu$rad at 45\,GeV and 1\,$\mu$rad at 80\,GeV and the effect on polarization is negligible.

The shift of $\nu_0$ is 1.6$\times$10$^{-6}$, corresponding to 0.71\,keV. It is worth noting that the tune shift
may be measured by comparing the spin tune with solenoidal fields
 on and off.


\section{Running Scheme for precise measurements of the Z and W masses}
\label{sec:running-scheme}

One of the main lessons of the LEP energy calibration program presented in Section~\ref{sec:LEP} was that a regular energy measurement is mandatory to improve by an order of magnitude or more the accuracy on the beam energy for FCC-ee. The LEP running scheme with isolated energy measurements at ends of coasts and interpolation of the beam energy between measurements was limited strongly by systematic effects.

Energy calibration through resonant depolarization can be performed at regular intervals during physics data taking if a large enough number of transversely polarized bunches are available for energy measurements. A typical interval between measurements could be 5 to 10 minutes since on that time scale the interpolation of the energy using orbit data, magnetic field and other independent monitoring should not pose a problem. Despite the fact that frequent energy measurements would be performed at FCC-ee, a model of the energy evolution over time must be constructed to gain confidence in the interpolation (even if only done on the time scale of a few minutes) and the underlying reasons for energy changes must be understood. The main components of such a model should include:
\begin{itemize}
    \item Changes of the main dipole magnetic field, for example due to ambient temperature;
    \item Changes of the machine circumference due to geological effects and Earth tides as observed at LEP and at LHC;
    \item Changes of orbit corrector settings induced by the evolving machine element alignment due to long term ground motion;
    \item Changes in the distribution, phases and calibrations of the RF system that affect the CM energy at each IP.
\end{itemize}
As a modern machine FCC-ee will be equipped with a real-time orbit feedback system that stabilize the orbit within the resolution and errors of the beam position monitor system. For example the long term stability of the beam orbits at the LHC is at the level of 50\,$\mu$m.

On the Z pole the polarization time is 29 hours for a 10\% polarization level. We will assume that a polarization level of 5\% is required for a good energy calibration. It is clear that the situation would be more comfortable if lower polarization levels could be used. Wigglers may be used to lower the polarization time on the Z pole as described in the previous section. Due to the power limitations of the RF system and very likely also due to local heat deposition, it is not possible to operate wigglers with the full beam. In order to generate sufficient polarization for the likely 100 to 200 non-colliding calibration bunches for energy measurements, wigglers could be switched on for two hours until a sufficient initial polarization were obtained on the calibration bunches. At this point the wiggler field would be ramped down and the main physics bunches would be injected over around 30 minutes. At regular intervals one of the calibration bunches would then be depolarized to determine the energy. With 100 calibration bunches that are depolarized every 10 minutes, each bunch has 17 hours to re-polarize which is sufficient to reach again the 5\% level. In this way it is possible to perform energy measurements for a period that is only limited by the lifetime of the calibration bunches and the number of beam particles that are burned away by the polarimeter (see sections below). To avoid limitations due to Toucheck losses described below, one should consider to top-up the a depolarized bunch immediately. With such a regular top-up after depolarization, there could be no limit to the length of a physics coast.

The lifetime of the bunches in FCC-ee is limited by the Touschek effect due to the extremely small transverse dimensions of the beams. A detailed evaluation of the lifetime is presented in Section~\ref{sec:ecal-touschek} below. For bunches with a populations of $10^{11}$ particles the lifetime is around 1~hour. Since the injected beam is not polarized, regular over-injection of the calibration bunches is not possible even if the injection process itself is clean enough not to depolarize the already existent transverse polarization.
The limitations due to the Touschek lifetime force the intensity of the calibration bunches to be relatively low, and it may limit the maximum length of machine fillings before the beam must be dumped and the process is started again.

\subsection{Touschek effect for pilot bunches} 
\label{sec:ecal-touschek}

The non-colliding pilot bunches with \num{4e10} particles will be used for continuous energy measurement via spin depolarization. Since these bunches are non colliding, energy spread and therefore bunch length are approximately a factor \num{4} smaller than for colliding bunches. Thus, intrabeam scattering and Touschek scattering rates are increased. Lifetime calculations for Touschek lifetime are based on the semi-analytical formulae in flat beam approximation given in \cite{LeDuff:1993jm}:
\begin{equation}\label{eq:touschekLeDuff}
\frac1{\tau_t} = \left\langle \frac{Nr_0c}{8\pi\gamma^2\sigma_x\sigma_y\sigma_s}\cdot\frac{D(\xi)}{\delta_{\text{acc}^3}} \right\rangle,
\end{equation}
where $N$ is the number of particles per bunch, $r_0$ the classical electron radius, $\sigma_i$ the rms beam sizes, $\gamma$ the relativistic factor.
Furthermore, $D$ is given as \cite{LeDuff:1993jm}:
\begin{equation}
D(\xi)=\sqrt{\xi}\left\{-\frac{3}{2}e^{-\xi}+\frac{\xi}{2}\int_\xi^\infty\frac{\ln ue^{-u}}{u}\text{d}u+\frac1{2}(3\xi-\xi\ln\xi+2)\int_\xi^\infty\frac{e^{-u}}{u}\text{d}u\right\},
\end{equation}
with 
\begin{equation}
\xi = \left(\frac{\delta_{\text{acc}}\beta_x}{\gamma\sigma_x}\right)^2.
\end{equation}
Finally, $\delta_{\text{acc}}$ is the energy acceptance which is dominated by the dynamic aperture. 
Since some quantities in Eq.~\ref{eq:touschekLeDuff} depend on the azimuth $s$, it is necessary to integrate along the circumference (which is indicated by the angle brackets in Eq.~\ref{eq:touschekLeDuff}). 

\begin{figure}[htbp]
\centering
\includegraphics[width=0.6\textwidth]{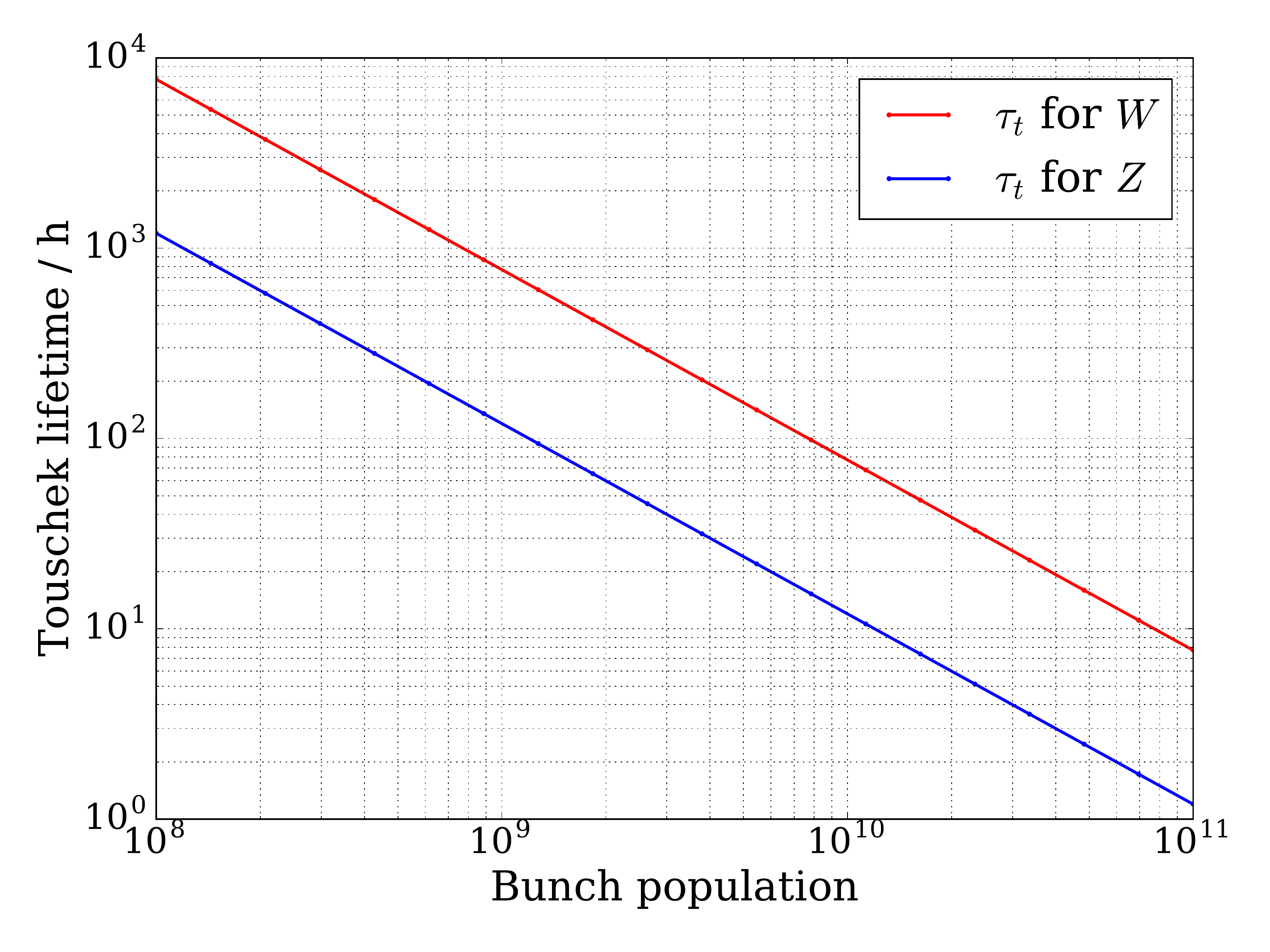}
\caption{Touschek lifetime as a function of number of particles per non-colliding bunch for $Z$ and $W$ energies.}
\label{fig:lifetimeVSBunchpopulation}
\end{figure}

\begin{figure}[htbp]
\centering
\includegraphics[width=0.6\textwidth]{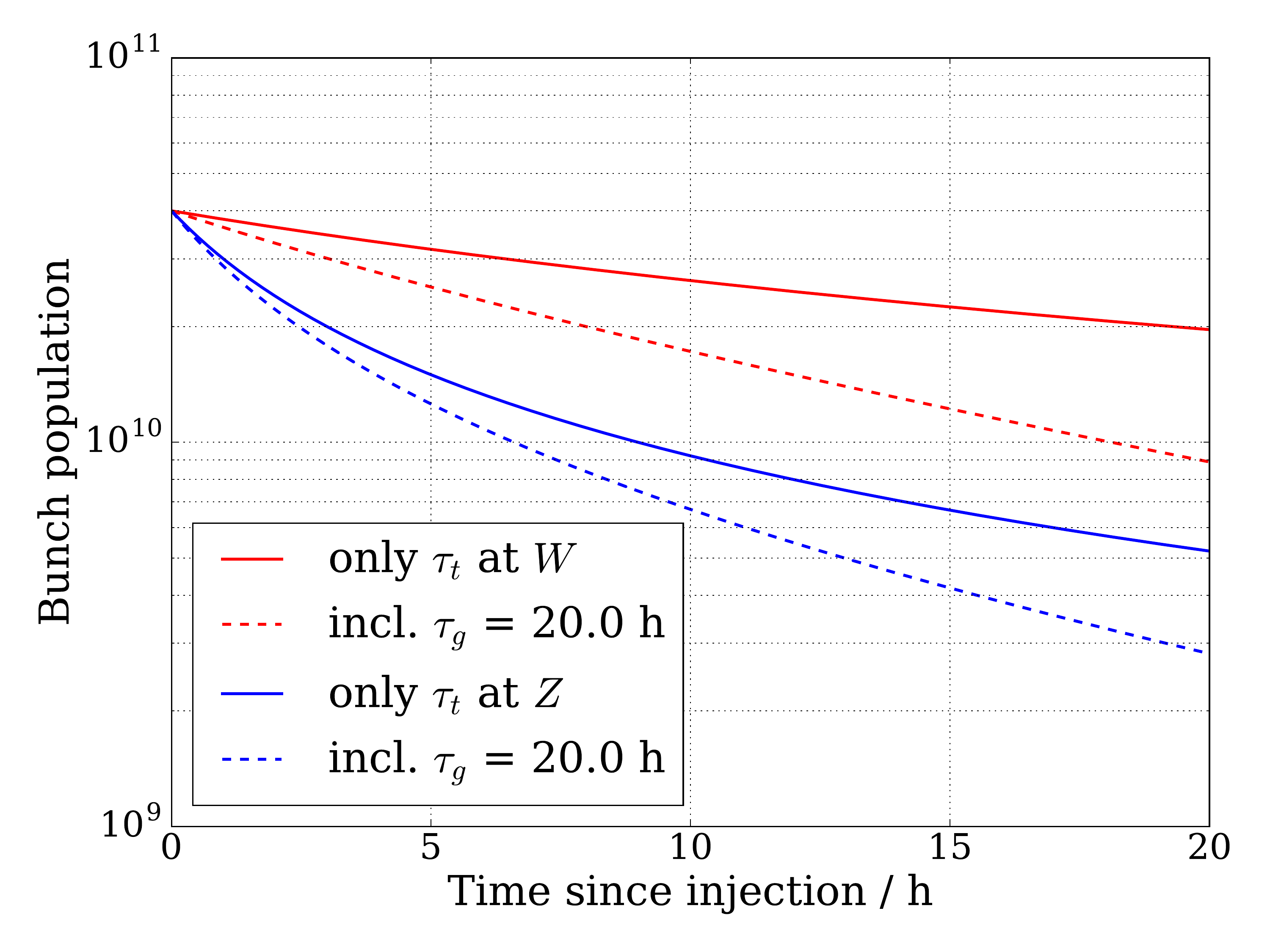}
\caption{Simulated intensity drop in non-colliding bunches due to Touschek lifetime and combination of Touschek and assumed gas scattering lifetime of $\tau_g=\SI{20}{\hour}$ for $Z$ and $W$ energies.}
\label{fig:intensityVStime}
\end{figure}

In Fig.~\ref{fig:lifetimeVSBunchpopulation} the evolution of Touschek lifetime as a function of bunch intensity for the non-colliding bunches is shown for the $Z$ and $W$ operating energies. The Touschek lifetime for a bunch intensity of \num{4e10} is estimated to be $\tau_t\approx\SI{3}{\hour}$ for the $Z$ energy and $\tau_t\approx\SI{19}{\hour}$ for the $W$ energy. In Fig.~\ref{fig:intensityVStime} the simulated intensity drop due to both Touschek lifetime and an assumed gas lifetime of \SI{20}{\hour} is presented for $Z$ and $W$ energies.

The Touschek scattering rate depends on the polarization of the particles in the bunch. To estimate the effect, the following formulae have been used from \cite{Lee:2005xx}:
\begin{equation}\label{eq:touschekVSpolarization}
\frac{\tau_t(P=0)}{\tau_t(P)} = 1 + \frac{F(\xi)}{C(\xi)}\cdot P^2,
\end{equation}
where $P$ is the level of polarization, $\xi$ is the same as before and
\begin{align}
F(\xi) & = -\frac{\xi}{2}\int_\xi^\infty\frac1{u^2}\ln\frac{u}{\xi}e^{-u}\text{d}u,\\
C(\xi) & = \xi\int_\xi^\infty\frac1{u^2}\left\{\left(\frac{u}{\xi}\right)-\frac1{2}\ln\left(\frac{u}{\xi}\right)-1\right\}e^{-u}\text{d}u.
\end{align}

\begin{figure}[htbp]
\centering
\includegraphics[width=0.6\textwidth]{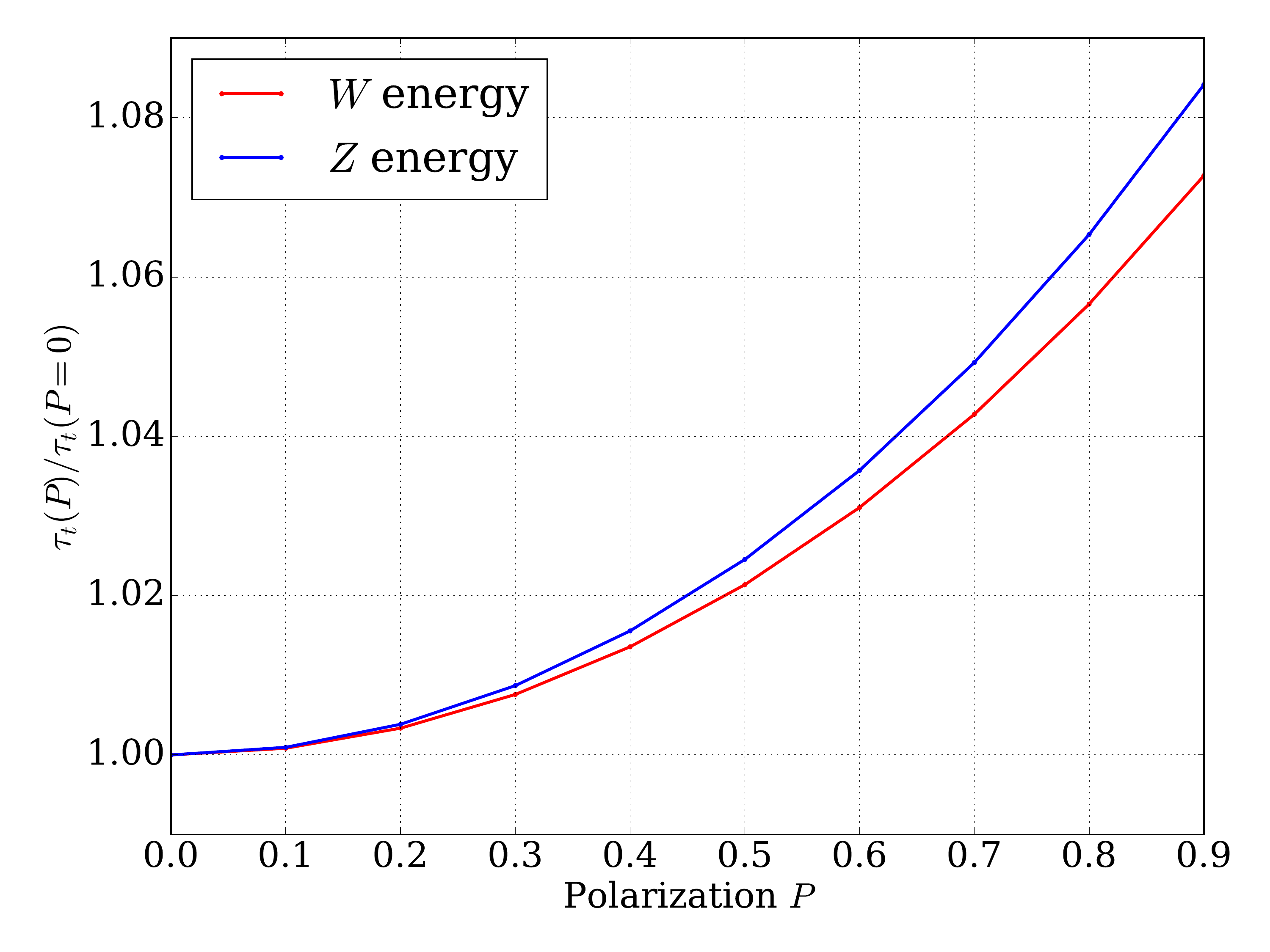}
\caption{Relative change of Touschek lifetime as a function of polarization for $Z$ and $W$ beam energies.}
\label{fig:touschekVSpolarization}
\end{figure}

Figure~\ref{fig:touschekVSpolarization} shows how the Touschek lifetime component varies with the level of polarization according to Eq.~\ref{eq:touschekVSpolarization}. 
Since the Touschek scattering rate for the $Z$ energy is comparably high, the drop of scattering rate due to depolarization of the bunch can be used as a possible additional indicator for depolarization. For the $W$ energy, the change in loss rate might not be as easily detectable since the Touschek lifetime is considerably larger.

\subsection{Polarization Wigglers} \label{sec:hardw-polarwigs}

Polarization wigglers are needed in FCC-ee due to the very long natural polarization times at the Z (more than 200 hours), however, they are not needed for running at the W pair 
threshold (or at higher energies).

The mode of operation is as follows: polarization wigglers will only be switched on at the beginning of every fill where only a small number of bunches would be circulating (the so-called ``pilot bunches'', which form around 2\% of the total number of bunches). These bunches (which will not be in collision) will eventually be used for depolarization measurements. With the current design adequate polarization levels (10\% transverse polarization) will be achieved after 100 minutes of wiggler operation\\

\begin{table}[htbp]
\caption{Specification for the FCC-ee wiggler scheme and LEP parameters shown for comparison. \vspace{3mm}}
\label{tabwiggler}
\centering
\begin{tabular}{|lcc|}
\hline\hline
 & FCC-ee & LEP \\
 \hline
 Number of units per beam	& $ 8 \times 3$ &	8 \\ 
Full gap height [mm] &	90	& 100 \\
Central field $B^+$ [T] & 	0.7	& 1.0 \\
Central pole length [mm]	 & 430	& 760\\
Asymmetry ratio $B^+/B^-$	& 6 & 2.5 \\
Critical energy of SR photons [keV] & 	900	& 1350 \\ 
\hline\hline
\end{tabular}
\end{table}


The number of wigglers is a compromise between the synchrotron radiation power produced by the system and the critical energy of the photons. The FCC-ee will use 8 wigglers per beam, each comprising three units, thus with   each three high field regions and six low field regions. The length of one wiggler unit is chosen so that the total orbit excursion in the horizontal plane is manageable, less than 0.8 mm for a B field of 0.7T  for 3~m units and a total wiggler length of around 9~m, see Fig.~\ref{eliana-f1}. 

\begin{figure}[htbp]
  
   \centering
   \includegraphics[width=0.75\textwidth]{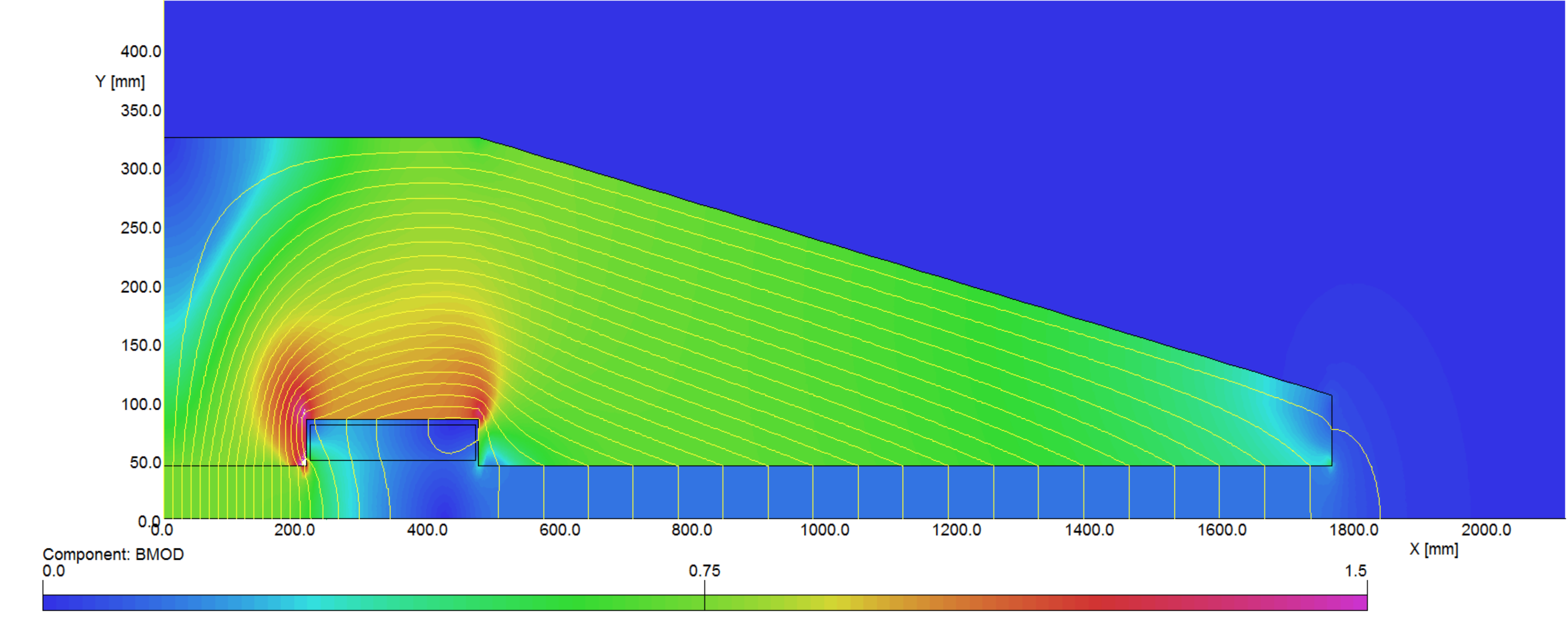} 
   
 \caption{Magnetic design of a half unit of the polarization wiggler. In this representation the beam will travel in the z-direction.}
   \label{fig:mag-desgn-polwig}
\end{figure}

When operated at their nominal settings at 45\,GeV they will produce a significant additional (about 40\%) SR power per bunch.
However, the total beam current in the machine would be low when the wigglers are on. 
The wigglers need to be located in a dispersion-free straight section; the straight sections around PH and PF are suitable. 
The LEP damping and emittance wiggler design was taken as a reference \cite{wigglerjowett} 
and modified to meet the FCC-ee requirements. 

The main parameters of the polarization wigglers for FCC-ee are given in Table~\ref{tabwiggler}. The 2D magnetic design for half of a unit is shown in Fig.~\ref{fig:mag-desgn-polwig}.

\subsection{Polarimetry}
\label{sec:polarimeter}

Inverse Compton scattering is the classical way to measure beam polarization in lepton machines.This technique has been successfully used at LEP~\cite{Placidi:1988nj,Knudsen:1991cu} and HERA~\cite{Barber:242702}. Fast measurement of the beam polarization allows to apply the resonant depolarization technique for precise beam energy determination~\cite{Skrinski1989,Arnaudon:1992rn}.


\subsubsection{Inverse Compton Scattering}

\begin{figure}[htbp]
\centering
\includegraphics[width=0.6\textwidth]{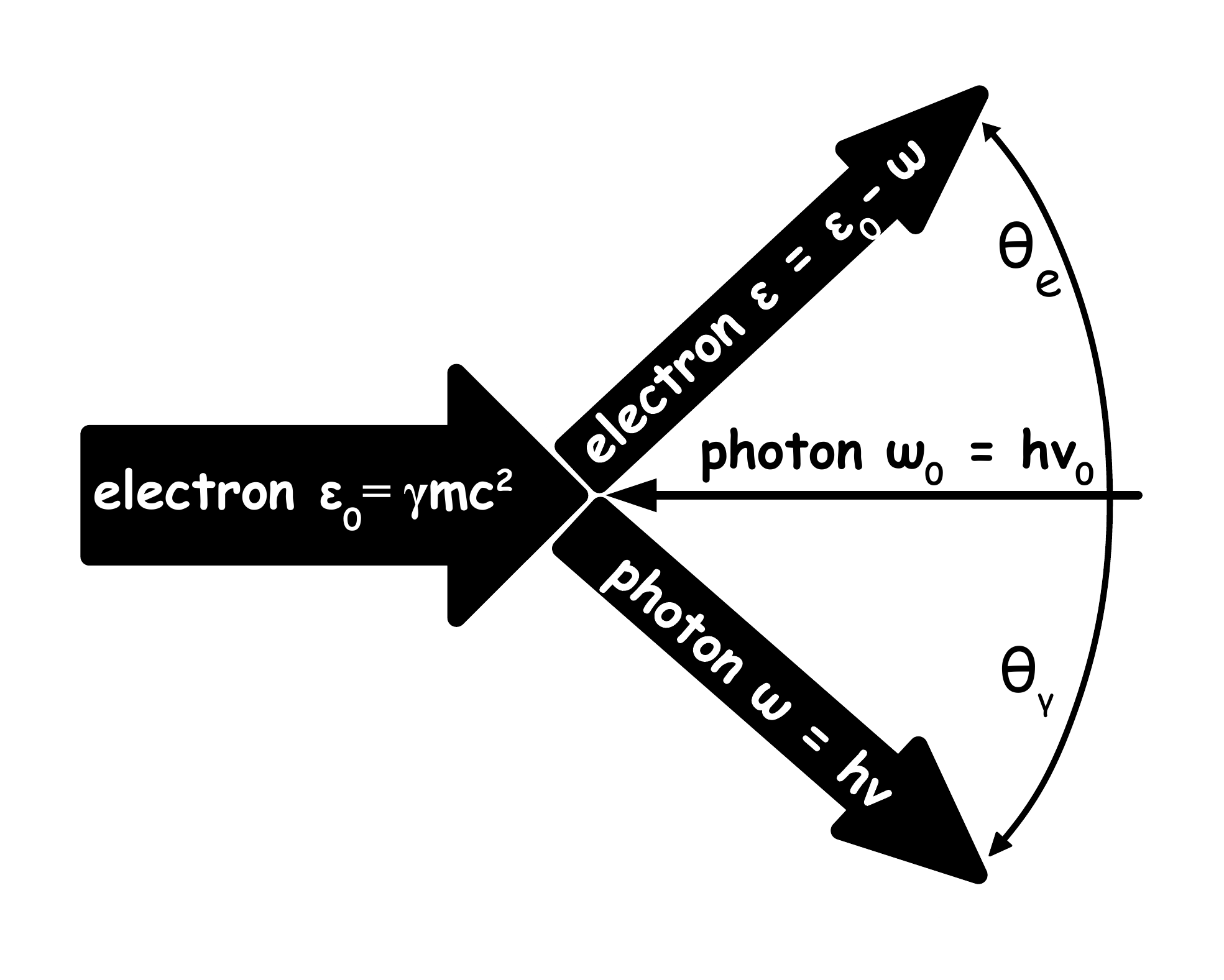}
\caption{Inverse Compton scattering: the thickness of every arrow qualitatively reflects the energy of each particle.
The values of $\omega_0, \varepsilon_0$ and $\omega, \varepsilon$ are the energies of the photon and electron in their initial and final states correspondingly, while $\theta_\gamma$ and $\theta_e$ are the scattering angles of photon and electron.}
\label{fig:kin}
\end{figure}

An illustration for the process of Inverse Compton Scattering (ICS) is presented in Fig.~\ref{fig:kin}.

Considering only the ultra-relativistic case ($\varepsilon_0, \varepsilon, \omega \gg \omega_0$) we introduce the universal scattering parameter
\begin{equation}
u
=\displaystyle\frac{\omega}{\varepsilon}
=\displaystyle\frac{\theta_e}{\theta_\gamma}
=\displaystyle\frac{\omega}{\varepsilon_0-\omega}
=\displaystyle\frac{\varepsilon_0-\varepsilon}{\varepsilon},
\label{u}
\end{equation}
bearing in mind the energy and transverse momenta conservation laws while neglecting the corresponding impacts of initial photon.
Parameter $u$ lies within the range $u \in \left[0,\kappa\right]$ and is limited from above by the longitudinal momentum conservation. $\kappa$ is twice the initial energy of the photon in the rest frame of the electron, expressed in units of the electron rest energy
\begin{equation}
\kappa=4\frac{\omega_0\varepsilon_0}{(mc^2)^2} = 2\times 2\gamma\frac{\omega_0}{mc^2}.
\label{kappa}
\end{equation}
If the electron-photon interaction is not head on, the angle of interaction $\alpha\neq\pi$ affects the initial photon energy seen by the electron, and $\kappa$ parameter becomes\footnote{this is correct when $\tan(\alpha/2) \gg 1/\gamma$.}
\begin{equation}
\kappa(\alpha)=4\frac{\omega_0\varepsilon_0}{(mc^2)^2} \sin^2\left(\frac{\alpha}{2}\right).
\label{kappa_any}
\end{equation}

\begin{figure}[htbp]
\centering
\includegraphics[width=0.75\textwidth]{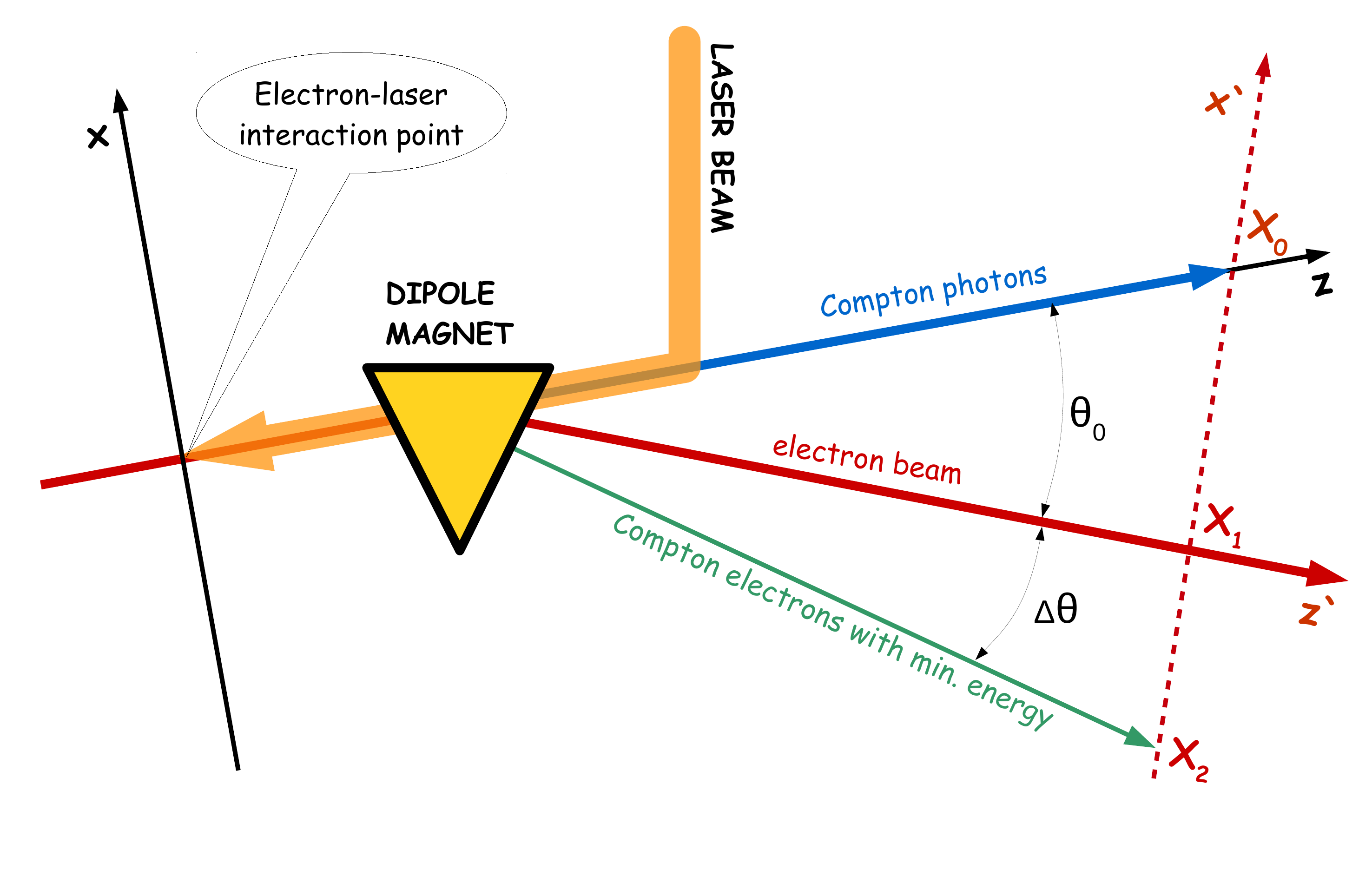}
\caption{Regular layout of ICS experiments realization.}
\label{fig:cxema}
\end{figure}

For the FCC-ee polarimeter we consider the interaction of laser radiation with the electrons in the electron beam energy range $\varepsilon_0 \in [45:185]$\,GeV.
The energy of the laser photon $\omega_0$ is coupled with the radiation wavelength in vacuum $\lambda_0$: $\omega_0 = hc/\lambda_0$, where $hc = 1.23984193$~eV$\cdot\mu$m.
For the particular case of $\lambda_0=1\;\mu$m, $\varepsilon_0=100$\,GeV and $\alpha=\pi$ one obtains the ``typical'' value of $\kappa$ parameter $\kappa \simeq 1.9$.
The maximum energy of the back-scattered photon $\omega_{max}$ obviously corresponds to the minimal energy of the scattered electron $\varepsilon_{min}$, both values are easily obtained from definitions (\ref{u}) -- (\ref{kappa_any}) when $u=\kappa$:
\begin{equation}
\omega_{max} = \frac{\varepsilon_0\kappa}{1+\kappa} \;\text{ and }\; \varepsilon_{min} = \frac{\varepsilon_0}{1+\kappa}.
\label{wmax_emin}
\end{equation}
Note that $\omega_{max}=\varepsilon_{min}$ when $\kappa=1$.
It's not hard to show that the scattering angles of photon $\theta_\gamma$ and electron $\theta_e$ (see Fig.~\ref{fig:kin}) depend on $u$ and $\kappa$ as:
\begin{equation}
\theta_\gamma = \frac{1}{\gamma}\sqrt{\frac{\kappa}{u}-1} \;\text{ and }\; \theta_e = \frac{u}{\gamma}\sqrt{\frac{\kappa}{u}-1}.
\label{thetas}
\end{equation}
The electron scattering angle $\theta_e$ can never exceed the limit $\max(\theta_e)=\kappa/2\cdot\gamma = 2\omega_0/mc^2$ and we see that this value does not depend on $\varepsilon_0$.
Almost any experimental application of laser radiation back-scattering on an electron beam implies the use of the minimal scheme shown in Fig.~\ref{fig:cxema}: the laser radiation is focused, inserted into the machine vacuum chamber and directed to the interaction point where scattering occurs. A dipole magnet is used to separate scattered photons (and electrons) from the non-interacting electrons that propagate further in the vacuum chamber.
$x$-axis and $z$-axis define the coordinate system in the interaction point, the plane of the figure is the plane of machine, the vertical $y$-axis is perpendicular to the plane of figure.
After the dipole, the coordinate system $(x',z')$ is rotated by the beam bending angle $\theta_0$.

\subsubsection{ICS cross section}

The ICS cross section is sensitive to polarization states of all initial and final particles~\cite{berestetskii1982quantum}.
It is common to average the polarization terms of the final states, then the cross section depends solely from the initial photon and electron polarizations.
To describe the polarization states of laser and electron beams in the coordinate system $x,y,z$, presented in Fig.~\ref{fig:cxema}, we introduce modified Stokes parameters.
\begin{itemize} \label{poldefs}
\item $\xi_\perp \in [0:1]$ and $\varphi_\perp \in [0:\pi]$ are the degree of laser linear polarization and its azimuthal angle.
\item $\xi_\circlearrowright \in [-1:1]$ is the sign and degree of circular polarization of the laser radiation: $\sqrt{\xi_\perp^2 + \xi_\circlearrowright^2}=1$.
\item $\zeta_\perp \in [0:1]$ and $\phi_\perp \in [0:2\pi]$ are the degree of transverse e$\pm$ beam polarization and its azimuthal angle.
\item $\zeta_\circlearrowright \in [-1:1]$ is the sign and degree of longitudinal spin polarization of the electrons: $\sqrt{\zeta_\perp^2 + \zeta_\circlearrowright^2} \in [0:1]$.
\end{itemize}
The ICS cross section may be described by the sum of three terms: $d\sigma = d\sigma_0 + d\sigma_\parallel + d\sigma_\perp$. The three terms correspond to unpolarized electrons ($d\sigma_0$) longitudinally polarized electrons ($d\sigma_\parallel$) and to transversely polarized electrons ($d\sigma_\perp$) and may be expressed by:
\begin{equation}
\begin{aligned}
d\sigma_0 & = & \frac{r_e^2}{\kappa^2(1+u)^3} &
\left( \kappa (1 + (1+u)^2) - 4\frac{u}{\kappa}(1+u)(\kappa-u)\Bigl[ 1-\xi_\perp \cos\bigl(2(\varphi-\varphi_\perp)\bigr) \Bigr] \right)  \;du\;d\varphi,\\
d\sigma_\parallel &  = & \frac{\xi_\circlearrowright \zeta_\circlearrowright r_e^2}{\kappa^2(1+u)^3} &
 u(u+2)(\kappa-2u)  \;du\;d\varphi ,\\
d\sigma_\perp  & = &  - \frac{ \xi_\circlearrowright \zeta_\perp r_e^2}{\kappa^2(1+u)^3} &
 2u\sqrt{u(\kappa-u)} \cos(\varphi-\phi_\perp)  \;du\;d\varphi 
\label{suall}
\end{aligned}
\end{equation}
In equations~(\ref{suall}) $r_e$ is the classical electron radius and $\varphi$ is the observer azimuthal angle. The third term $d\sigma_\perp$, which is the most important for an FCC-ee polarimeter, does not affect the total cross section, which in absence of longitudinal polarization of electrons is obtained by integration of $d\sigma_0$ only:
\begin{equation}
\sigma_0( \kappa)  =
\frac{2 \pi r_e^2}{\kappa} \left( \left[ 1-\frac{4}{\kappa}-\frac{8}{\kappa^2}\right] \mathrm{log}(1+\kappa)
+ \frac{1}{2}\left[ 1-\frac{1}{( 1+\kappa)^2}\right] +\frac{8}{\kappa}\right).
\label{tcs}
\end{equation}
For $\kappa \ll 1$ equation (\ref{tcs}) tends to the Thomson cross section $\sigma_0=\frac{8}{3}\pi r_e^2 \left( 1-\kappa\right)$.

The above expressions may be used to build a Monte-Carlo generator to allow further analysis of the scattered particle distributions.
The dimensionless parameter $u\in[0:\kappa]$ is obtained according to the initial values of  $\varepsilon_0$, $\omega_0$, $\alpha$ and polarization coefficients. The probability distribution of $u$ is defined by the cross section (\ref{suall}).
The required properties of the final state particles like $\omega$, $\varepsilon$, $\theta_e$ or $\theta_\gamma$ are obtained from equations (\ref{u}) and (\ref{thetas}).
The influence of the bending magnet shown in Fig.~\ref{fig:cxema} on the scattered electrons is not yet considered.

\subsubsection{Bending of electrons}

Let's describe the integrated dipole strength by the parameter $Bl$, assuming  that this quantity is proportional to the integral magnetic field along the electron trajectory.
The electron with energy $E$ will be bent to the angle $\theta=Bl/E$ under the assumption that $Bl$ is the same for all energies under consideration
\footnote{The validity of this assumption will be discussed on page \pageref{DeltaL}.}.
By equation (\ref{u}) we express the energy $\varepsilon$ of scattered electron through the ICS parameter $u$: $\varepsilon=\varepsilon_0/(1+u)$.
This electron is bent by the dipole to the angle
\begin{equation}
\theta = \frac{Bl}{\varepsilon} =
\frac{Bl}{\varepsilon_0} + u\frac{Bl}{\varepsilon_0}=
\theta_0 + u\theta_0,
\label{bend}
\end{equation}
i.~e. $\theta$ is the sum of the beam bending angle $\theta_0$ and the bending angle $\Delta\theta = u\theta_0$ induced by the electron energy loss from ICS.
Both $\theta_0$ and $\Delta\theta$ are shown in Fig.~\ref{fig:cxema} for the maximum possible $u$ value $u=\kappa$.
Note that $\kappa\theta_0$ does not depend on $\varepsilon_0$.
In Ref.~\cite{Muchnoi:2008bx} it was suggested to use the ratio $\Delta\theta/\theta_0=\kappa$ for the ILC beam energy determination.

Let us introduce the quantity $\vartheta \equiv \gamma(\theta-\theta_0) = u\vartheta_0$ which is the angle $\Delta\theta$, measured in units of $1/\gamma$.
The scattering angle of an electron due to ICS, expressed in the same units, is $\vartheta = u\sqrt{\kappa/u-1}$ as follows from eq.~(\ref{thetas}).
By splitting $\vartheta$ into $x$ and $y$ transverse components and gathering all angles together we obtain:
\begin{equation}
\begin{aligned}
\vartheta_x  = & \sqrt{u(\kappa-u)}\cos\varphi + u\vartheta_0, \\
\vartheta_y  = & \sqrt{u(\kappa-u)}\sin\varphi.
\end{aligned}
\label{txty}
\end{equation}
Since the back-scattered photons are not bent by the dipole, the photon transverse angles (eq.~(\ref{thetas})) in the same coordinates and in the same units as in eq.~(\ref{txty}), are given by
\begin{equation}
\begin{aligned}
\eta_x  = & - \sqrt{\kappa/u-1}\cos\varphi - \vartheta_0, \\
\eta_y  = & - \sqrt{\kappa/u-1}\sin\varphi.
\end{aligned}
\label{nxny}
\end{equation}

\subsubsection{Polarimeter location}

\begin{figure}[htbp]
\centering
\includegraphics[width=0.9\textwidth]{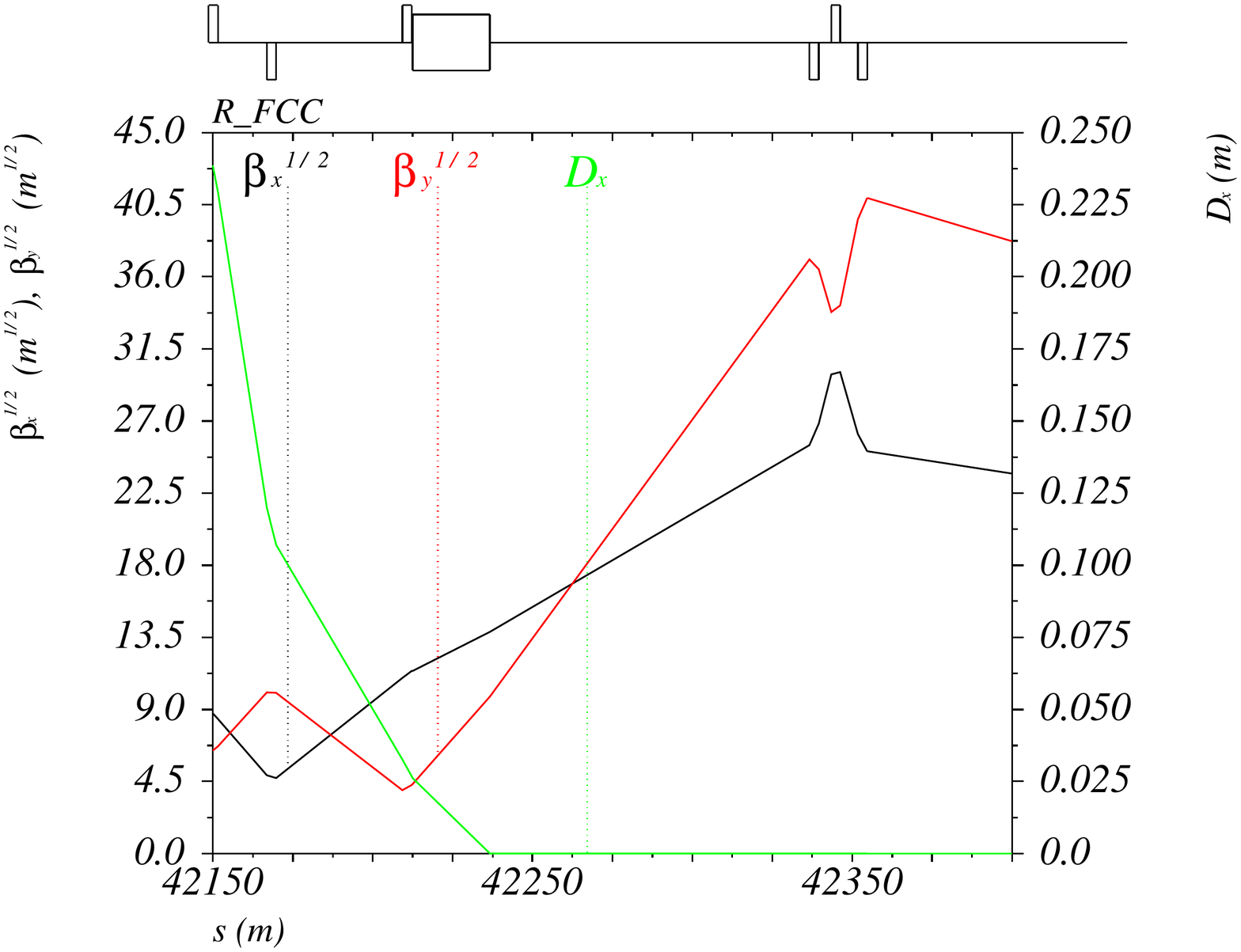}
\caption{Possible location of the polarimeter in the FCC-ee lattice between the quadrupole and the dipole to the right of the vertical betatron function minimum ($s$ coordinate $ \approx 42'212$).}
\label{fig:madx-polarimeter}
\end{figure}

A possible location for  the  polarimeter is in the FCC-ee section shown in Fig.~\ref{fig:madx-polarimeter}. After the dispersion suppressing dipole magnet roughly 100\,m of free drift space can be used to separate of the ICS photons and electrons from the beam. The interaction of the pulsed laser beam with the electron beam occurs just between the dipole and the preceding quadrupole, close to the local minimum of vertical $\beta$-function.

\begin{figure}[htbp]
\centering
\includegraphics[width=0.9\textwidth]{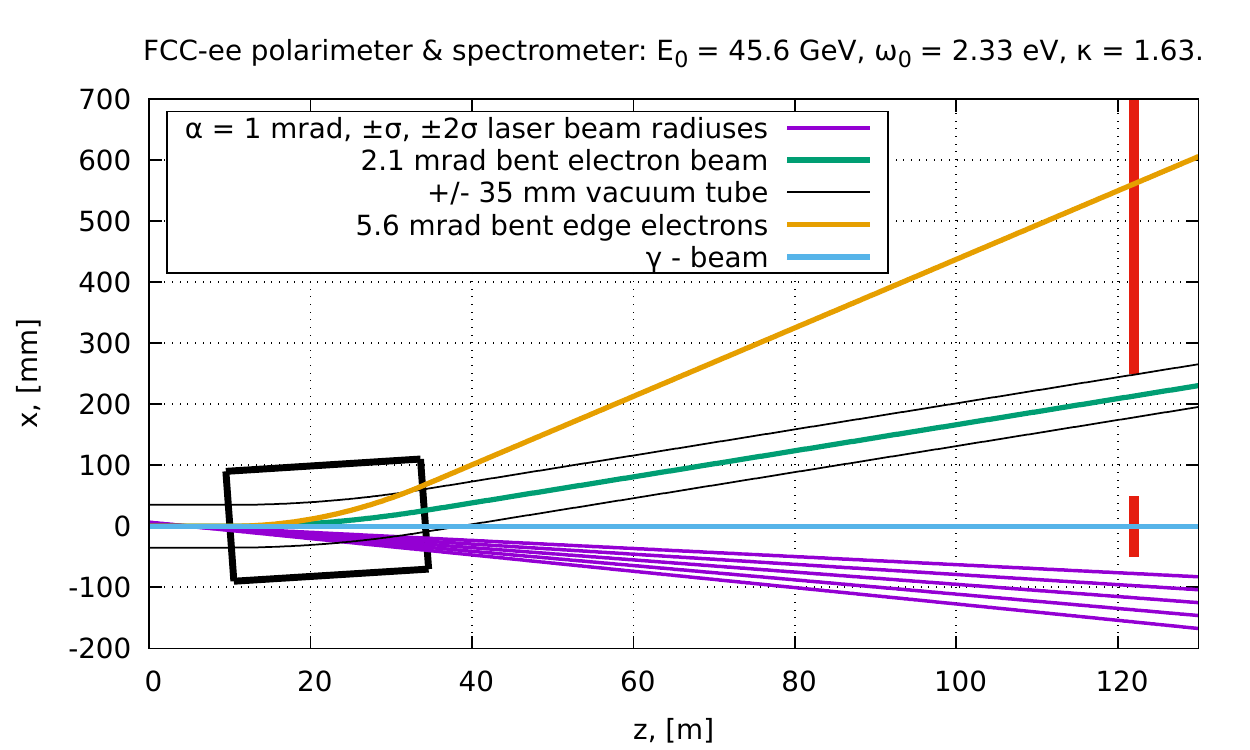}
\caption{Sketch of the polarimeter with the lattice dipole ($L=24.12$\,m, $\theta_0=2.13$\,mrad, $B=0.0135$\,T, $R_0=11302$\,m), the vacuum chamber and the particle trajectories. Red vertical bars on the right side indicate the location of the scattered particles detectors 100\,m away from the centre of the dipole.}
\label{fig:100m}
\end{figure}

\subsubsection{Polarimeter layout}

A sketch of the polarimeter is presented in Fig.~\ref{fig:100m}, the apparatus arrangement is in horizontal plane. The laser radiation $\lambda_0=532$~nm is inserted to the vacuum chamber from the right side and focused to the laser-beam interaction point LIR ($z=5$\,m), the laser spot transverse size at the LIR is $\sigma_0=0.25$\,mm.
The laser-electron interaction angle is $\alpha = \pi - 0.001$ and the relative difference between $\kappa$ of eq.~(\ref{kappa}) and $\kappa(\alpha)$ of eq.~(\ref{kappa_any}) is only $2.5\cdot10^{-7}$.

Figure~\ref{fig:100m} may be used to estimate the difference of the B-field integral seen by  electrons of different energies. All of the electrons enter the dipole of length $L$ along the same beam orbit. The radius of trajectory in the dipole will be dependent on the electron energy. Let $R_0$ to be the radius of an electron with energy $\varepsilon_0$ and $\theta_0 = L/R_0$ is the beam bending angle.
The minimal radius of an electron after scattering on the laser will be $R_0/(1+\kappa)$.
After passing the dipole these two electrons will have the difference $\Delta x \simeq \kappa L \theta_0 /2 $ in transverse horizontal coordinates.
With the parameters of Fig.~\ref{fig:100m} this difference is $\Delta x \simeq 43$\,mm.
The length of the trajectories of these two electrons inside the dipole will be also different, i.~e. even in case of absolutely uniform dipole their field integrals will not be the same. For a rectangular dipole shape the  expression for the relative difference of the  trajectory length is
\begin{equation}
\frac{\Delta L}{L} = \frac{2}{\theta_0}
\left[
\frac{1}{1+\kappa} \arcsin \left( \frac{\theta_0}{2} (1+\kappa) \sqrt{1+ \left(\frac{\kappa\theta_0}{2}\right)^2} \right)
- \arcsin \left( \frac{\theta_0}{2} \right) \right].
\label{DeltaL}
\end{equation}
The relative difference depends only on $\theta_0$ and $\kappa$.
With the set of parameters from Fig.~\ref{fig:100m}, i.~e. $\theta_0=2.13$\,mrad and $\kappa=1.63$, $\Delta L/L = 2.63\cdot10^{-6}$.

The assumption about the equality of the integrals of the magnetic field for the electron beam and scattered electrons is therefore valid to the level of a few parts per million for a homogeneous hard edge dipole. The difference could be reduced with a shorter dipole which on the other hand will harden and increase the synchrotron radiation photons emitted by the dipole.

\subsubsection{Distributions of scattered particles}

A MC generator was created to obtain the 2D ($x,y$) distributions of scattered photons and electrons at the detectors presented in Fig.~\ref{fig:100m}.
The ICS parameters are: $\varepsilon_0=45.6$\,GeV and $\omega_0=2.33$~eV.
The spectrometer configuration is described by the beam bending angle $\theta_0=2.134$\,mrad, the lengths of the dipole $L=24.12$\,m and two spectrometer arms.
The first arm with a length $L_1=117$\,m is the path from laser-electron IP to the detector, the second arm with a length $L_2=100$\,m is the path from the longitudinal centre of the dipole to the detector.
The impact of the electron beam parameters is accounted by introducing the angular spreads according to the transverse beam emittances $\epsilon_x=0.27$~nm and $\epsilon_y=1$~pm.
The horizontal and vertical beam electron angles $x'$ and $y'$ are described by normal distributions with means equal to zero and standard deviations $\sigma_x = \sqrt{\epsilon_x/\beta_x}$ and $\sigma_y = \sqrt{\epsilon_y/\beta_y}$, with 
 values of $\beta$-functions from Fig.~\ref{fig:madx-polarimeter}.

\begin{figure}[htbp]
\centering
\includegraphics[width=\textwidth]{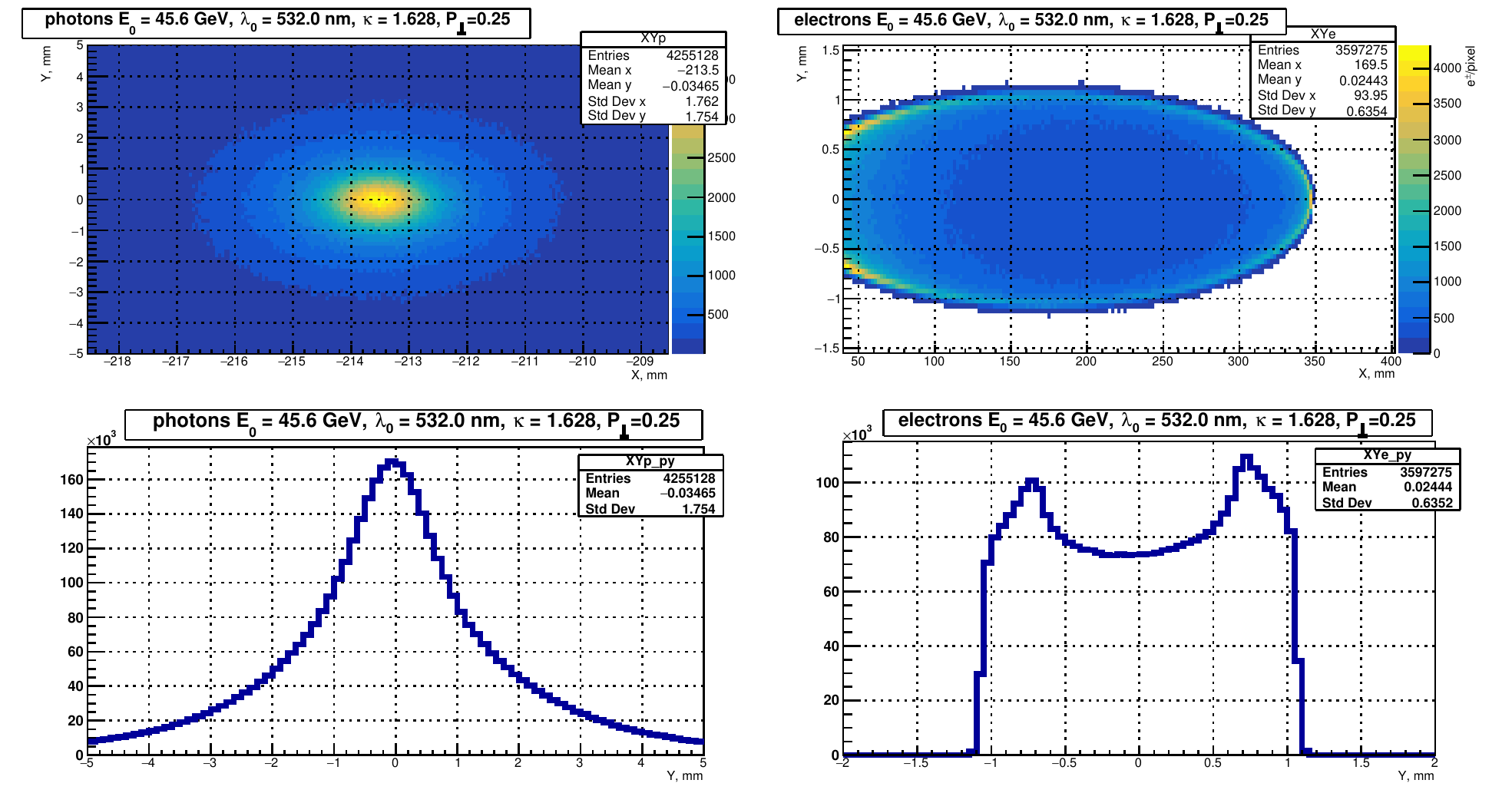}
\caption{MC generation of ICS photon and electrons at the detector plane for $P_\perp = \xi_\circlearrowright \zeta_\perp = 0.25$ and $\phi_\perp=\pi/2$. The bottom plots present the projection of the two dimensional distributions (top plots) onto the vertical (y) axis.}
\label{fig:pos25}
\end{figure}

\begin{figure}[htbp]
\centering
\includegraphics[width=\textwidth]{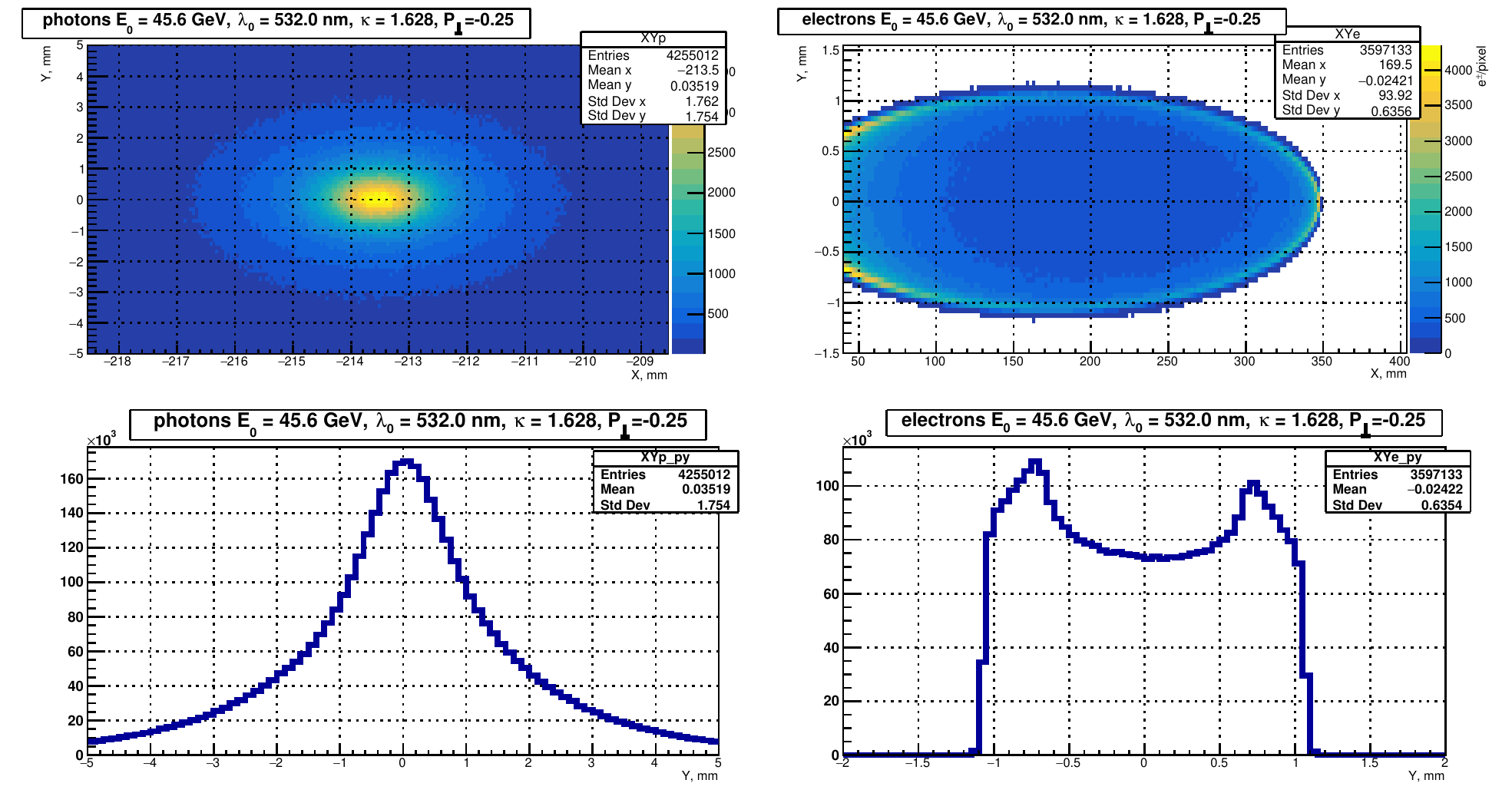}
\caption{MC results of ICS photon and electrons at the detector plane for $P_\perp = \xi_\circlearrowright \zeta_\perp = -0.25$ and $\phi_\perp=\pi/2$. The bottom plots present the projection of the two dimensional distributions (top plots) onto the vertical (y) axis.}
\label{fig:neg25}
\end{figure}

The MC generation procedure is as follows:
\begin{itemize}
\item generate $u\in[0,\kappa]$ and $\varphi\in[0:2\pi]$ according to 2D function $d\sigma(u,\varphi)$ (eq.~(\ref{suall})),
\item generate $x'$ and $y'$ according to corresponding normal distributions,
\item construct the photon $X_\gamma, Y_\gamma$ and electron $X_e, Y_e$ transverse coordinates on the detector plane:
\end{itemize}
\begin{equation}
\begin{aligned}
X_\gamma & =  x' L_1 - \frac{L_1}{\gamma}\sqrt{ \kappa/u-1} \cos \varphi - \theta_0 L_2, \\
Y_\gamma & =  y' L_1 - \frac{L_1}{\gamma}\sqrt{ \kappa/u-1} \sin \varphi, \\
X_e      & =  x' L_1 + \frac{L_1}{\gamma}\sqrt{u(\kappa-u)} \cos \varphi + u \theta_0 L_2, \\
Y_e      & =  y' L_1 + \frac{L_1}{\gamma}\sqrt{u(\kappa-u)} \sin \varphi.
\end{aligned}
\end{equation}
The results of such a simulation for an electron beam with $\zeta_\perp=25$\% vertical  spin polarization ($\phi_\perp=\pi/2$) are presented in Fig.~\ref{fig:pos25} and Fig.~\ref{fig:neg25}.
The difference between the figures is the laser polarization $\xi_\circlearrowright=+1$ (Fig.~\ref{fig:pos25}) and  $\xi_\circlearrowright=-1$ (Fig.~\ref{fig:neg25}).
The 2D distributions for both photons and electrons are plotted along the same horizontal axis $x$, where $x=0$ corresponds to the position of the electron beam.
The detectors for scattered particles are located outside the machine vacuum chamber. The scattered electron distribution starts around $x = 40$\,mm which is assumed to be the edge of the vacuum chamber.

\begin{figure}[t]
\centering
\includegraphics[width=\textwidth]{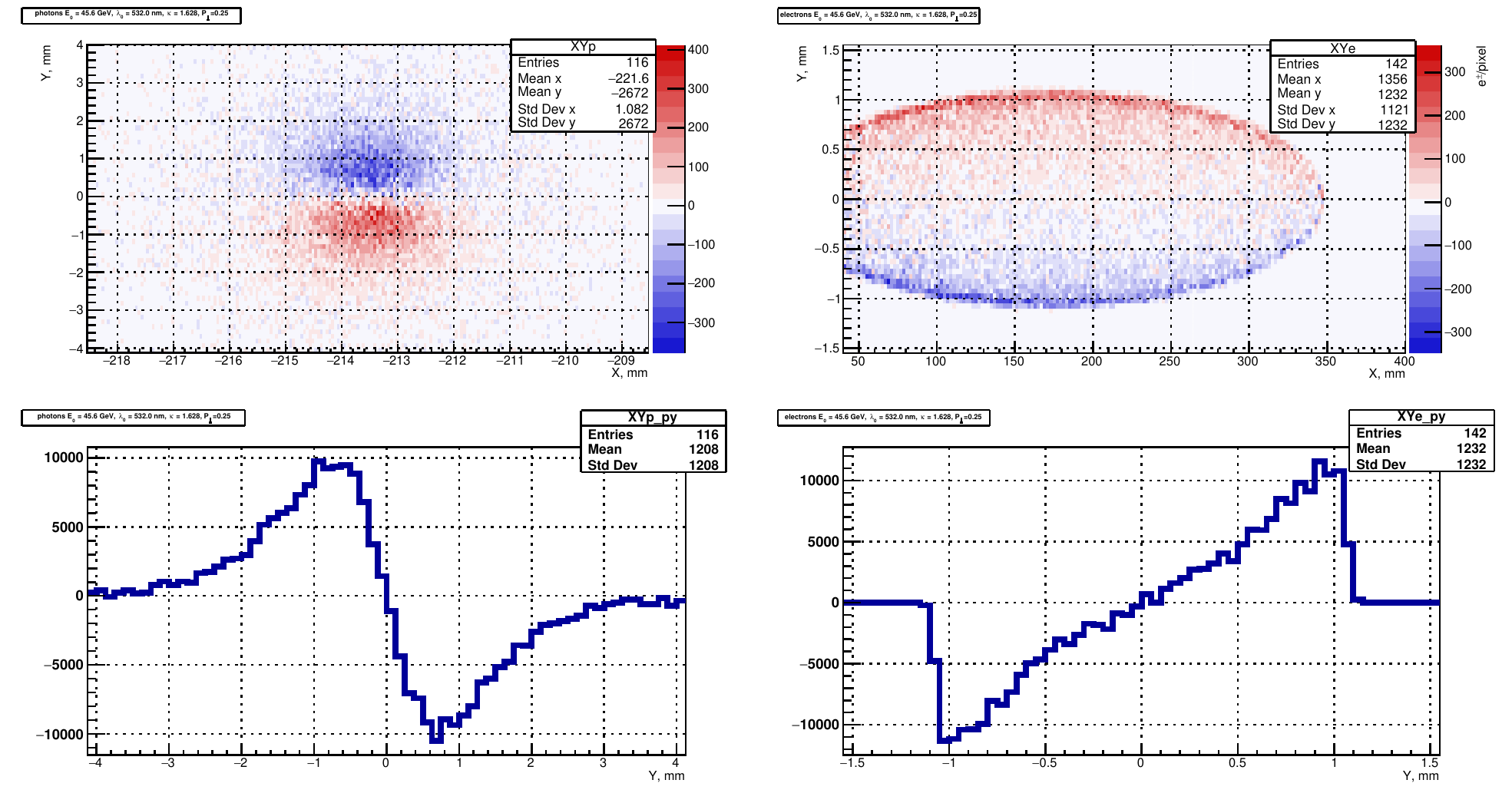}
\caption{The difference between corresponding distributions in Fig.~\ref{fig:pos25} and Fig.~\ref{fig:neg25}.}
\label{fig:thedifference}
\end{figure}

The 1D distributions at the bottom of each figure are the projections of the 2D distributions to the vertical axis $y$. The mean $y$-values of these distributions are shifted up or down with respect to zero according to the beam polarization and corresponding asymmetries in ICS cross section.
In Fig.~\ref{fig:thedifference} all distributions are obtained by subtraction of corresponding distributions from Fig.~\ref{fig:pos25} and Fig.~\ref{fig:neg25}.
Detecting the up-down asymmetry in the distribution of laser backscattered photons is a classical way to measure the transverse polarization of the electron beam.  In \cite{Mordechai:2013zwm} it was proposed to use the up-down asymmetry in the distribution of scattered electrons for the transverse polarization measurement at the ILC. It was suggested to measure the distribution of scattered electrons with a Silicon pixel detector.

The up-down asymmetry in the distribution of scattered electrons peaks at the scattering angles $\theta^*_e = \pm 2\omega_0/mc^2$, which corresponds approximately to $\pm 9\;\mu$rad at 45\,GeV (see eqs.~(\ref{thetas}) and (\ref{suall})).
The asymmetry is only observable if the distributions are not blurred by the electron beam emittance. On the other hand the maximum up-down asymmetry in the distribution of scattered photons occurs at the scattering angles of $\theta^*_\gamma \simeq 1/\gamma$ which is almost the same as $\theta^*_e$ in our particular case. 

The following benefits may be obtained by measuring the scattered electrons in addition to the scattered photons:
\begin{itemize}
\item Scattered electrons propagate to the inner side of the machine radius, i.~e. there is no direct background from high energy synchrotron radiation (only scattered photons).
\item Unlike photons, charged electrons are easily detected by ionization losses, while photons must be converted into electromagnetic showers.
\item Despite the fact that the fluxes of scattered photons and electrons are the same, the flux density of electrons is much lower due to bending and corresponding spatial separation by energies. Simultaneous detection of multiple scattered electrons thus is much easier.
\item Analysis of the scattered electrons distribution allows to measure both longitudinal and transverse beam polarizations.
\item As one can observe from Figs.~\ref{fig:pos25}--\ref{fig:thedifference}, the inversion of the laser circular polarization leads to a redistribution of the scattered electron density within a fixed elliptic shape. This fact potentially provides better systematic accuracy for beam polarization determination.
\item A downside of the electron detection is its dependence on stray electromagnetic fields over the very long lever arm between LIR and detector plane.
\end{itemize}
Eventually both photon and electron distributions should be measured by the FCC-ee polarimeter. The photon measurements will exploit the LEP and HERA experience, they must measure the centre of the photon distribution in both $x$ and $y$ dimensions. The latter is required for direct beam energy determination, which will be discussed below.

\subsubsection{Scattered electrons distribution}

The successful application of the method of direct electron beam energy determination by backscattering of laser radiation is based on the measurement of $\omega_{max}$ (see eq.~\ref{wmax_emin}) at sufficiently low energies to obtain a  good absolute accuracy. A positive experience of this method was obtained at the low energy colliders VEPP-4M, BEPC-II and VEPP-2000~\cite{Achasov:2017abc}.
Despite the fact that this method is not directly applicable to FCC-ee due to the high electron energies, this section highlights what may be learned from the shape of the scattered electron distribution.

From (\ref{txty}) the following quadratic equation is obtained for $u$:
\begin{equation}
(\vartheta_x-u\vartheta_0)^2 + \vartheta_y^2 = u(\kappa-u),
\label{nequality}
\end{equation}
with the roots
\begin{equation}
u^\pm = \frac{\kappa+2\vartheta_0\vartheta_x\pm\sqrt{\kappa^2-4(\vartheta_x^2+\vartheta_y^2(1+\vartheta_0^2)-\kappa\vartheta_0\vartheta_x)}}{2(1+\vartheta_0^2)}.
\label{upm}
\end{equation}
The average value of $u$ and its limiting value for the large values of $\vartheta_0$ do not depend on $\vartheta_y$:
\begin{equation}
\langle u \rangle
= \frac{u^+ + u^-}{2}
= \frac{\kappa/2+\vartheta_0\vartheta_x}{1+\vartheta_0^2} \;\;\xrightarrow{\vartheta_0\gg1}\;\; \frac{\vartheta_x}{\vartheta_0}.
\label{nutxty-one}
\end{equation}
In the $\vartheta_x,\vartheta_y$ plane all the scattered electrons are located inside an ellipse, see in Figs.~\ref{fig:pos25}, \ref{fig:neg25}), described by the solutions of eq.~(\ref{upm}).
The centre of the ellipse is located at $[\vartheta_x =\kappa\vartheta_0/2; \vartheta_y=0]$ with horizontal half-axis $A=\kappa\sqrt{1+\vartheta_0^2}/2$ and vertical half-axis $B=\kappa/2$.
This implies that
\begin{equation}
\vartheta_x^{max} = \frac{\kappa}{2}\Bigl(\vartheta_0+\sqrt{1+\vartheta_0^2}\Bigr)\xrightarrow{\vartheta_0\gg1}\kappa\vartheta_0.
\label{varthetamax}
\end{equation}
Following the notation introduced above, $\vartheta$-s are angles measured in units of $1/\gamma$, while $\theta$-s are angles in radians.
In radians expression (\ref{varthetamax}) becomes
\begin{equation}
\Delta\theta = \frac{\kappa}{2}\Bigl(\theta_0 + \sqrt{1/\gamma^2+\theta_0^2}\Bigr)\xrightarrow{\theta_0\gg1/\gamma}\kappa\theta_0,
\label{thetamax}
\end{equation}
where $\Delta\theta$ and $\theta_0$ are defined in Fig.~\ref{fig:cxema}.
The ICS cross section may be transformed from variables $u,\varphi$ to $\vartheta_x,\vartheta_y$ with the Jacobian matrix:
\begin{equation}
\mathbf{J}= \left(
\begin{matrix}
\displaystyle\frac{\partial\vartheta_x}{\partial u} & \displaystyle\frac{\partial\vartheta_x}{\partial\varphi}\\[1em]
\displaystyle\frac{\partial\vartheta_y}{\partial u} & \displaystyle\frac{\partial\vartheta_y}{\partial\varphi}
\end{matrix}
\right)
=
\left(\begin{matrix}[r]
\displaystyle \vartheta_0 + \frac{\kappa/2-u}{\sqrt{u(\kappa-u)}}\cos\varphi &
\displaystyle -\sqrt{u(\kappa-u)}\sin\varphi \\[1em]
\displaystyle  \frac{\kappa/2-u}{\sqrt{u(\kappa-u)}}\sin\varphi &
\displaystyle  \sqrt{u(\kappa-u)}\cos\varphi
\end{matrix}\right).
\label{nJacob}
\end{equation}
The matrix determinant is:
\begin{equation}
\det(\mathbf{J}) = \kappa/2 - u + \vartheta_0\sqrt{u(\kappa-u)}\cos\varphi = \sqrt{\kappa^2/4-\vartheta_x^2-\vartheta_y^2(1+\vartheta_0^2)+\kappa\vartheta_0\vartheta_x} .
\end{equation}
Hence, $du d\varphi = 2 d\vartheta_xd\vartheta_y/\det(\mathbf{J})$, where the factor ``2'' is due to the summing of the ``up'' and ``down'' solutions of eq.~(\ref{upm}).
We will now perform another change of variables from $\vartheta_x,\vartheta_y$ to  $x$ and $y$. With those new variables the cross section exists inside a circle of radius $R=1$ centred at $(x=0;y=0)$:
\begin{equation}
x = \frac{\vartheta_x-\kappa\vartheta_0/2}{\kappa/2\sqrt{1+\vartheta_0^2}},\;\;\;\;\;
y = \frac{\vartheta_y}{\kappa/2}.
\label{nxy}
\end{equation}
Then:
\begin{equation}
\begin{aligned}
du d\varphi & = & \frac{\kappa\,dx\,dy}{\sqrt{1 - x^2 - y^2}} ,\\
u = \langle u \rangle  &=& \frac{\kappa}{2} \left( 1 + \frac{x\vartheta_0}{\sqrt{1+\vartheta_0^2}}\right),\\
\sin(\varphi) & = & \frac{y\, \kappa}{2\sqrt{u(\kappa-u)}}.
\end{aligned}
\label{janduxy}
\end{equation}
In (\ref{janduxy}) vertical transverse electron polarization ($\phi_\perp=\pi/2$) is assumed for which $\cos(\varphi-\phi_\perp)=\sin(\varphi)$.
For backscattering of circularly polarized laser radiation ($\xi_\circlearrowright = \pm 1$) on the electron beam, where both vertical transverse ($\zeta_\perp \neq 0, \phi_\perp=\pi/2$) and longitudinal ($\zeta_\circlearrowright \neq 0$) polarizations are possible, the cross sections (\ref{suall}) may be expressed with the new variables:
\begin{equation}
\begin{aligned}
d\sigma_0  & = & \frac{r_e^2}{\kappa(1+u)^3\sqrt{1 - x^2 - y^2}} &
\left( 1 + (1+u)^2 - 4\frac{u}{\kappa}(1+u)(1-\frac{u}{\kappa}) \right) & dx\;dy,\\
d\sigma_\parallel & = & \frac{\xi_\circlearrowright \zeta_\circlearrowright  r_e^2}{\kappa(1+u)^3\sqrt{1 - x^2 - y^2}} &
\;\;\;\;\;\;\;\; u(u+2)(1-2u/\kappa) & dx\;dy ,\\
d\sigma_\perp & = & - \frac{ \xi_\circlearrowright \zeta_\perp r_e^2}{\kappa(1+u)^3\sqrt{1 - x^2 - y^2}} &
\;\;\;\;\;\;\;\; u  y & dx\;dy .
\end{aligned}
\label{xyall}
\end{equation}
Due to the presence of $\sqrt{1 - x^2 - y^2}$ in the denominator of (\ref{xyall}) the cross section exhibits a singularity at the edge of a circle (ellipse), which is however integrable.

\subsubsection{Electron and photon detection}

\begin{figure}[ht]
\centering
\vspace{-4mm}
\includegraphics[width=0.8\textwidth]{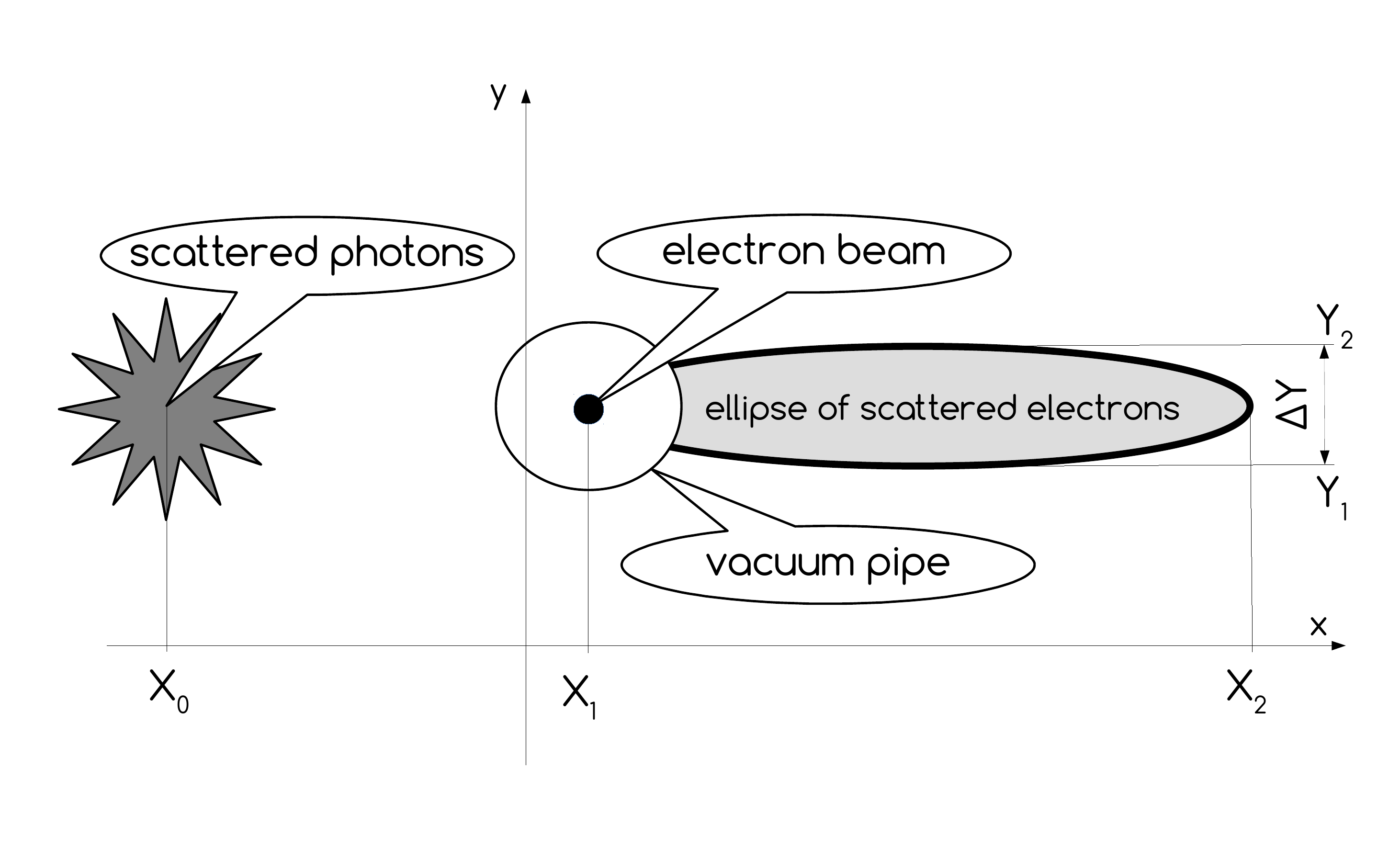}
\vspace{-4mm}
\caption{The $xy$ plane of particle detection. $X_0$ is the horizontal centre of gravity of the scattered photon distribution. $X_1$ is the unperturbed electron beam position and the extreme left edge of the scattered electron ellipse, while $X_2$ is the right side of the same ellipse. The vertical size is given by $\Delta Y = L_1\cdot(4\omega_0/mc^2)$.}
\label{fig:poldet-principle}
\end{figure}

The detectors for the scattered photons and electrons are proposed to be installed as shown in Fig~\ref{fig:100m}. These pixel detectors will measure the $x$ and $y$ positions of each particle according to the following scheme of Fig.~\ref{fig:poldet-principle}.

The photon detector should consist of a few pixel detector planes sandwiched with Tungsten (or another absorber material) plates to initiate electromagnetic showers. A detailed Geant simulation of the detector must be made for a proper design.

For the detection of the scattered electrons we consider a position measurement using a silicon pixel detector (as in \cite{Mordechai:2013zwm}) placed at a distance $L_1= 117$\,m from the LIR and $L_2 = 100$\,m from the centre of bending dipole.
The active dimension of the detector is 400$\times$4\,{\rm mm}$^2$, it is offset horizontally 40\,mm with respect to the beam axis.
The size of the pixel cell assumed to be 2$\times$0.05\,{\rm mm}$^2$, i.~e. there are 200 pixels in $x$ and 80 pixels in $y$.

To extract the desired information the MC distribution of scattered electrons will be fitted by the theoretical cross section of Eq.~(\ref{xyall}).
This cross section has a very sharp edge at $x^2+y^2=1$, so the integrals of (\ref{xyall}) over each pixel are required for fitting.
However, the dependency of the cross section on $u$ and $y$ is rather weak and it was found to be enough to evaluate the integral
\begin{equation}
I_{xy} = \int\limits_{x_0}^{x_1}\int\limits_{y_0}^{y_1}\frac{dx\,dy}{\sqrt{1-x^2-y^2}}
\label{Ixy}
\end{equation}
over a rectangular pixel limited by $[x_0,x_1]$ in $x$ and $[y_0,y_1]$ in $y$.

The result of integration is:
\begin{equation}
\begin{aligned}
I_{xy} & =
x_1 \arctan\!\left[\frac{y_1 D_{10}     - y_0 D_{11}}  {y_1 y_0       + D_{11}D_{10}}\right] -
    \arctan\!\left[\frac{x_1(y_1 D_{10} - y_0 D_{11})} {x_1^2 y_0 y_1 + D_{11}D_{10}}\right] +
y_1 \arctan\!\left[\frac{x_1 D_{01}     - x_0 D_{11}}  {x_1 x_0       + D_{11}D_{01}}\right] -  \\
& -
x_0 \arctan\!\left[\frac{y_1 D_{00}     - y_0 D_{01}}  {y_1 y_0       + D_{00}D_{01}}\right] +
    \arctan\!\left[\frac{x_0(y_1 D_{00} - y_0 D_{01})} {x_0^2 y_0 y_1 + D_{00}D_{01}}\right] -
y_0 \arctan\!\left[\frac{x_1 D_{00}     - x_0 D_{10}}  {x_1 x_0       + D_{00}D_{10}}\right],
\end{aligned}
\label{Irect1}
\end{equation}
where $D_{ij}=\sqrt{1-x_i^2-y_j^2}$ and $i,j=[0,1]$.
As a second step $I_{xy}$ is convoluted with the two-dimensional normal distribution of initial electrons, $P(x,y) = \frac{1}{2\pi\sigma_x\sigma_y}\exp\left(-\frac{x^2}{2\sigma^2_x}-\frac{y^2}{2\sigma^2_y}\right).$
Here $\sigma_x$ and $\sigma_y$ are the r.m.s. electron beam sizes projected to the detection plane. Finally the last step is to account for $u$ and $y$ in eq.~(\ref{xyall}).

\begin{figure}[htb]
\centering
\includegraphics[width=\textwidth]{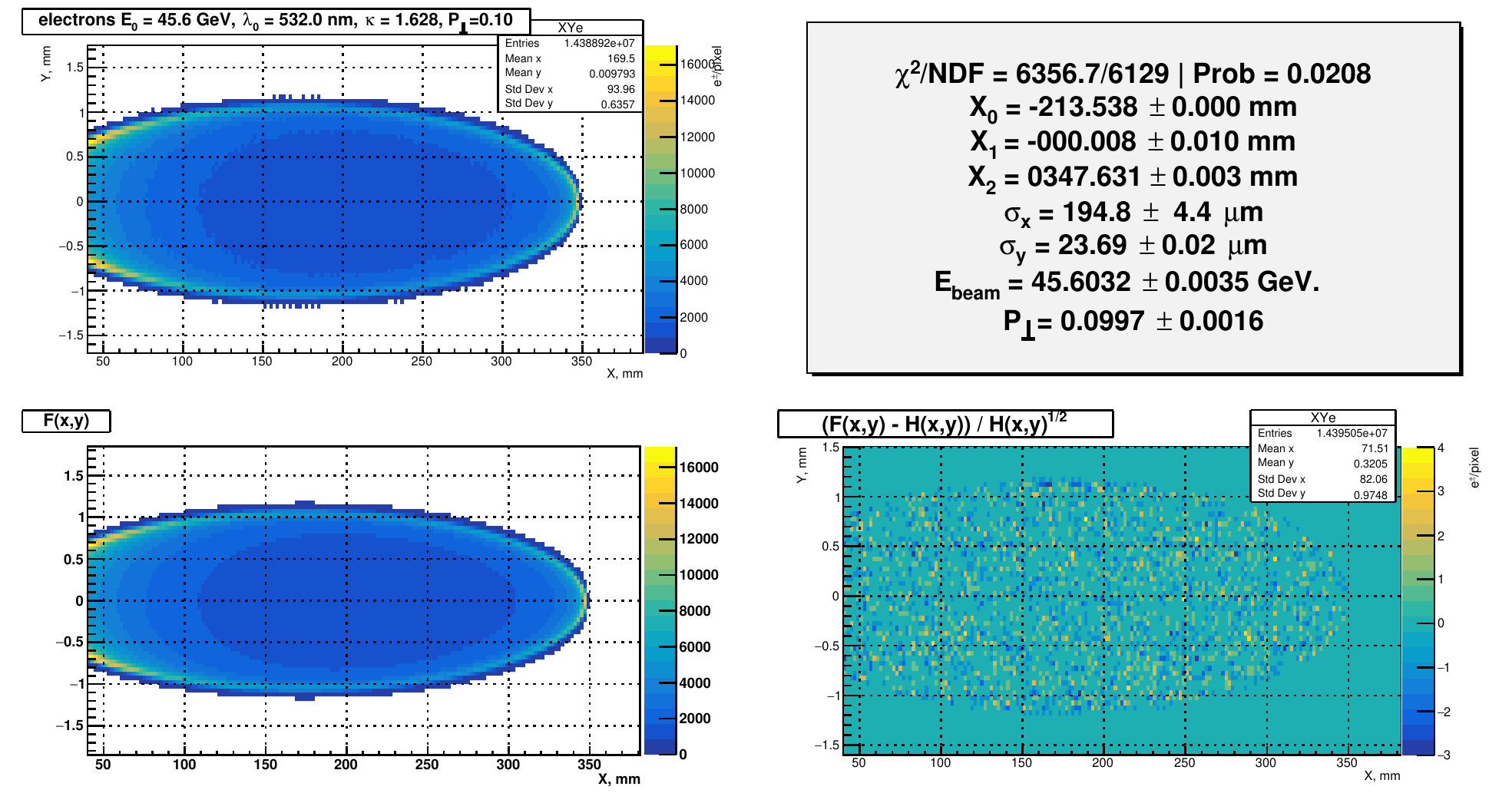}
\caption{{\em Top-left:} MC distribution of scattered electrons $H(x,y)$.
{\em Bottom-left:} function $F(x,y)$ after fitting to the MC distribution.
{\em Bottom-right:} normalized difference: $(F(x,y)-H(x,y))/\sqrt{H(x,y)}$.
{\em Top-right:}  $F(x,y)$ parameters obtained from the fit, only the mean $x$ value  of the scattered photons distribution ($X_0$) was fixed).}
\label{fig:TheResult}
\end{figure}

The function $F(x,y)$ was built based on these considerations to describe the shape of the scattered electrons distribution, see Fig.~\ref{fig:TheResult}.
The function has nine parameters (and a normalization factor):
\begin{itemize}
\item The first parameter is $\kappa$, defined in eq.~(\ref{kappa}). This parameter is fixed based on an approximate value of the beam energy because $F(x,y)$ has only a very week dependence on $\kappa$, variation of $\pm$1\% may be neglected in the fit.
\item The next four parameters are $X_1,X_2,Y_1,Y_2$ define the positions of the ellipse edges, see Fig.~\ref{fig:poldet-principle}.
\item The sixth and seventh parameters are sensitive to the polarizations $P_\perp = \xi_\circlearrowright \zeta_\perp$ and $P_\parallel = \xi_\circlearrowright \zeta_\circlearrowright$. In the example of Fig.~\ref{fig:TheResult} conditions were fixed to $\phi_\perp = \pi/2$ and $\zeta_\circlearrowright=0$.
\item The eighth and ninth parameters are the electron beam sizes $\sigma_x$ and $\sigma_y$ at the azimuth of the detector.
\end{itemize}

The results presented in Fig.~\ref{fig:TheResult} were obtained with $2 \cdot 10^7$ backscattered MC events, with about $1.44 \cdot 10^7$ inside the acceptance of the electron detector.
The physical parameters obtained directly from the fit results are the ellipse positions $X_1,X_2,Y_1,Y_2$, the electron beam transverse sizes $\sigma_x$ and $\sigma_y$ and the beam polarization  $P_\perp$, measured with 1.6\% relative accuracy (0.16\% absolute accuracy).
The beam energy is reconstructed using the following relation
\begin{equation}
E_{beam} = \frac{(mc^2)^2}{4\omega_0} \cdot \frac{X_2-X_1}{X_1-X_0}.
\end{equation}
The statistical accuracy of the energy reconstruction is around $10^{-4}$ fur such a statistics.

\subsubsection{The flux of back scattered photons}

For a CW TEM$_{00}$ laser radiation propagating along $z$-axis,
the optical radius $w(z)$ of the gaussian beam is by definition the transverse distance from the $z$-axis where the radiation intensity drops to $1/e^2$ ($\simeq$13.5\%) of the peak value.
We define the laser beam size as $\sigma(z) = w(z)/2$ which corresponds to an intensity of $1/e$ ($\simeq$36.8\%) of the peak.
For laser light of wavelength $\lambda_0$ focused at $z=0$ to a waist size $\sigma_0$, the beam size will evolve along $z$:
\begin{equation}
\sigma(z) = \sigma_0 \sqrt{ 1 + \left(\frac {z} {z_R}\right)^2},
\text{ where }
z_R = \frac{4 \pi \sigma_0^2}{\lambda_0} \text{ is the Rayleigh length}.
\end{equation}
The power density [W/cm$^2$] of a Gaussian beam with power $P$ [W] is
\begin{equation}
I(r,z) = \frac{P}{2 \pi \sigma(z)^2} \exp \left( - \frac {r^2} {2\sigma(z)^2}\right).
\end{equation}
The radiation power is related to the number of laser photons emitted per second,
\begin{equation}
P = dE/dt = h \nu \cdot dN/dt\;\;[\text{J s}^{-1}].
\end{equation}
The longitudinal density of laser photons along $z$ is 
$\rho_\parallel = dN/dz = P \lambda_0/hc^2$~[cm$^{-1}$].
The Rayleigh length corresponds to the distance along the $z$ direction from the beam waist where the on-axis intensity decreases by a factor 2, i.e. $I(0,z_R) = I(0,0)/2$. The far field divergence is $\theta = \sigma_0/z_R = \lambda_0/(4\pi\sigma_0)$.

\begin{figure}[htbp]
\centering
\includegraphics[width=0.5\textwidth]{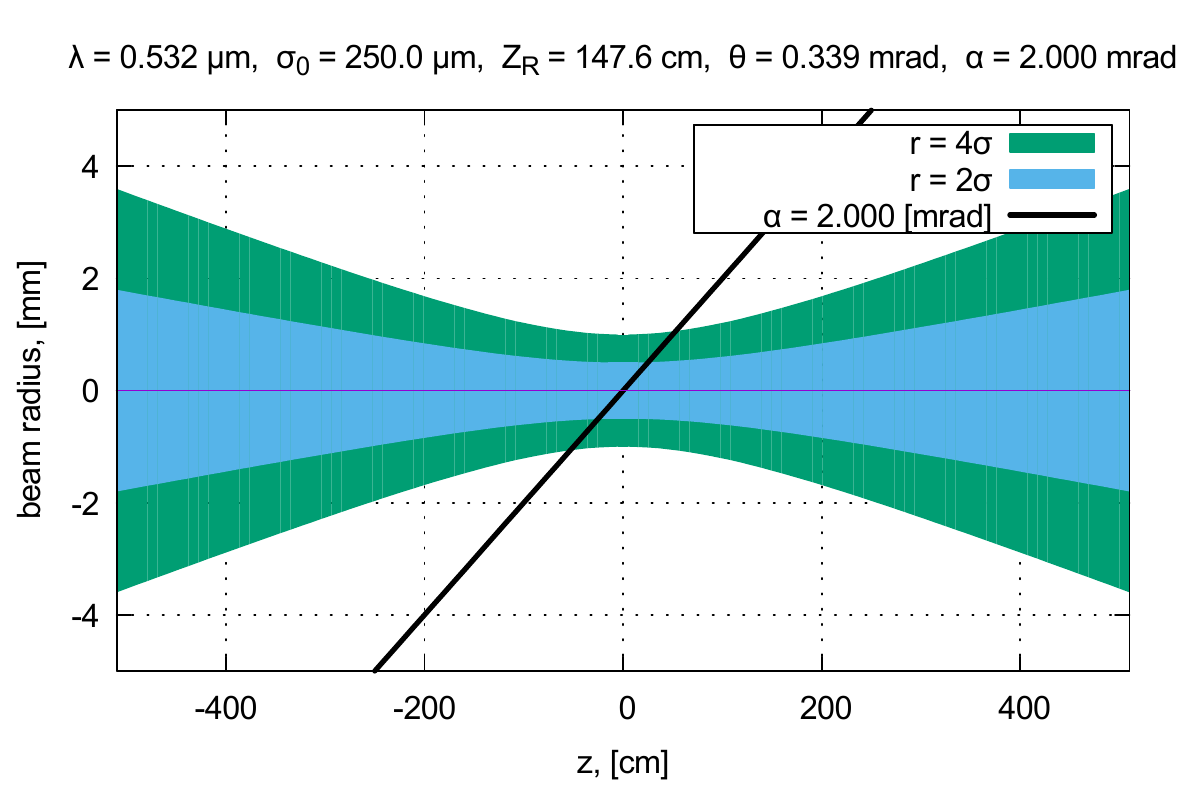}
\caption{An electron (the black sloping line) passing through the laser beam waist.}
\label{fig:laser-ebeam-schema}
\end{figure}

\begin{figure}[htbp]
\centering
\includegraphics[width=0.5\textwidth]{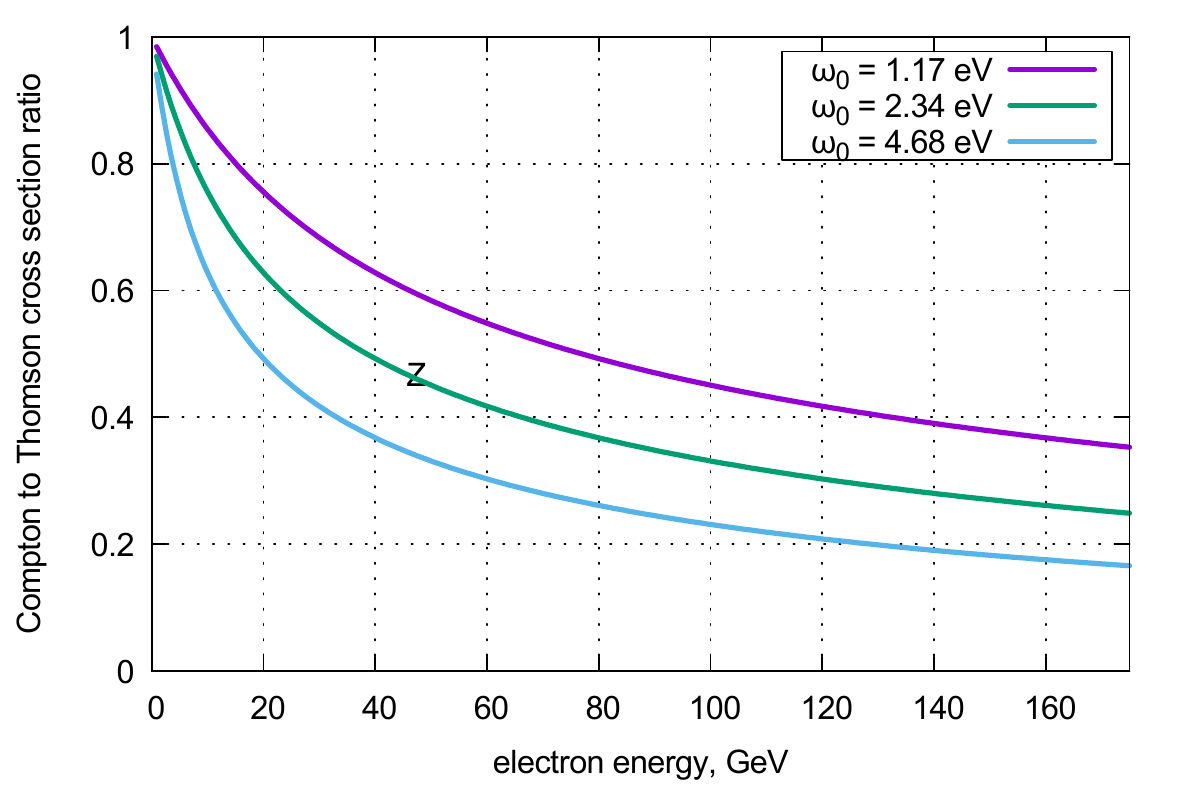}
\caption{The ratio of the ICS cross section to Thomson cross section vs FCC beam energy.}
\label{fig:ratio-ics-tcs}
\end{figure}

An electron ($v/c\simeq1$) propagating towards the LIR encounters the laser focus  with a small incident angle $\alpha$. The photon target density seen by this electron is given by
\begin{equation}
\rho_\perp =  \rho_\parallel\frac{(1+\cos\alpha)}{2\pi\sigma_0^2} \int \limits_{-\infty}^{\infty}
\frac{\exp \left( - \frac {z^2 \tan^2\alpha} {2\sigma(z)^2}\right)}{1+(z/z_R)^2} dz.
\end{equation}
The probability $W$ for Compton scattering of the electron is determined by the product of the density $\rho_\perp$ and the scattering cross section.
The latter is defined by eq.~(\ref{tcs}) and depends on parameter $\kappa$, see Fig.~\ref{fig:ratio-ics-tcs}.
The maximum scattering probability $W_{max}$ is reached for $\alpha=0$ and at low energy with Thomson cross section $\sigma_T=0.665$~barn,
\begin{equation}
W_{max} =  \frac{\sigma_T}{\pi\sigma_0^2}\frac{P \lambda_0}{ h c^2 } \int \limits_{-\infty}^{\infty}\frac{dz}{ 1 + (z/z_R)^2}
 = \frac{4 \sigma_T P}{ h c^2 } \int \limits_{-\infty}^{\infty}\frac{dx}{ 1 + x^2} = \frac{4 \pi \sigma_T P}{ h c^2 } = \frac{P}{P_c}
\end{equation}
where $ P_c = \hbar c^2/2\sigma_T \simeq 0.7124 \cdot 10^{11}$~[W] is the power of laser radiation required for 100\% scattering probability.
$W_{max}$ depends neither on the radiation wavelength $\lambda_0$ nor on the waist size $\sigma_0$, but only on the laser power because the two beams are parallel. A low energy electron bunch with a population of $0.7\cdot10^{11}$ particles colliding head-on with 1~W of laser radiation will therefore produce one Compton scattering event.

\begin{figure}[htbp]
\centering
\vspace{-0.5cm}
\includegraphics[width=0.45\textwidth]{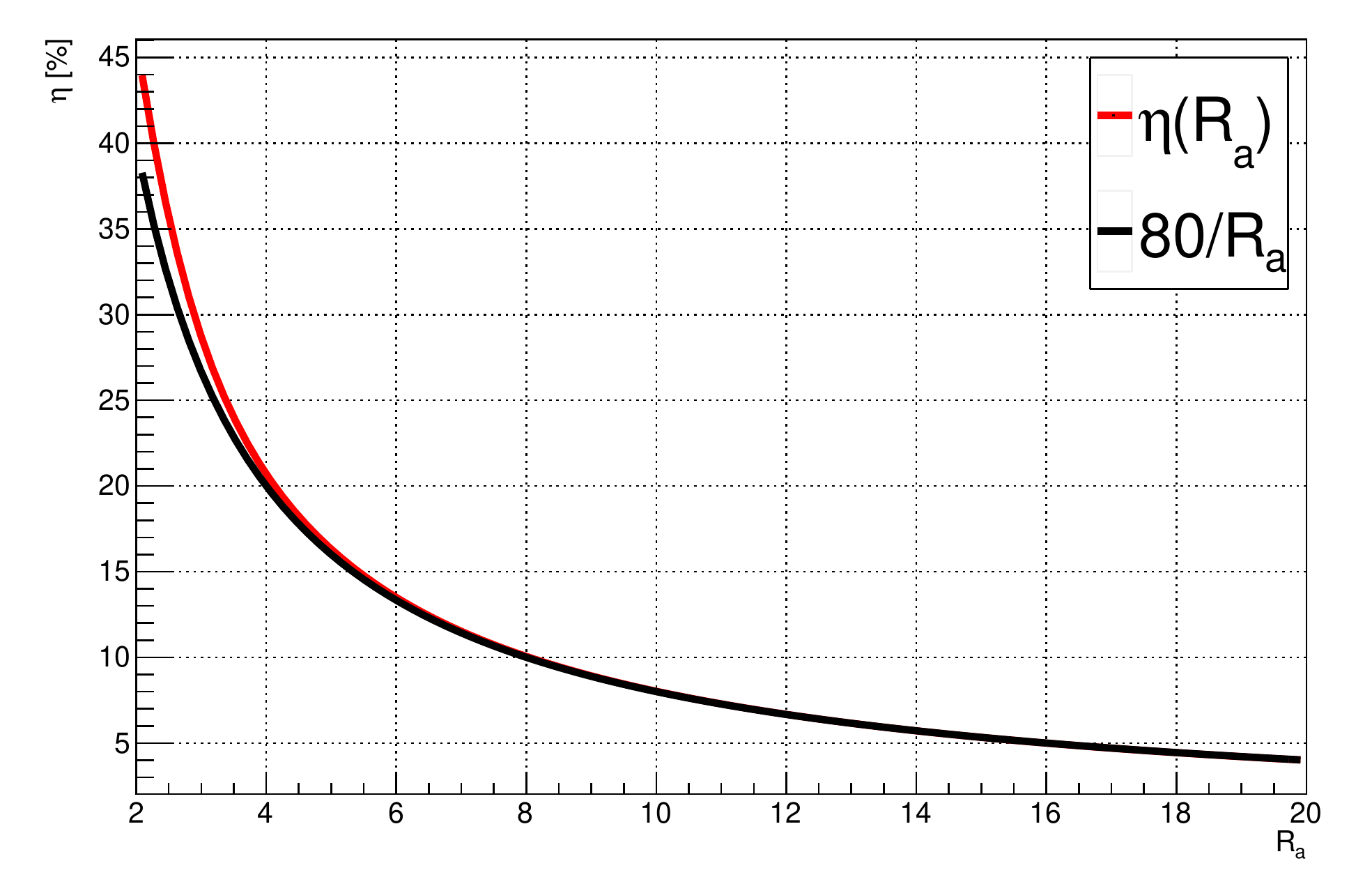}
\caption{$\eta(R_A)$ vs $R_A$ as defined by eq.~(\ref{etara}) and its simple approximation.}
\label{fig:eta1}
\end{figure}

The reduction of the scattering probability when $\alpha \neq 0$ is given by the ratio of angle $\alpha$ to the laser divergence angle $\theta=\sigma_0/z_R$.
Since a mirror is required to deliver the laser beam to the LIR, $\theta$ should be always smaller than $\alpha$; this ratio defines the separation laser and electron beams at the location of the mirror (see Fig.~\ref{fig:100m}).
Defining the ''angle ratio'' as $R_{A} = \alpha/\theta$, the reduction of the scattering probability loss may be expressed by
\begin{equation}
\eta(R_A) = \frac{W(\alpha)}{W_{max}} = \frac{1}{\pi} \!\!\int \limits_{-\infty}^{\infty} \!\!\exp \left( - \frac{x^2 R_A^2}{2(1+x^2)}  \right) \frac{dx}{ 1 + x^2}.
\label{etara}
\end{equation}
Fig.~\ref{fig:eta1} presents $\eta(R_A)$ as a function of $R_A$.

\subsubsection{In-vacuum light mirror}

The laser beam for ICS must be injected into the vacuum chamber through a transparent window. An in-vacuum mirror directs the photons towards the LIR to collide with the electrons.

The polarimeter layout sketched out in Fig.~\ref{fig:100m} is installed on the inner ring, the outer ring is positioned on the side of the ICS photon beam. To avoid interference with the outgoing e-/e+ beams and with the outer ring vacuum chamber and to minimize the impact of the SR photons on its surface, the mirror should be placed below the circulating and the light injected from above. For such a configuration direct impacts of SR photons can be minimized, an important factor since the mirror will remain inserted inside the vacuum chamber for very long periods.

\begin{figure}[bht]
\centering
\includegraphics[width=0.5\textwidth]{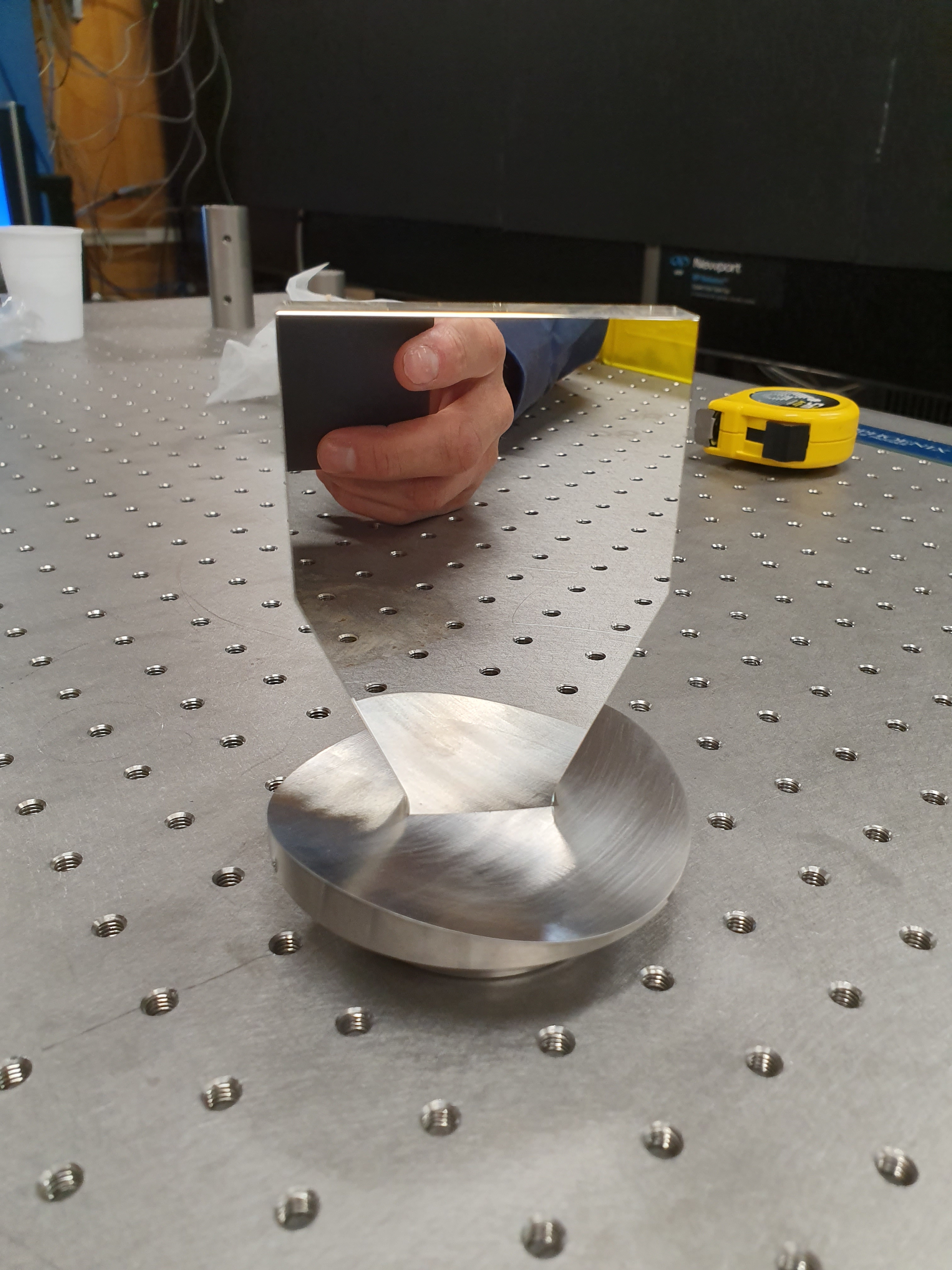}
\caption{The LHC synchrotron radiation monitor (BSRT) in-vacuum mirror including the support holding the dielectric mirror. The support follows exactly the shape of the vacuum chamber, RF contacts (not shown on image) ensure contact between vacuum chamber and support.}
\label{fig:lhc-sr-mirror}
\end{figure}

With the Ampere-level currents circulating inside the chamber at the Z pole and the short bunch length, beam induced heating of the mirror and its supports must be carefully avoided or at least minimized. The example of the LHC synchrotron radiation monitor extraction mirror highlights the damage that can be induced on mirror and mirror support by high intensity beams~\cite{Andreazza:1566631}. The initial design with non-shielded metallic (bulk Silicon) support and metallic mirror was damaged by intensities that were still below LHC design. An improved design of mirror and support, the later fully shielded, solved all problems and limited the temperature increase of the mirror and support to less then 1$^\circ$. The new glass mirror with thin dielectric coating is shown in Fig.~\ref{fig:lhc-sr-mirror}. It must however be noted that the r.m.s. bunch length at the LHC of 8\,cm is roughly an order of magnitude larger than at FCC-ee with a much lower frequency spectrum (up to a few GHz). At FCC-ee the situation will therefore be more difficult, and active cooling of the support may have to be considered.

For a vacuum chamber radius of 35\,mm the mirror could have a diameter of up to around 30 mm with an orientation of around 45$^\circ$ to the vertical. For a distance between mirror edge and beam of 20\,mm, the centre of the mirror would be roughly 30\,mm below the beam line and a slightly enlarged vacuum chamber could be designed for that location. With a spot size of 0.25\,mm the distance between mirror and LIR should not exceed 25\,m to contain most of the light (3-4$\sigma$ radius). The angle $\alpha$ between laser and beam would then be close to 1\,mrad. The laser light would be be injected from above the vacuum chamber with a transparent window. A rotation of the mirror around the vertical axis can be used to adjust laser light and electron beam overlap in the horizontal plane. To tune the laser beam angle in the vertical plane, either the mirror must  be adjustable in angle or the incident laser beam angle could be adjusted using a mirror system above the beam line and out of vacuum. The mirror must be extractable from the vacuum chamber in which case RF contacts (fingers) must be inserted to shield the retracted mirror from the beam.

\subsubsection{Pulsed laser}

\begin{figure}[htbp]
\centering
\includegraphics[width=0.8\textwidth]{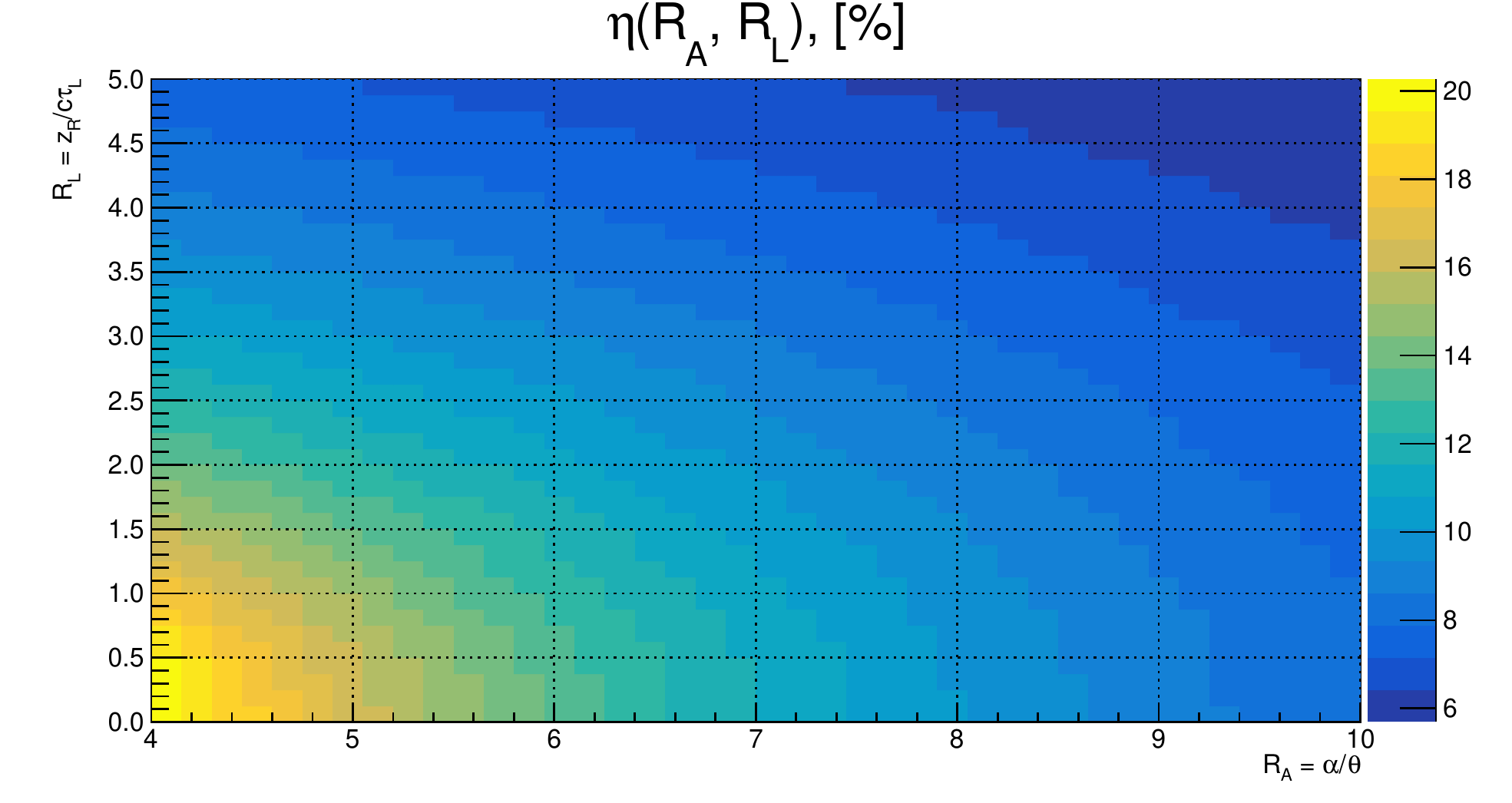}
\caption{Simulated efficiency map $\eta(R_L, R_A)$ for pulsed laser power.}
\label{fig:nlna-sim}
\end{figure}

The FCC-ee will operate with polarized pilot bunches for regular beam energy measurement by resonant depolarization and laser operation in CW mode is thus not possible.
The FCC-ee revolution frequency $\simeq 3$\,kHz is comfortable for solid-state lasers operating in a Q-switched regime.
The laser pulse propagation can be described as:
\begin{equation}
\rho_\parallel(s,t) = \displaystyle\frac{N_\gamma}{\sqrt{2\pi}c\tau_L}\exp\left(-\frac{1}{2}\left(\frac{s-ct}{c\tau_L}\right)^2\right),
\end{equation}
where $\tau_L$ and $E_L$ are pulse duration and energy, $N_\gamma = E_L\lambda/hc$.
The scattering probability for $\alpha=0$ is
\begin{equation}
W  =  \displaystyle\frac{(E_L/\sqrt{2\pi}\tau_L)}{P_c} \cdot
\frac{1}{\pi}\int\limits_{-\infty}^{\infty}\!\frac{\exp\{-2(x\,z_R/c\tau_L)^2\}}{1+x^2}
\,dx =
\displaystyle\frac{P_L}{P_c} \cdot \frac{1}{\pi}\int\limits_{-\infty}^{\infty}\frac{\exp\{-2(x R_L)^2\}}{1+x^2}dx,
\end{equation}
where $P_L = E_L/\sqrt{2\pi}\tau_L$ is the instantaneous laser power and $R_L=z_R/c\tau_L$ is the ``length ratio''.
The scattering probability for an arbitrary $\alpha$ is
\begin{equation}
W  =  \displaystyle\frac{P_L}{P_c} \cdot \frac{1}{\pi}
\int\limits_{-\infty}^{\infty}\frac{\exp\left(-x^2\left(2 R_L^2+\displaystyle\frac{R_A^2}{2(1+x^2)}\right)\right)}{1+x^2}\,dx =
\displaystyle\frac{P_L}{P_c} \cdot \eta(R_L, R_A)
\label{eq:nlna}
\end{equation}
where
$$
P_L = E_L/\sqrt{2\pi}\tau_L; \;\;\;\;\;\;
P_c \simeq 0.7124 \cdot 10^{11}~\text{[W]}; \;\;\;\;\;\;
R_L=z_R/c\tau_L; \;\;\;\;\;\;
R_A = \alpha/\theta_0 = 4\pi\sigma_0/\alpha.
$$
The efficiency map $\eta(R_L, R_A)$ obtained by numerical integration of eq.~(\ref{eq:nlna}) is presented in Fig.~\ref{fig:nlna-sim}.

With this information it is now possible to  estimate the flux of backscattered photons from one FCC-ee bunches and the associated bunch lifetime in the configuration of Fig.~\ref{fig:100m}:
\begin{itemize}
\item Laser wavelength $\lambda_0$ = 532~nm;
\item Compton cross-section correction  $R_\times\simeq$50\% (see Fig.~\ref{fig:ratio-ics-tcs});
\item Waist size $\sigma_0$ = 0.25\,mm, Rayleigh length $z_R$ = 148\,cm;
\item Far field divergence $\theta$ = 0.169\,mrad;
\item Interaction angle $\alpha$ = 1.0\,mrad (horizontal crossing);
\item Laser pulse energy $E_L$ = 1 [mJ], pulse length $\tau_L$=5~[ns] (sigma);
\item Instantaneous laser power $P_L$ = 80 [kW], $P_L/P_c = 1.1 \cdot 10^{-6}$;
\item Ratio of angles $R_A$ = 5.9, ratio of lengths $R_L$ = 0.98, leading to $\eta(R_L, R_A)$ $\simeq$13\% (see Fig.~\ref{fig:nlna-sim});
\item Scattering probability $W = P_L/P_c \cdot R_\times \cdot \eta(R_L, R_A) \simeq 7 \cdot 10^{-8}$;
\item For $N_e = 10^{10}$ electrons per bunch and laser repetition frequency $f_{ref} =3$\,kHz, the ICS phton rate is $\dot{N}_\gamma = f \cdot N_e \cdot W\simeq 2\cdot 10^{6}$ [s$^{-1}$];  
\item Average laser power is $P=f \cdot E_L\simeq3$~W.
\end{itemize}
The influence of the electron beam sizes on the above estimations was not considered because it is negligible. For those parameters the ICS photon rates correspond to the bunch lifetimes of only 1.4~hours which is quite low considering the long stable machine fillings that are aimed for at FCC-ee. It is therefore important to avoid continuous measurements on the polarized pilot bunches, limiting those periods to the time required to depolarize a bunch by observing the loss of polarization. There is potentially also margin for a reduction in laser power.

\subsubsection{Polarimeter Summary}

A ${\rm e^+e^-}$ polarimeter based on inverse Compton scattering with simultaneous detection of the scattered e+/e- and of the Compton photons provides a powerful and redundant way to measure the transverse polarization with an accuracy of 1\% every second. The suggested apparatus will be able to measure in addition the beam energy, the longitudinal polarization and the beam size at the location of the laser-beam interaction.

The statistical accuracy of direct beam energy determination from the e+/e- distribution is at the level of $\Delta E/E < 100$~ppm within a 10~s measurement time. Sources of systematic errors however require additional studies. 

Once the resonant depolarization (RDP) is performed regularly, frequent cross-calibrations of the spectrometer can be made by comparison with the RDP result; this, combined with the measurements of energy differences between e+ and e- at the interaction points would provide powerful  cross-checks of the energy model.

\subsection{Resonant depolarization process}
\label{sec:depol}
The value of the spin tune may be determined with a fast RF depolarizer when the depolarizer frequency is tuned to the value of the fractional spin tune. See Section 12 for a detailed discussion about the kind of spin tune that is measured.  In practice the depolarizer frequency is swept over a narrow frequency interval. For well tuned kick strength and frequency sweep rate, the polarization may be destroyed or even flipped in sign when the fractional spin tune is within the sweep range. This procedure was first discussed in~\cite{Arnaudon:1992rn,Arnaudon:1994zq}.

The synchrotron tune $Q_s$ plays an important role in the spin dynamics at FCC-ee. $Q_s=0.05$ at \PZ corresponds to a reasonable compromise between requirements for high luminosity and polarization.

The modulation of the spin precession frequency  by  synchrotron oscillations can be characterized by the synchrotron modulation index $\xi=\Delta \nu/Q_s = \nu_0 \sigma_\delta/Q_s$, where $\sigma_\delta$ is the r.m.s. relative energy spread $\delta=\Delta E/E$ of the beam. For the LEP parameters at \PZ, with $\nu_0=103.4$, $Q_s=0.065$ and $\sigma_\delta=0.0007$, the index was small, $\xi = 1.1$. Such a low value is well suited for resonant depolarization; it implies that the central resonance frequency peak and all the synchrotron side bands were narrow, not overlapping each other.

\begin{figure}[bth]
\centering
\includegraphics[width=0.48\linewidth]{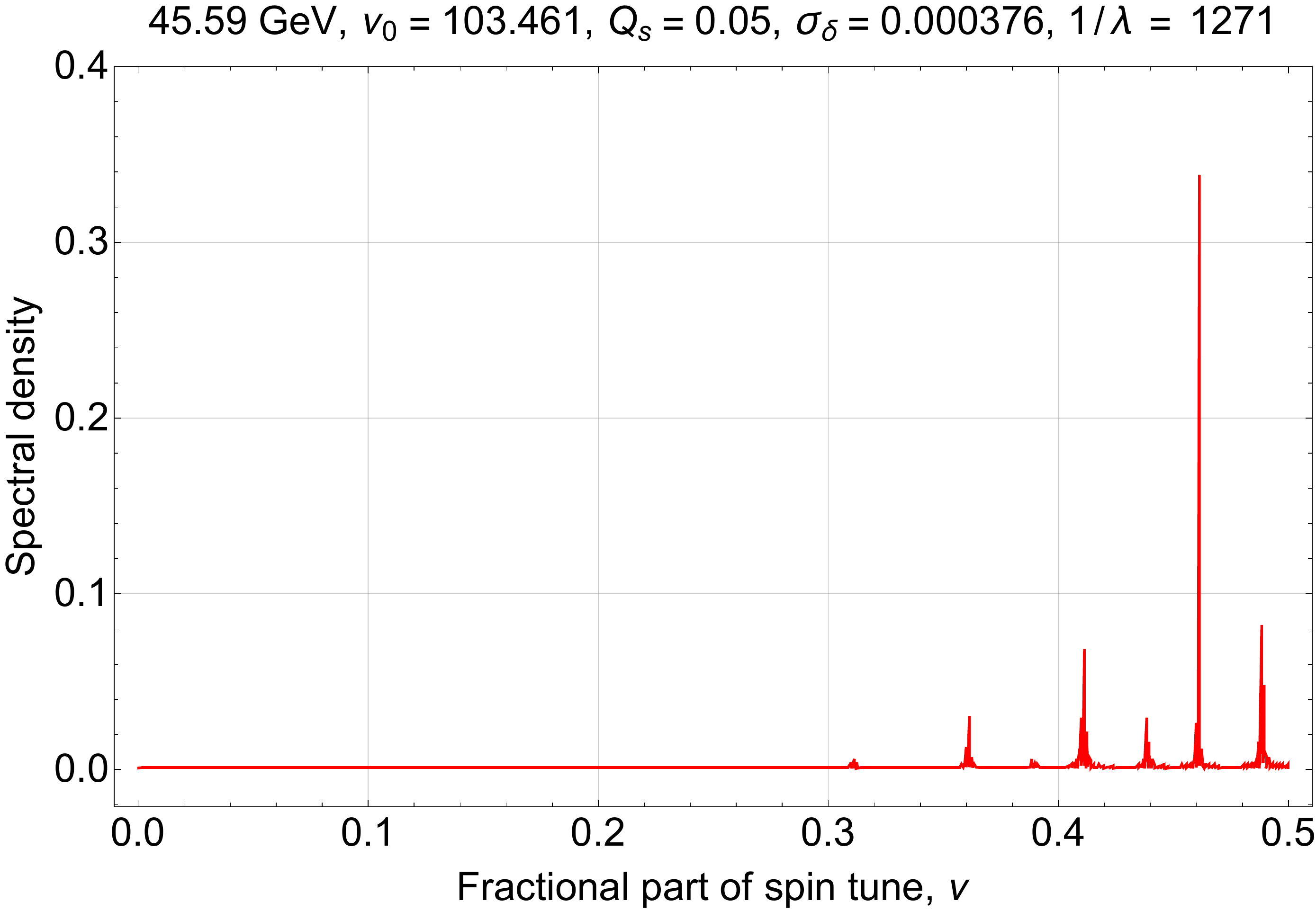}\hfill\includegraphics[width=0.48\linewidth]{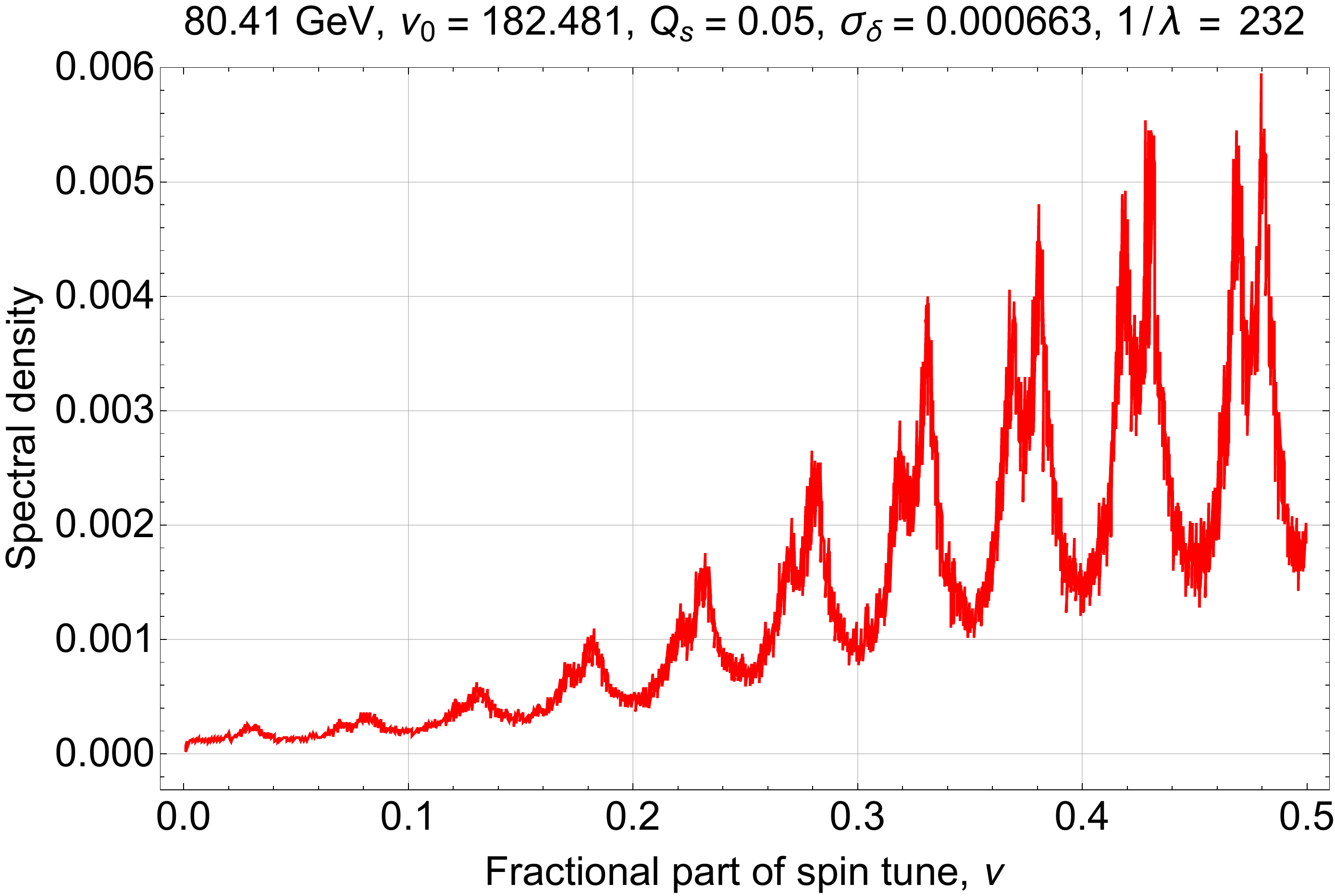}
\caption{Fourier spectra of the single particle spin motion at 45.5 (left) and \SI{80}{\giga\electronvolt} (right), obtained for 32768 turns of free spin precession, perturbed by random energy fluctuations due to emission of SR quanta. The highest peak corresponds to the fractional part of the spin tune {$\nu_0$}, it is surrounded by the first synchrotron side bands $\nu=\nu_0 \pm Q_s$. The side bands are due to the modulation of the spin-precession frequency by synchrotron oscillations. The double peaks in the figure on the right are due to the mirroring of the synchrotron sideband peaks around 0.5 by the Fourier transform.}\label{fig:depol_process_spectral_density}
\end{figure}

For FCC-ee at \PZ the modulation index is also favourable for pilot bunches. So with a relative energy spread of $\sigma_\delta = 4\times 10^{-4}$ the modulation index is $0.83$ which is even smaller than at LEP thanks to the lower energy spread. But the situation degrades significantly with increasing beam energy. Already at the \PW threshold, for the pilot bunch parameters of $\nu_0=182.41$, $\sigma_\delta = 6.6\times 10^{-4}$, $Q_s=0.05$, the synchrotron modulation index reaches $\xi=2.4$. The spin precession spectrum looses the sharp discrete line structure observed at the \PZ, and becomes more chaotic and continuous, with a wide central peak and sidebands. This is illustrated by the numerical simulation of the spin motion, produced by the spin tracking code of Ivan Koop~\cite{koop1,Koop:2019rjk}, with results presented in Fig.~\ref{fig:depol_process_spectral_density}.
At \SI{80}{\giga\electronvolt} the peaks in the spectrum are much wider than at \SI{45}{\giga\electronvolt}. This is the direct consequence of a large value of the synchrotron modulation index $\xi=2.4$ at \PW threshold.

Radiation damping of synchrotron oscillations is an essential ingredient of these simulations. At the \PW thrshold the damping rate is $\lambda=1/232$ per turn, and the width of the spin precession peak is proportional to the $\lambda$. The spectrum line shape can be described approximately by:
\begin{equation}
f(\nu) = A\cdot \frac{\Delta}{\sqrt{\Delta^2 + (\nu-\nu_0)^2}},
\end{equation}
with 3 free parameters $A$, $\Delta$ and $\nu_0$. From spin tracking results at various energies, one obtains an approximate power law dependence of $\Delta$ on the equilibrium beam parameters
\begin{equation}
\Delta = 0.0035 \frac{\lambda}{0.000686}\cdot\left( \frac{E [\si{\giga\electronvolt}]}{80}\cdot \frac{\sigma_\delta}{0.000663} \right)^{\!2.5}\!\cdot \left( \frac{0.05}{Q_s} \right)^{\!3}.
\end{equation}

At $\SI{80}{\giga\electronvolt}$ the peak width reaches $\Delta=0.0035$ which is approximately 10 times larger than what would be required for an accurate and reliable RDP. Taking into account the energy dependence of all input parameters, $\lambda \propto E^3$, $\sigma_\delta \propto E$, $Q_s \simeq \textnormal{const}$, line width $\Delta$ scales with beam energy as
\begin{equation}
\Delta \sim E^8.
\end{equation}
This steep scaling law prevents RDP at or above \SI{100}{\giga\electronvolt} per beam for FCC-ee. The spins will not resonate beyond this energy limit, at least with a synchrotron tune of $Q_s=0.05$. The higher order synchrotron side bands  overlap and beam depolarization will occur at any frequency.

\begin{figure}[tbh]
	\centering
	\includegraphics[width=1\linewidth]{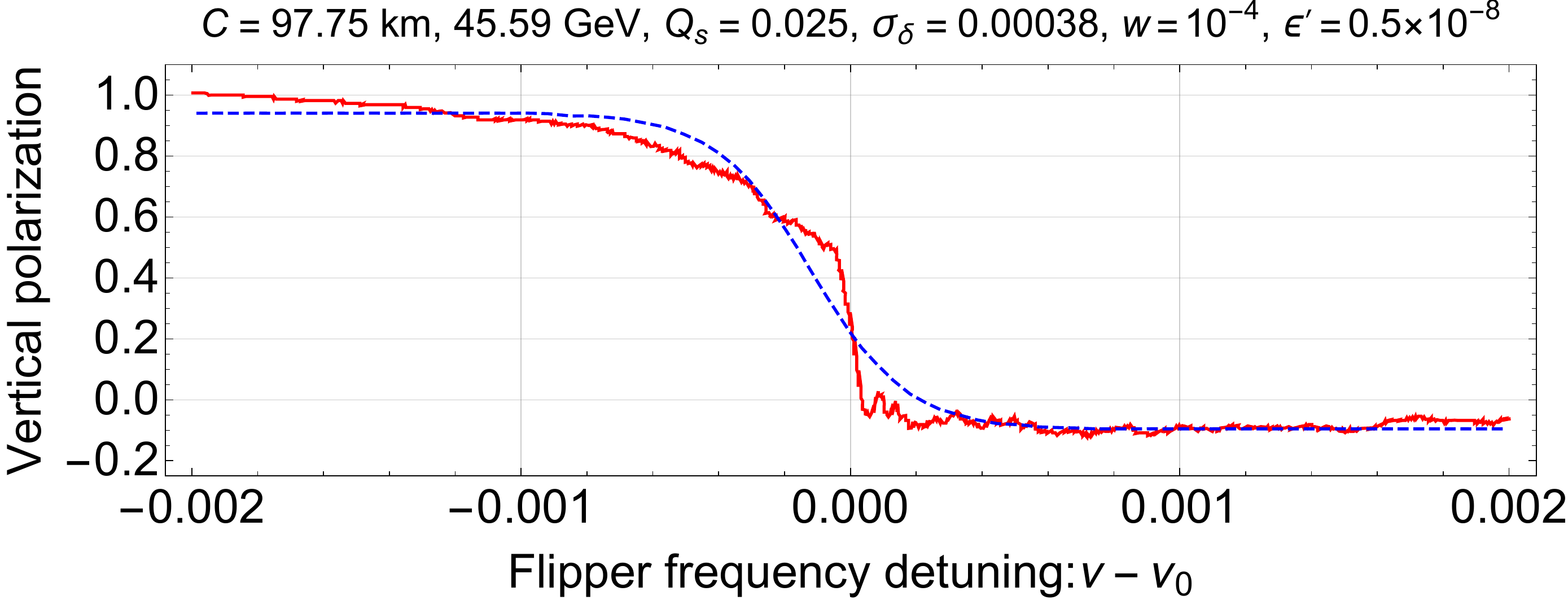}
	\caption{Simulation of a frequency sweep with the depolarizer on the \PZ pole showing a very sharp depolarization at the exact spin tune value.}\label{fig:depol_process_pol_depol_z}
\end{figure}

\begin{figure}[tbh]
	\centering
	\includegraphics[width=0.7\linewidth]{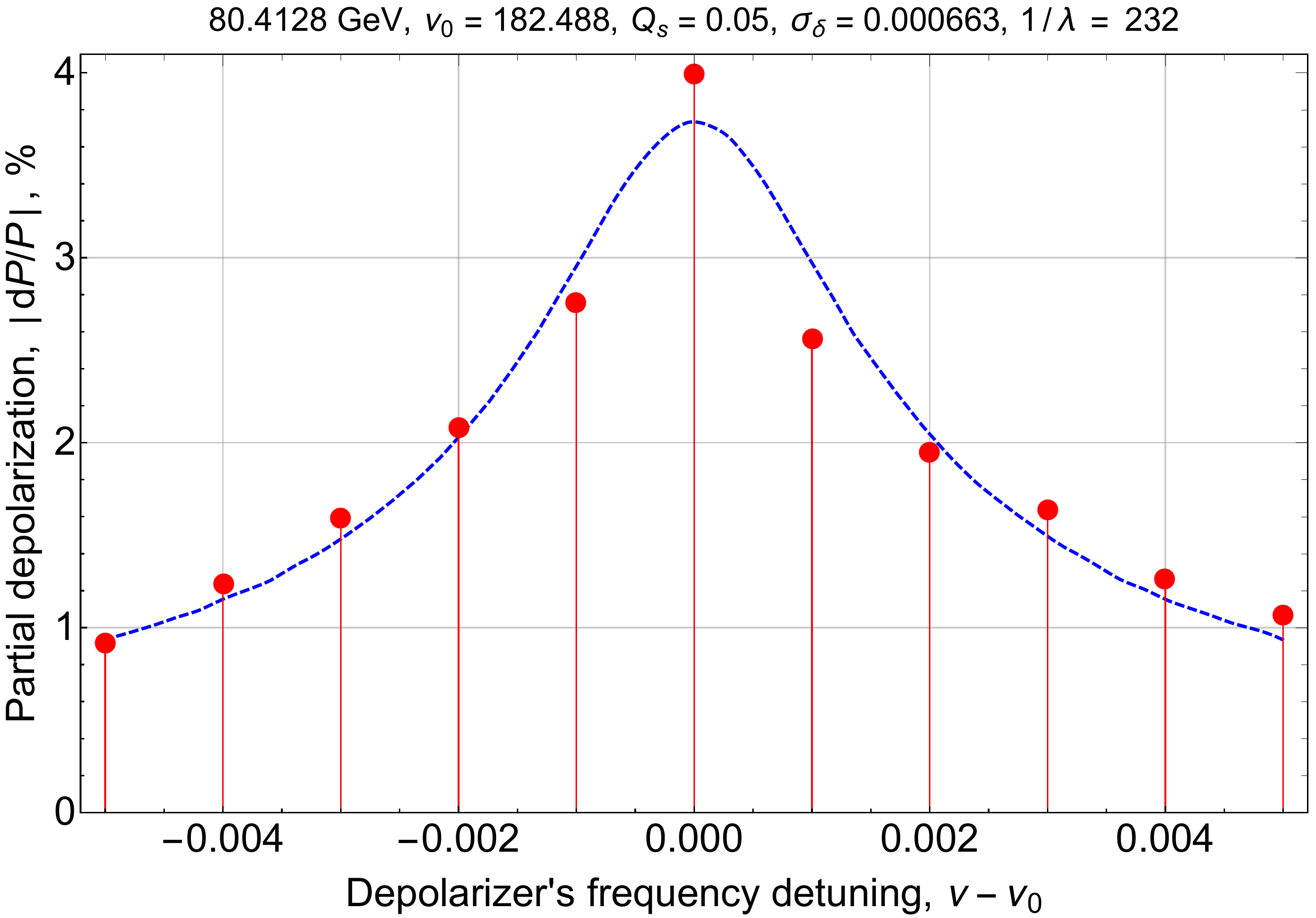}
	\caption{At the WW threshold the energy spread is too large to perform a wide frequency sweep, progressing in little steps is required.}\label{fig:depol_process_pol_depol_w}
\end{figure}

Figures~\ref{fig:depol_process_pol_depol_z} and \ref{fig:depol_process_pol_depol_w} present simulation results for the RDP process at the \PZ pole and at the \PW threshold.
On the \PZ pole the beam depolarization occurs sharply at the exact spin tune value. At the \PW threshold a partial beam depolarization with progressing small steps in the depolarizer's frequency was applied. The spin resonance can be located, but it is not clear if the depolarization is large enough to precisely locate the resonance. 

\subsubsection{RF Depolarizer}

The LEP RF depolarizer (RF magnet) consisted of a magnetic kicker with a maximum integrated field of $\sim 0.4 \times 10^{-3}$\,Tm. The typical field strength for good depolarization was $0.2 \times 10^{-3}$\,Tm, corresponding to an orbit kick of $1.2 \mu$rad. The kicker pulse had a FWHM of around 1\,$\mu$s. Individual LEP bunches could be kicked once per turn by the kicker.

The LHC transverse feedback system is a good example of a modern feedback system for high energy beams. Four 1.5\,m long electrostatic kicker magnets per beam and per plane are used to provide a kick of $2 \mu$rad at 450\,GeV with a useful bandwidth ranging from 1\,kHz to 1\,MHz. The system is capable of acting on single bunches spaced by 25~ns. This system would provide adequate strength and bandwidth over the entire FCC-ee energy range even with a fourth of the strength installed in the LHC. With a lower strength requirement it should be possible to increase the bandwidth to match the FCC-ee bunch spacing at the Z pole.

\section{Sources of beam energy changes and systematic errors}
\label{sec:energy-systematics}

\subsection{Spin precession and beam energy}
\label{sec:RDP}

The motion of the electron spin vector $\vec{S}$ in external electromagnetic fields $\vec{E}$, $\vec{B}$ obeys the Thomas-Bargmann-Michel-Telegdi equation \cite{Thomas:1927, BMT}(Gaussian unit system)
\begin{eqnarray}
\displaystyle\frac{\mbox{d}\vec{S}}{\mbox{d}t}&=&\vec{W}\times\vec{S}\,, \nonumber \\
\vec{W}&=&-\Biggl(\frac{q_0}{\gamma}+q'\Biggr)\vec{B}_\perp-\frac{q}{\gamma}\vec{B}_\parallel-
	\Biggl(\frac{q_0}{\gamma+1}+q'\Biggr)\frac{\vec{E}\times\vec{v}}{c}\,,
\label{eq:BMT}
\end{eqnarray}
where $t$ is the time in the laboratory frame, $\vec{B}_\parallel=(\vec{B}\cdot\vec{v})/\vec{v}$, $\vec{B}=\vec{B}_\perp+\vec{B}_\parallel$, $\vec{v}$ is the particle velocity, $\gamma=1/\sqrt{1-v^2/c^2}$ is the Lorentz factor, $q_0$ and $q'$  are the normal and anomalous parts of the gyromagnetic ratio $q=q_0+q'=e/mc+q'$, $\vec{W}$ is the spin precession frequency vector, $e$ and $m$ are the electron charge and mass respectively, $c$ is the speed of light -- this is similar to equation \ref{eq:Omega_Thomas} with the correspondence $q'/q_0= a = (g-2)/2$. The revolution frequency of the particle $\vec{\Omega}_L$ (Larmor frequency) in the external electromagnetic field is
\begin{equation}
\displaystyle
\vec{\Omega}_L=-\frac{q_0}{\gamma}\vec{B}_\perp-
	\frac{q_0\gamma}{\gamma^2-1}\frac{\vec{E}\times\vec{v}}{c}\,.
\label{eq:Larmor}
\end{equation}
In the absence of electric and longitudinal magnetic fields the difference between the Larmor and spin precession frequencies is defined by the anomalous part of gyromagnetic ratio. 

The resonant depolarization technique (RDP)~\cite{Bukin:1975db,Derbenev:472402} is based on the measurement of the spin precision frequency $|\vec{W}|$. Depolarization of the beam can be achieved when the frequency of the external electromagnetic field $\Omega_D$ satisfies the resonant condition
\begin{equation}
W\pm\Omega_D=\Omega_L\cdot n\,,
\label{resonant_condition}
\end{equation}
where $n$ is an integer.

\subsection{Average beam energy determination from the spin precession frequency}
In an accelerator with a perfectly flat orbit (no electric, radial and longitudinal magnetic fields on the reference trajectory) the spin precesses around the direction of the guiding field. The instantaneous (local) spin precession frequency is
\begin{equation}
\displaystyle
W_{inst}=\Omega_L\cdot(1+\gamma\frac{q'}{q_0})\,.
\label{eq:instant_spin_freq}
\end{equation}
The integrated spin precession over one revolution is
\begin{equation}
\displaystyle
W= \frac{1}{2\pi}\oint \left(\frac{q_0}{\gamma}+q'\right)B_\perp(\theta) d\theta
\displaystyle
 =\Omega_0\cdot\left(1+\frac{q'}{q_0}\frac{\left<B_\perp\right>}{\left<B_\perp/\gamma\right>}\right)\,,
\label{eq:spin_freq}
\end{equation}
where $\theta$ is azimuthal coordinate along the closed orbit, $\Omega_0=q_0 \left<B_\perp/\gamma\right>$ is the revolution frequency, the brackets $\left<\right>$ designate $(1/2\pi)\oint...d\theta$. To first order the ratio $\left<B\right>/\left<B/\gamma\right>$  equals the average Lorentz factor $\left<\gamma\right>$. Neglecting higher orders equation (\ref{eq:spin_freq}) becomes 
\begin{equation}
W=\Omega_0(1+\gamma q'/q_0) \;. 
\end{equation}
Introduction of the spin tune $\nu_0 = W_0/\Omega_0-1=\gamma q'/q_0$ leads to the expression for the average energy $E$
\begin{equation}
E=\nu_0\frac{mc^2}{q'/q_0}\,.
\end{equation}
The ratio of the anomalous and normal parts of the gyromagnetic ratio is $q'/q_0=1.15965218091\cdot10^{-3}\pm0.26\cdot10^{-12}$. The electron rest mass is $mc^2=0.5109989461\pm0.31\cdot10^{-8}$\,MeV \cite{PDG:2016}. Hence, the beam energy is given by
\begin{equation}
E[MeV]=440.64846\cdot\nu_0\,,
\label{eq:nu_E}
\end{equation}
with a relatve accuracy of
\begin{equation}
\displaystyle
\frac{\Delta E}{E}=\sqrt{\left(\frac{\Delta (mc^2)}{mc^2}\right)^2+
								 \left(\frac{\Delta (q'/q_0)}{q'/q_0}\right)^2}\simeq
								 \frac{\Delta (mc^2)}{mc^2}=7.8\cdot10^{-8}\,.
\end{equation}

Because the depolarization process requires many turns, the depolarization time is larger than the damping time of synchrotron oscillations, which itself is larger than the revolution and the synchrotron oscillation periods. Thus the resonant depolarization technique measures the spin precession frequency averaged over a large number of turns and over the ensemble of particles in a bunch. The measured spin precession frequency is defined by the integral of the guiding field.

\subsection{Sources of beam energy variation}\label{seq:sources-e-changes}

This section discusses the main sources of average energy variations that are measurable by RDP.

\subsubsection{Bending field drifts}\label{sec:energy-bending-drifts}

The momentum of a particle in a machine is defined by the integral of the transverse guiding field as 
\begin{equation}\label{eq:def-energy-bending}
P =  \frac{Z e}{2 \pi} \oint B_\perp(s) ds = Z \cdot 44.7 \mathrm{MeV/c/Tm}  \oint B_\perp(s) ds \; ,
\end{equation}
where $Ze$ is the particle charge, for e+e- colliders $Z = 1$.
For a  well aligned circular machine where the beam is  centred on the mean orbit (see below), the integral of the magnetic field is dominated by the bending dipole field. Any change of the dipole field generates therefore a change in momentum and energy of the beam. Such changes can be due to power supply drifts, tunnel and magnet temperature drifts affecting the integrated field or external perturbations as observed by LEP with earth currents generated by railways~\cite{Assmann:1998qb}. Those effects may have implications on the machine stability but do not impact the beam energy accuracy provided the beam energy is measured at sufficiently short time intervals. On the basis of the experience of LEP and the expected time scale of such bending field changes, measurement intervals of a few minutes as foreseen in the operational scenario of Section~\ref{sec:running-scheme} seem to be adequate.

\subsubsection{Circumference drifts}\label{sec:energy-circumf-drifts}

The average beam energy depends on the length of the beam orbit $C$ and on the actual machine circumference $C_c$ defined as the orbit length for which the beam is centred on average on the quadrupoles. The latter orbit is also referred to as the ``central orbit''. The relative energy change induced by a difference between $C$ and $C_c$ is
\begin{equation}\label{eq:en-change-orbitlength}
\frac{\Delta E}{E_0} = =-\frac{1}{\chi}\frac{C - C_c}{C_c} \,.
\end{equation}
For FCC-ee the momentum compaction factor $\chi$ may be as low as $\chi \simeq 10^{-5}$. If the orbit length $C$ corresponds to a RF frequency $f_{RF}$ and the central orbit to a frequency $f_{RFc}$ the energy change may also be expressed as
\begin{equation}\label{eq:en-change-frf}
\frac{\Delta E}{E_0} = =\frac{1}{\chi}\frac{f_{RF} - f_{RFc}}{f_{RFc}} \,.
\end{equation}
A similar expression is also obtained in terms of revolution frequency $f_{rev}$ since
\begin{equation}\label{eq:frf-frev}
f_{rev} = f_{RF}/h
\end{equation}
where $h$ is the RF harmonic number.

Large machines like FCC-ee will be subject to earth tides with  circumference changes by $\Delta C \simeq \pm 2$\,mm for $C$ of 100\,km~\cite{TIDENIM,Wenninger:387157, PhysRevAccelBeams.20.081003}. For a momentum compaction factor of $\chi \simeq 10^{-5}$ the corresponding energy changes reaches  $\pm 2\cdot 10^{-3}$ or $\pm 90$\,MeV around the Z resonance. Besides tides slow long-term geological deformations may lead to changes with a similar order of magnitude. For LEP such slow changes were even larger than for tides~\cite{Assmann:2004gc}.

The circumference changes can be compensated with a slow feedback on the RF frequency based on the mean radial orbit position measured using the arc BPMs to maintain the beam radially stable, as it is done at the LHC~\cite{PhysRevAccelBeams.20.081003}. This feedback should reduce the radial changes to the micrometer level (subject to the long term stability of the BPM electronics for example) or equivalently the relative energy changes of $< 10^{-5}$. Residual energy changes can be tracked using the regular energy calibrations, a periodicity of a few minutes should again be adequate.

\subsubsection{Vertical fields of orbit correctors and quadrupoles}

Horizontal orbit distortions due to vertical magnetic fields do not change  equation \eqref{eq:nu_E} between average beam energy and average spin precession frequency, but they affect the value of the average energy according to equation \eqref{eq:def-energy-bending}) and through the change of the orbit length~\cite{Assmann:2004gc}. Therefore, the stability of the energy between the calibrations depends on field stability in the elements defining the horizontal orbit, i.e. mainly the alignment of quadrupole magnets and the settings of horizontal orbit corrector magnets.

The orbit curvature distortion $\Delta k$ will change the energy by \cite{Assmann:2004gc,Chao:384825, Errors}
\begin{equation}
\displaystyle\frac{\Delta E}{E}=-\frac{1}{\chi C}\oint\Delta k\eta_x ds\,,
\label{eq:continuous_cor}
\end{equation}
where $\eta_x$ is the dispersion function, $C$ is the circumference, $\chi$ is the momentum compaction factor. A point-like corrector with deflection $\theta_1=\oint \Delta k ds=\oint (\Delta B_y/B\rho)ds$  changes the beam energy according to
\begin{equation}
\displaystyle \frac{\Delta E}{E_0}=-\frac{\theta_1\eta_{x}}{\chi C}\,.
\label{eq:hor-corrector-1}
\end{equation}
where $\eta_{x}$ is the dispersion function at the corrector position. For the full orbit the energy change can be estimated by summing over all correctors (or kicks),
\begin{equation}
\displaystyle \frac{\Delta E}{E_0}=-\frac{\sum_i \theta_{1,i} \eta_{x,i}}{\chi C}\,,
\label{eq:hor-corrector-all}
\end{equation}
This contribution may be estimated in terms of the rms orbit deviation  by converting the  corrector kicks into the standard deviation of the horizontal orbit $\sigma_x$, which yields an rms energy shift of
\begin{equation}
\sigma\left(\frac{\Delta E}{E_0}\right)=\frac{2\sqrt2\sin(\pi\nu_x)}{\chi C} \frac{\left<\eta_x\right>}{\left<\beta_x\right>} \sigma_x\,,
\end{equation}
where $\left<\eta_x\right>$ and $\left<\beta_x\right>$ are averages over the ring of the dispersion and horizontal beta function  respectfully.
Substitution of $\nu_x=269.138$, $C =97756$\,m, $\chi=1.5\cdot 10^{-5}$, $\left<\eta_x\right>=0.19$\,m and $\left<\beta_x\right>=143.59$\,m gives
\begin{equation}
\sigma\left(\frac{\Delta E}{E_0}\right)=-1.2\cdot10^{-3}[m^{-1}]\cdot \sigma_x[m]\,.
\end{equation}
To achieve an energy stability of $10^{-6}$ the horizontal orbit between calibrations must be stable within $\sigma_x=0.8$\,mm which is as well above the requirements for stable operation ($\sigma_x < 0.1$\,mm) and the achievements of a modern orbit feedback and BPM systems ($\sigma_x < 0.01$\,mm).

Quadrupole shifts in the horizontal plane also generate an effective horizontal correctors. The corresponding energy shift is obtained from equation \eqref{eq:continuous_cor}, where $\Delta k=K_1\Delta x$ with $K_1$ being the quadrupole strength and $\Delta x$ the quadrupole displacement.
Table \ref{tbl:QuadHorizontalStability} summarizes values of the quadrupole shifts in the horizontal plane to ensure an energy stability of $\Delta E/E_0=10^{-6}$ between the calibrations. The last line in the table is given for the 720 cell quadrupoles QF4.
\begin{table}[htbp]
\caption{Quadrupoles shifts for a relative energy change of $10^{-6}$.\vspace{3mm}}
\label{tbl:QuadHorizontalStability}
\centering
\begin{tabular}{|l|c|}
\hline
Quadrupole	& $\Delta x$ [$\mu$m]										\\ \hline
QC7.1		& $200$	 							\\ \hline
QY2.1		& $76$	 						\\ \hline
QFG2.4		& $160$	 						\\ \hline
QF4.1	& $140$	 						\\ \hline
QG6.1		& $35$	 						\\ \hline
QF4 family	& $\Delta x/\sqrt{720}=5$	 \\ \hline
\end{tabular}
\end{table}

The beam orbits of FCC-ee will be stabilized over many hours by an orbit feedback with target stabilization levels of tens of micrometers rms. Such requirements match the achieved orbit stability at the LHC or at synchrotron light sources. The deflections applied by the orbit feedback will effectively compensate the impact of the quadrupole misalignments and ensure sufficient energy stability between the regular minute-interval energy measurements.

\subsubsection{Vertical fields of sextupoles and betatron oscillations}

Horizontal orbit distortions also affect the beam energy through the contribution of the machine sextupoles to the integrated magnetic field. This effect is equivalent to the effect of the horizontal orbit correctors and quadrupoles described in the previous section. The kick due to a sextupole is given approximately by $-K_2 L_s x^2$ where $ K_2=\frac{1}{B\rho}\frac{\partial^2B_y}{\partial x^2}$ is the sextupole strength, $L_s$ the sextupole length and $x$ the orbit offset in the sextupole. For orbit excursions of 0.2\,mm, the kicks  reach values of 0.2\,$\mu$rad, more than one order of magnitude smaller that typical kicks used to correct the orbit.

Particles with horizontal betatron oscillations experience the transverse field in the sextupoles $B_\perp\propto x^2$. The average over the bunch particles does not necessarily cancel out, thereby inducing an energy (and spin tune) shift and a widening of the distribution of spin precession frequencies.
The simplest estimate for the precession-frequency or energy shift is \cite{Derbenev:1107974}
\begin{equation}
\frac{\Delta\nu}{\nu}=\frac{\Delta E}{E}=-\frac{1}{2\pi}\oint \left(\varepsilon_x\beta_x(s)+\eta_x(s)^2\sigma_\delta^2 \right)K_2(s) ds\,,
\end{equation}
For the lattice FCCee\_z\_213\_nosol\_18 the resulting energy shift is
\begin{equation}  \label{eq:sext-de-shift}
\frac{\Delta E}{E} \simeq -3\cdot10^{-6}
\end{equation}
which is clearly sizable given the target accuracy of $10^{-6}$.

A more accurate estimate~\cite{TuneSpread2} also includes the path lengthening which shifts the particle energies because the RF frequency is constant:
\begin{equation}
\begin{split}
-\chi\frac{\Delta E}{E}&= \varepsilon_x
\left<\left(K_0(K_0^2+2K_1)+\frac{K_2}{2}\right)\beta_x\eta_x+ \left(1+K_0\eta_x\right)\gamma_x
\right>+ \\
&+ \frac{\sigma_\delta^2}{2}
\left<\left(K_0(K_0^2+2K_1)+\frac{K_2}{2}\right)\eta_x^3+
\frac{1}{2}\left(1+K_0\eta_x\right)\eta_x'^2+\eta_x^2\left(\frac{2K_0}{R}+K_1\right)\right> \,,
\end{split}
\end{equation}
where $R=C/2\pi$ is the average orbit radius, $K_0=\frac{B_y}{B\rho}$ is the orbit curvature, $K_1=\frac{1}{B\rho}\frac{\partial B_y}{\partial x}$ is the quadrupole strength, $\beta_x, \gamma_x, \eta_x, \eta_x'$ are Courant-Snyder and dispersion functions.
For the parameters of lattice FCCee\_z\_213\_nosol\_18 this expression results in a systematic shift essentially identical to the result of equation~\eqref{eq:sext-de-shift}.

\subsection{Systematic errors of the average beam energy determination}\label{sec:energy-syst-errors}

This section discusses effects that break the correspondence between the average beam energy and the precession frequency determined by RDP. Such errors may constitute irreducible systematic errors on the  energy determination by RDP. In some cases the error can not be measurable directly but can be estimated theoretically with sufficient accuracy.

\subsubsection{Energy dependent momentum compaction}
The momentum compaction factor $\chi$ relates the  revolution frequency $\Omega$ to its energy $E=E_0(1+\delta)$ where $\delta=(E-E_0)/E_0$ is relative energy deviation according to $\Omega=\Omega_0(1-\chi\delta)$. The chromaticity of the momentum compaction factor introduces asymmetries into the particle revolution frequency distribution, and breaks relation \eqref{eq:nu_E}. For a momentum compaction chromaticity $\chi_1$, defined as $\chi=\chi_0+\chi_1\delta$, the synchrotron-oscillation equation is modified according to
\begin{equation}
\ddot{\delta}=-\omega_{syn}^2\delta-\omega_{syn}^2\displaystyle\frac{\chi_1}{\chi_0}\delta^2\,,
\label{eq:synch-1}
\end{equation}
The spin precession frequency may be expressed as 
\begin{equation}
W-\Omega_0=\gamma\Omega\frac{q'}{q_0}=
\gamma_0\Omega_0\frac{q'}{q_0}(1+\delta(1-\chi_0)-\delta^2(\chi_0+\chi_1))\,,
\end{equation}
where $\Omega_0$ is revolution frequency of the particle with equilibrium energy and the term $\gamma\Omega$ was expanded in terms of $\delta$ neglecting terms beyond $\delta^2$.

Solving \eqref{eq:synch-1} with the help of perturbation theory  yields
\begin{equation} \label{eq:spin-freq-nl-alphac}
\begin{array}{l}
\displaystyle
\delta=a\cos(\omega t)-\frac{\chi_1}{\chi_0}\frac{a^2}{2}+
       \frac{\chi_1}{\chi_0}\frac{a^2}{6}\cos(2\omega t)+
       \left(\frac{\chi_1}{\chi_0}\right)^2\frac{a^3}{48}\cos(3\omega t)\,, \\
\displaystyle
\omega=\omega_{syn}-\frac{5}{12}\omega_{syn}\left(\frac{\chi_1}{\chi_0}\right)^2a^2\,,
\end{array}
\end{equation}
where $\omega_{syn}$ and $a$ are the unperturbed frequency and amplitude of the synchrotron oscillations. 
Assuming a Gaussian beam energy distribution with standard deviation $\sigma_\delta E_0$, the average amplitude values in the first order are
$\left<a^2\right>=2\sigma_\delta^2$, $\left<a^4\right>=8\sigma_\delta^4$.
Then the expected value and dispersion of the momentum deviation are 
\begin{align} \label{eq:syncho-osc-perturbation-means}
\left<\delta\right>=-\frac{\chi_1}{\chi_0}\sigma_\delta^2\,, & \quad \left<\delta^2\right>=\sigma_\delta^2\,,
\end{align}
and by substituting in Eq.~(\ref{eq:spin-freq-nl-alphac}) the average spin precession frequency becomes
\begin{equation}
\left<W-\Omega_0\right>_\delta=\gamma_0\Omega_0\frac{q'}{q_0}\left(1-\chi_0\sigma_\delta^2-\frac{\chi_1}{\chi_0}\sigma_\delta^2\right)\,.
\label{eq:spin-fr-2}
\end{equation}
Hence the energy obtained using the measured spin precession frequency (Eq.~\eqref{eq:nu_E}) is
\begin{equation}
E_{meas}=E_0\left(1-\frac{\chi_1}{\chi_0}\sigma_\delta^2-\chi_0\sigma_\delta^2\right)\,,
\label{eq:e-meas-1}
\end{equation}
while the average beam energy is
\begin{equation}
\left<E\right>=E_0\left(1-\frac{\chi_1}{\chi_0}\sigma_\delta^2\right)\,.
\end{equation}

\begin{figure}[htb]
\centering\includegraphics[width=0.8\linewidth]{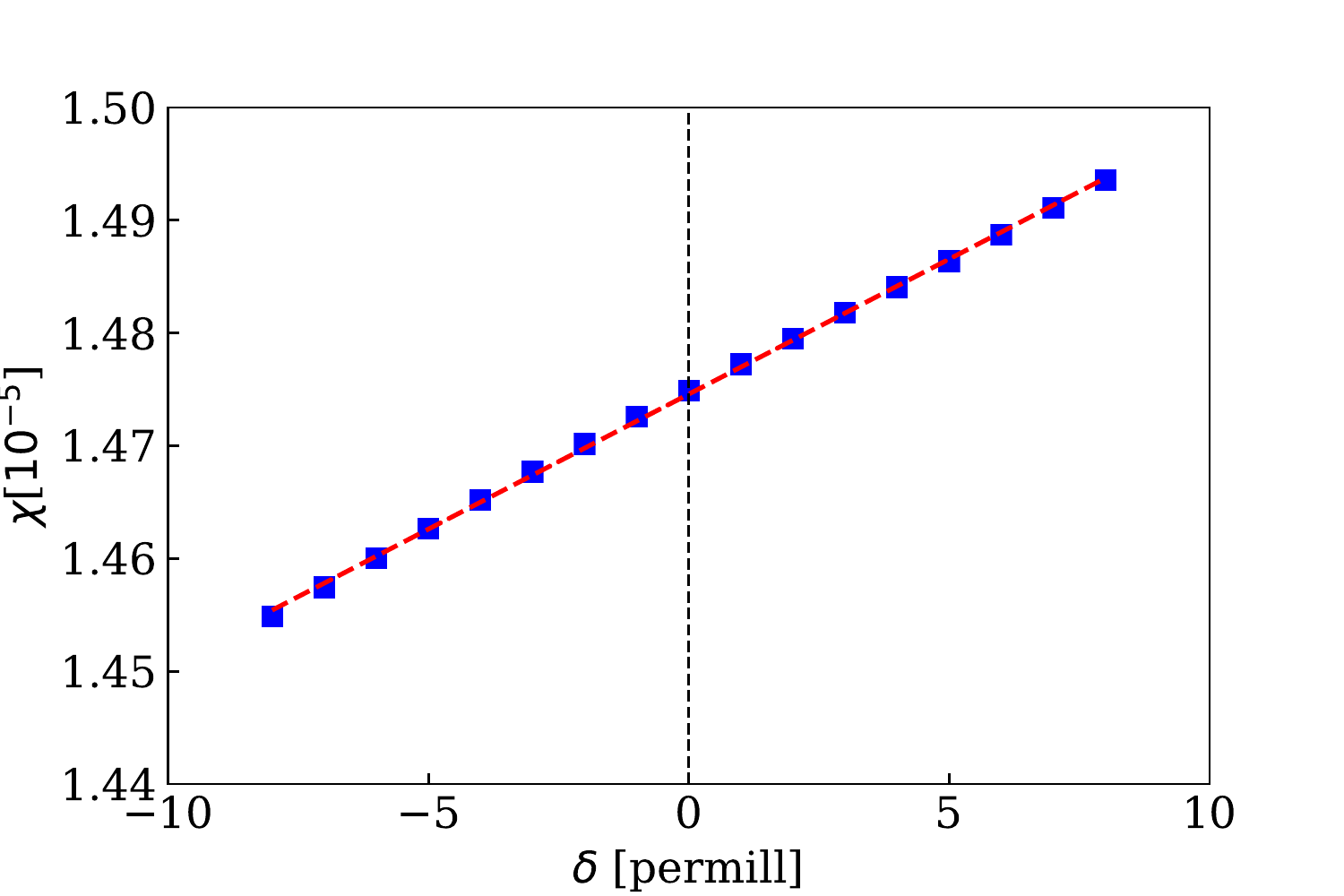}
\caption{ \label{fig:alpha-dp} Dependence of the momentum compaction factor $\chi$ on the relative energy offset $\delta$ for the the  Z lattice FCCee\_z\_213. The slope $\chi_1 = d\chi/d\delta$ is $2.4 \cdot 10^{-5}$ and $\chi_1/\chi_0 \simeq 1.6$.}
\end{figure}

The difference between average and  measured energies is
\begin{equation}
\frac{\left<E\right>-E_{meas}}{E_{meas}}=\displaystyle\frac{\chi_0\sigma_\delta^2}{1-\frac{\chi_1}{\chi_0}\sigma_\delta^2-\chi_0\sigma_\delta^2}=2\cdot 10^{-12}\,.
\end{equation}
where we used  $E_0=45.6$\,GeV, $\chi_0=1.5\cdot10^{-5}$, $\chi_1=2.4\cdot10^{-5}$, $\sigma_\delta=3.8\cdot10^{-4}$ (parameters of FCCee\_z\_213 lattice) for the non-colliding bunches used for energy calibration. The dependence on $\chi$ on the relative energy offset is presented in Fig~\ref{fig:alpha-dp}. Colliding bunches have a larger energy spread $\sigma_{\delta,bs}=1.3\cdot10^{-3}$ due to beamstrahlung at the IP \cite{Bogomyagkov:beamstrauhlung}. This leads to an energy difference between colliding and non-colliding bunches of
\begin{equation}
\displaystyle\frac{\left<E\right>_{col}-E_{meas}}{E_{meas}}=\displaystyle\frac{\chi_0\sigma_\delta^2+\frac{\chi_1}{\chi_0}(\sigma_\delta^2-\sigma_{\delta,bs}^2)}{1-\frac{\chi_1}{\chi_0}\sigma_\delta^2-\chi_0\sigma_\delta^2} \simeq 2.5\cdot 10^{-6}\,.
\end{equation}
This uncertainty can be constrained by a measurement or a prediction of the $\chi_1$ derivative term. Most likely a model may have to be used as the change of $\chi$ of a reasonable interval of $\delta \simeq \pm 2 \cdot 10^{-3}$ seems too small to be easily measurable.

\subsubsection{Vertical orbit distortions}
Radial magnetic fields break relation \eqref{eq:nu_E} between beam energy and spin precession frequency. The main source of such fields are vertical alignment errors of quadrupoles which induce vertical orbit distortions, and the vertical orbit correctors that are powered to compensate the orbit errors. Several authors estimated this effect in \cite{Assmann:262009,Bogomyagkov:2018nst}. Following the later publications the average beam energy shift $\overline{\Delta\nu}$ and its uncertainty $\sigma_{\overline{\Delta\nu}}$ are given by
\begin{align}
\label{eq:vshift-1}
\overline{\Delta\nu}&=\frac{\nu_0^2}{2}\frac{\overline{\left<z^2\right>}}{Q}\sum_{k=-\infty}^{\infty} \frac{k^4}{(\nu_z^2-k^2)^2(\nu_0 - k)}\,, \\
\sigma_{\overline{\Delta\nu}}&=\frac{\nu_0^2\sqrt{3}}{2}\frac{\overline{\left<z^2\right>}}{Q}
\sqrt{2\nu_0\sum_{k=-\infty}^{\infty} \frac{k^8}{(\nu_z^2-k^2)^4(\nu_0 - k)^2(\nu_0 + k)}}\,, \\
Q&= \frac{\pi}{2\nu_z^3}\cot\pi\nu_z+\frac{\pi^2}{2\nu_z^2}\csc^2\pi\nu_z\,,
\end{align}
where $\left<\right>$ corresponds to an average over the circumference, and $\bar{\,}$ represents the average over orbits, $\nu_z$ is the vertical betatron tune and $\nu_0$ is the unperturbed spin tune. To derive this expression the following assumptions were made: the spin rotation angle is given by $\Phi(\theta) = a \gamma \theta$ where $\theta$ is the azimuth angle (i.e. no straight section) and the vertical beta function is assumed to be constant
$\beta_z=const=\left<\beta_z\right>$.

Table \ref{tbl:EnergyShiftVertical} presents tolerable vertical orbit distortions for a relative energy bias of $10^{-6}$. The latest machine simulations yield a typical vertical orbit r.m.s. $\sqrt{\overline{\left<z^2\right>}}$ of $\simeq 0.2$\,mm, well inside the tolerance limit of $10^{-6}$.
\begin{table}[htbp]
\caption{Beam energy shift and uncertainty due to vertical orbit distortions for an uncertainty of $\Delta E/E = 10^{-6}$.\vspace{3mm}}
\label{tbl:EnergyShiftVertical}
\centering
\begin{tabular}{|l|c|c|c|}
\hline
$E\,,$ GeV									& 45.6							& 78.65							& 81.3							\\ \hline	
$\sqrt{\overline{\left<z^2\right>}}$ (mm) &	0.6		&	0.28							&	0.27							\\ \hline
$\nu_z$									&	269.22						&	269.2						&	269.2						\\ \hline
$\Delta E$ (keV)							& -31							& -54							& -56							\\ \hline
$\sigma_{\Delta E}$ (keV)			&  46							& 82								& 85								\\[3pt] \hline
\end{tabular}

\end{table}

\subsubsection{Longitudinal fields}
Uncompensated longitudinal detector fields shift the spin recession frequency without affecting the beam energy\cite{Longitudinalfield} and break the relation \eqref{eq:nu_E} between energy and spin tune. The spin tune shift $\Delta\nu_0$ of an uncompensated solenoid is
\begin{equation}
\Delta\nu_0=\frac{\varphi^2}{8 \pi}\cot(\pi\nu_0)\approx\frac{1}{8\pi}\cot(\pi\nu_0)\left(\frac{\Delta B_c}{B_0}\frac{2B_0L_c}{B\rho}\right)^2\approx 2\times 10^{-9}\,,
\end{equation}
where $\varphi=(B_0L_0+2B_cL_c)(1+q'/q_0)/B\rho$ is the spin rotation angle by an imperfectly compensated longitudinal field of $B_0=2$\,T and where $\Delta B_c=0.1$\,T is the error of the compensating solenoid field. $L_c=0.75$\,m is the length of the compensating solenoid, $B\rho=152.105\,\mbox{T}\cdot\mbox{m}$ is the beam rigidity for a beam energy of 45\,GeV and $\nu_0=103.484$ is the unperturbed spin tune.
The corresponding beam energy error is
\begin{equation}
\frac{\Delta E}{E_0}=\frac{\Delta \nu\cdot 440.65}{E_0}\approx 2\times 10^{-11}
\end{equation}
which can be safely neglected.

\subsection{Width of the spin tune distribution}

Since the spin precession frequency is proportional to the particle energy and magnetic field \eqref{eq:spin_freq} synchrotron and betatron oscillations generate or increase the width of the spin tune distribution.

A particle with energy $E=E_0(1+\delta)$, different from the equilibrium value $E_0$, is subject to synchrotron oscillations with frequency $\omega_{syn}$. The revolution frequency of the particle oscillates around the equilibrium value $\Omega_0$ according to $\Omega=\Omega_0(1-\chi_0\delta)$, where $\chi_0$ is the momentum compaction factor and $\delta = a\cos{\omega_{syn} t}$ is the time varying energy shift. Substituting the particle energy and revolution frequency in \eqref{eq:spin_freq} yields
\begin{equation}
W=\Omega_0\left(1+\nu_0-\chi_0\nu_0\frac{a^2}{2}\right)+
  \Omega_0\left(\nu_0(1-\chi_0)-\chi_0)\sin(\omega_{syn} t\right)+
 +\chi_0\Omega_0\nu_0\frac{a^2}{2}\cos(2\omega_{syn} t)\,,
\label{eq:synchr_oscl}
\end{equation}
where $\nu_0=\gamma_0 q'/q_0$. The first term in \eqref{eq:synchr_oscl} is responsible for the shift and the widening of the spin precession frequency distribution. The second and third terms are responsible for the side bands and could be neglected if the expected shift and width of the spin tune distribution are smaller than the synchrotron oscillation frequency. Averaging over synchrotron oscillations and noting that $\left<a^2\right>=2 \sigma_\delta^2$ yields
\begin{equation}
\left<\frac{W-\Omega_0(1+\nu_0)}{\Omega_0(1+\nu_0)}\right>=
\left<-\frac{\chi_0\nu_0\displaystyle\frac{a^2}{2}}{1+\nu_0}\right>=
-\frac{\chi_0\nu_0\sigma_{\delta}^2}{1+\nu_0}=-2\cdot 10^{-12}
\end{equation}
which can be safely neglected.

\section{Centre-of-mass energy corrections and errors}
\label{sec:cm-energy}

Colliding beam experiments require precise knowledge of the average centre-of-mass energy $\langle \sqrt{s} \rangle$, which depends on the individual beam energies $E_+$ and $E_-$, on the average beam crossing angle $\alpha$, on optical parameters at the IP, and on the beam distributions:
\begin{equation}
\left<\sqrt{s}\right>_{x',y'}=2 \sqrt{E_+E_-}
\cos\sfrac{\alpha}{2} \left( 1
 - \frac{1}{4}\sigma_{x'}^2 - \frac{1}{4}\sigma_{y'}^2 \left[ 1-\tan^2\sfrac{\alpha}{2} \right] + \dots \right)\,,
\label{eq:InvariantMass-5}
\end{equation}
where  $\sigma_{x'}^2 = \varepsilon_x/\beta^*_x$ and $\sigma_{y'}^2 = \varepsilon_y/\beta^*_y$ correspond to the angular spreads ($p_{x'}/p_0$ and $p_{y'}/p_0$), assumed to be equal for both beams. For the typical FCC-ee beam emittances and betatron functions at the IPs, both $\sigma_{x'}^2$ and $\sigma_{y'}^2$ are at the level or smaller than $10^{-9}$ and can be neglected. Equation \ref{eq:InvariantMass-5} can therefore be simplified to
\begin{equation}
\langle \sqrt{s} \rangle = 2 \sqrt{E_+ E_-} \cos{\sfrac{\alpha}{2}}.
\label{eq:InvariantMass-6}
\end{equation}

\subsection{Distributed energy loss}
\label{sec:sr-loss-along-ring}

Local energy deviations due to the combination of energy loss from synchrotron radiation or collective effects (impedances) and energy gains by the RF system are a fundamental and important ingredients that must be taken into account when propagating the average beam energy measured by RDP to the IPs to reconstruct the CM energies.

\subsubsection{RF system and synchrotron radiation energy loss}

The originally proposed  FCC-ee RF system for the Z pole and WW operation  consists of RF superconducting cavities delivering around 100\,MV of accelerating voltage to each beam. The RF cavities are installed in two groups on opposite sides of the ring at points D and J. The RF systems of electron and positron beams are physically separated for Z and W operation. 
At higher energies more cavities will be added, and for operation at the top threshold the RF system will be shared by the two beams to save on RF cavities.

\begin{figure}[htb]
\centering\includegraphics[width=.9\linewidth]{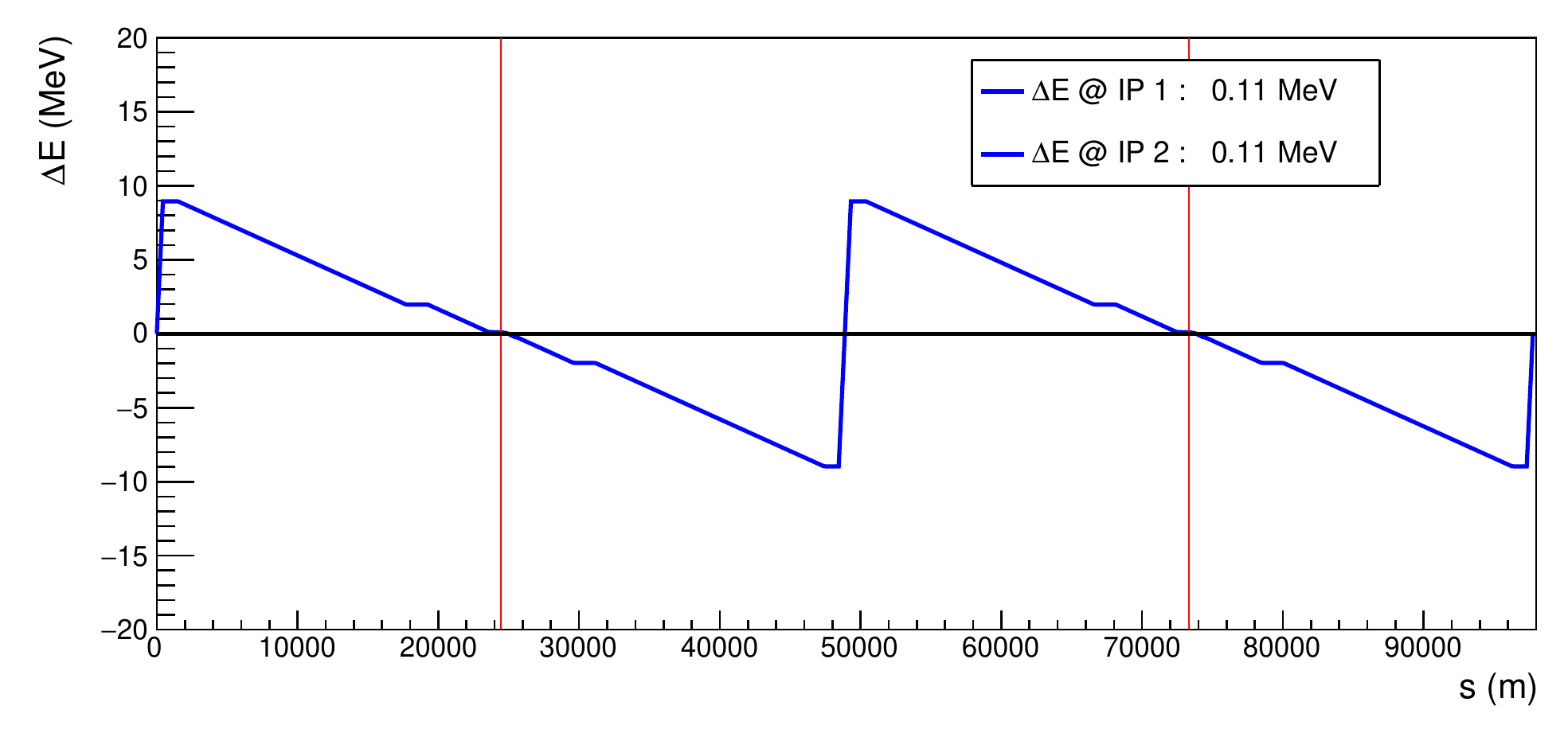} 
\centering\includegraphics[width=.9\linewidth]{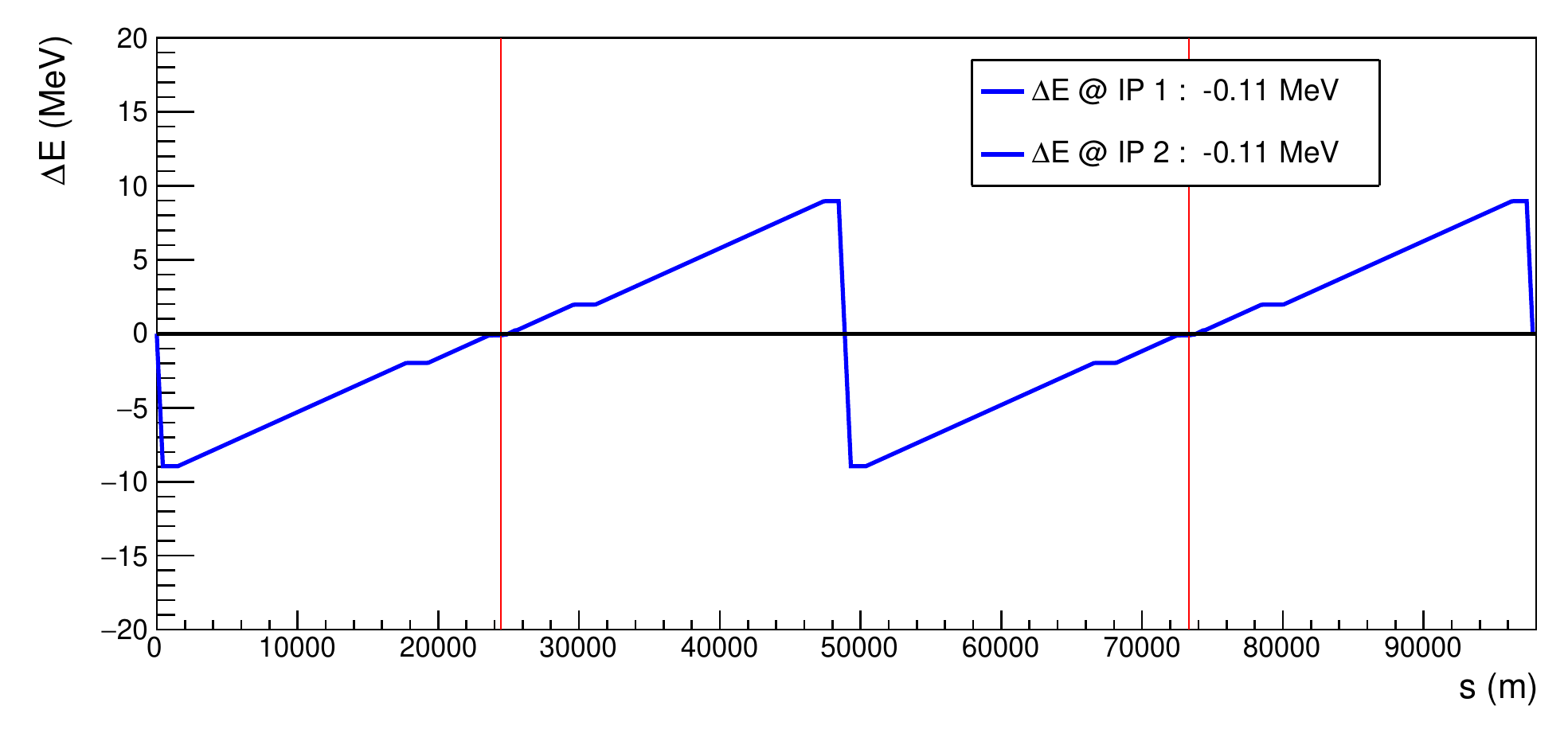}
\caption{\label{fig:sawtooth-Z-2rf} Energy sawtooth at the Z pole for the two beams (top: beam direction left to right, bottom: beam direction right to left), the vertical axis corresponds to the relative energy offset and the horizontal axis to the longitudinal coordinate. The two IPs are indicated by the red vertical lines. The energy gains in the two RF stations are exactly equal in this ideal configuration.}
\end{figure}

Figure~\ref{fig:sawtooth-Z-2rf} presents the energy sawtooth for the nominal configuration at the Z pole. The total energy loss is 36\,MeV. Due to the asymmetry of the straight sections around the experiments the beams have an energy offset of around $\delta_{1,2} \simeq \pm 0.1 \mathrm{MeV}$ in the ideal configuration. For perfectly symmetric rings the offsets have opposite signs, i.e. $\delta_{1} = - \delta_{2}$. 

The local energy loss by SR, $U(s)$, in a magnet is proportional to $E(s)^2 B(s)^2$ where $E(s)$ and $B(s)$ are the local energy and the local transverse magnetic field. The relative uncertainty on the energy loss $\delta U/U$ is therefore proportional to $2 \delta B/B$. Over one quarter of the ring, i.e. the typical distance between an RF station and an IP, the energy loss at the Z pole is around 10\,MeV. The uncertainty on the energy loss is around 10\,keV provided the magnetic field is known to $\delta B/B \sim 5 \cdot 10^{-4}$ which is a realistic target. 

Imperfect phasing of the two RF stations, one with respect to the other, will lead to unequal energy gains in the two RF sections and a shift of the local energy all around the ring including of course the IPs. Because the total energy gain in the two RF stations must remain constant, a change in energy gain of one RF station must be exactly compensated by the other RF station. The same argument applies to voltage errors. Because the RF systems of the two beams will be based on different cavities the phase and voltage calibration errors will be independent.

For the ideal RF system both RF stations D and J are perfectly phased and provide the same RF voltage $V_o$ such that
\begin{equation}\label{eq:vrf-ideal-rf}
\Delta E_{SR} = 2 V_o \sin{\Phi_s}
\end{equation}
where $\Delta E_{SR}$ is the energy loss per turn from SR and $\Phi_s$ is the stable phase angle. In the presence of voltage errors $\delta V_{J,D}$ and phase errors $\phi_{J,D}$ we obtain
\begin{equation}\label{eq:vrf-imperfect-rf}
\Delta E_{SR} = (V_o + \delta V_J) \sin(\Phi_s + \phi_J) + (V_o + \delta V_D) \sin(\Phi_s + \phi_D) \;.
\end{equation}
This expression can be rewritten in the form
\begin{equation}\label{eq:vrf-imperfect-rf2}
\Delta E_{SR} = 2 V_o \sin{\Phi_s} + \delta E_D + \delta E_J
\end{equation}
where $\delta E_{D,J}$ are the changes of energy gains in stations D and J. To first order
\begin{equation}
\delta E_D \simeq \delta V_D \sin(\Phi_s) + V_o \cos(\Phi_s) \sin(\phi_D)
\end{equation}
with a similar expression for $\delta E_J$. From relations \ref{eq:vrf-imperfect-rf2} and \ref{eq:vrf-ideal-rf} one notes that $\delta E_D = -\delta E_J$ i.e. the loss of effective accelerating voltage in one station must be compensated by the other station. As a consequence the local energy of the beam changes by $\delta E_D$ in one of the two IPs and by $-\delta E_D$ in the other IP. The energy changes of the two IPs are therefore fully anti-correlated. This argument applies to the RF systems of both beams. Errors in the evaluation of the effective energy gains of the two RF stations manifest themselves in anti-correlated change of the CM energies at the experiments. At the Z pole a continuous monitoring of the reconstructed Z masses by the two experiments provides precious indications of uncontrolled energy shifts at the IPs due to RF system errors.

\begin{figure}[htb]
\centering\includegraphics[width=.9\linewidth]{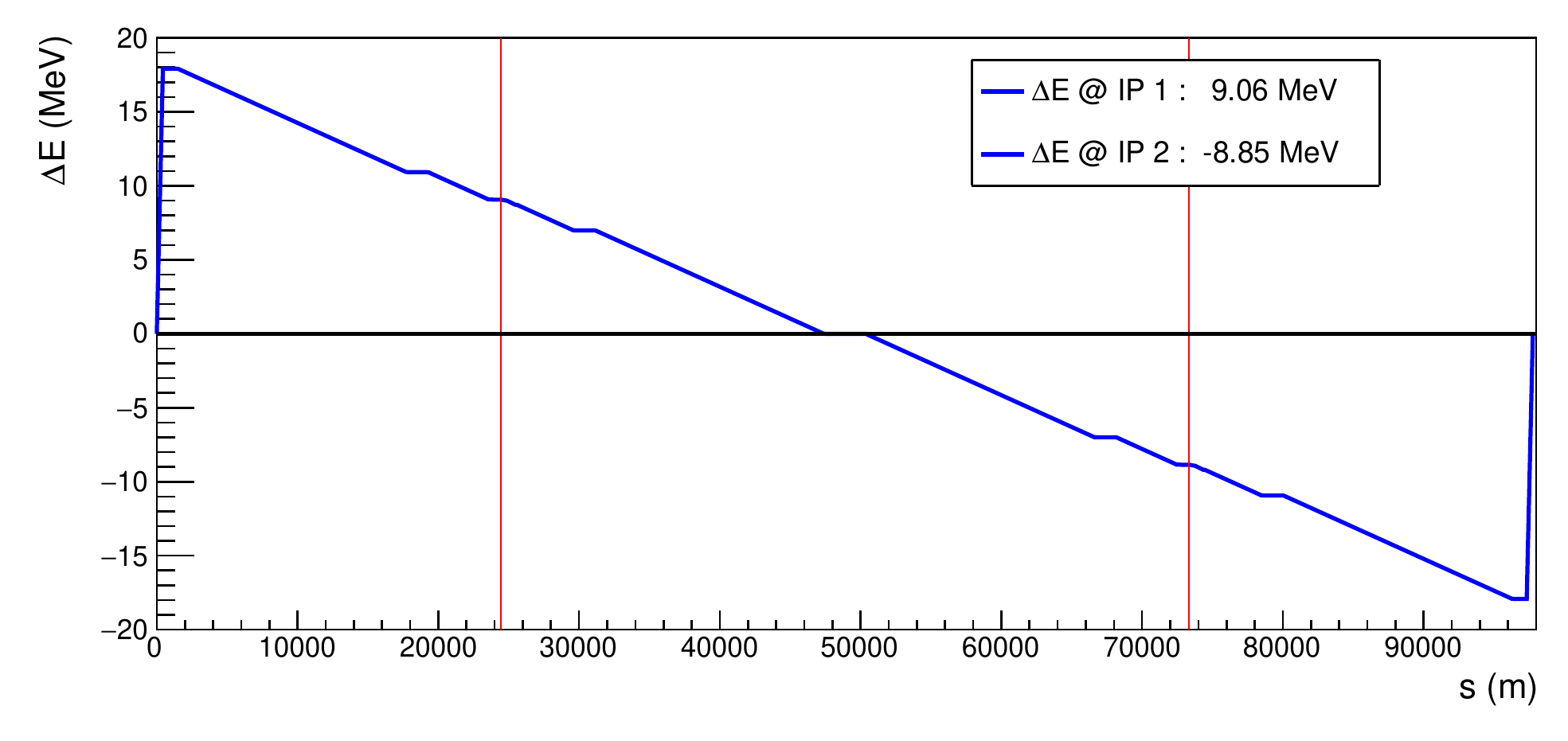}
\centering\includegraphics[width=.9\linewidth]{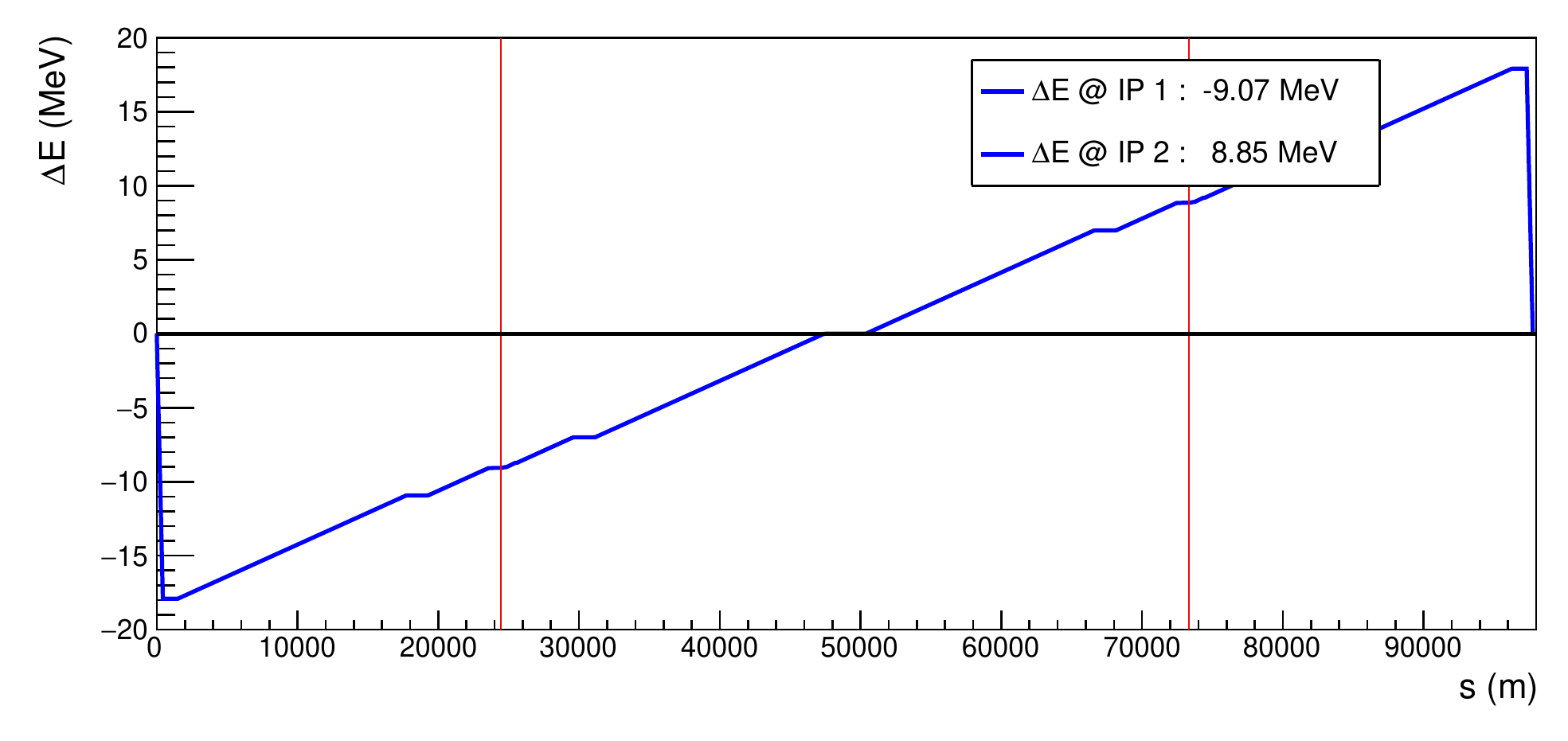}
\caption{\label{fig:sawtooth-Z-1rf} Energy sawtooth at the Z pole for the two beams with a single RF station per beam in the same location (top: beam direction left to right, bottom: beam direction right to left), the vertical axis corresponds to the relative energy offset and the horizontal axis to the longitudinal coordinate. The two IPs are indicated by the red vertical lines.}
\end{figure}

It was noted that, for the Z pole or WW  operation, it is possible to operate with a single RF station without reduction of the nominal power.  As becomes clearer below, such a layout is highly  beneficial for understanding and controlling systematic uncertainties due to the synchrotron-radiation energy losses around the ring. The energy saw-toothing for  the machine operated with a single RF system  is shown in  Fig.~\ref{fig:sawtooth-Z-1rf}.  By design the energy gain of the RF station of each ring must exactly compensate the energy loss $E_{\rm SR}$,  and the cavity automatically compensates voltage errors by shifting its phase. This design also ensures that the colliding beams and the pilot bunches have the same average energy. The beams have energy offsets of around $\pm$9\,MeV at the IPs, still with opposite signs for the two beams, i.e. $\delta_1 = - \delta_2$. In such a configuration,  RF-related uncertainties can be avoided, which removes one of the important contributions to the centre-of-mass energy uncertainty. 

For example, if point D is used as single RF station for one of the beams and point J for the other (or the reverse), then the centre-of-mass energy is shifted by around $+18$\,MeV at one IP, and by around $-18$\,MeV at the other, with respect to the average beam energy as determined by RDP. It is better, from the point-of-view of energy calibration, and probably more practical, from the point-of-view of infrastructure and operation, to have all the RF in a single point. In that case the beam energies of the two beams are different at the IPs, but the centre-of-mass energies are identical. 

In either configuration, the comparison between the Z masses measured by the two experiments provides a test of the energy calibration chain at the level of 8\,keV precision overall. In addition, the energy difference between the ${\rm e^+}$ and ${\rm e^-}$ beams results in a boost of the centre-of-mass system, which can be inferred from ${\rm e^+ e^- \rightarrow \mu^+ \mu^-}$ events, with the same method as for the centre-of-mass energy spread measurement (Section~\ref{sec:sigmaE-muons}) and with similar precision. This measurement leads to two other inclusive and independent tests (one for each IP) of the energy calibration chain at the keV level. These numbers concern the Z run, and are correspondingly larger for the W run, with a precision worse by two orders of magnitude larger, given the available statistics.  

With a high precision BPM system it is possible to monitor energy gain variations of the RF stations since those appear as radial offsets in the two machine halves. For a horizontal dispersion in the arc of 30\,cm and a long-term accuracy of the BPMs of 3\,$\mu$m it is possible to observe relative energy gains of $10^{-5}$. In controlled experiments such an orbit monitoring of the energy sawtooth was used with success at LEP to calibrate the effective RF voltage distribution around the ring~\cite{LEPrfcal.SL2002}.

\subsubsection{Longitudinal impedance losses}

Due to its large size and high number of accelerator components the longitudinal impedance budget is rather high. Section 2.6.9 of the FCC-ee CDR~\cite{Abada2019} summarizes the estimated FCC-ee longitudinal impedance budget and the associate power loss which is dominated, with 2/3 of the total budget, by resistive wall effects. For a short bunch length at the Z pole of 3.5\,mm the total power loss amounts to almost 14\,MW compared to the SR power loss of 50\,MW. This corresponds to a distributed energy loss per turn of around 9\,MeV compared to the 36\,MeV due to SR. For the nominal bunch length with beamstrahlung which is around 13\,mm, the power loss will be significantly reduced, probably by a factor between 3 and 4. However, the residual power loss remains significant and must be accounted for to reconstruct the local beam energy at the IPs. Spreads in bunch lengths and bunch intensities will have to be accounted for in the reconstruction of the local beam energy with time.

The longitudinal power loss can be measured by injecting bunches of different intensities (colliding or non-colliding) and measuring their orbit differences. The intensity or bunch-length dependent power loss will induce orbit shifts between the different bunches that can in principle be measured rather accurately as  was done at LEP~\cite{LEPImpedanceLoc}.

\subsection{Dispersion at the IP}

For beams colliding with an offset at the IP,
the CM energy spread and shift are affected by the local dispersion at the IP.
For a total IP separation of the beams of $2u_0$ the expressions for the CM energy shift and spread are~\cite{Jowett:806265}
\begin{equation}
\Delta \sqrt{s} = - 2 u_0 \frac{\sigma_E^2 (D_{u1} - D_{u2})}{E_0 (\sigma_{B1}^2+ \sigma_{B2}^2)}
\end{equation}
\begin{equation}
\sigma_{\sqrt{s}}^2 =  \sigma_E^2
  \left[ \frac{\sigma_{\epsilon}^2 (D_{u1}+D_{u2})^2 + 4\sigma_{u}^2}
        {\sigma_{B1}^2 + \sigma_{B2}^2} \right]
\end{equation}
$D_{u1}$ and $D_{u2}$ represent the dispersion at the IP for the two beams labelled by 1 and 2. $\sigma_E$ is the beam energy spread assumed here to be equal for both beams and $\sigma_{\epsilon} = \sigma_E/E$ is the relative energy spread. $\sigma_{Bi}$ is the total transverse size of beam (i) at the IP,
\begin{equation}
\sigma_{Bi}^2 = \sigma_u^2 + \left( D_{ui} \sigma_{\epsilon} \right)^2
\end{equation}
with $\sigma_u$ the betatronic component of the beam size. These expressions assume that the dispersion does not depend on the beam energy offset, an assumption that is not correct for the FCC-ee lattice and that will be discussed later.

If the beam sizes at the IP are dominated by the betatronic component which is rather likely, the energy shift simplifies to
\begin{equation}
\Delta \sqrt{s} = - u_0 \frac{\sigma_E^2 \Delta D^*}{E_0 \sigma_u^2}
\end{equation}
where $\Delta D^* = D_{u1} - D_{u2}$ is the difference in dispersion at the IP between the two beams. This effect applies to both planes ($u$ = x,y). In general due to the very flat beam shapes the most critical effect arises in the vertical plane.

At the Z pole for the nominal beam parameters at the IP and for $\Delta D^* = 1$\,$\mu$m the shift will be
\begin{equation}
|\Delta \sqrt{s}| = 96 \; |u_0|  \; [\mathrm{keV/nm}] 
\end{equation}
in the vertical plane with full beamstrahlung where $\sigma_{\epsilon} = 0.13$\%. Current simulations of machines errors and their corrections~\cite{Charles} results in a spread of $D_{y} \approx 10 \mu\mathrm{m}$ with extreme values of $\approx 30 \mu\mathrm{m}$. The beam offset $2 u_0$ must be controlled to 0.1~nm which is below $\sigma_u$/100 to ensure that the systematic error does not exceed 100\,keV. 
The horizontal dispersion $D_{x}$ may reach up to 0.2\,mm which requires a control of the beam offsets to 300~nm or roughly 5\% of the horizontal beam size. 

To minimize the systematic uncertainties arising from dispersion, the beams have to be scanned one against the other (separator scans as mentioned in Section~\ref{LEPEcmcorr}) at regular intervals, to optimize their overlap with high accuracy. The scan frequency depends on the stability of the offset, which itself depends on ground motion, the thermal stability of the ring, and the performance of the orbit feedback systems. At very high beam-beam tune shift, the beam distributions themselves may change during the scan, which requires even more care in evaluating experimental systematic errors, and small amplitude  separator scans, potentially reducing the intrinsic precision of the method.

Colliding-beam offsets also result in measurable beam-beam deflection at the beam collision frequency, which can be picked up with the beam position monitors. Another possibility is to measure high-energy photons from radiative Bhabha scattering or beamstrahlung at the collision point: calorimeters could be situated in the first bending magnets downstream of the IP. Furthermore, shifts in the position of the high-energy photon spot,  or in the amplitude of the beam-beam deflection, would indicate residual vertical dispersion each time the RF frequency would be modified to follow the ground motion. Further study of these intrinsically passive methods should be further investigated.  

The horizontal plane poses a particular challenge due to the large crossing angle which couples the transverse and the longitudinal planes: the beams cannot be scanned in the transverse plane for the optimum, as a scan also shifts the longitudinal position of the collision point. In the horizontal plane the beam overlap must be scanned with the relative RF phase of electron and position beams (i.e. arrival time at the IP). It must also be noted that in the crossing plane the bunch head, centre and tail collide with counterparts of the opposing bunch. Any shape distortion of difference in the energy distributions of heads and tails may introduce subtle systematic errors that must be investigated in the future.  

The difference in dispersion $\Delta D^*$ can be obtained by measuring the change in beam overlap optimum as a function of an applied RF frequency change. For $\Delta D^*$ of 10\,$\mu$m an energy change of 0.1\% applied through an RF frequency change will generate a beam separation of 10~nm. Since sub-nm accuracy is required for the beam overlap to control $\Delta \sqrt{s}$, it should be possible to measure and monitor $\Delta D^*$ to well below 1\,$\mu$m.

The average dispersion can be measured by the experiment by observing the IP position shift due to an applied energy offset. This information can be obtained together with the measurement of $\Delta D^*$. The required resolution depends on the impact on the energy spread.

\begin{figure}[htb]
\centering\includegraphics[width=0.8\linewidth]{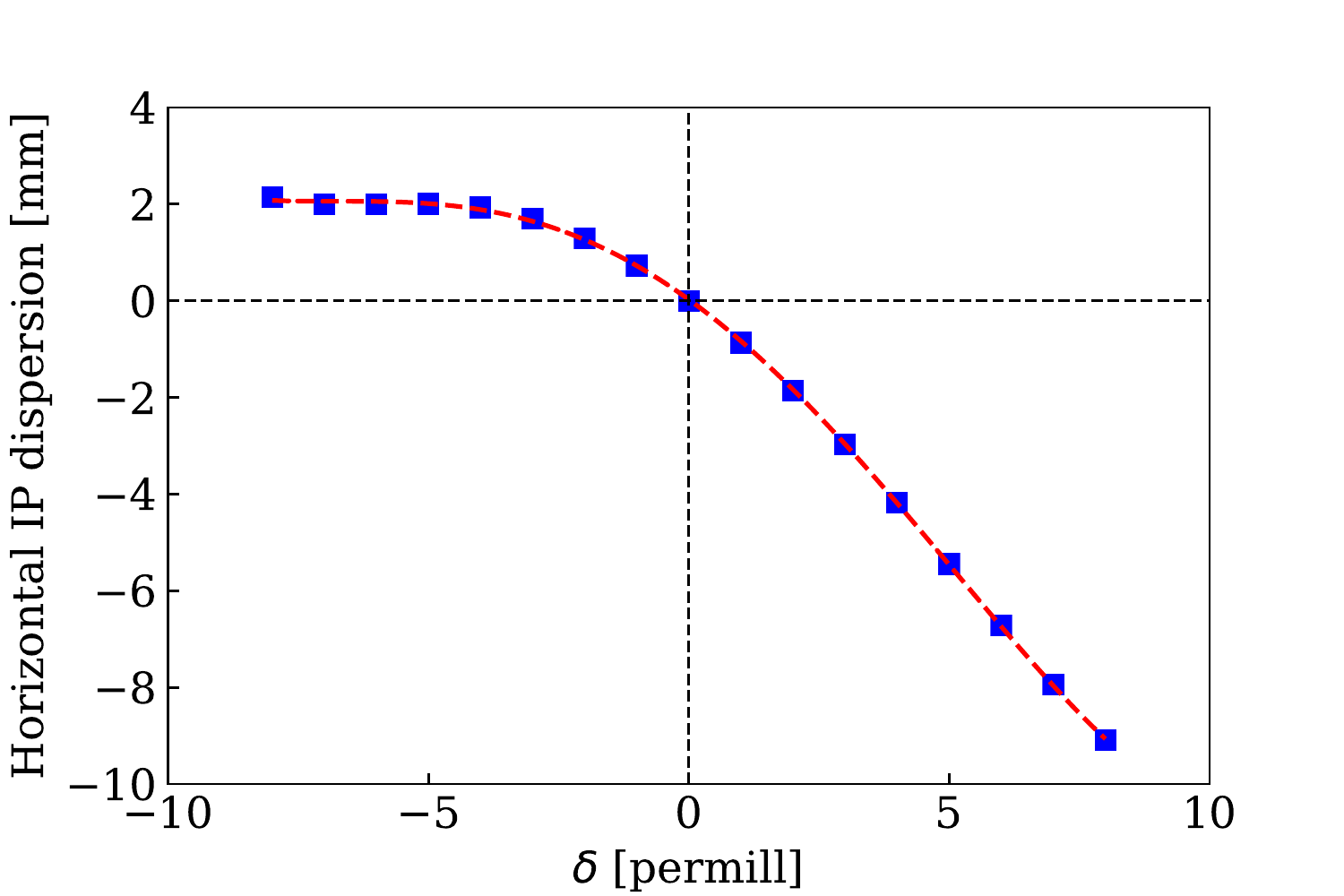}
\caption{ \label{fig:dxip-dp} Dependence of the horizontal dispersion at the IP on the relative energy offset $\delta$ for the the  Z lattice FCCee\_z\_213. The line is a fourth order polynominal fit to the points.}
\end{figure}

Figure~\ref{fig:dxip-dp} presents the dependence of the horizontal dispersion at the IP on the relative momentum offset for a typical FCC-ee Z lattice. There is a large derivative term, leading to dispersions of $\simeq 1$\,mm for for offsets of $10^{-3}$ that correspond to the energy spread of colliding bunches. This value is significantly larger than the residual dispersion of a perturbed lattice after correction at $\delta = 0$. The assumption that the dispersion is independent of the energy offset, which is used to derive the equations presented in this section, are clearly not correct for this machine, and higher order  effects will have to be included in the evaluation of the uncertainty.

The criticality of the dispersion at the IP may be reduced by operating in a regime where beamstrahlung is weak since in that case the energy spread may be a factor of 3 smaller, gaining around one order of magnitude on the sensitivity to $\Delta D^*$. Operating for some time in such a regime may provide insights into the systematic errors due to dispersion at the IPs.

\subsection{Chromaticity of the betatron function at the IP}

\begin{figure}[htb]
\centering\includegraphics[width=0.49\linewidth]{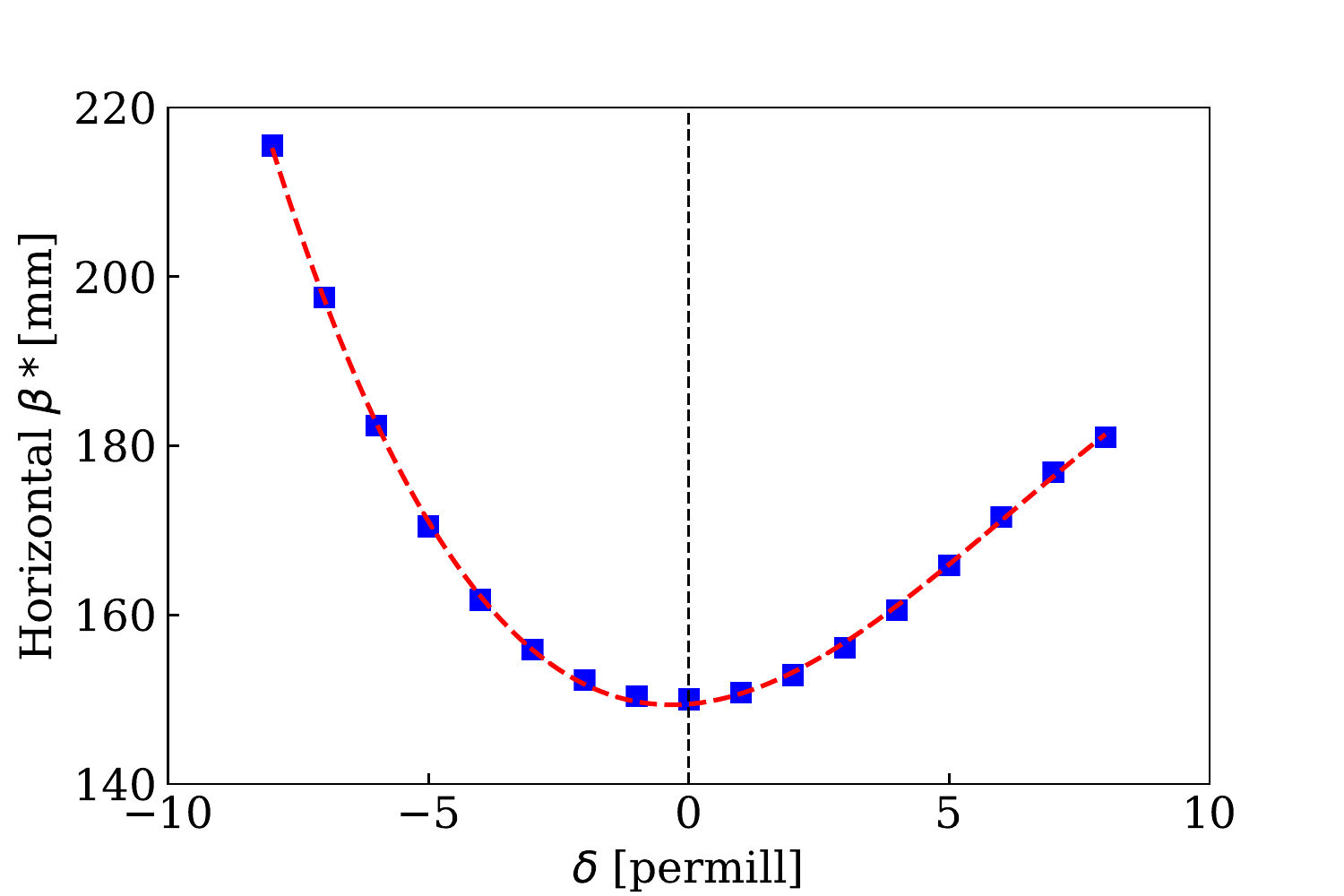}
\centering\includegraphics[width=0.49\linewidth]{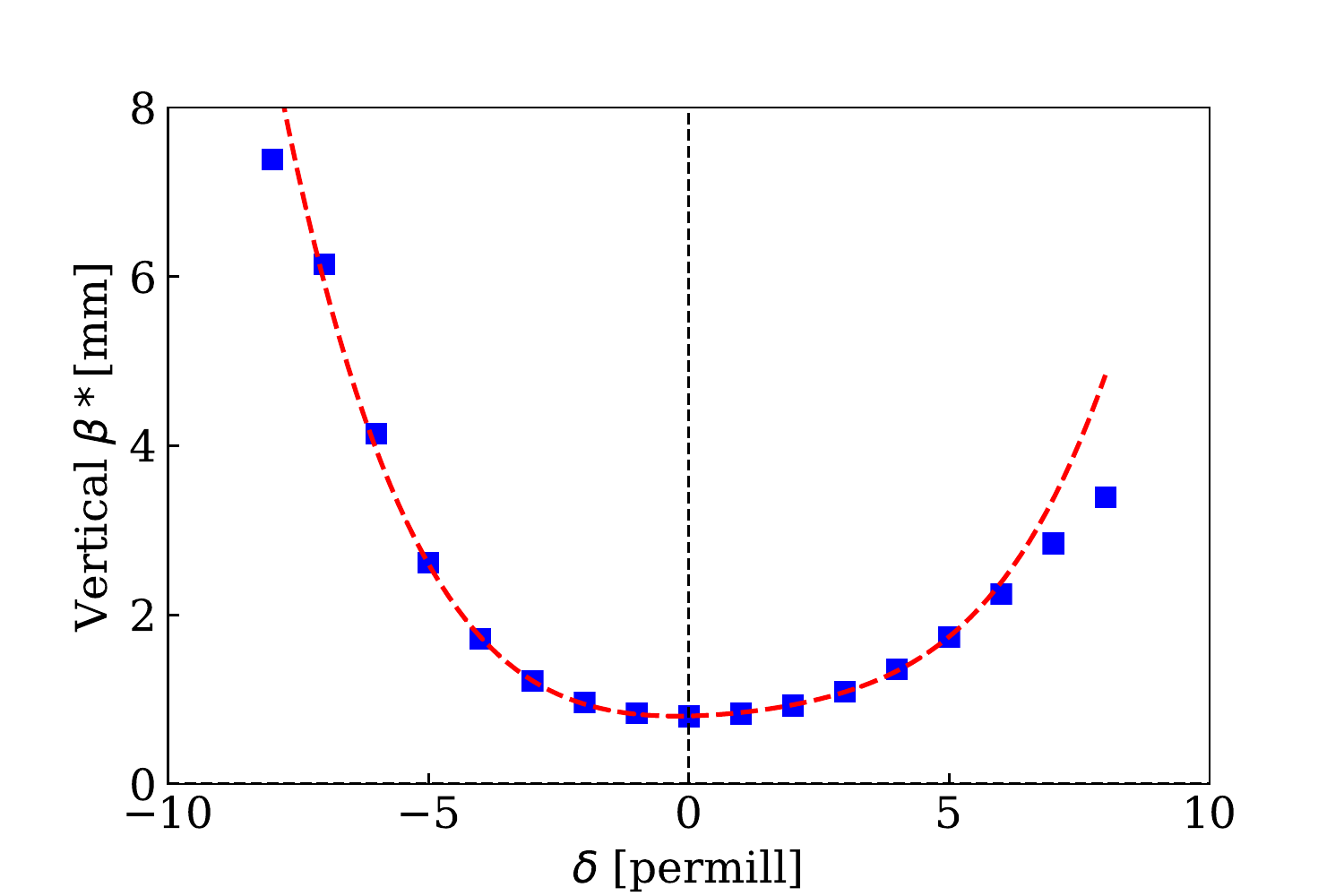}
\caption{\label{fig:betastar-dp} Dependence of the horizontal and vertical $\beta^*$ on the relative momentum offset $\delta$ for the Z lattice FCCee\_z\_213. The point are simulated with MADX, the dashed line is a 4$^\mathrm{th}$ order polynomial fit to the points (restricted to the range $\delta = \pm 0.005$ for the vertical $\beta^*$). The first order derivatives $(1/\beta^*) d\beta^*/d\delta$ are 3.5 for the horizontal and 17.5 for the vertical plane which corresponds roughly to the entries in Table~\ref{tbl:beta_chromaticity}.}
\end{figure}

The chromaticity of $\beta^*$ results in different particle densities as a function of the particle energies. As a consequence the luminosity distribution over the CM energy may become asymmetric leading to a potential bias of the CM energy. For a particle density for the $i$th bunch
\begin{equation}
\label{eq:density-1}
n_i(x_i,y_i,s_i,t,\delta_i)=\frac{N_i}{(2\pi)^{4/2}\sigma_x\sigma_y\sigma_z\sigma_\delta}
\exp\!\left[
-\frac{x_i^2}{2\sigma_{x,i}^2} 
-\frac{y_i^2}{2\sigma_{y,i}^2}
-\frac{(s_i\mp ct)^2}{2\sigma_z^2}-\frac{\delta_i^2}{2\sigma_\delta^2}
 \right]\,,
\end{equation}
the luminosity is calculated according to \cite{Herr2013,ref:Suzuki:luminosity} as
\begin{equation}
\label{eq:luminosity-1}
\mathcal{L}=2f_0
\int n_1(x,y,s,t,\delta_1)n_2(x,y,s,t,\delta_2)dx dy ds dct d\delta_1 d\delta_2\,,
\end{equation}
where the crossing angle has been neglected since its influence is small. It is assumed that the average energies of both beams are equal to $E_0$, $\sigma^2_{x,y}=\varepsilon_{x,y}\beta_{x,y}^*$ is the beam size in the corresponding plane and  $\varepsilon_{x,y}$ is the emittance. The betatron function at the IP $\beta_{x,y}^*$ depends on the energy offset $\delta$,
\begin{equation}
\beta_{x,y}^*(\delta)=\beta_{0\,x,y}+\beta_{1\,x,y}\delta+\beta_{2\,x,y}\delta^2 + \mathrm{higher\; order\; terms} \;\;,
\end{equation}
where the linear term $\beta_{1\,x,y}$ is responsible for the chromaticity of the beta function. Figure~\ref{fig:betastar-dp} presents $\beta^*$($\delta$) for the FCC-ee Z lattice.  The importance of higher order terms is apparent. 

\begin{table}[htbp]
\caption{Beta function chromaticity and corresponding bias of the invariant mass for a centre-of-mass energy of 45\,GeV. The statistical uncertainty of the MC simulation correspond to 2.4 keV. \vspace{3mm}}
\label{tbl:beta_chromaticity}
\centering
\begin{tabular}{|l|c|c|c|}
\hline
$\frac{1}{\beta_x}\frac{d\beta_x}{d\delta}$	& $\frac{1}{\beta_y}\frac{d\beta_y}{d\delta}$	& $\Delta \sqrt{s}$ (keV) & $\frac{\Delta \sqrt{s}}{\sqrt{s}}$				\\ \hline
0			&	15	&	$-49$		&	$-1.1\cdot10^{-6}$					\\ \hline
200			&	0		&	$-26$		&	$-5.7\cdot10^{-7}$					\\ \hline
200			&	15	&	$-75$		&	$-1.6\cdot10^{-6}$					\\ \hline
\end{tabular}
\end{table}

Asymmetries in the dependence of $\beta^*$ on $\delta$ may induce biases of the centre-mass energy. Since the is no analytical solution for the overlap integral for energy dependent $\beta^*$ (and transverse beam size) the impact on the CM was estimated for the first order chromaticity  term $\beta_{1,x,y}$ with the help of a Monte-Carlo simulation.
Table \ref{tbl:beta_chromaticity} presents the CM energy bias for different values of beta function chromaticities. The chromaticity of $\beta^*$ must be measured during operation, and the energy shift estimated using the full dependence of $\beta^*$ on $\delta$ by numerical integration of the integrals.

The main issue with this effect is the impact of the strong beam-beam forces on the $\beta^*$ chromaticity. The curves presented in Fig.~\ref{fig:betastar-dp} correspond to the situation of non-colliding beams, and it is likely that a measurement of the chromaticity may only be obtained without collisions. The impact of the beam-beam interaction on the $\beta^*$ chromaticity  requires detailed studies.

\subsection{Modification of particle energies due to collective fields}
\label{sec:collective}

In the resonant depolarization process, the whole bunch transverse polarization is measured by its interaction with the polarized laser before and after its excitation with the depolarization kicker.  
The individual particles are affected by the electric charge of the surrounding bunch and are therefore subject to both a static potential and to additional inter-bunch motion, implying longitudinal and transverse motions, and the resulting changes of kinetic energy. At the interaction point however, the surrounding field is largely compensated locally by that of the counter-rotating bunch.  These effects will eventually require a full relativistic field treatment. An indicative first order-of-magnitude estimate based on field potentials is given in the following, with two main results: (i)~the modification of particles total energy due to the collective potential is estimated between 120 and 400\,keV; (ii)~the cancellation at the IP results in essentially a full cancellation of the effect. These effects are expected to modify the relationship between the spin-tune and centre-of-mass energy in a way which depends on the bunch populations, and can be tested with one of the methods, momentum measurement in the polarimeter or reconstruction of the centre-of-mass energy in the detectors, which are described in Section\ref{sec:point-to-point}. These measurements can be performed with different bunch intensities, and may not require a significant loss of luminosity if higher bunch intensities are used with correspondingly smaller number of bunches.

\subsubsection{Collective fields of the bunches}

The potential energy of the particles depends on the dimensions of the  bunch it belongs to. It varies along the orbit following the change of the bunch dimensions. Therefore the electron energy $E=\gamma mc^2$ at the IP is different from that in the arcs. Electrons in the fields of the their own bunches will have potential energies \cite{Zimmermann:1997:SpaceCharge}
\begin{equation}
U[eV]=\frac{N_p e^2[Gs]}{\sqrt{2\pi}\sigma_z[cm]}\left(\gamma_e+\ln(2)-2\ln\left(\frac{\sigma_x+\sigma_y}{r}\right) \right)\frac{10^{-7}}{e[C]}\,,
\label{eq:OwnBeamPotential}
\end{equation}
where $\gamma_e=0.577$ is Euler constant, $N_p=4\cdot 10^{10}$ is the bunch population, $r_{ip}=15$\,mm and $r_{arc}=20$\,mm are vacuum chamber radius at IP and in the arcs respectively, $\sigma_{x,IP}=6.2\cdot10^{-6}$\,m and $\sigma_{y,IP}=3.1\cdot10^{-8}$\,m , $\sigma_{x,arc}=1.9\cdot10^{-4}$\,m and $\sigma_{y,arc}=1.2\cdot10^{-5}$\,m are horizontal and vertical beams sizes at the IP and in the arcs.
The potential energies of electrons at the IP and in the arcs for $E_0=45.6$\,GeV are
\begin{align}
\frac{U_{ip}}{E_0}  &=\frac{192\,{\rm keV}}{45.6\,{\rm GeV}}=4.2 \cdot 10^{-6}\,, \\
\frac{U_{arc}}{E_0}&=\frac{120\,{\rm keV}}{45.6\,{\rm GeV}}=2.6 \cdot 10^{-6}\,.
\end{align}

In the vicinity of the interaction point, the particles will experience the fields of the counter-rotating bunch. The longitudinal projection of the transverse field will accelerate the particles and change their energies The absolute value of the potential energy of the opposite bunch reaches its maximum when an electron reaches the centre of the bunch. Estimating the potential energy in the centre of the opposite bunch $\{x,y,s,z=s-ct\}=\{0,0,0,0\}$ according to
\begin{equation}
U(x,y,s,ct)=-\frac{\gamma N_p r_e mc^2}{\sqrt{\pi}}\int_0^\infty dq
\frac{\exp\left[-\frac{(x+s\cdot \alpha)^2}{2\sigma_{x}^2+q}-\frac{y^2}{2\sigma_{y}^2+q}-\frac{\gamma^2(s+ct)^2}{2\gamma^2\sigma_{s}^2+q}\right]}{\sqrt{2\sigma_x^2+q} \sqrt{2\sigma_y^2+q} \sqrt{2\gamma^2\sigma_s^2+q}}\,,
\end{equation}
for beam energy $E_0=45.6$\,GeV yields
\begin{equation}
\frac{U(0,0,0,0)}{E_0}=-\frac{0.4\,{\rm MeV}}{45.6\,{\rm GeV}}=-9.3\cdot 10^{-6}\,.
\end{equation}

\subsubsection{Invariant mass in the external field}

For calculation of the invariant mass in the presence of external fields we need to remember the definition of the four-momentum
\begin{equation}
P^\mu=(E-e\varphi,\vec{p})=(E-e\varphi,\vec{\mathcal{P}}-\frac{e}{c}\vec{A})\,,
\end{equation}
where $\vec{p}$ is the kinematic momentum, $\vec{\mathcal{P}}$ is the generalized momentum, $E$ is the particle energy, $e\varphi$ is the particle potential energy, and $\varphi$ and $\vec{A}$ are the scalar and vector potentials of the external field.
The energy-momentum relation is $(E-e\varphi)^2=m^2c^4+c^2(\vec{p})^2$.
The invariant mass (centre-mass energy) $s$ of two colliding particles is
\begin{equation}
s =(P_1^\mu+P_2^\mu)^2=2E_1e_1\varphi+2E_2e_2\varphi+2E_1E_2-(e_1\varphi)^2-(e_2\varphi)^2-2\vec{p^{(1)}}\vec{p^{(2)}}\,.
\end{equation}
Introducing normalized coordinates $\delta=(E_i-E_0)/E_0$ and $u=e_i\varphi/E_0$, the longitudinal projection of the momentum is
\begin{equation}
p_{i,s}=\sqrt{(E_i-e_i\varphi)^2-p_{i,x}^2-p_{i,y}^2} 
           =E_0\sqrt{(1+\delta_i-u)^2-\left(\frac{p_{i,x}}{E_0}\right)^2-\left(\frac{p_{i,y}}{E_0}\right)^2}\,.
\end{equation}
Assuming that $\delta_i$, $p_{i,x,y}/E_0$ obey normal distributions one gets the average invariant mass
\begin{align}
\left<s \right>&=4E_0^2\cos^2\sfrac{\alpha}{2}(1-u^2)-2E_0^2\sigma_{px}^2\cos\alpha-2E_0^2\sigma_{py}^2\cos\alpha\,, \\
\left< \sqrt{s}\right>&=2E_0\cos\sfrac{\alpha}{2}\left( 1-\frac{u^2}{2}\right) -\frac{E_0}{2}\left(\sigma_\delta^2\cos\sfrac{\alpha}{2}+\sigma_{px}^2\cos\sfrac{\alpha}{2}+\sigma_{py}^2\frac{\cos\alpha}{\cos\sfrac{\alpha}{2}}\right)\,,
\end{align}
The variance of the centre-of-mass energy is
\begin{equation}
\left<s\right>-\left<\sqrt{s}\right>^2=2E_0^2\cos^2\sfrac{\alpha}{2}\left(\sigma_\delta^2+\sigma_{px}^2\tan^2\sfrac{\alpha}{2}\right)\,.
\end{equation}
Substituting values of the potential energy from the previous sections we obtain, for beam energies $E_0=45.6$\,GeV, the shift due to beam potentials
\begin{equation}
\frac{\left<\sqrt{s}\right>-2E_0\cos\sfrac{\alpha}{2}}{2E_0\cos\sfrac{\alpha}{2}}=\left( 1-\frac{(e\varphi)^2}{2E_0^2}\right)\approx 4\times 10^{-10}\,.
\end{equation}

\subsection{Effects from interaction with the counter-rotating beam}
\label{sec:oncoming_beam}

The particles in colliding bunches, unlike in pilot bunches used for energy calibration,  interact with the electromagnetic field from the counter-rotating bunches. The particle energies are affected by beamstrahlung (BS) and by the ``kick" that, because of the beam crossing angle $\alpha$, they feel from the opposite charge bunch. The magnitude of the effect depends on the bunch length $\sigma_z$, or equivalently the energy spread $\sigma_\delta$ in the bunch, which in turn strongly depends on BS. Therefore, the  problem is best solved in a consistent way with multi-turn beam-beam tracking codes, such as  {\tt Lifetrac}~\cite{Valishev:927046}.

\subsubsection{Effect of beamstrahlung}

\begin{figure}[htb]
\centering
\includegraphics[width=0.95\columnwidth]{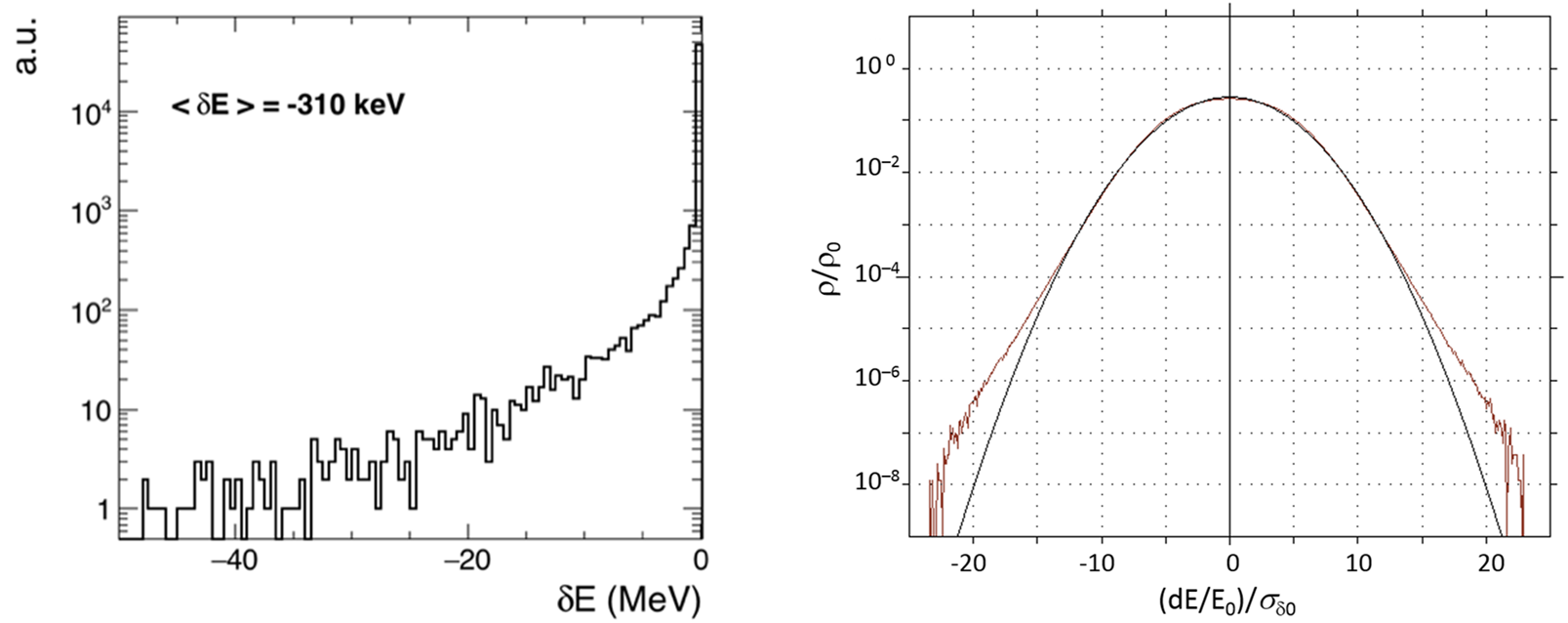}
\caption{\label{fig:de_bs} Left: Distribution of the energy that the beam particles lose in the interaction region at Z peak, as obtained from the {\tt Guinea-Pig} simulation program~\cite{Schulte:1999tx}. Right: Equilibrium energy distribution in a bunch, obtained from the {\tt Lifetrac} code. The black curve is a Gaussian with $\sigma_{\delta} = 3.4 \sigma_{\delta_0}$}
\end{figure}

At each crossing, the energy loss due to BS, the distribution of which is displayed in the left panel of Fig.~\ref{fig:de_bs} for the Z-pole running parameters, amounts to 310\,keV on average. Beamstrahlung also leads to a significant increase (by a factor $\sim 3.4$ at Z peak) in energy spread and, accordingly, in bunch length. The skewed energy loss distribution causes the equilibrium energy distribution, shown in the right panel of Fig.~\ref{fig:de_bs}, to be not strictly Gaussian. Because the equilibrium $\sigma_{\delta}$ and $\sigma_z$ strongly depend on the bunch population, scales are normalized, hereafter and in all figures, to their unperturbed (without BS) values, $\sigma_{\delta_0}$ and $\sigma_{z_0}$.

\begin{figure}[htb]
\centering
\includegraphics[width=0.95\columnwidth]{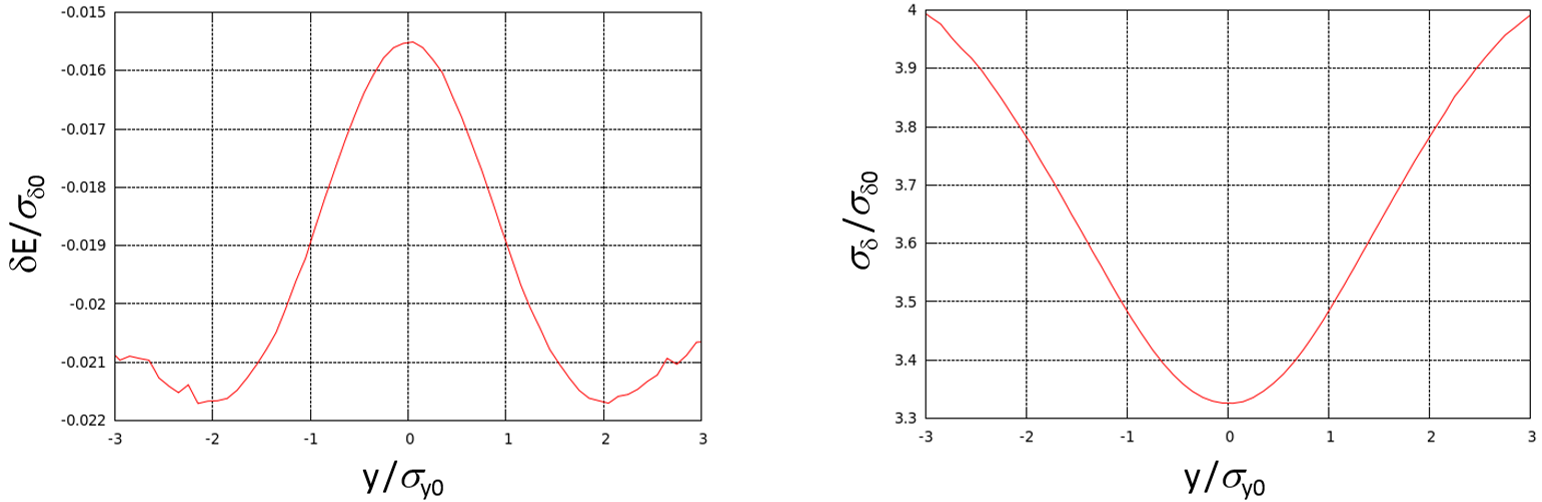}
\caption{\label{fig:de_yy} Energy loss per collision (left) and energy spread (right) at Z peak versus the vertical coordinate.}
\end{figure}

The beamstrahlung experienced by a given particle also depends on the particle coordinates. This dependence generates correlations between energy loss, energy spread and coordinates. For example, Fig.~\ref{fig:de_yy} shows that the largest losses are experienced by particles with a vertical coordinate $y$ such that $\left|y\right| / \sigma_y > 2$ (left panel), and that the energy spread $\sigma_{\delta}$ depends on $y$ accordingly (right panel). The dependence on the horizontal transverse coordinate is softer. The dependence on the longitudinal coordinate is stronger, and comes together with another source of energy change, that will be described in Section~\ref{sec:oncomingbeam_kick}.

The average energy loss per crossing due to beamstrahlung, calculated as the difference between the average beam energies before and after the crossing with the oncoming bunch ($\Delta E \approx - 310$\,keV at the Z pole, and $\approx - 1.42$\,MeV at the WW threshold) causes the equilibrium RF phase to change: the bunches are displaced in the longitudinal direction by approximately \mbox{1 mm} at the Z pole and \mbox{0.6 mm} at the WW threshold,  providing an increase in the energy transmitted to the beam in the RF cavities, to compensate the energy loss $\Delta E$. 
If $\Delta E_1$ denotes the energy shift of colliding bunches with respect to pilot bunches just before the IP, this shift amounts to $\Delta E_1 - \Delta E$ just after the IP.
In first approximation, if the energy losses in the arcs do not depend on $\Delta E$ (for small $\Delta E$), these shifts extend to the entire sections from RF cavity to IP, and from IP to RF cavity, respectively. Since the revolution frequency for all bunches is the same, they must have equal path lengths. It follows that, if the IPs are located symmetrically with respect to the RF cavities, $\Delta E_1 = \Delta E / 2$. In other words, 
the particles in colliding bunches have an energy larger (smaller) than the particles in pilot bunches by $\Delta E / 2$ prior to (after) the crossing, but equal on average during the crossing to the pilot bunch mean particle energy.

Another difference in energy loss with respect to pilot bunches, $\delta E_{\rm arc}$, arises in the arcs because of the difference in orbits (due to dispersion) and of the direct dependence of losses on the energy. At first order, this effect is linear in $\Delta E$ and therefore has the same values, but opposite signs in the sections before and after IP. This difference therefore leads to a shift in energy by $\delta E_{\rm arc}$ at the location of RF cavities, but there is no additional longitudinal displacement of the bunch in this place, since the total losses per revolution do not change. Instead, there is no additional energy shift at the IP, but the longitudinal displacement of the bunch relative to the pilot bunches slightly increases. But in the following orders of approximation there is no full compensation. Besides, the interaction region is not quite symmetrical with respect to the IP. In general, the collision energy shift due to $\Delta E$ should be quite small, but this question requires further study and clarification.

\subsubsection{Energy kick induced by the crossing angle}
\label{sec:oncomingbeam_kick}

\begin{figure}[htbp]
\begin{center}
\includegraphics[width=0.7\columnwidth]{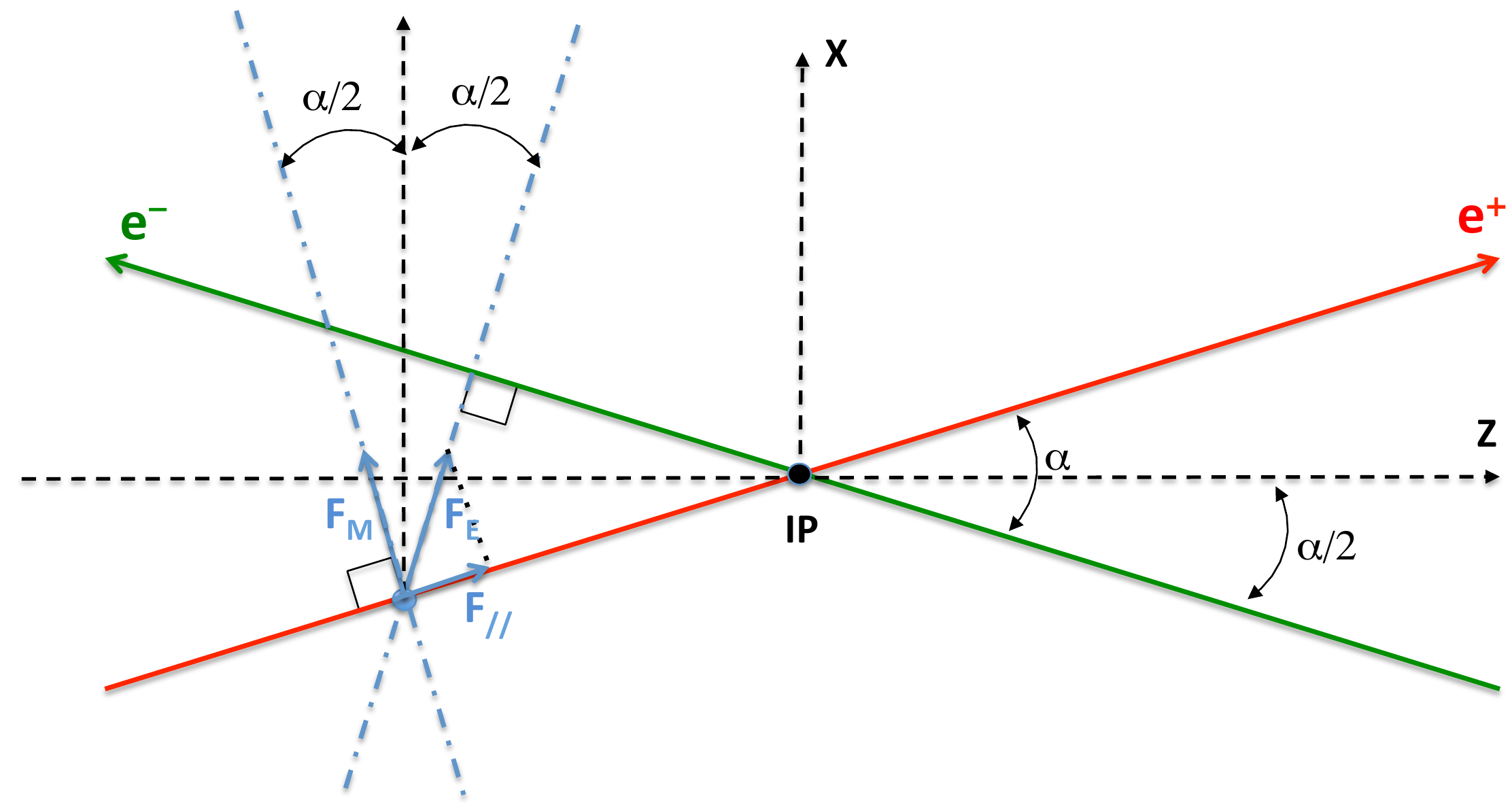}
\caption{\small Schematic view of the electric and magnetic attractive Lorentz forces $\vec{F}_E$ and $\vec{F}_M$ acting on each positron from the opposite electron bunch, upon bunch crossing at the interaction point (IP). Similar forces from the positron bunch affect each electron. The beam crossing angle is denoted $\alpha$. The $Z$ axis is the bisecting line of the two beam axes at the interaction point, the $X$ axis is orthogonal to the $Z$ axis such that the horizontal ($X,Z$) plane contain the two beam axes. 
}
\label{fig:energykick-schema}%
\end{center}
\end{figure}

In addition to beamstrahlung, the particles in colliding bunches experience an attractive force from the counter-rotating bunch with an electric component $\vec{F}_E$ orthogonal to the opposite bunch trajectory (since its field is compressed into a plane) and a magnetic component $\vec{F}_M$ orthogonal to the particle trajectory, as illustrated in Fig.~\ref{fig:energykick-schema}. The electric and magnetic forces have equal strengths, $F_E = F_M = F$, in the ultrarelativistic case and in the laboratory frame. Because of the beam crossing angle $\alpha$, the compound force has a component $F_\parallel$ along the particle trajectory -- accelerating the particles in the region before the interaction point (IP) and decelerating them in the region after the IP -- and a component $F_\perp$ orthogonal to the particle trajectory -- changing the particle direction in the $(X,Z)$ plane subtended by the two beam axes, and therefore increasing the crossing angle $\alpha$ with the opposite bunch before the IP (and reducing it after the IP). A quick examination of Fig.~\ref{fig:energykick-schema} gives the following relations: 
\begin{eqnarray}
F_\parallel & = & F_E \sin\alpha = F \sin\alpha, \label{eq:Fpar}\\
F_\perp & = &  F_M + F_E \cos\alpha = F \left(1 + \cos\alpha\right),\label{eq:Fperp}
\end{eqnarray}
and similarly, projecting onto the $X$ and $Z$ axis:
\begin{eqnarray}
F_X & = F_E \sin\sfrac{\alpha}{2} + F_M \sin\sfrac{\alpha}{2} & = 2F \sin\sfrac{\alpha}{2}, \label{eq:Fx}\\
F_Z & = F_E \cos \sfrac{\alpha}{2} - F_M \cos\sfrac{\alpha}{2} & = 0 \label{eq:Fz}.
\end{eqnarray}

The longitudinal kick $F_\parallel = F \sin\alpha$ (Eq.~\ref{eq:Fpar}) changes the energy of the particles for a nonzero crossing angle. Because $F$ is positive before the IP and negative after the IP, particles located in the centre of their bunch, crossing the IP at the centre of the oncoming bunch, see their energy unchanged by $F_\parallel$ when they exit the interaction region. However, for particles with a nonzero longitudinal coordinate, this is no longer the case: the impact from the oncoming beam is equivalent to the appearance of a nonlinear RF cavity, which reduces the synchrotron tune and distorts the shape of the potential well, and in turns affects the shape of the longitudinal particle distribution. This effect exists even in a head-on collision (because of hour-glass), but it becomes much stronger with large Piwinski angle and was experimentally observed at the DA$\Phi$NE collider \cite{Drago:2011zza}.

\begin{figure}[htb]
\centering\includegraphics[width=.7\linewidth]{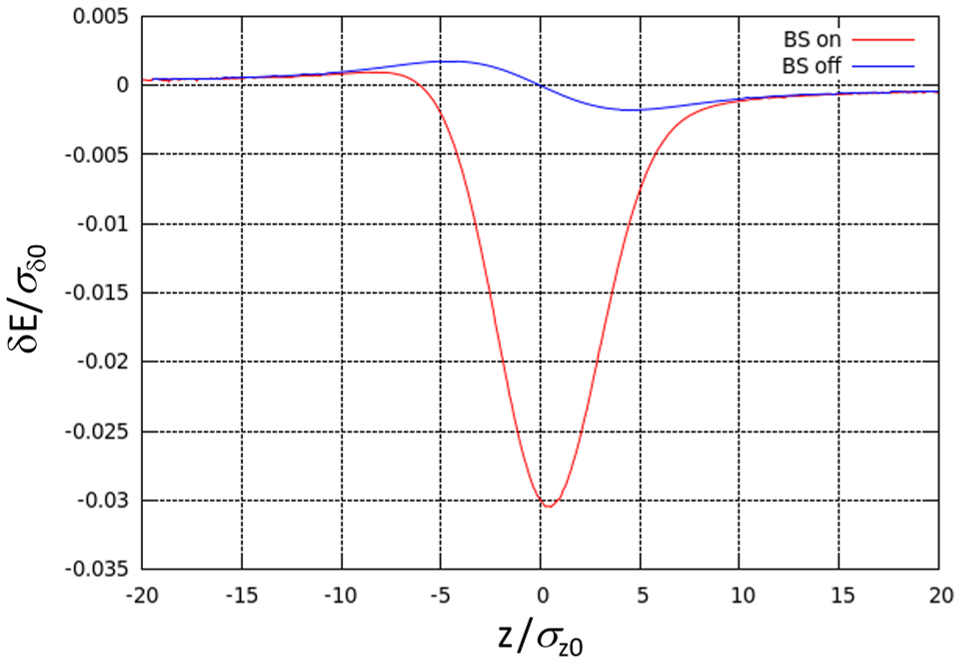}
\caption{\label{fig:de_zz} Particle energy change per crossing at the Z pole versus the particle longitudinal coordinate $z$. The red curve includes the energy loss due to BS, while the blue curve shows the dependence without BS, where everything is determined only by the crossing angle.}
\end{figure}

For particles that exit the interaction region (i.e., that do not collide), the dependence of the energy change on the particle longitudinal coordinate for each crossing is presented in Fig.~\ref{fig:de_zz} (in this context ``crossing'' extends from ``well before IP'' to ``well after IP''), without BS (blue curve) and with the additional energy loss due to BS (red curve).  

\begin{figure}[htb]
\centering
\begin{tabular}{cc}
\includegraphics[width=0.45\columnwidth]{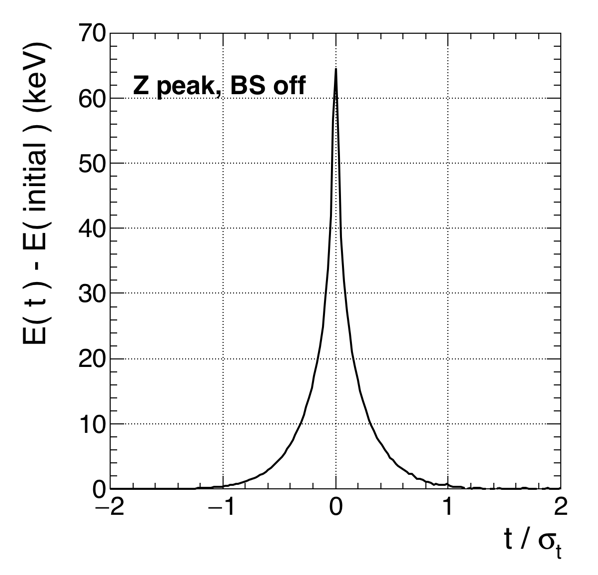} & 
\includegraphics[width=0.45\columnwidth]{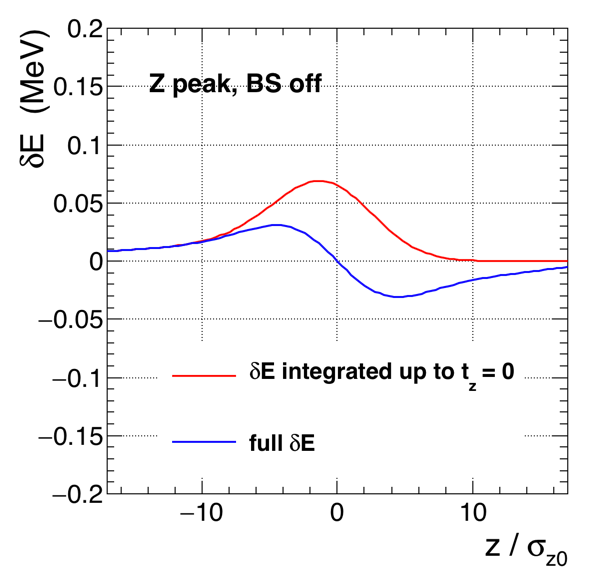}
\end{tabular}
\caption{\label{fig:deltaE_buildup} Left: Variation of the energy induced by the crossing angle for a particle in the centre of the bunch, as a function of time, as it moves through the interaction region. The time is expressed in units of $\sigma_t = \sigma_z / c$ where $\sigma_z$ denotes the bunch length at equilibrium, and $t = 0$ defines the time at which the  particle crosses the IP. Right:  Variation of the energy of a beam particle as a function of its longitudinal position $z$ within the bunch. The blue curve shows the total energy change and is equivalent to the blue curve shown in Fig.~\ref{fig:de_zz}. For the red curve, the energy kicks are integrated only up to the time when the particle reaches the IP. The particles in the head of the bunch experience less acceleration before the collision than particles in the tail, which makes this red curve asymmetric.}
\end{figure}

A more detailed picture for the case without BS is presented in Fig.~\ref{fig:deltaE_buildup} where it can be seen by how much the  energy of a particle located in the centre of its bunch changes during crossing (left panel) and how the total change depends on the particle longitudinal coordinate (right panel, blue curve). The total change, averaged over all particles in the bunch, is exactly zero\footnote{This is strictly true in the symmetrical case, when the populations of the counter bunches are equal. If the bunch populations deviate from the nominal value by $\pm 5\%$ (i.e. the relative difference between colliding bunches is about $10\%$), the energy spreads and the bunch lengths differ by almost a factor of two, and $\Delta E$ and longitudinal displacements differ by slightly more than a factor of two. This means that the centres of bunches no longer meet at the IP: the weak (less populated) bunch decelerates and the strong one accelerates by $\sim 1$\,keV due to the beam-beam kicks, with negligible effect on the centre-of-mass energy. In any case, the bunches ultimately find the correct equilibrium phases, taking into account SR losses in the arcs, energy gain in RF cavities, BS losses at the IPs and the energy changes (gain or loss, which contributes to $\Delta E$) due to the crossing angle.}. In contrast, for particles that collide, the positive energy kick before the IP is not  compensated by the negative energy kick after the IP, as shown by the red curve of Fig.~\ref{fig:deltaE_buildup} (right panel). 
This net effect causes an average increase $\delta E$ of the colliding particle energies with respect to the non-colliding particle energies, and with respect to the particles in the pilot bunches. 

Numerically, the most accurate method to predict the average collision energy (and therefore $\delta E$) is a direct counting in beam-beam simulations:
\begin{equation}
< E > = \frac{\sum E_c L_c}{\sum L_c},
\end{equation}
where $E_c$ is the particle energy at the moment of collision with a thin slice of the opposite bunch, $L_c$ is the luminosity produced by this elementary collision, and the sum is taken over all the particles of the bunch, all the slices of the opposite bunch, and all the turns -- about $10^9$ particle-turns in total. The average particle energy shift (with Z-pole running parameters) amounts to $\delta E = 1.3 \times 10^{-6} E_0 \approx 60.5$\,keV for particles that collide. \\

\noindent When it comes to the average centre-of-mass energy of the collision:
\begin{equation}
    \sqrt{s} = 2 \sqrt{E_1 E_2} \cos\sfrac{\alpha}{2} = 2 \sqrt{\vert p_{Z,1} p_{Z,2}} \vert,
\end{equation}
where $E_{i}$ are the average energies of the two particles in collision, and $p_{Z,i}$ their average momentum components along the Z axis, the increase of $E_i$ is exactly compensated by the increase of the beam crossing angle $\alpha$ (i.e., the decrease of $\cos\sfrac{\alpha}{2}$). Indeed, because the forces along the $Z$ axis are exactly 0 (Eq.~\ref{eq:Fz}), $p_{Z,i}$ is not modified, and the centre-of-mass energy does not change. The determination of the average centre-of-mass energy therefore requires the measurement of the average beam energy from pilot bunches with resonant depolarization, the measurement of the crossing angle in collision, and the determination of the crossing angle increase due to beam-beam effects. Methods to measure the average crossing angle and its average increase due to beam-beam effects are discussed in Section~\ref{sec:EnergySpreadMuons}.

\section{Centre-of-mass energy spread and beam crossing angle determination}
\label{sec:EnergySpreadAndCrossingAngle}

All precision measurements performed at FCC-ee (cross sections, asymmetries) depend on the centre-of-mass energy of the collisions, $\sqrt{s} = 2\sqrt{E_{\rm e^+}E_{\rm e^-}}\cos\sfrac{\alpha}{2}$. Their interpretations in terms of electroweak precision pseudo observables (the Z mass and width $m_{\rm Z}$ and $\Gamma_{\rm Z}$, the Weinberg angle $\sin^2\theta^{\rm eff}_\ell$, the W mass and width $m_{\rm W}$ and $\Gamma_{\rm W}$, the top quark mass and width $m_{\rm top}$ and $\Gamma_{\rm top}$~\cite{Gomez-Ceballos:2013zzn}, the electromagnetic coupling constant $\alpha_{\rm QED} (m^2_{\rm Z})$~\cite{Janot:2015gjr}, etc.) therefore benefit from an accurate knowledge of the centre-of-mass energy spectrum during collisions. Should the electron and positron beam energy profiles be Gaussian, this knowledge ``just'' amounts to the determination of the average beam energies in collision, their relative asymmetry and spread, and the beam crossing angle. The average beam energies is measured with resonant depolarization for non-colliding bunches. The other parameters are dealt with in this section.

\subsection{Determination from \texorpdfstring{${\rm e^+e^-} \to  \mu^+\mu^- (\gamma)$}{mmg} events}
\label{sec:EnergySpreadMuons}

\subsubsection{Total energy and momentum conservation}

\noindent Dimuon events, ${\rm e^+e^-} \to \mu^+\mu^- (\gamma)$, accompanied by one initial-state-radiation (ISR) photon ($\gamma$) along one of the two beam directions, are fully constrained by total momentum-energy ($p_x, p_y, p_z, E$) conservation: 

\begin{eqnarray}
E^+\sin\theta^+\cos\varphi^+ + E^-\sin\theta^-\cos\varphi^- + | p_z^\gamma | \tan\alpha/2 & = &\sqrt{s} \tan\alpha/2, \label{eq:xconv} \\
E^+\sin\theta^+\sin\varphi^+ + E^-\sin\theta^-\sin\varphi^- \ \phantom{+ | p_z^\gamma | \tan\alpha/2} \ \ & = & 0, \label{eq:yconv}\\
E^+\cos\theta^+\phantom{\cos\varphi^+} + E^-\cos\theta^-\phantom{\cos\varphi^-} + \phantom{| }p_z^\gamma\phantom{ | \tan\alpha/2} \  & = & 0, \label{eq:zconv}\\\
E^+\phantom{\sin\theta^+\cos\varphi^+} + E^-\phantom{\sin\theta^-\cos\varphi^-} + \ | p_z^\gamma |/\cos\alpha/2 & = &\sqrt{s}/ \cos\alpha/2, \label{eq:econv} 
\end{eqnarray}
where $\sqrt{s}$ is the centre-of-mass energy of the collision, $E^\pm$ are the $\mu^\pm$ energies, $\alpha$ is the beam crossing angle, the $Z$ axis is the bisecting line of the two beam axes at the interaction point, the $X$ axis is orthogonal to the $Z$ axis such that the ($X,Z$) plane contain the two beam axes, and the $Y$ axis such that ($X,Y,Z$) is an orthonormal system. The polar angles $\theta^\pm$ are measured with respect to the $Z$ axis, and the azimuthal angles $\varphi^\pm$ with respect to the $X$ axis. The centre-of-mass energy of the collision $\sqrt{s}$ is spread around its average by the beam energy spread $\sigma_\delta^\pm$ arising from synchrotron radiation (SR) and beamstrahlung (BS). Beam energy spread therefore plays a role similar to that of ISR photons in these equations. 

Under these hypotheses, Eqs.~\ref{eq:xconv} to~\ref{eq:econv} are straightforwardly solved event-by-event for $\alpha$ and $x_\gamma = p_z^\gamma/\sqrt{s}$ as a function of the four muon polar and azimuthal angles:
\begin{equation}
\alpha = 2\arcsin\left[ \frac{\sin\left( \varphi^- - \varphi^+\right)\sin\theta^+\sin\theta^-}{\sin\varphi^-\sin\theta^- - \sin\varphi^+\sin\theta^+} \right], {\rm \ \  and}
\label{eq:alpha}
\end{equation}
\begin{equation}
x_\gamma = -\frac{x_+\cos\theta^+ + x_-\cos\theta^-}{\cos(\alpha/2) +\lvert x_+\cos\theta^+ + x_-\cos\theta^- \rvert },
\label{eq:reduced}
\end{equation}
where $x_\pm = E^\pm \cos(\alpha/2) / (\sqrt{s} - \lvert p_Z^\gamma \rvert )$ are the reduced muon energies, and can be expressed as well as a function of the muon polar and azimuthal angles:
\begin{equation}
x_\pm = \frac{\mp \sin\theta^\mp\sin\varphi^\mp}{\sin\theta^+\sin\varphi^+-\sin\theta^-\sin\varphi^-}.
\label{eq:boost}
\end{equation}

\subsubsection{Measurement of the crossing angle in collisions}
\label{sec:crossinganglemeasurement}

\begin{figure}[ht]
\begin{center}
\includegraphics[width=0.72\columnwidth]{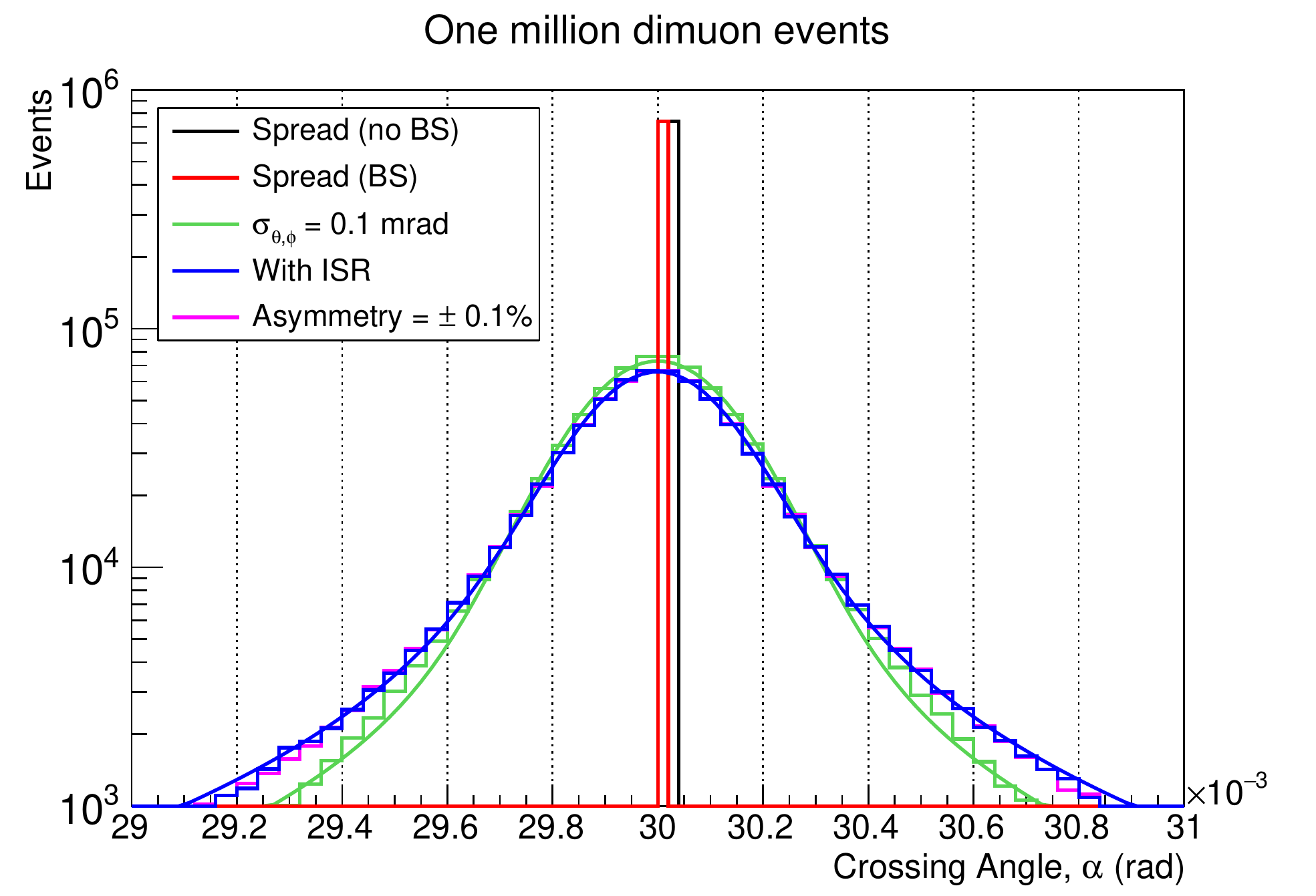}
\includegraphics[width=0.72\columnwidth]{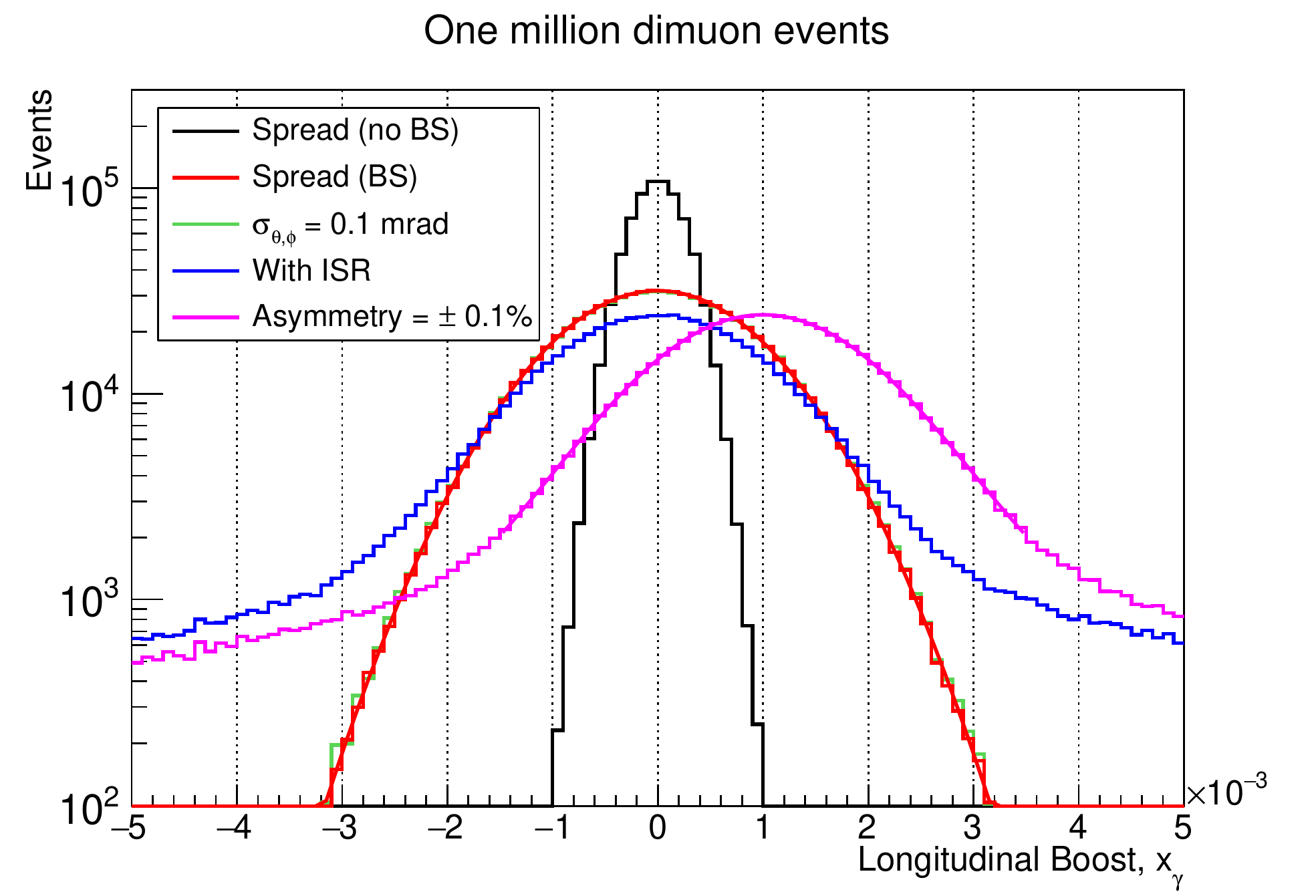}
\caption{\small Distributions of the crossing angle $\alpha$ (top) and the longitudinal boost $x_\gamma$ (bottom) for $10^6$ dimuon events generated at the Z pole, as determined event by event from the muon polar and azimuthal angles. The crossing angle kick due to beam-beam interactions ($+0.178$\,mrad on average with nominal parameters) is not included in the top plot. For an event to enter the distributions, both muons are required to satisfy $\lvert \cos\theta ^\pm\rvert < 0.9$ and $\lvert \sin\varphi^\pm \rvert > 0.2$. Pink: most realistic distribution, with nominal Gaussian beam energy spread (0.132\%), beam energy asymmetry of $\pm 0.1\%$, muon angular resolution of 0.1\,mrad, and ISR simulated up to second order in $\alpha_{\rm QED}$ (with the possibility of photon emission by the two beams at nonzero angle with respect to the beams). Blue: same as pink, but with no asymmetry between the electron and positron energies. Green: same as blue, but without ISR. Red: same as green, but with perfect angular resolution. Black: same as red, but without beamstrahlung (energy spread of 0.038\%). The green and blue $\alpha$ histograms are fitted to a Voigtian to guide the eye. The red and pink $x_\gamma$ histograms are fitted to a Gaussian and a Voigtian, respectively.} 
\label{fig:boostalpha}%
\end{center}
\end{figure}

The value of the beam crossing angle can be determined for each event (Eq.~\ref{eq:alpha}), together with the reduced muon energies, without the use of the $p_z$ conservation equation (Eq.~\ref{eq:zconv}). This value is therefore independent of whatever happens along the $z$ axis, such as beam energy spread or ISR/BS photon emission along that axis, as is illustrated in Fig.~\ref{fig:boostalpha} (top). In this figure, the distribution of the value of $\alpha$ as reconstructed from the muon angles is displayed for $10^6$ dimuon events generated at the Z pole ($\sqrt{s} = 91.2$\,GeV) with $\alpha=30$\,mrad and with the nominal Gaussian beam energy spread: 0.132\% (0.038\%) with (without) BS. If the muon angles are perfectly measured, the reconstructed crossing angle is strictly equal to the nominal crossing angle for all events, independently of the energy spread (black and red histograms), as previously inferred. An uniform polar and azimuthal muon angular resolution of 0.1\,mrad, typical of the detector designs studied for FCC-ee and other ${\rm e^+e^-}$ colliders, causes the distribution to acquire a Voigtian shape with a Breit-Wigner width and a Gaussian sigma, of about 0.1\,mrad each. Initial state radiation strictly along the $z$ axis would not modify this distribution, but the Breit-Wigner width is quite sensitive to the angular distribution of ISR photons -- and thus of the colliding electrons and positrons -- and increases to 0.17\,mrad. The beam vertical divergence at the interaction point ($\sim 0.045$\,mrad at the Z pole) and the crossing angle kick due to beam-beam interaction (Section~\ref{sec:oncoming_beam}), both not simulated in this figure, would add in quadrature to this width, slightly increasing it to 0.19\,mrad. With $10^6$ dimuon events, expected to be recorded in 5 minutes at the Z pole, the crossing angle (taken as the peak of the fitted Voigtian function) can be determined with a sub-$\mu$rad statistical precision:
\begin{equation}
\langle \alpha \rangle = 29.9998 \pm 0.0003 {\rm \,mrad}.
\end{equation}

\subsubsection{Measurement of the centre-of-mass energy spread in collisions}
\label{sec:sigmaE-muons}

\begin{figure}[htbp]
\begin{center}
\includegraphics[width=0.75\columnwidth]{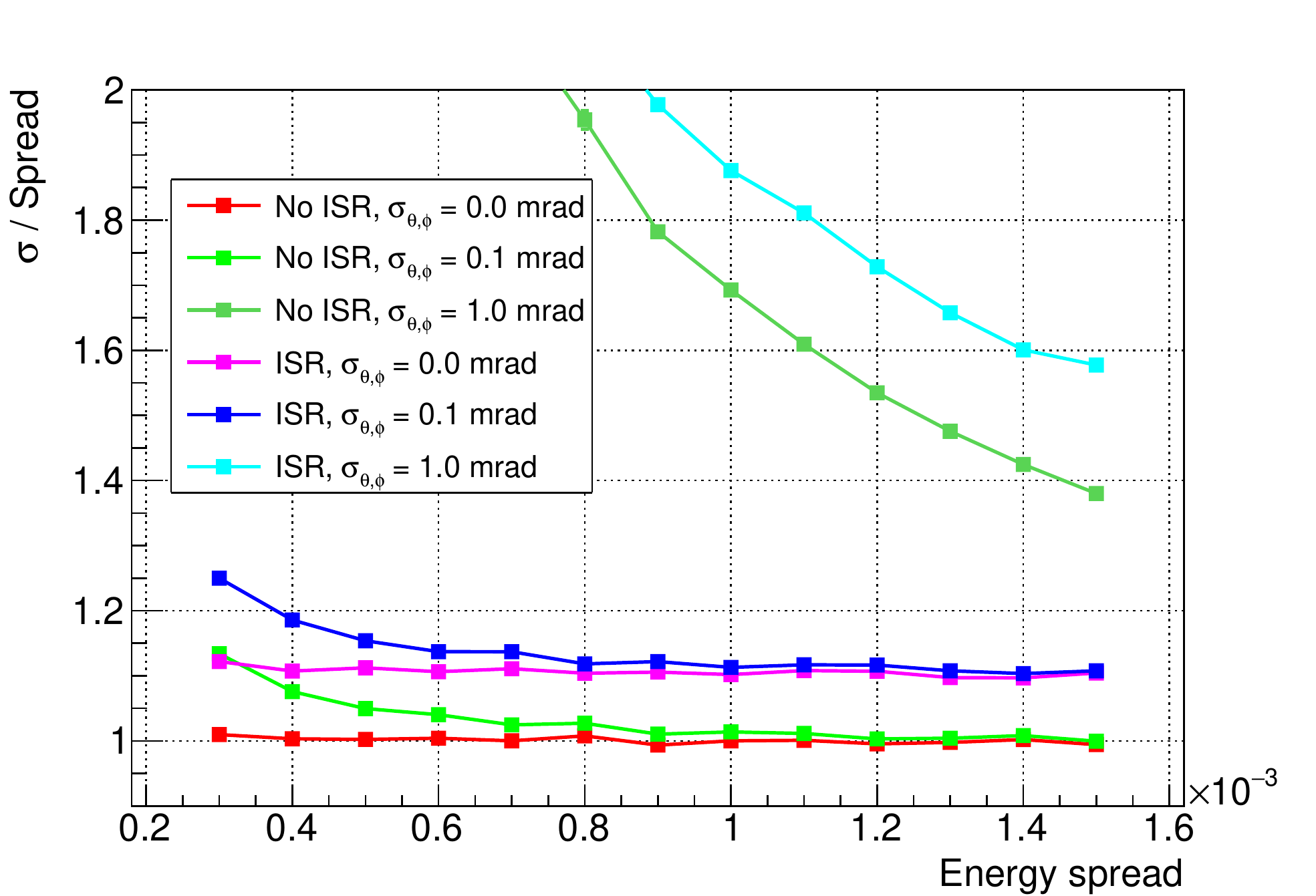}
\caption{\small Ratio of the fitted $x_\gamma$ distribution Gaussian width $\sigma$ to the true centre-of-mass energy spread, for a beam-energy spread varying from 0.03\% to 0.15\%, reconstructed from dimuon events generated at the Z pole. Red/Pink: without/with, perfect muon angular resolution; Green/Blue: without/with ISR, muon angular resolution of 0.1\,mrad; Dark green/light blue: without/with ISR, muon angular resolution of 1\,mrad. Each dot uses a sample of $10^5$ dimuon simulated events.}
\label{fig:boostcalib}%
\end{center}
\end{figure}

The resulting value of $x_\gamma$ (Eq.~\ref{eq:boost}), determined for each event by injecting Eqs.~\ref{eq:alpha} and~\ref{eq:reduced} in the $p_z$ conservation equation (Eq.~\ref{eq:zconv}), can be reinterpreted as the relative difference between the energies of the colliding electron and positron for this very event. In the absence of initial state photon emission (ISR), and if the muon angles are perfectly measured, the distribution of $x_\gamma$ is therefore nothing but the distribution of the relative centre-of-mass energy spread, centred around the mean asymmetry between the electron and positron beam energies, as illustrated by the black (no BS) and red (BS) histograms of Fig.~\ref{fig:boostalpha} (bottom). In this figure, the energy spread is assumed to be Gaussian and identical for both beams, but the distribution of $x_\gamma$ remains that of the relative centre-of-mass energy spread even if it were not the case. Angular resolution (green histogram) and photon emission (blue histogram) do not affect the average energy asymmetry (assumed to be 0.1\% in the pink histogram), which can be determined with an excellent precision from the $10^6$ dimuon event sample,
\begin{equation}
\langle x_\gamma \rangle = (0.9991 \pm 0.0015)\times 10^{-3}.
\end{equation}
They do, however, broaden and alter the shape of the $x_\gamma$ as displayed in the figure, and therefore need to be known/predicted with some accuracy to be unfolded towards the extractions of the centre-of-mass energy spread distribution. To estimate the precision required on the knowledge of the angular resolution and photon emission spectrum, the $x_\gamma$ distribution is fit with a Gaussian shape in the $\pm 2 \sigma$ interval, with or without ISR, with either perfect or finite angular resolution, and for a Gaussian beam energy spread varying from 0.030\% to 0.150\%, representative of all the possible FCC-ee conditions expected at the Z pole. The ratio of the fitted $\sigma$ to the true centre-of-mass energy spread is displayed in Fig.~\ref{fig:boostcalib} in the different hypotheses. 

As already alluded to, the fitted $\sigma$ in absence of ISR and with perfect angular resolution (red dots) equals exactly the value of the relative centre-of-mass energy spread $\sigma_\delta/\sqrt{2}$. With $10^6$ dimuon events, the statistical precision reached on $\sigma_\delta$ is at the level of one part in a thousand:
\begin{eqnarray}
\sigma_\delta = (0.03804\pm 0.00004 )\% & { \rm for\  an\  energy\  spread\  of}  & 0.038\% {\rm \ (no\ BS),} \label{eq:SPnoBS}\\
\sigma_\delta = (0.13185\pm 0.00011 )\% & { \rm for\  an\  energy\  spread\  of}  & 0.132\% {\rm \ (nominal\ BS).} \label{eq:SPBS} 
\end{eqnarray}
Initial state radiation increases $\sigma$ by about 10\%, irrespective of the beam energy spread (pink dots). To maintain the precision achieved without ISR (Eqs.~\ref{eq:SPnoBS} and~\ref{eq:SPBS}), the ISR spectrum must therefore be known to 1\% or better, a figure that was already surpassed by one order of magnitude at the time of LEP for the Z lineshape determination, and that is expected to improve by at least another order of magnitude for the Z lineshape fit at FCC-ee. A muon angular resolution of 0.1\,mrad causes $\sigma$ to further increase by 0.5\% in nominal beamstrahlung conditions (green and blue dots), and must therefore be known to about 10\% over the whole detector acceptance to maintain the per-mil precision on the centre-of-mass energy spread. It is beyond the scope of the present analysis to develop the complete strategy for unfolding the known effects of ISR and angular resolution towards the extraction of the full relative centre-of-mass energy spread spectrum. 

Figure~\ref{fig:boostcalib} also shows the devastating effect of a 10-times worse muon angular resolution (1\,mrad instead of 0.1\,mrad, dark green and light blue dots), which then becomes the dominant component of the $x_\gamma$ distribution Gaussian width and would have to be known to 0.1\% over the whole detector acceptance to preserve a per-mil precision on the centre-of-mass energy spread in nominal beamstrahlung conditions. A muon angular resolution of 0.1\,mrad or better is therefore an inescapable requirement for the future FCC-ee detectors.
It is beyond the scope of this analysis to document detailed methods to measure the muon angular resolution with a modest 10\% accuracy for each point of the ($\theta,\varphi$) plane: it can be trivially extracted for each track from the comparison of the angles reconstructed with the odd ($1^{\rm st}, 3^{\rm rd}, \dots, (n-1)^{\rm th}$) and even ($2^{\rm nd}, 4^{\rm th}, \dots, n^{\rm th}$) hits along that track. The result for the $\varphi$ resolution can even be cross-checked with the spread of the $\alpha$ distribution (dominated by this resolution). 

\subsubsection{Absolute angle determination}
It is, however, of the utmost importance to explain how and with what systematic precision the \underline{absolute} values of the angles $\theta^\pm$  and $\varphi^\pm$ are determined. Any unknown systematic bias would indeed modify the $\alpha$ and $x_\gamma$ distributions in an unpredictable manner, and jeopardize the accuracy on the centre-of-mass energy spread. Systematic biases on the muon angles arise from the fact that the ``local'' reference frame is in general different from the ``natural'' reference frame used in Eqs.~\ref{eq:xconv} to~\ref{eq:econv}. In the local reference frame, the $Z_{\rm local}$ axis coincides with the detector solenoid axis and might differ from the natural bisecting line between the two beam axes and the horizontal ($X_{\rm local},Z_{\rm local}$) plane might not naturally contain the two beams (with a direct subsequent difference between the $Y_{\rm local}$ and $Y$ axes).

\begin{figure}[htbp]
\begin{center}
\includegraphics[width=0.49\columnwidth]{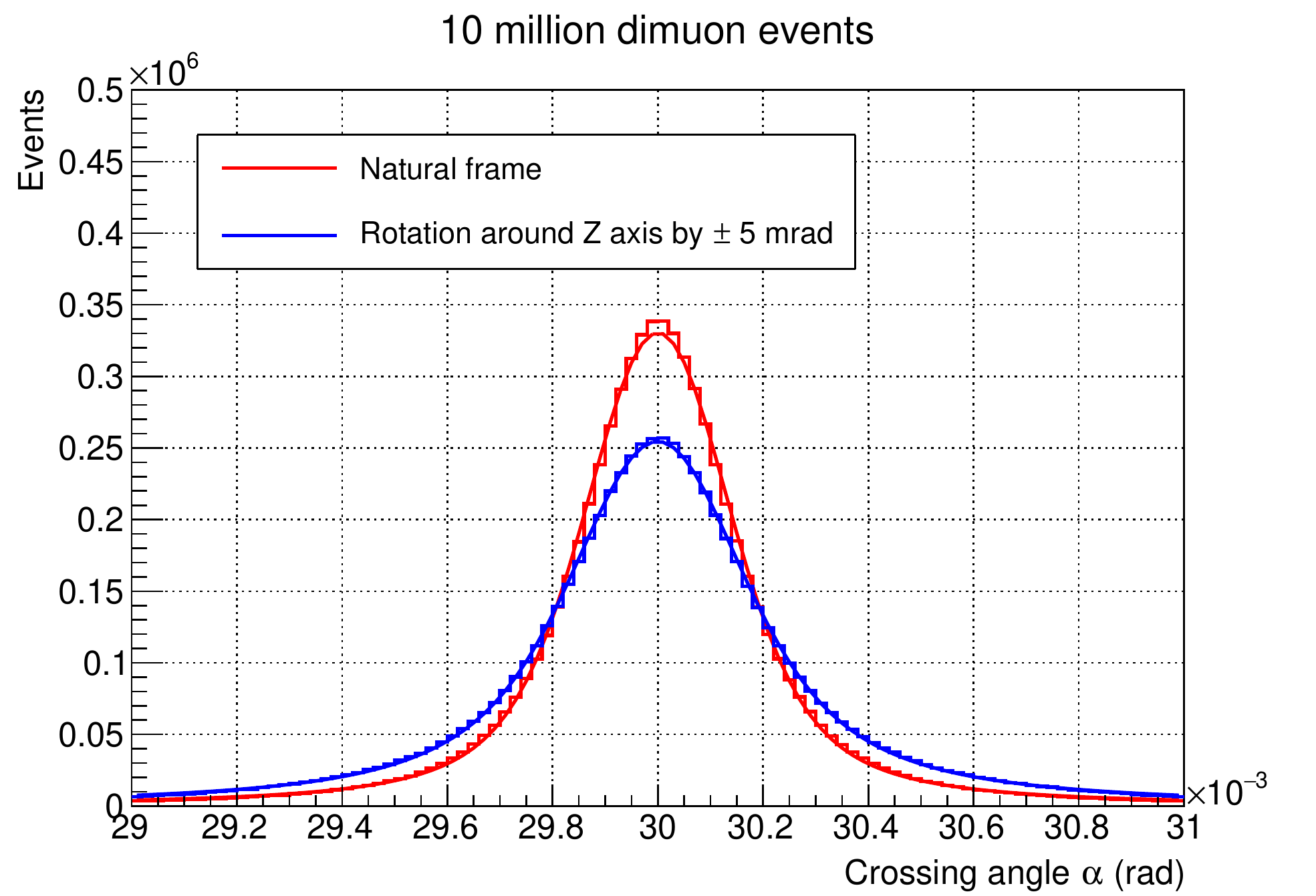}
\includegraphics[width=0.49\columnwidth]{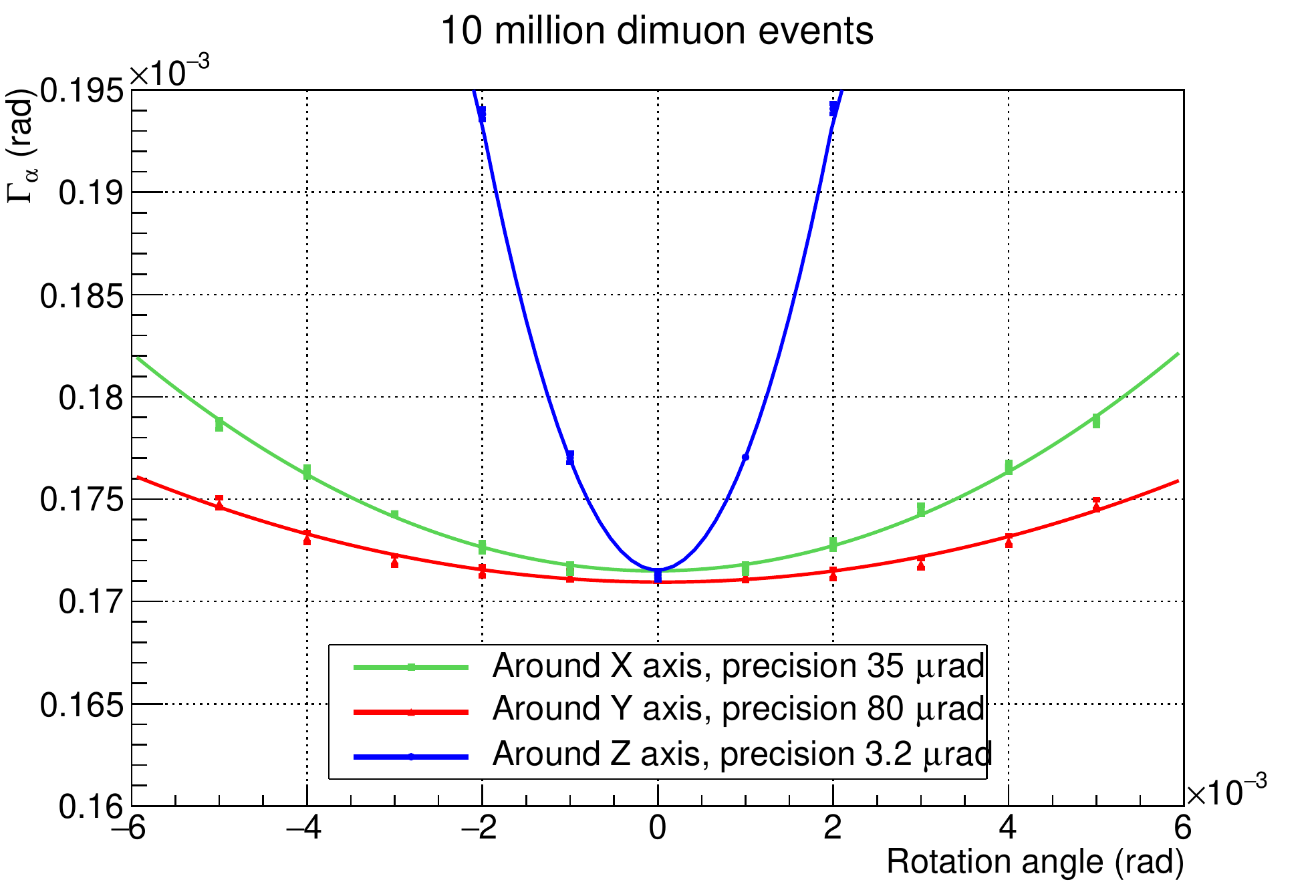}
\caption{\small (Left) Distribution of the crossing angle $\alpha$ reconstructed from the muon directions, in a sample of $10^7$ dimuon events generated at the Z pole, with ISR, angular resolution of 0.1\,mrad, maximal beamstrahlung, and beam energy asymmetry of $\pm 1\%$. The red histogram assumes that the local and natural frames coincide, while the blue histogram is obtained when the natural frame is rotated around the $Z$ axis by $\pm 5$\,mrad. Both histograms are fitted to a Voigtian. (Right) Variation of the Voigtian width $\Gamma_\alpha$ with a rotation of the natural frame around the $X$ (green), $Y$ (red), and $Z$ (blue) axes, as a function of the rotation angle from $-6$ to $+6$\,mrad, and fitted to parabolas to minimize the width. In the simulation used to make this figure, the local and natural frames were chosen to coincide, and the minimum of the Voigtian width is indeed found for rotation angles compatible with 0\,$\mu$rad.}
\label{fig:alpharotation}%
\end{center}
\end{figure}

As exemplified in Fig.~\ref{fig:boostalpha} (top), however, the Voigtian width of $\alpha$ distribution increases with anything happening in the transverse plane (e.g., angles of the ISR photons, $\varphi$ angular resolution, etc.). A modification of any of the three axis directions would therefore have a direct impact on this width. All possible modifications can be parametrized with three successive Euler rotations, around each of the three axes $X$, $Y$, and $Z$. The largest effect is expected from a rotation around the $Z$ axis, as it changes both the $X$ and $Y$ transverse axes simultaneously. A smaller but still measurable effect is expected from the rotations around the $X$ and $Y$ axes, as they change the $Y$ and $X$ directions, respectively. The $\alpha$ distribution is shown in Fig.~\ref{fig:alpharotation} (left) when all local axes coincide with those of the natural frame, and with a rotation of the local frame around the $Z$ axis by $\pm 5$\,mrad, for a sample of $10^7$ dimuon events. The right panel of the same figure displays the variation of the Voigtian width with a rotation of the natural frame around the $X$, $Y$, and $Z$ axes, as a function of the rotation angle. The rotation angles of the natural frame with respect to the local frame around the $X$, $Y$, and $Z$ axes can be determined by minimizing the Voigtian width, with a precision of 35, 80, and 3.2\,$\mu$rad in less than one hour at the Z pole, and of 10, 25, and 1\,$\mu$rad within a ten-hours fill. These figures get quadratically better with smaller $\varphi$ resolution, which dominates the natural width of the $\alpha$ distribution. 

\begin{figure}[htbp]
\begin{center}
\includegraphics[width=0.49\columnwidth]{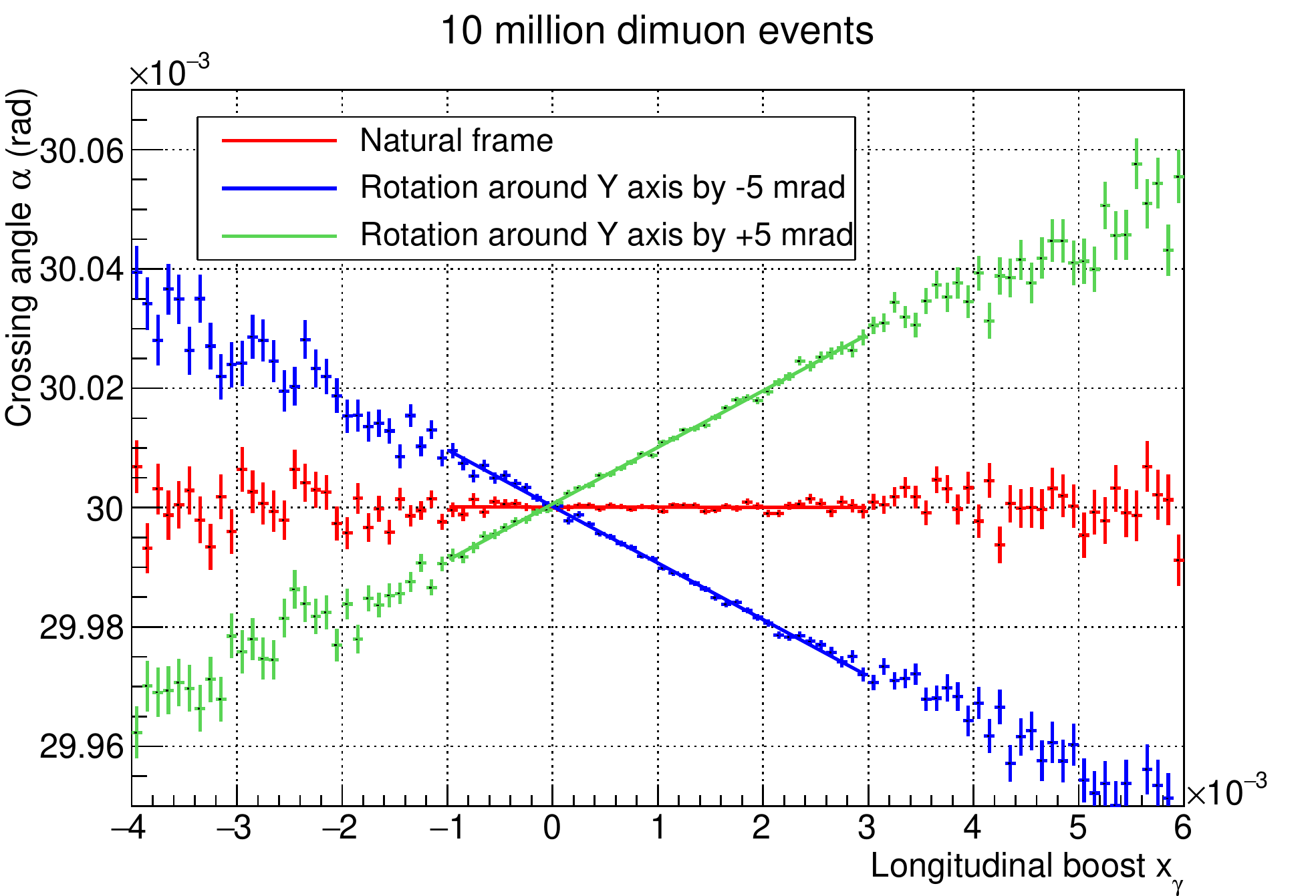}
\includegraphics[width=0.49\columnwidth]{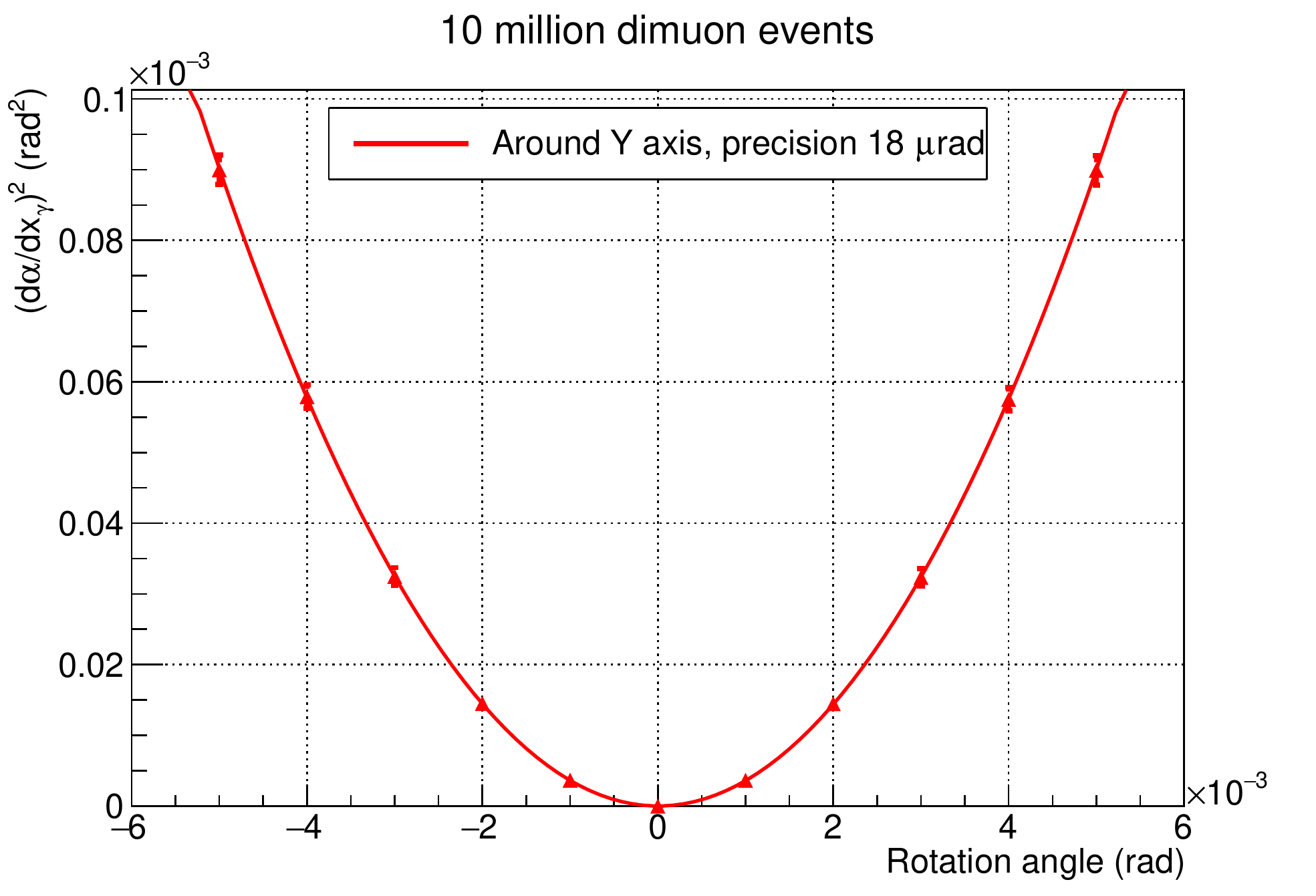}
\caption{\small (Left) Distributions of $\langle \alpha \rangle$ vs $\langle x_\gamma \rangle$ in a sample of $10^7$ dimuon events generted at the Z pole, with ISR, angular resolution of 0.1\,mrad, maximal beamstrahlung, and beam energy asymmetry of $\pm 1\%$, when the natural frame is rotated with respect to the local frame by $-5$ (blue), $0$ (red), and $+5$\,mrad (green) around the $Y$ axis. The distributions are fitted to a linear function. (Right) Variation of the square of the slope of this linear fit as a function of the rotation angle around the $Y$ axis from $-6$ to $+6$\,mrad, and fitted to a parabola to minimize the slope. In the simulation used to make this figure, the local and natural frames were chosen to coincide, and the minimum is indeed found for a rotation angle compatible with 0\,$\mu$rad.}
\label{fig:alphaxslope}%
\end{center}
\end{figure}

The determination of the rotation angle around the $Y$ axis can be improved by exploiting the observation that the $X$ and $Z$ measurements get mixed with such a rotation, resulting in a strong correlation between the reconstructed $\alpha$ and $x_\gamma$ values, as displayed in Fig.~\ref{fig:alphaxslope} (left) for rotation angles of $-5$, $0$, and $+5$\,mrad with a sample of $10^7$ dimuon events. The minimization of this slope allows the determination of the rotation angle of the natural frame with respect to the local frame around the $Y$ axis with a precision of 18\,$\mu$rad in less than one hour, and of 5\,$\mu$rad within a 10-hours fill, thus improving the precision by a factor five with respect to the $\alpha$ Voigtian width minimization. 

With such accuracies, the systematic biasses on $\langle \alpha \rangle$ and on $\langle x_\gamma \rangle$ are found to be smaller than $0.1\,\mu$rad and $10^{-7}$, respectively. For all practical purposes, the variations of the $x_\gamma$ distribution around $\langle x_\gamma \rangle$ are already insignificant with 100 times smaller event samples. There might be even more precise methods to determine with the data the orientation of the natural frame with respect to the local frame. As the precision reached with the method presented here is sufficient, it is left to the future FCC-ee scientists to find and optimize them.  

The importance of a nonzero crossing angle for the success of the method cannot be overemphasized. In all figures of this section, the muon directions are required to satisfy $\lvert \cos\theta^\pm \rvert < 0.9$ and $\lvert \sin\varphi^\pm \rvert > 0.2$. The former requirement is imposed by the acceptance of the detector tracker, but the latter requirement rejects events close to the horizontal plane in which the two muons are back-to-back and for which the numerator and the denominator of Eq.~\ref{eq:alpha} are both close to zero. It is only when $\varphi$ is nonzero that the finite crossing angle generates a nontrivial angle between the two muon directions, which in turns allows $\alpha$ and $x_\gamma$ to be determined with precision, even when $x_\gamma=0$. When the beams collide head on instead, the muons are produced back-to-back, and the vanishing crossing angle can never be determined with precision from Eq.~\ref{eq:alpha}, with the consequence of an inaccurate alignment of the local frame to the natural frame and of uncontrollable systematic biasses on the muon angles. 

\subsubsection{Number of events and time needed at the various running points}

To conclude, the number of events required (and the time needed to collect them) to satisfy the precision requirements on the beam energy spread measurement at the various centre-of-energies, with nominal FCC-ee parameters, as given in Sections~\ref{sec:Zrequirements}, \ref{sec:Wrequirements} and~\ref{sec:TOPrequirements}, are summarized in Table~\ref{tab:time} for the method described above. The beam energy spread can be monitored with the required precision within a couple minutes at and around the Z pole for the Z width and the $\alpha_{\rm QED}(m^2_{\rm Z})$ determination; within four minutes at the WW threshold for the W width determination; and within 30 minutes at the top-pair threshold for the top-quark width determination. Other measurements, such as the Z, W, Higgs-boson and top-quark masses or the effective Weinberg angle $\sin^2\theta^{\rm eff}_{\ell}$, are much less affected by the knowledge of beam energy spread (or the lack thereof). 

\begin{table}[htbp]
  \centering
\caption{\small Requirements on the precision of the beam energy spread measurement for nominal FCC-ee parameters. The first line indicates the precision electroweak pseudo observables relying on measurements that are significantly biassed by the beam energy spread, and the third line gives the corresponding ``acceptable'' uncertainty arising from the precision with which the beam energy spread is determined. The value of the acceptable uncertainty is chosen so that the FCC-ee target uncertainty on the pseudo observable determination (100\,keV for the Z width, $3\,10^{-5}$ on the electromagnetic coupling constant, 1.3\,MeV on the W width, and 45\,MeV on the top-quark width) is increased by about $\sim 5\%$. The centre-of-mass energies at which the measurement is performed are shown in the third line, and the precision of the energy spread measurement required to reach the acceptable uncertainty in the fourth line. The number of ${\rm e^+e^-} \to \mu^+\mu-$ events needed to reach this precision is given in the fifth line. The dimuon rate determined from the luminosity (sixth line) and the $\mu^+\mu-$ production cross section (seventh line) is displayed in the eighth line. The time needed to reach the required precision on the beam energy spread is deduced in the last line. \vspace{0.2cm}}
\label{tab:time}
\begin{tabular}{|c|c|c|c|c|c|c|c|}
\hline
Pseudo Observable & \multicolumn{3}{|c|}{$\Gamma_{\rm Z}$} & \multicolumn{2}{|c|}{$\alpha_{\rm QED}(m^2_{\rm Z})$} & $\Gamma_{\rm W}$ & $\Gamma_{\rm top}$ \\
\hline
Acceptable error & \multicolumn{3}{|c|}{35\,keV} & \multicolumn{2}{|c|}{$10^{-5}$} & 0.5\,MeV & 18\,MeV\\
\hline\hline
$\sqrt{s}$ (GeV) & 87.9 & 91.2 & 93.8 & 87.9 & 93.8 & 161 & 350 \\
\hline
$\sigma(\delta E)/\delta E$ & $0.8\%$ & $0.2\%$ & $0.8\%$ &\multicolumn{2}{|c|}{$0.7\%$} & 11\% & 35\% \\
\hline
$N_{\rm e^+e^- \to \mu^+\mu^-}$ & $5\, 10^4$ & $8\, 10^5$ & $5\, 10^4$ & \multicolumn{2}{|c|}{$6.5\, 10^4$} & 260 & 25 \\
\hline\hline
{\cal L} ($10^{34}\,{\rm cm^{-2}s^{-1}}$) & \multicolumn{5}{|c|}{230} & 28 & 1.8 \\
\hline
$\sigma_{\mu\mu}$ (pb) & 185 & 1450 & 460 & 185 & 460 & 4.0 & 0.8\\
\hline
Dimuon rate (Hz) & 425 & 3325 & 1050 & 425 & 1050 & 1.1 & 0.015 \\
\hline\hline
Time needed & 2\,min & 4\,min & $< 1$\,min & 3\,min & 1\,min & 4\,min & 30 min \\
\hline
\end{tabular}
\end{table}

If the collider operations called for smaller than nominal luminosities (e.g., due to smaller electrical power available, or for beam stability reasons), the dimuon rate would reduce proportionally. Because the energy spread would decrease as well, however, the precision requirements would become quadratically less stringent than displayed in Table~\ref{tab:time}. For example, if the relative energy spread at the Z pole decreases by a factor of two from 0.132\% to 0.066\% (thus still beamstrahlung dominated), the requirement on the relative precision is relaxed by a factor of four from 0.2\% to 0.8\%, and the number of $\mu^+\mu^-$ events needed to reach this precision decreases by a factor of 16 from $8\times10^5$ to $5\times 10^4$. In the same time, if halving the beam energy spread is the result of reducing the bunch population from $1.7 \times 10^{11}$ to $0.47 \times 10^{11}$ particles, with a corresponding increase in the number of bunches, the luminosity gets smaller by a factor 1.83.  In that case, the time needed to collect $5\times 10^4$ dimuon events becomes close to 30 seconds. The time periods indicated in the last line of the table are therefore to be considered as absolute maxima, irrespective of the (beamstrahlung-dominated) running conditions. 

\subsubsection{Measurement of the beam crossing angle increase in collisions}

As studied in Section~\ref{sec:oncoming_beam}, the beam crossing angle causes the beam-beam interactions at the IP to increase the ${\rm e^\pm}$ energies $E^0_\pm$ by a quantity $\delta E_\pm$, and to increase the beam crossing angle $\alpha_0$ by a quantity $\delta \alpha$. These beam-beam interactions, the effects of which were not included so far in this section, do not modify the centre-of-mass energy: 
\begin{equation}
    \sqrt{s} = 2 \sqrt{E_+^0 E_-^0} \cos\sfrac{\alpha_0}{2} = 2 \sqrt{E_+E_-} \cos\sfrac{\alpha}{2}, 
    \label{eq:Ecomwobeambeam}
\end{equation}
where $E_\pm = E_\pm^0 +\delta E_\pm$, and $\alpha = \alpha_0 + \delta\alpha$. Resonant depolarization of non-colliding bunches, unaffected by beam-beam interactions, allows the measurement of $E_\pm^0$. On the other hand, the method described in Section~\ref{sec:crossinganglemeasurement} delivers a measurement of $\alpha$, in the presence of beam-beam interactions. To determine the centre-of-mass energy from the left part of Eq.~\ref{eq:Ecomwobeambeam}, the unaltered crossing angle $\alpha_0 = \alpha - \delta\alpha$ is needed. It is therefore necessary to determine the crossing angle increase $\delta\alpha$ caused by beam-beam interactions. From the right part of Eq.~\ref{eq:Ecomwobeambeam}, the crossing angle increase $\delta\alpha$ is directly related to the beam energy increase $\delta E_\pm$:
\begin{equation}
\delta \alpha = \frac{1}{\tan\sfrac{\alpha}{2}} \left( \frac{\delta E_+}{E_+} + \frac{\delta E_-}{E_-}  \right),
\label{eq:crossangleincrease}
\end{equation}
which gives numerically, for the nominal FCC-ee paramaters at the Z pole, with $\alpha = 30$\,mrad, $E_\pm = 45.6$\,GeV, and $\delta E_\pm = 60.5$\,keV (as predicted by the {\tt LifeTrac} simulation code):
\begin{equation}
\delta \alpha = 0.177\, {\rm mrad}, 
\end{equation}
corresponding to a crossing angle relative increase of 0.58\%, to be measured experimentally. If $\Delta\delta\alpha$ is the precision with which the crossing angle increase can be measured, the impact of this precision on the $\sqrt{s}$ accuracy (as obtained from the left part of Eq.~\ref{eq:Ecomwobeambeam}) amounts to 
\begin{equation}
    \frac{\Delta \sqrt{s}}{\sqrt{s}} \simeq \frac{1}{4}\alpha\delta\alpha \ \frac{\Delta\delta\alpha}{\delta\alpha} \approx 1.3 \times 10^{-6} \  \frac{\Delta\delta\alpha}{\delta\alpha}.
\end{equation}
In other words, a measurement of $\delta\alpha$ with a moderate precision of 10\% would lead to a contribution $\Delta\sqrt{s}$ of 12\,keV on the centre-of-mass energy uncertainty, adequately small when compared to the uncertainty originating from the beam energy measurement with resonant depolarization. 

In principle, the ``filling'' period with the bootstrapping method~\cite{Ogur:IPAC2018-MOPMF001} is ideal for the measurement of the crossing angle increase: at the Z pole, half of the nominal intensity is first injected from the booster to the collider ring in electron and positron bunches, which are then alternately topped up by steps of 10\% every 52 seconds as indicated in Table~\ref{tab:bootstrapping} until the nominal intensity is reached for both. During the whole operation, these bunches collide with the nominal optics parameters. The corresponding luminosity, beam energy and centre-of-mass energy spreads, beam energy shifts (as determined by the {\tt Lifetrac} code), and crossing angle (as determined from Eq.~\ref{eq:crossangleincrease}), typically stabilize to the values indicated in Table~\ref{tab:bootstrapping} within less than 5 seconds, which leaves more than 40 seconds to record data at each step. The number of ${\rm e^+ e^-} \to \mu^+\mu^-$ events produced at each interaction point during these 40 seconds, with which the crossing angle and the centre-of-mass energy spread can be measured as explained above, is also indicated in the same table. 

\begin{table}[htbp]
  \centering
\caption{\small Number of particles $N_{\rm part}^\pm$ in each bunch of the collider ring (normalized to the nominal value at the Z pole) at each step of the FCC-ee filling period. Also given are the luminosity ${\cal L}$, the beam energy spreads $\sigma_\delta^\pm$ and the centre-of-mass energy spread $\sigma_{\sqrt{s}}$, all normalized to their nominal FCC-ee values; the energy kicks $\delta E^\pm$ (in keV); as computed by the {\tt Lifetrac} code. Finally, the crossing angle $\alpha$ (in mrad) and the number of ${\rm e^+ e^-} \to \mu^+\mu^-$ events produced at each interaction point during 40 seconds in each configuration are indicated in the last two columns. The {\tt LifeTrac} statistical uncertainties are of the order of 1\% for the luminosity and the spreads, and of the order of 3\% for the kicks.\vspace{0.2cm}}
 \label{tab:bootstrapping}
\begin{tabular}{|c|c|c|c|c|c|c|c|c|r|}
\hline
$N_{\rm part}^+$ & $N_{\rm part}^-$ & ${\cal L}$ & $\sigma_\delta^+$ & $\sigma_\delta^-$ & $\sigma_{\sqrt{s}}$ & $\delta E_+$ & $\delta E_-$ &  $\alpha$ & $N_{\mu^+\mu^-}$ \\
\hline\hline
0.50  &  0.50  &  0.37  &  0.68  &  0.68  &  0.680  &  39.2  &  39.2  & 30.1147 & 49210 \\ 
0.50  &  0.55  &  0.38  &  0.79  &  0.61  &  0.705  &  47.9  &  33.7  & 30.1193 & 50540 \\ 
0.60  &  0.55  &  0.44  &  0.64  &  0.84  &  0.747  &  35.5  &  51.5  & 30.1273 & 58250 \\ 
0.60  &  0.65  &  0.50  &  0.87  &  0.68  &  0.781  &  52.9  &  39.2  & 30.1347 & 66500 \\
0.70  &  0.65  &  0.56  &  0.69  &  0.93  &  0.819  &  40.1  &  56.5  & 30.1413 & 74480 \\ 
0.70  &  0.75  &  0.62  &  0.94  &  0.74  &  0.846  &  57.5  &  43.8  & 30.1480 & 82460 \\
0.80  &  0.75  &  0.68  &  0.76  &  0.99  &  0.883  &  44.7  &  61.6  & 30.1553 & 90440 \\
0.80  &  0.85  &  0.74  &  1.02  &  0.80  &  0.917  &  63.4  &  45.6  & 30.1593 & 98420 \\
0.90  &  0.85  &  0.81  &  0.82  &  1.04  &  0.936  &  49.2  &  65.2  & 30.1673 & 107730 \\
0.90  &  0.95  &  0.87  &  1.09  &  0.84  &  0.973  &  67.5  &  49.2  & 30.1707 & 115710 \\
1.00  &  0.95  &  0.91  &  0.86  &  1.12  &  0.998  &  49.2  &  67.5  & 30.1707 & 121030 \\ 
1.00  &  1.00  &  1.00  &  1.00  &  1.00  &  1.000  &  60.2  &  60.2  & 30.1760 & 133000 \\ 
\hline\hline
\end{tabular}
\end{table}

\begin{figure}[htbp]
\begin{center}
\includegraphics[width=0.75\columnwidth]{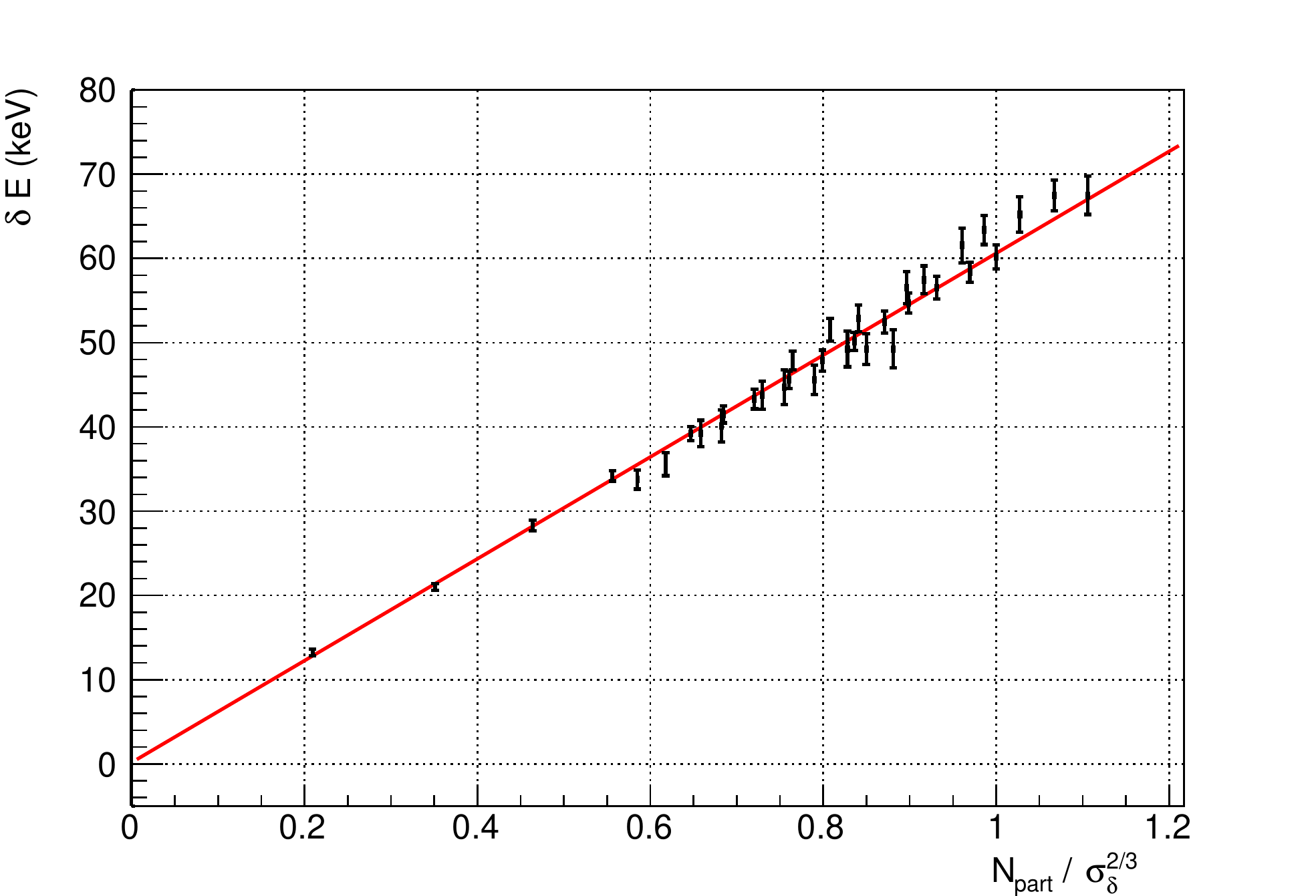}
\caption{\small Energy kick $\delta E$ (in keV) as a function of the opposite bunch population $N_{\rm part}$ divided by its energy spread $\sigma_\delta$ to the power 2/3, as obtained from {\tt Lifetrac} for varying bunch populations. Both $N_{\rm part}$ and $\sigma_\delta$ are normalized to their nominal FCC-ee values at the Z pole.  All other parameters are fixed to their nominal values at the Z pole. The uncertainties arise from the limited MC statistics. The line shows the result of a linear fit to the simulated points: the fitted energy shift is $0.2 \pm 1.0$\,keV for empty bunches, and amounts to $60.5 \pm 0.8$\,keV for nominal parameters. The uncertainty on these two parameters includes the limited MC statistics and the exponent variation in Eq.~\ref{eq:shift}.}
\label{fig:energykick}%
\end{center}
\end{figure}

Because the energy kicks $\delta E_\pm$ are directly proportional to the opposite bunch population $N_{\rm part}^\mp$, an extrapolation of the measured crossing angles with these varying bunch populations to $N_{\rm part}^\mp = 0$ should directly give the value of the unaltered crossing angle $\alpha_0$ and of the crossing angle increase $\delta\alpha$. The energy kicks also decrease when the opposite bunch length increase. A fit to a numerical integration of the analytical expression of the Lorentz force~\cite{Keil:269336, Placidi:353204} shows that these kicks are, all other parameters being equal, proportional to the opposite bunch population divided by the opposite bunch length (and therefore its energy spread) to the power 2/3:\footnote{The value of the power slightly depends on the bunch length range chosen for the fit, with a variation in the interval $0.67 \pm 0.05$. This dependence is taken as a systematic uncertainty in the rest of this section.}
\begin{equation}
    \delta E^\pm \propto \frac{N_{\rm part}^\mp}{{\sigma_\delta^\mp}^{2/3}}. 
    \label{eq:shift}
\end{equation}
This dependence was checked with the independent results from {\tt LifeTrac}, most of them listed in Table~\ref{tab:bootstrapping}, and complemented with four points at smaller bunch populations. As displayed in Fig.~\ref{fig:energykick}, the extrapolation of the linear fit to the {\tt Lifetrac} data gives a energy shift compatible with zero ($0.2 \pm 0.2 {\rm {\small (stat.)}} \pm 1.0 {\rm {\small (syst.)}}$\,keV) for empty bunches, and of $60.5 \pm 0.5 {\rm {\small (stat.)}} \pm 0.6 {\rm {\small (syst.)}}$\,keV for nominal parameters.

\begin{figure}[htbp]
\begin{center}
\includegraphics[width=0.49\columnwidth]{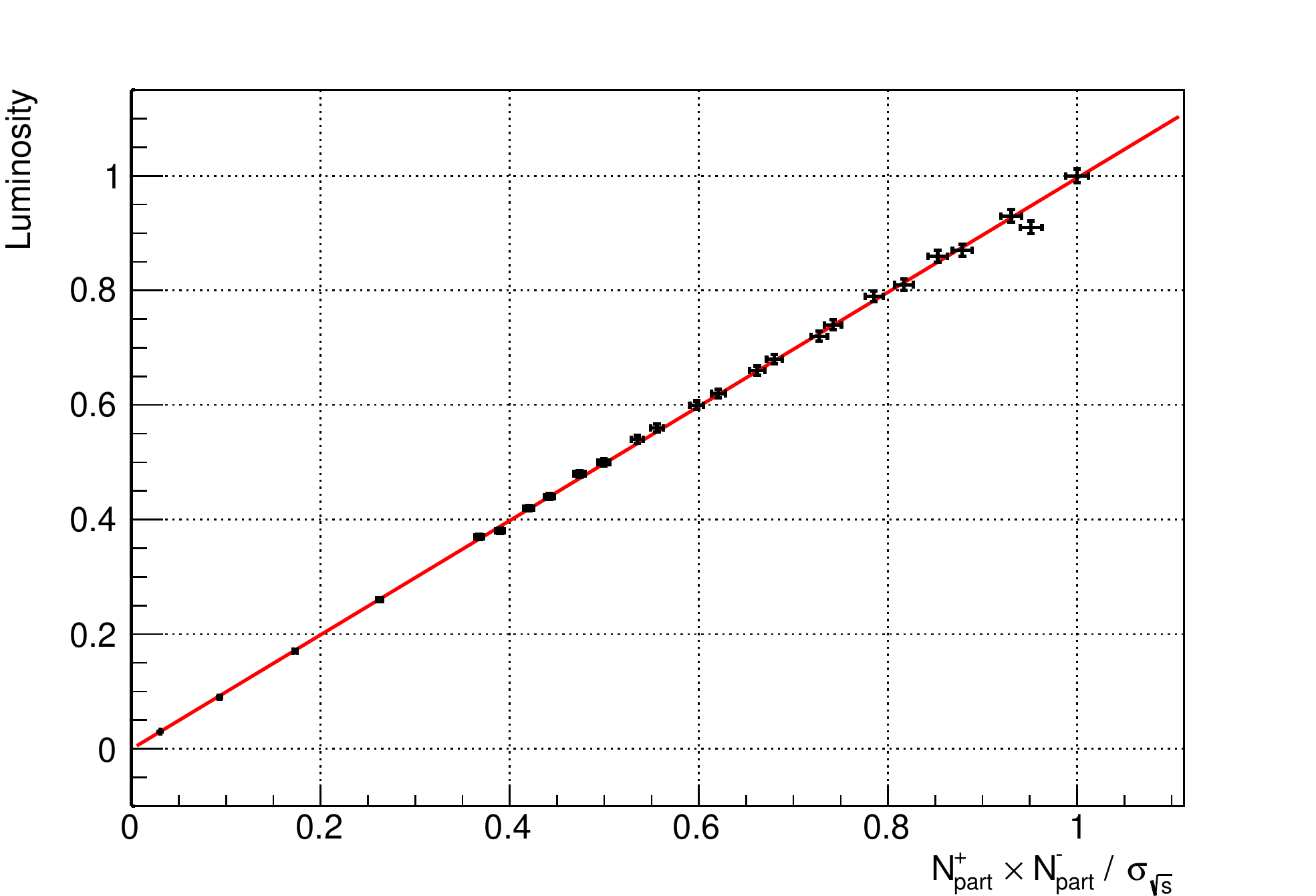}
\includegraphics[width=0.49\columnwidth]{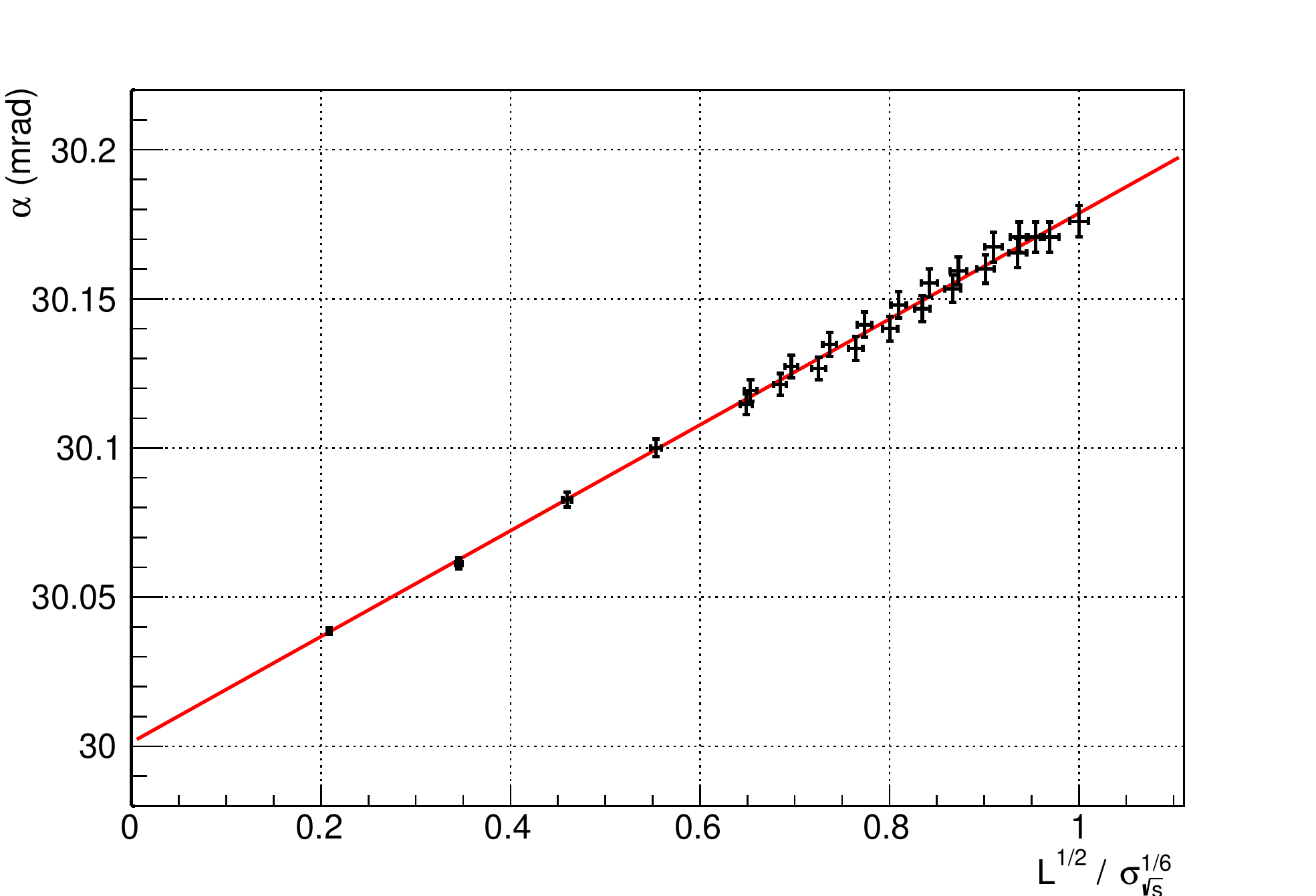}
\caption{\small Left: Luminosity ${\cal L}$ as a function of $N_{\rm part}^+ \times N_{\rm part}^- / \sigma_{\sqrt{s}}$. Right: Beam crossing angle $\alpha$ (in mrad) as a function of ${\cal L}^{1/2} / \sigma_{\sqrt{s}}^{1/6}$. Both plots are obtained from the {\tt Lifetrac} simulation code for bunch populations varying from 10\% to 100\% of the nominal FCC-ee value at the Z pole (keeping ${\rm e}^\pm$ bunch populations within $\pm 5\%$ from each other). The luminosity ${\cal L}$, the ${\rm e}^\pm$ bunch populations $N_{\rm part}^\pm$, and the centre-of-mass energy spread $\sigma_{\sqrt{s}}$ are normalized to their nominal values.  All other parameters are fixed to their nominal FCC-ee values at the Z pole. The uncertainties arise from the limited MC stastistics. The lines show the linear fits to the simulated points: for example, the fitted crossing angle is $30.0013 \pm 0.0031$\,mrad for empty bunches, and amounts to $30.1775 \pm 0.0032$\,mrad for nominal parameters.}
\label{fig:powerlaws}%
\end{center}
\end{figure}

For equal electron and positron bunch populations and lengths, the crossing angle increase is proportional to the (common) beam energy shift (Eq.~\ref{eq:crossangleincrease}), and  is therefore proportional to
\begin{equation}
\label{eq:comshift}
    \delta\alpha \propto \frac{N_{\rm part}}{\sigma_{\sqrt{s}}^{2/3}}.
\end{equation}
Under the same conditions, the luminosity ${\cal L}$ is proportional to
\begin{equation}
\label{eq:lumipopspread}
    {\cal L} \propto \frac{N_{\rm part}^2}{\sigma_z} \Leftrightarrow {\cal L} \propto \frac{N_{\rm part}^2}{\sigma_{\sqrt{s}}} .
\end{equation}
Equations~\ref{eq:comshift} and~\ref{eq:lumipopspread} can be merged into the following remarkable power law: 
\begin{equation}
\label{eq:remarkablepowerlaw}
    \delta\alpha \propto \frac{ {\cal L}^{1/2}} {\sigma_{\sqrt{s}}^{1/6}}.
\end{equation}
A small population asymmetry between electron and positron bunches would cause the bunch with smaller (larger) population and  larger (smaller) length to be more (less) kicked , with a vanishing effect, to first order, on $\delta \alpha$.  As displayed in Fig.~\ref{fig:powerlaws}, Eqs.~\ref{eq:lumipopspread} and~\ref{eq:remarkablepowerlaw} are indeed verified to hold with an asymmetry of up to $\pm 5\%$ with the {\tt Lifetrac} simulation code, for the same ${\rm e}^\pm$ bunch populations as in Fig.~\ref{fig:energykick}. In these two plots, $N_{\rm part}^2$ is replaced by $N_{\rm part}^+ N_{\rm part}^-$, and $\sigma_{\sqrt{s}} = \sigma_{\delta}^+ \oplus \sigma_{\delta}^-$.

The measurement of $\delta\alpha$ therefore requires the measurement of the luminosity, the centre-of-mass energy spread, and the crossing angle, with colliding beams and increasing bunch populations, up to the nominal FCC-ee value, while keeping the other machine parameters to their nominal value. These three quantities can be measured in situ by ${\rm e^+ e^-} \to \mu^+\mu^-$ events: the absolute measurements of $\alpha$ and $\sigma_{\sqrt{s}}$ are described earlier in this section, and the luminosity is simply proportional to $N_{\mu\mu}$, the number of $\mu^+\mu^-$ events, with the following statistical precision: 
\begin{equation}
    \frac{\Delta{\cal L}}{{\cal L}} = \frac{\Delta\sigma_{\sqrt{s}}}{\sigma_{\sqrt{s}}} = \frac{1}{\sqrt{N_{\mu\mu}}}; \ \ {\rm and} \ \ 
    \Delta\alpha = \frac{0.3\,{\rm mrad}}{\sqrt{N_{\mu\mu}}}.
\end{equation}

\begin{figure}[htbp]
\begin{center}
\includegraphics[width=0.80\columnwidth]{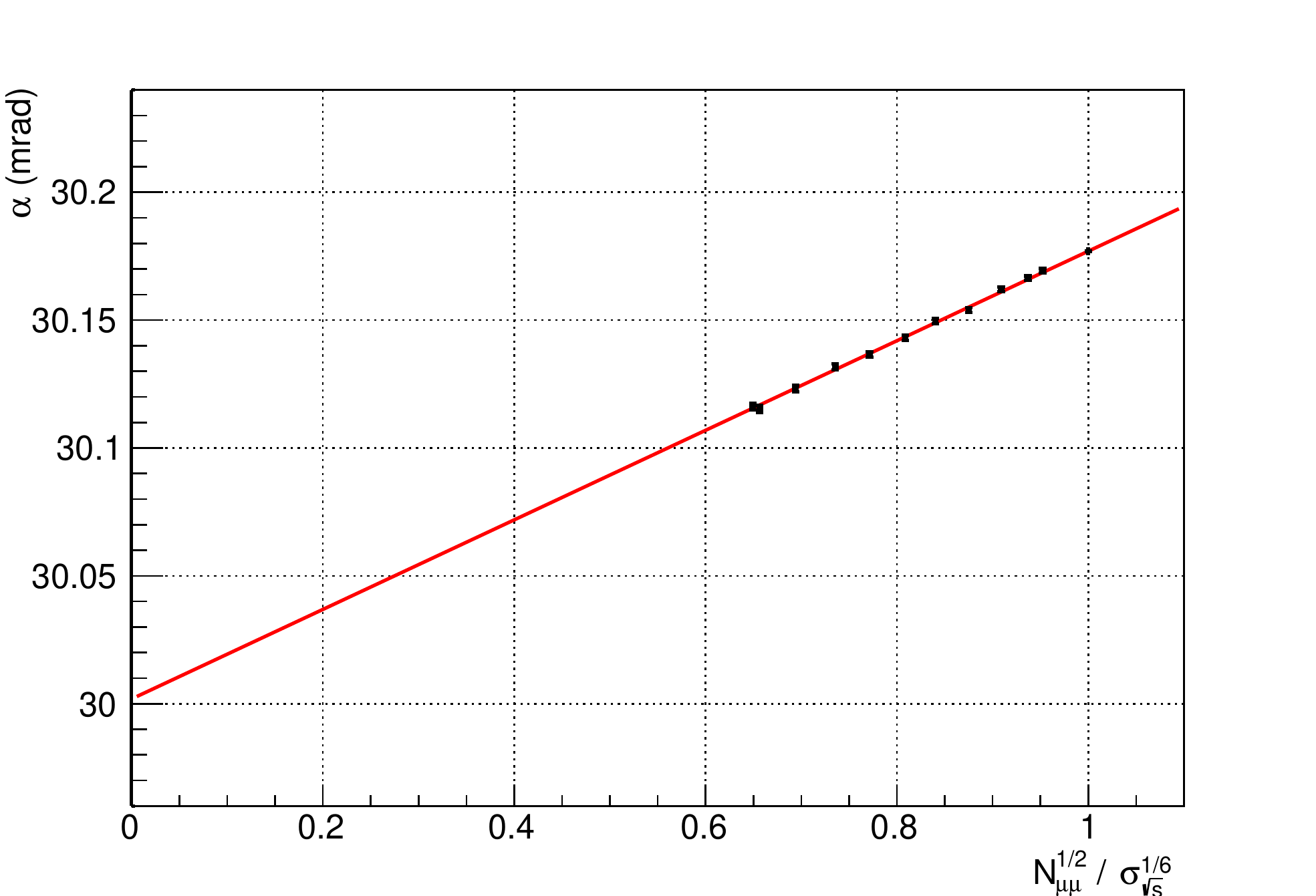}
\caption{\small Measurements of the crossing angle while increasing the ${\rm e}^\pm$ bunch populations during the FCC-ee filling period at the Z pole, as of function of the square root of the number of dimuon events collected in 40 seconds, divided by the measured centre-of-mass energy spread to the power 1/6 (normalized to their nominal values). The line shows a linear fit to the measurements. In this toy MC example, the fitted crossing angle well before the IP is $30.0008 \pm 0.0016$\,mrad, and its increase at the IP for nominal parameters amounts to $0.1761 \pm 0.0016$\,mrad (statistical errors only). Details can be found in the text.}
\label{fig:anglekickmeasurement}%
\end{center}
\end{figure}

The precision with which $\alpha$ and $N_{\mu\mu}^{1/2}/\sigma_{\sqrt{s}}^{1/6}$ can be measured at each step is illustrated in Fig.~\ref{fig:anglekickmeasurement}. A linear fit through the twelve measurements displayed in Fig.~\ref{fig:anglekickmeasurement} gives the following result: 
\begin{eqnarray}
\alpha_0 & = & 30.0008 \pm 0.0016 {\rm (stat.)} \pm 0.0031 {\rm (syst.)}\,{\rm mrad}, \\
\delta\alpha & = & 0.1761 \pm 0.0016 {\rm (stat.)} \pm 0.0032 {\rm (syst.)}\,{\rm mrad},
\end{eqnarray}
which represents a measurement of the crossing angle increase with a relative statistical accuracy of about 2\%, corresponding to an uncertainty on the centre-of-mass energy well within the requirements, of the order of 2.5\,keV. The systematic uncertainty due to non-equality of the electron and positron bunch populations during the filling period is of the same order. 

This measurement relies on the assumptions that, during this period, {\it (i)} the beam instabilities can be kept under control; and {\it (ii)} the detector high-voltages can be safely turned on -- as they will during regular top-up injection in stable collisions. Should any one of these two assumptions be not upheld by reality, a similar measurement could be done during stable collisions from the natural bunch population spread. It may be, however, that all bunch populations end up being strictly identical. Even in this case, it is still possible to exploit the fact that  all bunch populations vary between 101\% and 99\% of the nominal value in 104 seconds, with 2\% top-up injection every 52 seconds, alternately for electron and positron bunches. The measurement of the crossing angle, centre-of-mass energy spread, and luminosity from dimuon events in the 26 seconds following each top-up injection and the 26 seconds preceding the next one allow a precision of 0.016\,mrad on $\delta\alpha$ (10\,keV on the centre-of-mass energy), in one hour of nominal luminosity data taking at the Z pole. If necessary, this precision can be improved by a factor two in a setup where the population of half of the ${\rm e}^\pm$ bunches is only 99\% of the nominal value (inducing a 0.75\% luminosity loss).

Because the dimuon rate is smaller by factors 3.2 and 7.8 for the off-peak points ($\sqrt{s} = 93.8$ and $87.9$\,GeV) than at the Z peak, the precision on $\delta\alpha$ slightly degrades to 0.0043 and 0.0053\,mrad (3.0 and 3.6\,keV on $\sqrt{s}$) during the filling period, and to 0.028 and 0.047\,mrad/$\sqrt{{\rm h}}$ (19 and 32\,keV/$\sqrt{{\rm h}}$ on $\sqrt{s}$) during stable collisions, still within requirements. Due to the much smaller dimuon rate, this method cannot be used at higher energies. On the other hand, the uncertainty on the centre-of-mass energy due to the beam energy measurement (300\,keV at the WW threshold, more than 1\,MeV above) is expected to be significantly larger than the bias due to the crossing angle increase (less than 100\,keV), which  can anyway be predicted with reasonable precision from the calibration of {\tt Lifetrac} from the measurements at the Z pole. 

\subsection{Determination from beam measurements}

There are no instruments able to measure directly the energy spread of a beam. The most ``direct'' measurement of the beam energy spread $\sigma_E$ is based on the measurement of the bunch length $\sigma_s$ that may be converted into energy spread based on the knowledge of the momentum compaction factor $\chi$ and the synchrotron tune $Q_s$ through the relation
\begin{equation}
    \frac{\sigma_{E}}{E} = \frac{\sqrt{2}}{\chi R} Q_s \sigma_s
\end{equation}
where $R = C/2 \pi$ is the average machine radius.

The bunch length measurement can be performed using the synchrotron radiation emitted in the visible range and detected by a streak camera. Such measurements may provide sub-ps time resolution. As an alternative electro-optical crystals may provide similar resolutions. Reaching a 100\,fs accuracy over a large intensity range remains a challenge and will require significant R\&D, see Section 3.6 of the CDR~\cite{Abada2019}.

The beam crossing angle prior to the interaction point can be determined from beam position monitors (BPMs) attached to the final quadrupoles, about 2.1\,m from the IP. The relative alignment error of these BPMs is expected to be around $100\,\mu$m, which translates to an uncertainty on the absolute value of $\alpha_0$ of about 0.1\,mrad. For long-term tracking of the crossing angle variation with time, the BPM intrinsic resolution of $1\,\mu$m  yields a precision on $\alpha_0$ of the order of $1\,\mu$rad. The absolute value of $\alpha$ measured by the BPMs can be calibrated with the dimuon events measurements presented in the previous section. 

\section{Sum-up and monitoring of centre-of-mass energy uncertainties}
\subsection{Absolute uncertainty}

An overview of the effects impacting the centre-of-mass energy in a variety of ways has been presented in Sections~\ref{sec:energy-systematics} and~\ref{sec:cm-energy}. These effects lead, on the one hand, to energy variations that can be tracked by regular RDP measurements, and on the other, to local energy changes (at each IP) invisible in RDP measurements, or even to systematic biases of the RDP measurements. 

Table~\ref{tab:cm-error-summary} shows a summary of the expected energy shifts $\Delta \sqrt{s}/\sqrt{s}$ due to the various effects, as well as the estimated residual systematic uncertainty $\delta \sqrt{s}/\sqrt{s}$. Means to control either the amplitude of the shift $\Delta \sqrt{s}/\sqrt{s}$ or the systematic error are given in the last column. These numbers represent the sum of uncertainties to be assigned to a given energy point taken in isolation, and represent the ``absolute'' centre-of-mass energy uncertainty. The (smaller) point-to-point uncertainty is treated in Section~\ref{sec:point-to-point}. 

While most contributions to the total uncertainty are under control, two sources have not been estimated for the time being and will require further investigation:
\begin{itemize}
    \item The impact of the IP dispersion in the horizontal plane. 
    \item The $\beta^*$ chromaticity effect that results from the  beam-beam interaction.  
\end{itemize}

\begin{table}[htbp]
\caption{Summary of CM energy uncertainties for Z pole operation.  $\Delta \sqrt{s}/\sqrt{s}$ is the estimated energy shift due to the various effects and $\delta \sqrt{s}/\sqrt{s}$ the residual contribution to the systematic error on the CM energy. Entries labelled with NE indicate that the impact cannot be estimated at the current time.\vspace{3mm}}
\label{tab:cm-error-summary}
\centering
\begin{tabular}{|l|c|c|c|l|}
\hline
Source & $\Delta \sqrt{s}/\sqrt{s}$ & RDP & $\delta \sqrt{s}/\sqrt{s}$ & Error control \\
& ($10^{-6}$) & & ($10^{-6}$) & \\ \hline \hline
Dipole field drifts & 100 & Y & & Tunnel T, PC control \\ 
Circumference drifts & 2000 & Y & & Radial feedback \\
Hor. orbit distortions & 100 & Y & & Orbit feedback \\ 
Sextupoles, $\beta$-tron oscil. & 3 & Y &  & Orbit feedback, machine model \\ \hline
Energy dependence of $\chi$ & 2 & N & $< 0.2$ & Machine model  \\
Vert. orbit distortions & 0.3 & N & 0.3 & Orbit control, alignment  \\
Longitudinal fields & 1 & N & $< 0.3$ & Magnetic model \\ \hline
SR losses & 200 & N & 0.2 & Magnetic model, one RF station \\
Collective effects & 100 & N & 0.2 & Machine model \\
IP dispersion (vertical) & 100 & N & 1 & Beam overlap, $D^*$ measurement \\
IP dispersion (horizontal) & 100 & N & NE & Beam overlap \\
$\beta^*$ chromaticity & 1-5 & N & NE & Machine model, Beam-beam \\
 Collective field & 10 & N & & Real? \\
 Crossing angle & & N & & Muon measurements \\
\hline
\end{tabular}
\end{table}

\subsection{Relative monitoring of point-to-point uncertainties}
\label{sec:point-to-point}
As seen in Section~\ref{sec:Zrequirements}, the systematic uncertainties in the relative calibrations from one energy point to the other, in the centre-of-mass energy scans of the Z line shape and the WW threshold, yield most important systematic effects on the EW observable measurements. These uncertainties are labeled for instance $$\left \{\frac{\Delta \left(\sqrt{s_+}  - \sqrt{s_-}\right)}{\sqrt{s_+} - \sqrt{s_-}}\right \}_{\rm ptp-syst}$$ in Eq.~\ref{eq:ECM-errors}. Two potentially very precise methods have been identified to measure and control these uncertainties: 
\begin{itemize}
    \item The measurement of the scattered electron and positron energy end-point in their respective polarimeter;
    \item The direct measurement of the centre-of-mass energy using muon pairs.  
\end{itemize}

As described in Section~\ref{sec:polarimeter}, the envisaged FCC-ee polarimeter-spectrometer (P-S)  will contain a measurement of the end-point of the electron spectrum. The statistical power amounts to a precision of 4\,MeV every 10 seconds, or 40\,keV every day, or 4\,keV over the data taking of each of the off-peak points. This provides an independent relative test at the level of precision similar to that of the measurement by resonant depolarization (RD): for instance a sampling of 100 independent comparisons of the beam energy measured by RD and by the P-S with a statistical precision of 40\,keV at each of the Z scan energy points will be possible, allowing an evaluation of possible biases and uncertainties at this level of precision. The complement of instrumentation that is needed to ensure the stability of the device remains to be evaluated and designed -- temperature and magnetic probes, possibly in-situ geometric monitoring, come to mind. Past experience from the LEP spectrometer tells that this might be  a difficult challenge. 

\begin{figure}[htb]
\centering\includegraphics[width=1.0\linewidth]{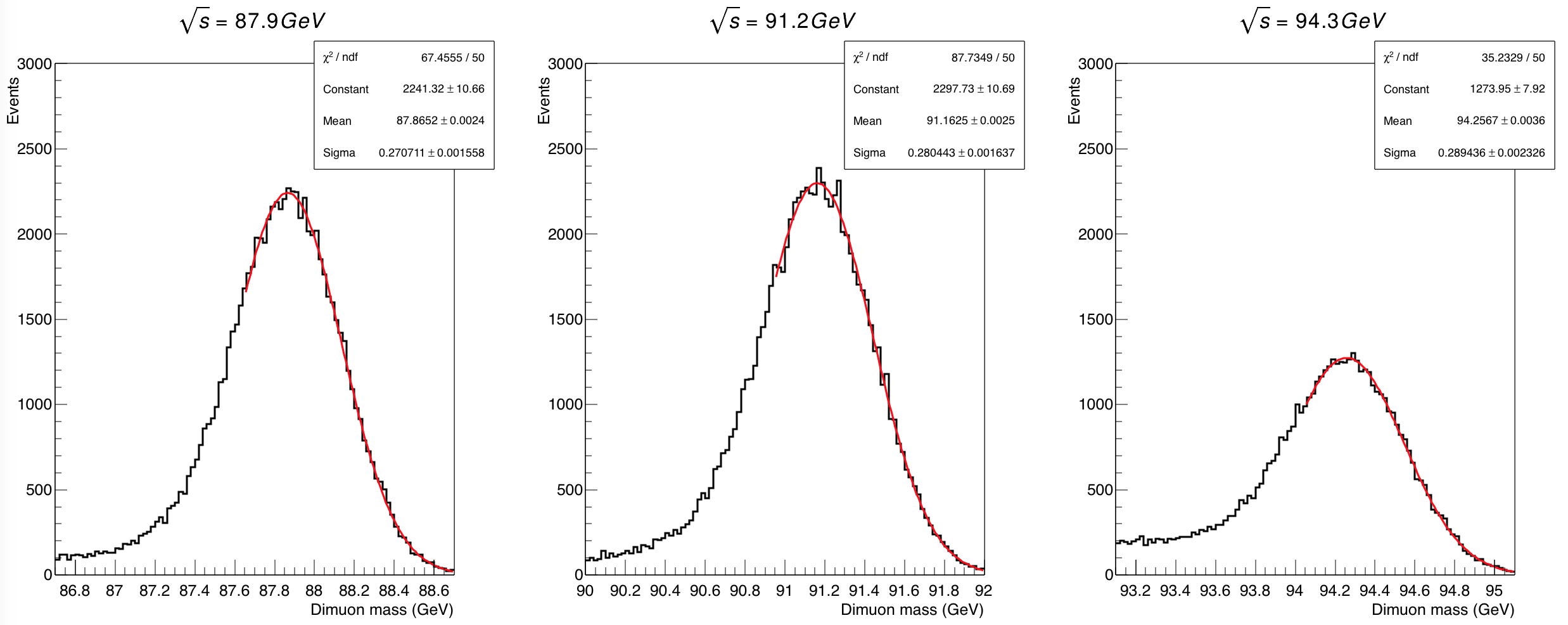}
\caption{\label{MuMu-masses-CLD} Invariant mass distribution of ${10^5}$ muon pairs in the CLD detector, at centre-of-mass energies of (left-to-right) 87.9, 91.2 and 94.3\,GeV respectively; the width of the distribution is dominated by the muon momentum measurement uncertainty. The data correspond to $521\,{\rm pb}^{-1}$, $69\,{\rm pb}^{-1}$, and $257\,{\rm pb}^{-1}$, which can be acquired in 4 minutes, 35 seconds and 2 minutes respectively}
\end{figure}

The muon pairs collected from the detectors offer a direct measurement of the centre-of-mass energy with high statistical power. Figure~\ref{MuMu-masses-CLD} shows the dimuon mass distribution for 100'000 muon pairs acquired at each of the Z resonance scan energies, measured in the CLD detector for muon reconstructed with an angle in excess of 20 degrees with respect to the $z$ axis. With the total integrated luminosities of the foreseen scan, the purely statistical precision of the measurement amounts to  5.5\,keV, 1.1\,keV, et 3.8\,keV for the three energy points of the scan. Here again, the breakdown of the data in more than 100 samples allows the systematic variations and the stability of the centre-of-mass energy measurement to be measured and controlled with a precision of 40\,keV or better. 

The effect of QED initial- and final-state radiation is readily visible in a shift of the fitted invariant mass with respect to the input centre-of-mass energy which amounts to $34.8$, $37.5$, and $43.3$\,MeV for the three points. Such a large shift -- added to the absolute calibration of the detector for the muon momentum measurement -- precludes the use of this method for absolute energy calibration. A careful estimate of the QED effects is needed to quantify and correct the expected dependence of this shift on the centre-of-mass energy (closely related to the QED effects on the line-shape itself). The possibility to ensure the long term stability of the magnetic field in the detector is the main instrumental challenge. 

In summary, at least two independent methods allow a verification of the stability of the energy calibration at the 40\,keV level of precision, which provides a good justification for assigning a value of 40\,keV for the point-to-point uncertainties. 

\subsection{Additional machine and beam monitoring tools}

\subsubsection{Orbit monitoring}

Earth tides induce roughly 1\,mm peak-to-peak amplitude circumference changes of the LEP/LHC ring~\cite{TIDENIM,Wenninger:387157,PhysRevAccelBeams.20.081003}, while longer term geological deformations induced seasonal circumference variations of around 2\,mm~\cite{Assmann:1998qb}. Due to the infrequent energy calibrations at LEP, which left many coasts un-calibrated, it was essential to be able to correct for such circumference changes that could affect the LEP energy by up to 20\,MeV at the Z pole (and more than twice as much at higher energies). While a model is available for earth tides with an accuracy of a few per cent, the long-term geological variations must be tracked with beam instrumentation. The mean radial beam position in the dispersive arc sections proved to be a very accurate monitoring of the circumference of LEP that was operated at a constant RF frequency. 

At FCC-ee, the RF frequency will have to follow the circumference variations to maintain the beams centred as it is done at the LHC~\cite{PhysRevAccelBeams.20.081003}. The full energy swing due to tides is predicted to be around 120\,MeV at the Z pole and 275\,MeV at the W threshold (for a momentum compaction factor $\chi$ of $1.5 \times 10^{-5}$). The accuracy of the circumference correction is defined by the accuracy of the arc beam position monitors. At the LHC, for example, the fill-to-fill reproducibility corresponds to a few $\mu$m. For FCC-ee, a much higher accuracy will be required. At the Z pole, for $\chi = 1.5 \times 10^{-5}$, an accuracy of 0.1\,MeV on the energy requires a measurement of the mean radius to 0.5\,$\mu$m, which is achievable with modern acquisition electronics. At the WW threshold, the mean radius must be known to 0.25\,$\mu$m for the same accuracy. The main uncertainty may well arise from the mechanical accuracy of the BPM blocks and accelerator component due to temperature variations induced by the synchrotron radiation load and stability of the air conditioning of the tunnel. 

Monitoring orbit differences of bunches with different intensities circulating in the ring at the same time is a powerful tool to determine, for example, energy losses due to impedances that can affect the local beam energy at the IP~\cite{LEPImpedanceLoc}. A resolution in the range of 0.1\,$\mu$m is desirable to observe and localize sub-MeV local energy losses. Since the bunches can be arranged to circulate together at the same time, this resolution only applies to the orbit or trajectory of different bunches measured at the same time. 

\subsubsection{Magnetic field monitoring}

The magnetic field of the LEP dipole magnets was not monitored in situ until 1994 when the first NMR probes were installed in a few dipole magnets. Later more than a dozen probes were distributed around the ring to obtain a reasonable sampling of the bending field variations~\cite{Assmann:1998qb}. The monitoring of the local field proved to be essential to understand the evolution of the beam energy during collisions due to tunnel temperature and train leakage currents. For LEP, this monitoring was essential because the energy calibration was normally only performed once at the end of the coasts: the energy had thus to be interpolated back by many hours and even days in case of a few coasts without calibration.

At FCC-ee, the frequent energy calibration does not make a local monitoring of the field as important as at LEP. It could nevertheless be a precious source of information to model the evolution of the beam energy  for operation at higher energies, for which energy calibration by RDP might not be available. It is therefore proposed to install field probes in a sample of magnets distributed all over the ring. The LEP bending field ranged between 40 and 120\,mT, a range that could be covered by NMR probes inserted between the vacuum chamber and the magnet yoke with the help field plates to homogenize the field locally. 

The FCC-ee bending field will be roughly a factor 4 lower for the same beam energy, and the operating range of field probes must cover the range of 10 to 40\,mT. At such low fields, NMR probes will most likely not be able to operate. Electron Spin Resonance (ESR) probes could be an alternative as they can cover magnetic field ranges of some mT. Hall probes are not recommended as an alternative due to longer term field accuracy and reproducibility.

\subsubsection{Energy calibration with proton beams}

A cross-calibration of the energy obtained with resonant depolarization can be performed by injecting protons into the rings and by comparing the RF frequencies of the proton and positrons beams. Such a technique was used at LEP to determine the energy at 20\,GeV~\cite{Bailey:212667}, to calibrate the protons beam energy at 450\,GeV in the SPS~\cite{Arduini:702743} and  to calibrate the LHC beam energy with high accuracy at 450\,GeV/c~\cite{PhysRevAccelBeams.20.081003}.

This technique is based on the fact that the revolution (and RF frequency) depends strongly on the charge over mass ratio of the circulating beam particles. For FCC-ee, the interest of that method is very limited for two main reasons:
\begin{itemize}
\item A proton injector chain is required to bring the protons into FCC-ee which seems to be a very complex and costly task.
\item Such a calibration cannot be performed in parallel to FCC-ee operation, it only provides isolated cross-calibration measurements.  
\end{itemize} 

Given those limitations, such a cross-calibration with protons does not seem to provide much added value.

\section{Integrated Simulation Tools}
\label{sec:sim-tools}
 
In this section, we emphasize the possible need of tool (or tools) that would allow a sound, yet realistic, calculation of the relationship between the measured frequency at which beam depolarization occurs, on the one hand, and the centre-of-mass energy that enters the physics measurements, on the other.  The interest of a good theoretical description of the motion of electrons and their spin in a well-modelled machine, is to be able to evaluate the impact on the energy calibration process of the various imperfections, of the various optimization procedures, and also to be able to prepare and execute the experimental verifications.  As can be seen in Sections~\ref{sec:polar}, ~\ref{sec:running-scheme}, 
and ~\ref{sec:energy-systematics}, the estimation of feasibility and the calculations of systematic uncertainties are based on a series of estimates of individual effects, based on often simplified models, ignoring their mutual dependence and possible interference.  
                  
The following material proposes, as preparation for future work, the concept of a framework, based on spin-orbit tracking simulations starting with the effect of a radio-frequency dipole, while at the same time estimating the ${\rm e^+e^-}$ centre-of-mass  energy at the interaction points (IP) in a given machine. 
\subsection{Spin Tune}
The resonant depolarization by means of a radio-frequency (RF) magnet nominally provides a measurement of a  spin precession frequency, which is equal to a  spin tune times the revolution frequency.  In order to use this number to work back to the centre-of-mass energies, an understanding of the meaning of the term spin tune is required. 
For a particle on the 6-D closed orbit  of a ring (which will always differ from the design orbit), the closed-orbit spin tune can be trivially extracted from the complex eigenvalues of the 1-turn 3x3 spin-rotation matrix on the closed orbit. The closed-orbit spin tune is denoted $\nu_0$ in modern literature.
The closed-orbit spin tune takes account of the ``energy saw tooth'' and resulting transverse closed-orbit shape arising from the loss of energy by radiation and its replenishment by the RF cavities.
Individual particles (or in simulation, groups of particles) undergo synchro-betatron motion away from the closed orbit; the instantaneous rate of spin precession and the axis of precession varies around the ring. An instantaneous rate of precession is not a spin tune. An ``eigentune'' can be extracted from a 1-turn 3x3 spin rotation matrix on a syncho-betatron trajectory, but this number depends not only  on the orbital amplitudes $\vec J \equiv (J_1,J_2,J_3)$, but also on the orbital phases $\vec \Phi \equiv (\Phi_1,\Phi_2,\Phi_3)$ and is therefore also not unique, and is therefore not a spin tune.  Instead, the concept of the Amplitude Dependent Spin Tune, ${\nu}(\vec J)$ must be introduced. An extensive survey on this topic can be found in Refs.~\cite{Hoffstaetter:2006zu}, \cite{Vogt:2000qj}, and~\cite{Barber:2004bi}.

References~\cite{Hoffstaetter:2006zu} (p. 163) and~\cite{Vogt:2000qj} contain an example for  protons at very high energy for single amplitude vertical motion, calculated with methods that are described therein. However, ${\nu}(\vec J)$ can also be discovered with a fast Fourier transform  (FFT) of the spin motion, as is done for instance in the numerical simulation of the spin motion, produced by the spin tracking code described in Refs.~\cite{koop1,Koop:2019rjk}, with results presented in Fig.~\ref{fig:depol_process_spectral_density}.  
The ${\nu}(\vec J)$  is, in fact, an equivalence class consisting of a single tune, the ``preferred spin tune'' ${{\nu}_P}(\vec J)$, which reduces to $\nu_0$ as the amplitudes go to zero, together with a countable infinity of tunes obtained by adding linear combinations of integer multiples of orbital tunes and integers to it. 
Thus ${\nu}(\vec J)$ is of the form
${{\nu}_P}(\vec J) + k_0 + k_1 Q_1 + k_2 Q_2 + k_3 Q_3$ with integers $k$ and tunes $Q$. Energy oscillations due to synchrotron motion contribute oscillations in the instantaneous precession rate, but the ${\nu}(\vec J)$ depends only on the amplitude of the synchrotron and betatron motions.
The  FFT exposes lines corresponding to the ${{\nu}_P}(\vec J)$
and, among other things, satellites separated from it by multiples of the synchrotron tune, the so-called synchrotron sideband resonances.
Spin flipping or resonant depolarization with the RF kicker can take place when the RF tune matches any of the ${\nu}(\vec J)$ displayed by the FFT -- not when it matches the closed-orbit spin  tune $\nu_0$.  Reference~\cite{Barber:2011zza} shows how the so-called single resonance model for RDP still applies on synchro-betatron trajectories. To discover the preferred spin-tune in presence of synchrotron motion, one should check that one is not sitting at the energy of one of the side-band. This was done at LEP by performing resonant depolarization after varying the synchrotron tune by an amount larger that the range of the frequency sweep of the kicker. 
Of course, in a beam with a smooth distribution of amplitudes, there is a distribution of ${{\nu}_P}(\vec J)$. The FFT will display a spread of lines centred on a line associated with an average of the orbital amplitudes. Furthermore, at higher order, the instantaneous rate of spin precession might acquire a quadratic dependence on the fractional energy deviation due to synchrotron  motion. For electrons, orbital damping and noise due to radiation must be included. Nevertheless, the FFT exposes a tune spectrum for the long-term spin motion, which in general is not centred on $\nu_0$.

\subsection{Simulation requirements}

A detailed simulation, incorporating both the spin motion on synchro-betatron trajectories and exposing the synchro-betatron resonances and the action of the RF kicker, the various energy shifts due to energy loss, beamstrahlung as well as the effect of collisions, is necessary to evaluate the relation between the observed resonant depolarization and the average beam energy, and, {\it in fine}, the centre-of-mass energy distribution in one given machine in a consistent way. 
To complete this program, a versatile spin-orbit tracking code is needed.  Among the several codes currently  available, a possible very suitable example could be {\tt Bmad}~\cite{BMAD,Sagan:Bmad2006}, which has been central to the continual improvement of the performance of CESR at Cornell. In addition to providing spin-orbit tracking, this toolkit is instrumental for a wide range of studies. The {\tt Bmad} code is written in modern Fortran, and so far includes the following features: high-order symplectic tracking (before damping); radiation effects; full 3-D spin motion; misalignments; crab crossing; space charge effects; wake fields; and wigglers.

With such a program, the spin tune distribution ${\nu}(\vec J)$ or its generalization can be obtained with an FFT. The Monte-Carlo simulations with {\tt Bmad} could include all known effects on the orbital and spin motion. This inclusion would probably entail a careful extension of the basic spin-orbit tracking code, and therefore require the attention of an expert. In any case, a proper study would involve a deep understanding of particle dynamics in storage rings and would probably take a couple of years. In fact it would provide material for one or more substantial PhD theses. Any correlation between effects would be automatically taken into account and it would become straightforward to check the sensitivity of the depolarizing RF frequency to machine parameters and imperfections and compare them with the estimates appearing in this document. The inclusion of the effects of the collective fields of an electron's own bunch and of the oncoming bunch, will be necessary. 
With this approach, using Monte-Carlo simulation, we would know what the RF magnetic field exposes/measures and, more importantly, the relation between the energy determined from the measurement of the rf-depolarizing frequency and the true centre-of-mass energy distribution.


\section{Summary}
\subsection{Present status}
\label{sec:status}
The present study has come so far to the following conclusions. 
\begin{enumerate}
    \item A workable scheme has been proposed to ensure high-precision knowledge of the average centre-of-mass energy and of the energy spread for the high precision measurements around the Z pole and the W pair threshold, based on resonant depolarization operated frequently on pilot bunches during physics runs. 
    \item The parameters of wigglers necessary to obtain, at the Z energies, a sufficient level of polarization in about 90 minutes, have been given. 
    \item The parameters of the depolarizing kicker have been defined and are well feasible. Especially at the WW energies, the depolarization imposes constraints on the synchrotron tune (${\rm Q_s}$) which have been specified and can be implemented. 
    \item The principle of polarimeter using a back-scattered laser and measuring both the recoil photons and the recoil electron or positron has been presented. It has been shown to ensure an excellent statistical and systematic precision, particular for the measurement of the depolarization of the beams. The polarimeter also acts as spectrometer allowing interpolation of energy measurements and independent stability checks.
    \item The determination of beam energy by resonant depolarization can be extrapolated to centre-of-mass energy given the knowledge of the distribution of synchrotron radiation loss around the ring, average beamstrahlung energy loss, and beam crossing angle, with satisfactory precision. 
    \item The compound interplay of collision offsets and residual dispersion can lead to significant centre-of-mass energy shifts, which constitute one of the more difficult uncertainty to mitigate. A scenario has been devised using frequent beam-beam offset scans. It has been noted that horizontal plane beam offsets as well and beam-beam induced $\beta^*$ chromaticity will require special treatment, as it might be more difficult to control. Further studies of the instrumentation needed to make the process more efficient (e.g. beamstrahlung monitors) should be studied.
    \item Systematic logging of all operation parameters will be essential to ensure that all corrections can be performed successfully without loss of data. 
    \item A number of critical parameters can be determined using muon pairs in the particle physics detectors themselves. In particular, the beam energy spread and  the beam crossing angle at the IP can be determined and monitored continuously. The centre-of-mass energy itself can be reconstructed and used as a means to reduce the point-to-point energy uncertainties. 
    \item At higher energies (Higgs factory, top pair threshold and above), it is unlikely that beam polarization will be available. Extrapolation from lower energies is possible using physics processes (Z$\gamma$, WW and ZZ events) and/or the polarimeter/spectrometer. The achieved precision will be sufficient not to affect significantly the measurements of the masses of the Higgs boson and of the top quark.  
\end{enumerate}

Consequently, we now are able to repeat, in Table~\ref{tab:Z-Ecal-errors-final} the Table~\ref{tab:Z-Ecal-errors} of errors stemming from the beam energy calibration work, with improved estimates for the point-to-point and energy-spread uncertainties.    

\begin{table}[htb]
\centering
\caption{Calculated uncertainties on the quantities most affected by the centre-of-mass energy uncertainties, under the final systematic assumptions. \vspace{3mm}
   \label{tab:Z-Ecal-errors-final}}
\begin{tabular}{|l|c|c|c|c|c|}
\hline\hline
 & statistics & $\Delta \sqrt{s}_{\rm abs} $ & $\Delta \sqrt{s}_{\rm syst-ptp}$ & calib. stats. & $\sigma_{\sqrt{s}}$ \\
Observable & & 100\,keV & {\bf 40\,keV} & 200\,keV/$\sqrt{N^i}$ &    $85  \pm {\bf 0.05 }$\,MeV \\ \hline
$\rm  m_Z$ (keV) & 4  & 100 & {\bf 28} & 1 & --  \\
$\rm  \Gamma_Z$ (keV) & 4  & 2.5 & {\bf 22} & 1 & {\bf 10}   \\
$\sin^2{\theta_{\rm W}^{\rm eff}}\times 10^6 $  from  $A_{\rm FB}^{\mu \mu}$ & 2  & -- & {\bf 2.4} & 0.1  & --\\
$\frac{\Delta \alpha_{\rm QED} ({\rm m_Z^2})}{\alpha_{\rm QED} ({\rm m_Z^2})} \times 10^5$ &  3 & 0.1 & {\bf 0.9} & -- & {\bf 0.1}\\ 
\hline\hline
\end{tabular} 
\end{table}

\subsection{Further studies and R\&D to be recommended}
\label{sec:future-studies}
The determination of centre-of-mass energies at a precision akin to the statistics available at FCC-ee is a major enterprise. In addition to beam equipment and diagnostics, it will eventually require a detailed model and simulation of the  accelerator and of its optimization procedures, as well a set of in situ studies carried out with active collaboration between the particle-physics-experiment teams and the accelerator-operation team. Success of this enterprise will be highly rewarding, with a potential legacy of historical measurements, and the exciting possibility that they could reveal the existence of new physics hitherto unknown. 

A few issues were revealed by this work. \begin{enumerate}
    \item The tools used for simulation of the orbit correction process, and of the simultaneous optimization of luminosity and polarization in  realistic machines should be integrated. This is essential to confirm the feasibility and operability of the proposed data taking scheme.  
    \item A critical point to address is the design of the diagnostics allowing control of the beam-beam offsets and the measurement of residual dispersion and the interaction point. This will allow the centre-of-mass energy shifts to be reduced and monitored, but should also benefit the optimization of luminosity. 
    \item The resonant depolarization process and its sensitivity to the energy spread and synchrotron tune should be further studied to optimize the procedures and the machine settings. 
    \item A detailed design of wigglers including proper management of the radiation is required. 
    \item Further reduction of the point-to-point uncertainties would still be welcome. This should involve development of an energy model, a thorough design of the monitoring devices, and of the data recording strategy.  
    \item Given that the Z run is scheduled at the beginning of the life of FCC-ee, all procedures, instrumentation, data processing and analysis should be ready well before the commissioning of the machine. 
\end{enumerate}
\bibliographystyle{jhep}
\bibliography{biblio.bib}


\end{document}